\documentclass[11pt,a4paper]{article}
\usepackage[T1]{fontenc}
\usepackage{lmodern}
\usepackage{microtype}
\usepackage{amsmath,amssymb,mathtools,bm}
\usepackage{slashed}
\usepackage[hyperfootnotes=false]{hyperref}
\hypersetup{hidelinks,pdftitle={Scherk--Schwarz Compactifications: From Circles to Strings},pdfauthor={Karim Benakli}}
\usepackage{placeins}
\usepackage[backend=biber,style=numeric,sorting=none,doi=true,url=true,eprint=true,giveninits=true,maxbibnames=50]{biblatex}
\usepackage{styles/ssreview-arxiv-refined-3}
\SSshownumbersfalse
\numberwithin{equation}{reviewquestion}
\usepackage{mathrsfs}
\usepackage{tabularx}
\usepackage{dsfont}

\usepackage[
  margin=2.5cm,
  top=2.8cm,
  bottom=2.8cm
]{geometry}

\DeclareMathOperator{\Tr}{Tr}

\title{\bfseries     Scherk–Schwarz Compactifications: From Circles to Strings}
\author{Karim Benakli\\\small LPTHE, Sorbonne Universit\'e and CNRS, \\ 4 place Jussieu, 75005 Paris, Paris, France}
\date{November 2026}
\begin{document}
\maketitle
\begin{abstract}
This review is written with a simple question in mind: what does one
need to know to enter the literature on Scherk--Schwarz
compactifications? It is organized as a sequence of questions that a
graduate student might naturally ask, together with the concepts,
calculations, and formal tools needed to work through the literature.
The discussion starts at an elementary field-theory level and
gradually develops toward more advanced aspects of supergravity,
orbifolds, quantum effects, and string theory. Phenomenological
applications are deliberately left aside.
\end{abstract}
\newpage
\begingroup
\small
\makeatletter
\renewcommand{\@pnumwidth}{3.5em}
\renewcommand{\@tocrmarg}{4.0em}
\makeatother
\tableofcontents
\endgroup

\newpage

\section*{Introduction}

The journey through these pages begins with a simple idea. Upon going once
around a circle of radius $R$, a field need not return to itself, but may return
to its image,
\begin{equation}
\Phi(y+2\pi R)
=
U \Phi(y),
\label{eq:intro-master-formula}
\end{equation}
under a transformation $U$. Applied in increasingly elaborate theories,
this simple idea has many ramifications, which have taken decades to explore.

Before turning to the story, let us first set the stage and introduce its
main characters. The first ingredient is extra dimensions. Their use in
attempts to relate gravity to other interactions goes back to the earliest
developments of Kaluza--Klein theory
\cite{Kaluza:1921tu,Klein:1926tv,Klein:1926fj,Einstein:1938fk}
(see also Ref.~\cite{Witten:2014fma} for a historical discussion).
Many of the explicit examples considered in this review will involve the
simplest internal spaces: the circle and the interval. They are sufficient
to illustrate many of the main ideas discussed below, while remaining
simple enough that several calculations can be carried through explicitly.

The second ingredient is supersymmetry. Supersymmetric theories have an
undeniable appeal to a high-energy theorist. \emph{Nolens volens}, the
question of supersymmetry breaking then imposes itself upon us: first, if
such theories are ever to describe our world, and second, out of
intellectual curiosity about what becomes of their remarkable structures
once supersymmetry is broken.

The third ingredient is string theory, our best-developed framework for a
quantum theory of gravity, in which both extra dimensions and
supersymmetry arise naturally. Familiar as these ingredients have become
in theoretical physics, neither has so far received experimental support.
Our discussion will therefore remain deliberately theoretical. In
particular, we shall largely refrain from phenomenological considerations
whose details are inevitably more sensitive to the passage of time.

If $U$ acts differently on bosons and fermions, then
Eq.~\eqref{eq:intro-master-formula} may break supersymmetry. But can one
choose $U$ arbitrarily? What properties must it satisfy? How does
supersymmetry breaking manifest itself, and in what sense is it
spontaneous? What scalar potential does it induce? What can be said about
the gravitino mass? Do all states feel supersymmetry breaking already at
the classical level? If not, how is the breaking transmitted to them by
radiative corrections? And is there any obstruction to embedding the
whole construction into string theory?

These are among the questions this review tries to answer. The novice
reader will already have guessed, from these questions and from the title
of the review, that the mechanism we shall be discussing is what is
commonly called Scherk--Schwarz compactification
\cite{Scherk:1978ta,Scherk:1979zr}. In their original construction,
symmetries of the higher-dimensional theory were used to generate masses
through a non-trivial dimensional reduction and, when the transformation
acts differently on the members of a supersymmetry multiplet, to break
supersymmetry spontaneously. Early field-theoretic implementations in
higher-dimensional supersymmetric theories, including matter and gauge
multiplets and the resulting supersymmetry-breaking mass spectra, were
developed in detail by Fayet \cite{Fayet:1985ua,Fayet:1985kt}.

String theory adds another layer to the problem. If supersymmetry is
unbroken at tree level, ordinary perturbative radiative corrections do
not provide a mechanism for breaking it, as follows from the
non-renormalization properties of supersymmetric string vacua
\cite{Dine:1986vd}. At the same time, classical supersymmetric and
non-supersymmetric string vacua cannot in general be connected by an
arbitrarily small deformation at finite distance in moduli space
\cite{Dine:1987vf,Banks:1988yz}. The alternatives are therefore sharply
constrained: supersymmetry breaking must already be present in the
classical string background, or arise from genuinely non-perturbative
physics.

This made Scherk--Schwarz breaking particularly attractive in perturbative
string theory, since it provides a tree-level and calculable realization
of supersymmetry breaking through the global compactification data.
Starting with the early string constructions
\cite{Rohm:1983aq,Kounnas:1988ye,Ferrara:1988jx,Kounnas:1989dk},
this possibility was explored in increasingly explicit settings.
At the same time, string theory made clear that the apparent freedom of
the field-theory construction is reduced by microscopic consistency
conditions. In particular, a small gravitino mass in perturbative string
constructions is tied to a tower of light states and to a
decompactification limit rather than to an arbitrary small deformation of
a compact vacuum \cite{Antoniadis:1988jn}. These observations will
reappear later in the review in a form that separates the rigid
monodromy from the physical scale associated with it.

\subsection*{How this review is organized}

There are many ways of writing a review on a subject such as this one.
I have chosen not to follow the most traditional route, in which a
sequence of chapters surveys and summarizes the literature on the
successive developments of the subject. Instead, I have adopted a more
personal approach.

I went back to several of my own papers and asked myself which questions
had led me to them, and what questions a reader would have to understand
in order to grasp them in depth. I then tried to write the corresponding
questions and answers in the form in which I would have liked to find
them in the literature. My hope is that the result will be useful to a
newcomer entering any of the subjects discussed here.

One consequence of this approach is that the review is, inevitably,
rather personal. A few of my own works appear repeatedly along the way.
They are not, however, the subject of the review, and the reader will
quickly see that they constitute only a small fraction of the material
covered. For me they have served mainly as guides: to understand them, a
student has to acquire a number of notions, both explicit and implicit,
and the discussion has been developed outward from there.

This choice has drawbacks. The treatment of the bibliography is
necessarily partial and is biased by the way in which each question is
posed. Many detours that would have made the bibliography more complete
have deliberately been avoided because they would also have changed the
nature of the review.

The scope of the review is similarly limited. It is neither a historical
account nor an exhaustive survey of Scherk--Schwarz constructions. In
particular, I shall not attempt a systematic catalogue of heterotic,
type~II, or orientifold constructions, nor a systematic review of flux
compactifications or gauged supergravities in their own right, nor a
review of supersymmetry phenomenology. Such constructions will enter only
when they illuminate one of the questions addressed here. The aim is to
understand the mechanism and the structures underlying it, rather than to
cover every context in which it has been used.

For the same reason, the discussion is not organized around a catalogue of
currently open problems. This is largely intentional. I have tried to
limit discussions whose content is particularly likely to become obsolete
as the subject develops, and to concentrate instead on questions and
structures that I hope will remain useful for longer.

The review starts at a deliberately elementary level, but the technical
detail increases as the discussion progresses and becomes, in a few
places, rather specialized. Roughly the first two thirds can be read
entirely within quantum field theory and supergravity, while the final
third brings string theory increasingly to the foreground.

Although the questions are arranged in a definite progression, individual
questions, or small groups of questions dealing with closely related
issues, have been written so as to remain reasonably self-contained. A
reader interested in a particular aspect of the subject should therefore
be able to enter the review at several different points without having to
read everything that precedes it. This partial independence is also one
reason why some notation and arguments are occasionally recalled.

A certain discipline has guided the questions and answers. Every notion
should first be accessible at an intuitive level. The review is intended
for physicists with rather different backgrounds, although familiarity
with quantum field theory, supergravity, and the basic ingredients of
string theory is assumed. Concepts that are more specific to the subject
are introduced and defined when they first become necessary.

Whenever possible, an abstract statement is accompanied by an explicit
example or calculation. Quite often, however, the answer admits a more
general formulation once the appropriate mathematical language is
introduced.\footnote{I would like to thank N.~Benakli for the effort spent giving
me intuitive explanations of mathematics that was not, at first,
intuitive to me.}
The purpose is not to pursue an elegant formalism for its own sake. The
hope is rather that the formalism provides a broader view of the question
and makes it easier for a newcomer to recognize the same structures in
related subjects and in parts of the literature not discussed explicitly
here. This choice sometimes makes the exposition heavier and introduces a
certain amount of repetition; I hope that the price is worth paying.

The discussion follows a corresponding progression. We begin with fields
on a circle and an interval, where the kinematics of the twist can be made
fully explicit. We then ask how the same data act on supersymmetry, how
they enter supergravity, and how the resulting breaking is transmitted
through quantum corrections. The last part of the review asks which of
these effective-field-theory constructions survive the additional
consistency conditions of string theory and how the same ideas are
realized when a simple classical internal geometry is no longer the most
useful description.

The primary purpose of a review is not to present new calculations or new
results in the same way as a research paper. I have nevertheless tried to
avoid producing a catalogue of summaries and, whenever possible, to bring
something original to the discussion. Sometimes this lies mainly in the
way the material is organized or explained; sometimes it lies in a
different way of looking at familiar physical results. Some formulations may be less familiar, and a few may not have
appeared in quite this form in the literature. I make no claim of
priority for these observations; their purpose is to clarify the
underlying structures.

Let me end this introduction by trying to answer a more personal
question: why write this review now?

The first answer is very simple: I was invited to write a review, and if
one is going to write one, why not this one? I first worked on
Scherk--Schwarz compactifications about thirty years ago, and in a sense I
began writing this review more than twenty years ago. On several
occasions I intended to write a review or a broader account, with the
scope oscillating between a general introduction to theories with extra
dimensions and a more focused discussion of Scherk--Schwarz
compactification. None of these projects was ever completed.

Over the years, however, a number of ideas, questions, and remarks
accumulated and remained, so to speak, in the drawers of my mind. Some of
them resurfaced when I prepared a workshop talk in Corfu; more recently,
others found their way into a paper. At that point I thought that I had enough notes, and perhaps a
sufficiently broad view of the subject, to write the review. I had
underestimated the challenge.

\begin{reviewquestion}{What happens to fields in higher-dimensional spaces when they move along a circle?}
\label{sec:why-periodic}

\begin{essentialidea}
On a circle, the points labelled by $y$ and $y+2\pi R$ are not distinct:
they represent the same point of the compact space. Requiring all fields to
be strictly periodic is the simplest way to implement this identification,
but it is not the only one. A field may instead return after one circuit of
the circle to its image under a transformation $U$ acting on the field
space. This gluing transformation is the monodromy and provides the basic
geometric ingredient of a Scherk--Schwarz compactification.
\end{essentialidea}

\narrativeheading{Periodicity and compactification}

Let us consider a field theory on a spacetime
\begin{equation}
\mathcal{M}_d\times S^1 ,
\end{equation}
where $\mathcal{M}_d$ is $d$-dimensional Minkowski space and $S^1$ is a
one-dimensional compact manifold. Let $y$ denote a coordinate on the
covering space $\mathbb{R}$, with the identification
\begin{equation}
y\sim y+2\pi R .
\end{equation}
The simplest possibility is to regard a field $\Phi$ as an ordinary
single-valued function on $S^1$, taking values in a fixed vector space.
Its lift to the covering line then satisfies
\begin{equation}
\Phi(x,y+2\pi R)=\Phi(x,y).
\end{equation}
For a scalar field this gives the Fourier expansion
\begin{equation}
\Phi(x,y)
=
\sum_{n\in\mathbb Z}
\Phi_n(x)e^{iny/R},
\end{equation}
and hence the familiar tower of compact momenta
\begin{equation}
p_y=\frac{n}{R},
\qquad n\in\mathbb Z .
\end{equation}

Historically, Kaluza--Klein theories are distinguished by the emergence
of lower-dimensional gauge interactions from higher-dimensional gravity
\cite{Kaluza:1921tu,Klein:1926tv,Klein:1926fj,Einstein:1938fk};
see also Ref.~\cite{Witten:2014fma} for a historical discussion.
Following standard usage in high-energy physics, we shall nevertheless
refer loosely to the modes generated by compactification as KK states,
even when gravity plays no role. Strict periodicity gives the simplest
KK spectrum, with momenta separated by integer units of $1/R$. This result,
however, already contains an additional assumption: the fibre above $y$
has been identified with the fibre above $y+2\pi R$ by the identity
transformation. Compactification identifies the points of the base space,
but it does not by itself require this particular identification of the
field fibres.

For visualization, the value of $\Phi$ in each fibre will be represented
in Fig.~\ref{fig:fibre-gluing} by an arrow. This should not be understood
as implying that $\Phi$ is a vector field: the arrow is merely a graphical
representation of an element of the fibre, on which the gluing
transformation $U$ acts.

%\newpage
\begin{figure}[t]
\centering
\includegraphics[width=\textwidth]{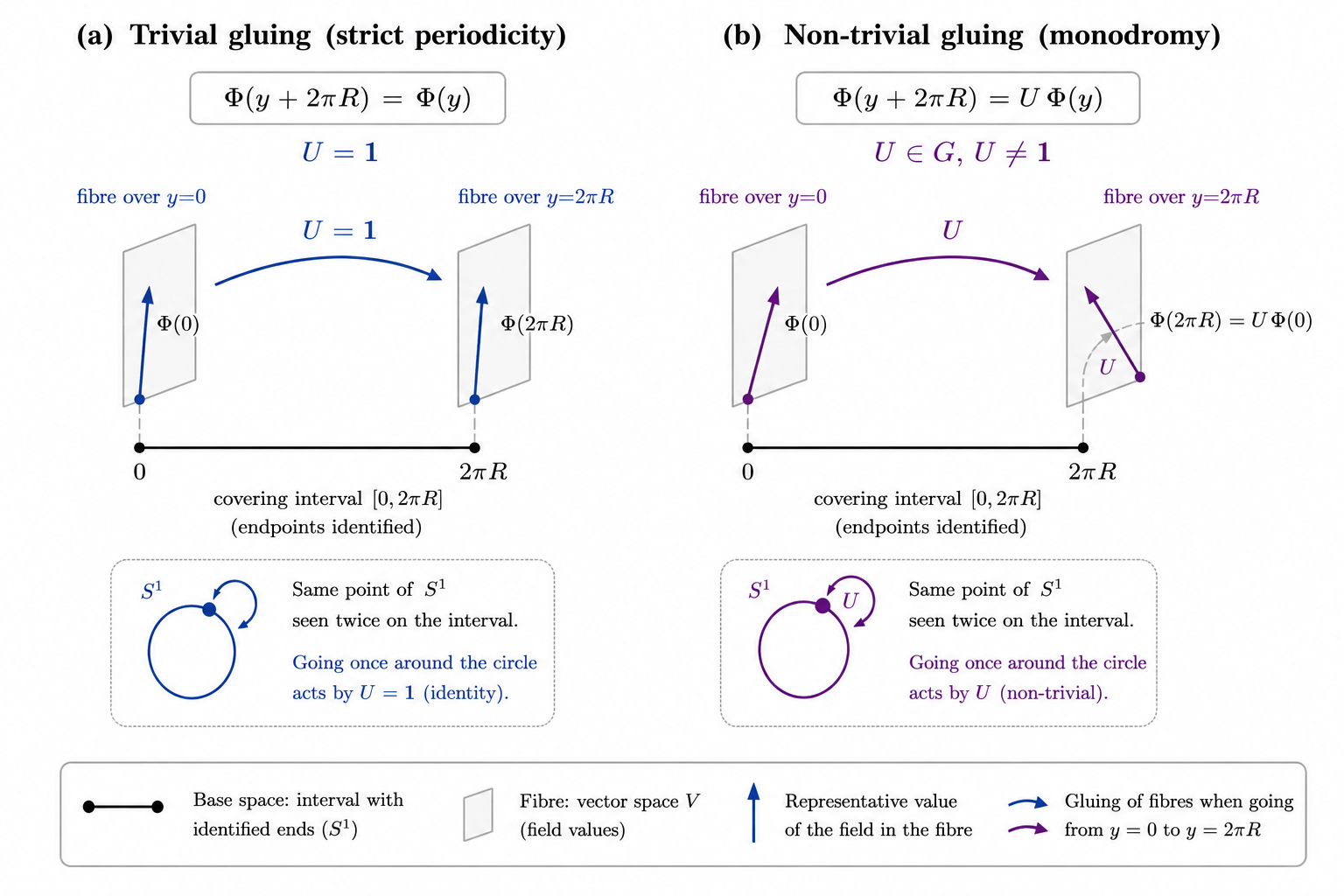}
\caption{%
Compactifying on a circle identifies $y$ and $y+2\pi R$; the
interval is only a convenient way of drawing $S^1$ (insets).
The arrows representing $\Phi$ are schematic elements of the fibre,
not spacetime vectors. In (a) the fibre is glued to itself by the
identity, $U=\mathbf{1}$, whereas in (b) it is glued by a non-trivial
monodromy $U$.
}
\label{fig:fibre-gluing}
\end{figure}

\narrativeheading{The more general global identification}

A more general description regards a field as a section of a fibre bundle
over the compact space. Imagine cutting the circle at one point and
representing it as the interval $0\leq y\leq2\pi R$. Reconstructing the
circle requires us to identify the two endpoints, but the corresponding
fibres need not be glued by the identity map. They may instead be related
by a transformation $U$,
\begin{equation}
\Phi(x,2\pi R)=U \Phi(x,0).
\end{equation}
As we shall discuss in the next question, consistency requires $U$ to be
an admissible symmetry transformation of the higher-dimensional theory.
Unrolling the circle onto its covering space gives the equivalent
condition
\begin{equation}
\Phi(x,y+2\pi R)=U \Phi(x,y).
\end{equation}
For $U=\mathbf{1}$ one recovers strict periodicity. For
$U\neq\mathbf{1}$ the lifted field on $\mathbb{R}$ is no longer periodic
in the naive sense, even though it defines a perfectly good global field
on the compact space once the fibres are glued by $U$. The relevant
global datum is therefore not the periodicity of a representative on the
covering line, but the monodromy specifying how the field is identified
after one circuit of the circle.

This distinction between the identification of points in the base space
and the gluing of the fibres above them is the elementary geometric
ingredient of a Scherk--Schwarz compactification
\cite{Scherk:1978ta,Scherk:1979zr}. Much of what follows in this review
will consist of determining which transformations $U$ are allowed,
understanding how the same $U$ acts in different representations, and
tracing the physical consequences of that action.

\narrativeheading{Monodromy and the fundamental group}

The same statement can be formulated intrinsically in terms of the
fundamental group. The group $\pi_1(S^1)$ consists of equivalence classes
of closed loops in $S^1$, two loops being equivalent when one can be
continuously deformed into the other while remaining on the circle.
Composition is defined by concatenation of loops, and for the circle
\begin{equation}
\pi_1(S^1)\simeq\mathbb Z ,
\end{equation}
where the integer labels the algebraic winding number around the circle.
There is therefore a single independent generator, and specifying the
monodromy associated with this generator determines the gluing around
every loop.

More generally, for a compact space $K$, flat monodromy data can be
described by a homomorphism
\begin{equation}
\rho:\pi_1(K)\longrightarrow G ,
\end{equation}
where $G$ acts on the fibres in the representation appropriate to the
field under consideration. The circle is the simplest possible example:
$\pi_1(S^1)$ has one generator, so the entire global datum is encoded in
one transformation $U=\rho(1)$. Which such homomorphisms define
physically distinct backgrounds, and in particular the role played by
conjugation in $G$, will be discussed in the next question.

A familiar example of non-trivial global boundary conditions is provided
by spinors. On $S^1$ there are two spin structures, corresponding to
\begin{equation}
\psi(y+2\pi R)=+\psi(y),
\end{equation}
and
\begin{equation}
\psi(y+2\pi R)=-\psi(y).
\end{equation}
This elementary example already shows that the geometry of the base space
does not uniquely determine the global identification of every field.
The precise relations between spin structures, spacetime fermion parity
$(-1)^{F_s}$, and genuine $R$-symmetry twists are subtler and will be
discussed later in the review.

\narrativeheading{Bibliographical remarks}

The higher-dimensional origins of Kaluza--Klein compactification are
represented here by the original works of Kaluza, Klein, and
Einstein--Bergmann
\cite{Kaluza:1921tu,Klein:1926tv,Klein:1926fj,Einstein:1938fk};
Ref.~\cite{Witten:2014fma} provides a useful modern discussion of part of
this early history. The generalized dimensional-reduction construction
based on a non-trivial gluing by a symmetry transformation was developed
by Scherk and Schwarz
\cite{Scherk:1978ta,Scherk:1979zr}. For the mathematical language of
fibre bundles, transition functions, and gluing data, a standard reference
is Husemoller \cite{Husemoller:1994}, while spin structures and spin
geometry are treated systematically by Lawson and Michelsohn
\cite{Lawson:1989}. We shall use these mathematical notions only to the
extent needed to formulate the physical questions that arise in the
compactification.

\begin{takehome}
\begin{itemize}
\item Fields on a compact space are naturally regarded as sections of
fibre bundles. Identifying points of the base space does not by itself
require the fibres to be identified trivially.
\item On a circle, strict periodicity corresponds to trivial gluing and
gives the compact momenta $p_y=n/R$.
\item More generally, a field may return to itself up to a monodromy
$U$ after going around a non-contractible cycle.
\item For a general compact space, flat monodromy data are encoded by
the action of its fundamental group on the relevant field fibres.
\end{itemize}
\end{takehome}

\nextquestion
{Why must the twist be associated with a symmetry of the higher-dimensional theory?}
\end{reviewquestion}

%%%%%%%%%%%%%%%%%%%%%%%%%%%%%%%%%%%%%%%%%%%%%%%%%%%%%%%%%%%%%%%
\begin{reviewquestion}
{Why must the twist be associated with a symmetry of the higher-dimensional theory?}
\label{sec:twist-symmetry}

\begin{essentialidea}
Compactification identifies different representatives of the same point
of the internal space. The gluing transformation must therefore assign
the same local physics to all such representatives. In the standard
field-theoretic Scherk--Schwarz construction, this requires the twist to
act as a symmetry of the higher-dimensional theory.
\end{essentialidea}

\narrativeheading{Why is this condition necessary?}

As discussed in the previous question, fields need not be strictly
periodic on the covering space: one may impose
\begin{equation}
    \Phi(x,y+2\pi R)=U\,\Phi(x,y).
    \label{eq:general_twist}
\end{equation}
This freedom is not unlimited. The coordinates $y$ and $y+2\pi R$
represent the same point of the compact space, and the two
representatives cannot be assigned different local dynamics. The
restriction becomes particularly transparent if we deliberately try to
violate it.

Consider, for example, a real scalar field with Lagrangian
\begin{equation}
    \mathcal L
    =
    \frac12(\partial\phi)^2
    -
    a\left[1-\cos\left(\frac{\phi}{f}\right)\right],
\end{equation}
where $2\pi f$ is the periodicity of the scalar potential. Suppose that
we try to impose the boundary condition
\begin{equation}
    \phi(y+2\pi R)
    =
    \phi(y)+\Delta .
\end{equation}
The kinetic term is invariant under a constant shift, but the potential
transforms according to
\begin{equation}
    \cos\left(\frac{\phi+\Delta}{f}\right)
    =
    \cos\left(\frac{\phi}{f}+\frac{\Delta}{f}\right).
\end{equation}
For a generic value of $\Delta$, this is not the original potential.
The two representatives of the same point would therefore be assigned
different local dynamics. The gluing is consistent only when
\begin{equation}
    \Delta=2\pi k f,
    \qquad k\in\mathbb Z,
\end{equation}
for which the shift is an actual symmetry of the theory. The example is
elementary, but it displays the general principle: the compact
identification may act non-trivially on the fields, but it cannot change
the local theory as one goes around the circle.

\narrativeheading{The field-theoretic criterion}

For an ordinary internal symmetry in a theory admitting a Lagrangian
description, the simplest form of the condition is
\begin{equation}
    \mathcal L(U\Phi)=\mathcal L(\Phi).
    \label{eq:lagrangian_invariance}
\end{equation}
More generally, it is sufficient that the action be invariant; the
Lagrangian density may change by a total derivative, provided that this
change is itself compatible with the global identification and gives no
contribution to the action. Equivalently, one may formulate the
requirement directly in terms of the equations of motion and the physical
observables: $U$ must map the higher-dimensional theory into itself. The
boundary condition \eqref{eq:general_twist} can then identify two
representatives of the same point without creating any local
inconsistency.

At the quantum level, the same statement must hold for the full quantum
theory. A classical transformation spoiled by an anomaly, for example,
cannot in general be used as an exact Scherk--Schwarz twist. Thus the
twist is not an arbitrary choice of quasi-periodic boundary condition:
around a non-contractible one-cycle it is a monodromy valued in the
appropriate symmetry group of the higher-dimensional theory. Which
elements of that group act non-trivially on the supercharges is a
separate question, to which we shall turn next.

\narrativeheading{Geometric interpretation}

The geometric formulation expresses the same requirement in the language
introduced in the previous question. A field is a section of a fibre
bundle, and the transition functions specify how local descriptions of
the fibre are related on overlapping patches. For a circle, the
transition function associated with going once around the compact
direction is precisely the monodromy $U$. A mathematically well-defined
bundle can therefore be constructed with non-trivial gluing.

Mathematical consistency of the bundle alone, however, is not enough for
a physical compactification. Its transition functions must also preserve
the structures that define the higher-dimensional theory: its action,
equations of motion, gauge structure, and, at the quantum level, its
exact operator content. In the standard Scherk--Schwarz construction
this is the physical statement that the monodromy takes values in an
exact symmetry group $G$. On a circle one may therefore write
\begin{equation}
    \rho:\pi_1(S^1)\simeq\mathbb Z\longrightarrow G,
\end{equation}
with $U=\rho(1)$. The abstract homomorphism introduced in the previous
question thus acquires a direct physical meaning: it specifies how an
allowed symmetry of the higher-dimensional theory is used to glue the
fields globally around the compact direction.

\narrativeheading{Beyond ordinary symmetries}

The preceding formulation can be extended beyond transformations acting
locally on a conventional set of fields. What is fundamentally required
is an exact automorphism of the complete theory. This distinction becomes
important in string theory, where two apparently different local
descriptions may be related by an exact duality, and such a duality can
itself be used as part of the gluing. This leads to duality-twisted and
non-geometric compactifications, in which no globally defined
higher-dimensional field configuration of the conventional kind need
exist.

Those possibilities will be encountered only much later in the review.
For the field-theoretic Scherk--Schwarz construction developed in the
next questions, the relevant twists will be ordinary symmetries of the
higher-dimensional theory. Keeping this simpler setting for the moment
will allow us to separate two logically distinct issues: first, whether
a twist is admissible at all, and second, whether an admissible twist
actually breaks supersymmetry.

\narrativeheading{Bibliographical remarks}

The requirement that the non-trivial dependence on the compact
coordinates be controlled by a symmetry of the higher-dimensional theory
is already explicit in the original construction of Scherk and Schwarz
\cite{Scherk:1978ta,Scherk:1979zr}. Their generalized dimensional
reduction uses precisely such a symmetry to factor the dependence on the
internal coordinates and to generate masses in the reduced theory. For
the underlying bundle language and the role of transition functions, we
refer again to the standard treatment in
Ref.~\cite{Husemoller:1994}. More general duality-valued gluings will be
discussed, with their appropriate literature, when we turn to string
compactifications.

\begin{takehome}
\begin{takehomeitems}

\item
A monodromy is not automatically an admissible Scherk--Schwarz twist.
Since compactification identifies different representatives of the same
spacetime point, the gluing must preserve the local higher-dimensional
physics.

\item
In the standard field-theoretic construction, the twist therefore acts
as a symmetry of the higher-dimensional theory. For a Lagrangian
invariant without boundary terms, this reduces to
$\mathcal L(U\Phi)=\mathcal L(\Phi)$.

\item
The symmetry must be exact at the level at which the compactification is
defined. In particular, a classical symmetry obstructed quantum
mechanically cannot in general define an exact quantum twist.

\item
More generally, the gluing may be an exact automorphism of the complete
theory, including suitable dualities in string theory.

\end{takehomeitems}
\end{takehome}

\nextquestion
{Among the admissible twists, which ones break supersymmetry?}

\end{reviewquestion}
%%%%%%%%%%%%%%%%%%%%%%%%%%%%%%%%%%%%%%%%%%%%%%%%%%%%%%%%%%%%%%%
\begin{reviewquestion}
{Among the admissible twists, which ones break supersymmetry?}
\label{sec:twist-supersymmetry}

\begin{essentialidea}
A twisted compactification preserves precisely those supersymmetry
transformations whose parameters are globally compatible with the
monodromy. For the flat compactifications considered here, the surviving
supersymmetries are therefore the invariant subspace of the monodromy
induced on the supersymmetry-parameter space.

This criterion tells us which supersymmetries survive the compactification.
It does not determine whether their absence should be interpreted as
spontaneous breaking. That is a separate dynamical question.
\end{essentialidea}

\narrativeheading{An admissible twist need not preserve all the original symmetries}

The previous question determined when a twist is admissible: it must act
as a symmetry, or more generally as an exact automorphism, of the
higher-dimensional theory. This requirement does not imply that every
symmetry of the parent theory survives after compactification. In
particular, an admissible monodromy may act non-trivially on the
supersymmetry generators. Admissibility and preservation are therefore
two distinct questions: the first asks whether the twist can consistently
be used to define the compactification, while the second asks which
symmetries remain globally well defined once that twist has been imposed.

\narrativeheading{The supercharge invariant-subspace criterion}

Consider a supersymmetric higher-dimensional theory with supersymmetry
generators $Q^I$ and corresponding spinorial transformation parameters
$\boldsymbol{\epsilon}^I$. Schematically, an infinitesimal
supersymmetry transformation acts as
\begin{equation}
    \delta_{\epsilon}\Phi
    \sim
    \bar{\boldsymbol{\epsilon}}_I\,[Q^I,\Phi],
\end{equation}
where the bracket is understood as graded whenever appropriate. Let
$\mathcal S$ denote the vector space of supersymmetry parameters. If
the fields obey the twisted identification
\begin{equation}
    \Phi(x,y+2\pi R)=U\,\Phi(x,y),
\end{equation}
the same transformation induces an action on $\mathcal S$, which we
denote by $R_{\epsilon}(U)$.

For the flat compactification considered here, a supersymmetry of the
lower-dimensional theory is generated by a parameter that can be chosen
independent of the compact coordinate. Compatibility with the global
identification then requires
\begin{equation}
    R_{\epsilon}(U)\,\boldsymbol{\epsilon}
    =
    \boldsymbol{\epsilon}.
\end{equation}
The preserved supersymmetries are consequently
\begin{equation}
    \mathcal S_{\rm preserved}
    =
    \ker\!\left(
        R_{\epsilon}(U)-\mathbf 1
    \right).
    \label{eq:q3-preserved-subspace}
\end{equation}
With several supersymmetry parameters, $R_{\epsilon}(U)$ is simply a
matrix acting on the full parameter space, and partial supersymmetry
breaking corresponds to a non-zero invariant subspace of dimension
smaller than $\dim\mathcal S$.

This criterion isolates the global part of the problem. In a more general
background, preserved supersymmetries must also satisfy the appropriate
local Killing-spinor equations, and the monodromy condition alone need
not be sufficient. For the flat backgrounds and flat twists that will
occupy us first, however, the invariant-subspace criterion contains the
relevant information.

\narrativeheading{When does a twist break supersymmetry?}

If the twist acts trivially on all supersymmetry generators, then
\begin{equation}
    R_{\epsilon}(U)=\mathbf 1,
\end{equation}
and all supersymmetries survive. Notice that this does not require the
twist itself to be trivial: $U$ may act non-trivially on other quantum
numbers and modify the Kaluza--Klein spectrum while leaving every
supercharge invariant. The existence of a non-trivial monodromy and the
breaking of supersymmetry are therefore logically distinct statements.

An $R$-symmetry is distinguished precisely by its non-trivial action on
the supersymmetry generators. For a continuous $R$-symmetry twist
$U_R(\omega)$, let the induced action on the supersymmetry parameters be
\begin{equation}
    R_{\epsilon}\!\left(U_R(\omega)\right)
    =
    \exp\!\left(
        2\pi i\omega\,\mathbb R_{\epsilon}
    \right),
\end{equation}
where $\mathbb R_{\epsilon}$ denotes the $R$-symmetry generator in the
representation carried by $\mathcal S$. In a basis in which this
generator is diagonal,
\begin{equation}
    R_{\epsilon}\!\left(U_R(\omega)\right)\epsilon_A
    =
    e^{2\pi i\omega q_A}\epsilon_A ,
\end{equation}
with $q_A$ the corresponding $R$-charge. The supersymmetry generated by
$\epsilon_A$ survives if and only if
\begin{equation}
    e^{2\pi i\omega q_A}=1.
\end{equation}
Depending on the spectrum of $R$-charges and on the value of the twist,
the compactification may therefore preserve all, part, or none of the
supersymmetry of the parent theory.

\narrativeheading{Several cycles}

Nothing in the invariant-subspace argument is specific to a single
circle. For a compact space $K$, let
\begin{equation}
    \rho_{\epsilon}:
    \pi_1(K)
    \longrightarrow
    \operatorname{GL}(\mathcal S)
\end{equation}
denote the monodromy representation induced on the supersymmetry
parameters. In the absence of additional local Killing-spinor
constraints, the globally preserved supersymmetries are
\begin{equation}
    \mathcal S_{\rm preserved}
    =
    \left\{
    \boldsymbol{\epsilon}\in\mathcal S
    \;\middle|\;
    \rho_{\epsilon}(\gamma)\boldsymbol{\epsilon}
    =
    \boldsymbol{\epsilon}
    \quad
    \text{for every }\gamma\in\pi_1(K)
    \right\}.
    \label{eq:q3-general-preserved-subspace}
\end{equation}
A supersymmetry parameter must thus belong to the common invariant
subspace of all monodromies. For a torus, the generators of
$\pi_1(T^n)\simeq\mathbb Z^n$ commute, and so must their monodromies.
For a more general compact space, the monodromies instead obey the
relations of its fundamental group.

\narrativeheading{What this criterion does not tell us}

The invariant-subspace criterion is kinematical. If
$\mathcal S_{\rm preserved}=\{0\}$, no global supersymmetry survives,
but this alone does not determine whether the breaking is spontaneous,
whether a Goldstino exists, or whether a super-Higgs mechanism occurs.
Those questions require additional dynamical information and will be
addressed later.

\narrativeheading{Bibliographical remarks}

The use of a higher-dimensional symmetry that acts non-trivially on the
supersymmetry generators is already central to the original construction
of Scherk and Schwarz
\cite{Scherk:1978ta,Scherk:1979zr}. Their first explicit supergravity
example uses a chiral symmetry of the higher-dimensional theory to give
mass to the gravitino while leaving the graviton massless. Applications
in higher-dimensional supersymmetric theories with matter and gauge
multiplets, including explicit $R$-symmetry twists and the corresponding
patterns of partial or complete supersymmetry breaking, were developed
further by Fayet
\cite{Fayet:1985ua,Fayet:1985kt}. We shall return to these constructions
when discussing the mass spectrum, the gravitino, and the dynamical
interpretation of Scherk--Schwarz breaking.

\begin{takehome}
\begin{takehomeitems}

\item
Admissibility and supersymmetry preservation are distinct questions:
an allowed twist need not preserve all the symmetries of the
higher-dimensional theory.

\item
For the flat compactifications considered here, the preserved
supersymmetries form the invariant subspace of the monodromy acting on
the supersymmetry-parameter space.

\item
A non-trivial twist may preserve all supersymmetries if it acts
trivially on the supercharges, while an $R$-symmetry twist may preserve
all, part, or none of them. With several cycles, the relevant subspace
is the common invariant subspace of all monodromies.

\item
This criterion is kinematical: the absence of invariant supersymmetry
parameters does not by itself establish spontaneous supersymmetry
breaking or a super-Higgs mechanism.

\end{takehomeitems}
\end{takehome}

\nextquestion
{How does a monodromy modify the Kaluza--Klein spectrum?}

\end{reviewquestion}
%%%%%%%%%%%%%%%%%%%%%%%%%%%%%%%%%%%%%%%%%%%%%%%%%%%%%%%%%%%%%%%
\begin{reviewquestion}
{How does a monodromy modify the Kaluza--Klein spectrum?}
\label{sec:twisted-kk-spectrum}

\begin{essentialidea}
A monodromy changes the global spectral problem. Fields become sections
of the flat bundle defined by the monodromy, and the Kaluza--Klein
masses are determined by the spectrum of the corresponding twisted
internal operator. The familiar shift $(n+\omega)/R$ is the special
result obtained for a phase monodromy on a flat circle.
\end{essentialidea}

\narrativeheading{The twisted spectral problem}

Consider a higher-dimensional spacetime locally of the form
$\mathcal M_d\times K$, with $K$ compact. For a scalar field of
higher-dimensional mass $m_0$, the Klein--Gordon operator separates
into a $d$-dimensional d'Alembertian and an internal Laplacian
$\Delta_K$. Writing
\begin{equation}
    -\Delta_K\psi_\alpha
    =
    \lambda_\alpha\psi_\alpha ,
\end{equation}
the corresponding lower-dimensional masses are
\begin{equation}
    m_\alpha^2
    =
    m_0^2+\lambda_\alpha .
\end{equation}
This is the ordinary Kaluza--Klein spectral problem.

Now equip the compactification with monodromy data
\begin{equation}
    \rho:\pi_1(K)\longrightarrow G ,
\end{equation}
where $G$ acts in the representation carried by the field. The field is
then a section of the associated flat bundle $E_\rho$, and the internal
spectral problem becomes
\begin{equation}
    -\Delta_\rho\psi_\alpha
    =
    \lambda_\alpha(\rho)\psi_\alpha ,
\end{equation}
so that
\begin{equation}
    m_\alpha^2
    =
    m_0^2+\lambda_\alpha(\rho).
\label{eq:q4-general-twisted-masses}
\end{equation}
Locally, on a contractible patch, the twisted and untwisted operators
may have exactly the same differential expression. What changes is
their global domain: only sections obeying the prescribed monodromy
conditions are admissible. A twist can therefore modify the spectrum
without modifying the local field equations. This distinction will
recur repeatedly, for example when supersymmetry can be defined locally
along a compact cycle although no supersymmetry parameter is globally
compatible with its monodromy.

\newpage
\narrativeheading{The familiar example of a circle}

For a flat circle $S^1_R$ with
\begin{equation}
    y\sim y+2\pi R ,
\end{equation}
consider a field satisfying the phase-twisted boundary condition
\begin{equation}
    \Phi(x,y+2\pi R)
    =
    e^{2\pi i\omega}\Phi(x,y).
\label{eq:q4-circle-monodromy}
\end{equation}
An internal plane wave $\psi_p(y)=e^{ipy}$ is allowed only if
\begin{equation}
    e^{2\pi iRp}=e^{2\pi i\omega},
\end{equation}
which gives
\begin{equation}
    p_n
    =
    \frac{n+\omega}{R},
    \qquad
    n\in\mathbb Z.
\label{eq:q4-shifted-circle-momentum}
\end{equation}
The internal eigenvalues are therefore
\begin{equation}
    \lambda_n(\omega)
    =
    \frac{(n+\omega)^2}{R^2},
\end{equation}
and the lower-dimensional masses become
\begin{equation}
    m_n^2
    =
    m_0^2
    +
    \frac{(n+\omega)^2}{R^2}.
\label{eq:q4-circle-masses}
\end{equation}

Thus the familiar replacement $n\to n+\omega$ is simply the spectrum
required by a phase-twisted boundary condition on a flat circle.
Figure~\ref{fig:q4-kk-shifted-levels} displays this special case. The
uniform translation of the momentum lattice visible there should not
be mistaken for a universal property of twisted compactifications: it
is the particularly simple realization of the more general spectral
problem described above.

\begin{figure}[t]
    \centering
    \includegraphics[width=0.95\linewidth]{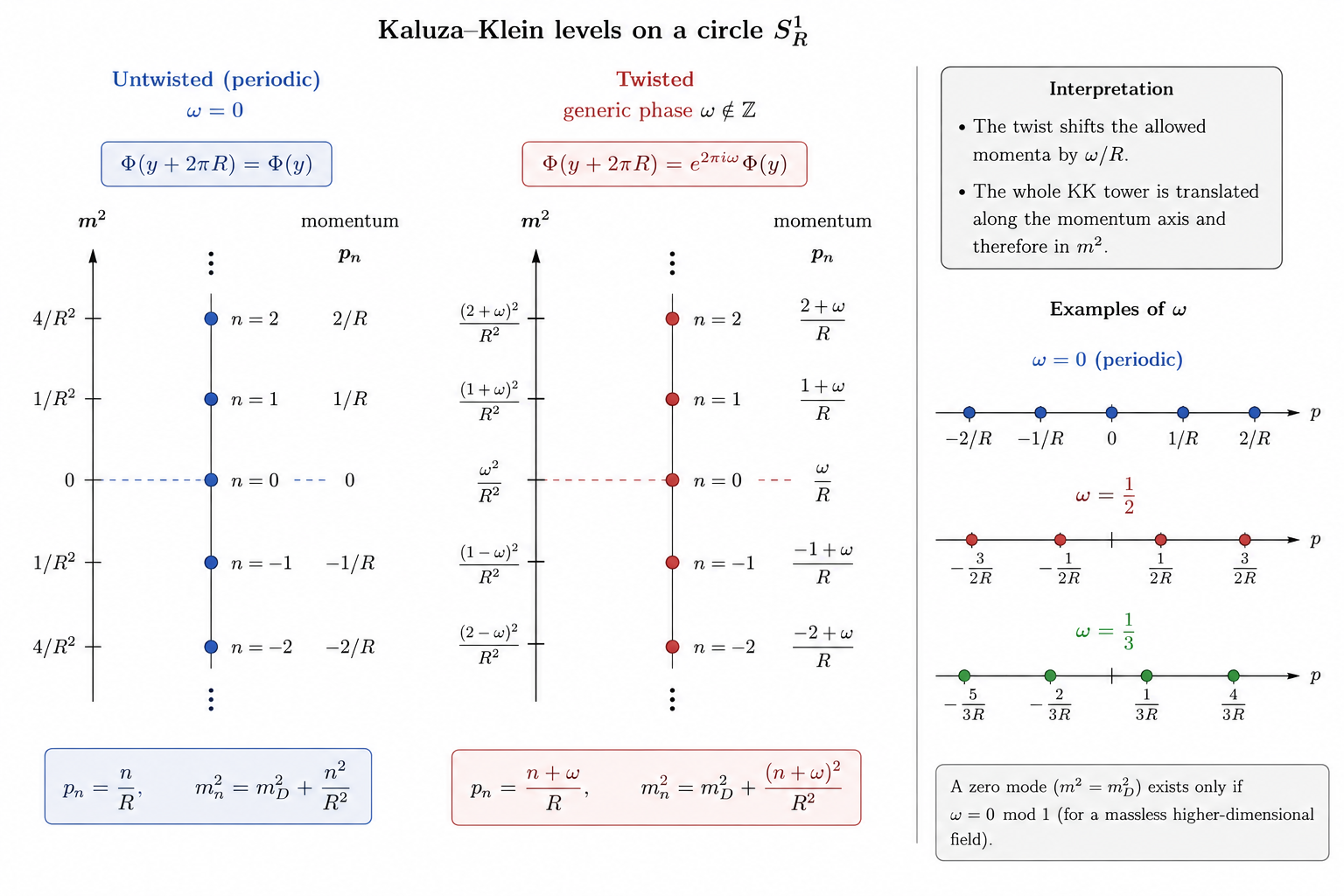}
    \caption{%
    Kaluza--Klein levels for a phase monodromy on a flat circle
    $S^1_R$. In the untwisted case the allowed momenta are
    $p_n=n/R$, whereas
    $\Phi(y+2\pi R)=e^{2\pi i\omega}\Phi(y)$ shifts them to
    $p_n=(n+\omega)/R$. The tower is therefore translated in momentum
    space. This rigid shift is special to the flat-circle problem; on a
    general compact space the spectrum is determined by the appropriate
    twisted internal operator.
    }
    \label{fig:q4-kk-shifted-levels}
\end{figure}

\narrativeheading{Twisted fields or periodic fields}

The same compactification can be described using the periodic field
\begin{equation}
    \widetilde\Phi(x,y)
    =
    e^{-i\omega y/R}\Phi(x,y),
\end{equation}
for which
\begin{equation}
    \widetilde\Phi(x,y+2\pi R)
    =
    \widetilde\Phi(x,y).
\end{equation}
The price of making the field periodic is that the internal derivative
acts as
\begin{equation}
    \partial_y
    \longrightarrow
    \partial_y+\frac{i\omega}{R}.
\label{eq:q4-shifted-derivative}
\end{equation}
The same global information may therefore be encoded either in twisted
boundary conditions or in a flat connection. This equivalence will
become particularly useful later when Scherk--Schwarz breaking is
described in terms of an $R$-symmetry connection or Wilson line.

\newpage
\narrativeheading{Several field components and several cycles}

If the field takes values in a representation space $V$ of the symmetry
group, a single monodromy need not act in the same way on every
component. For
\begin{equation}
    \Phi(y+2\pi R)=U\,\Phi(y),
\end{equation}
suppose that $U$ is diagonalizable and choose a basis
$\{v_\alpha\}$ satisfying
\begin{equation}
    Uv_\alpha
    =
    e^{2\pi i\omega_\alpha}v_\alpha .
\end{equation}
Writing
\begin{equation}
    \Phi(x,y)
    =
    \sum_\alpha\Phi_\alpha(x,y)v_\alpha ,
\end{equation}
each component obeys its own phase-twisted boundary condition and has
momenta
\begin{equation}
    p_{n,\alpha}
    =
    \frac{n+\omega_\alpha}{R},
    \qquad
    n\in\mathbb Z .
\end{equation}
Different components of the same higher-dimensional multiplet may
therefore generate differently shifted Kaluza--Klein towers. In particular, a zero mode with vanishing internal momentum can survive only in a monodromy-invariant component.

The same reasoning applies when the compact space contains several
independent cycles. For a rectangular $r$-torus,
\begin{equation}
    y^i\sim y^i+2\pi R_i,
    \qquad
    i=1,\ldots,r ,
\end{equation}
let $U_i$ denote the monodromy around the $i$th cycle. Since
$\pi_1(T^r)\simeq\mathbb Z^r$, these monodromies commute. When they can
be simultaneously diagonalized, common eigenvectors $v_\alpha$ may be
chosen such that
\begin{equation}
    U_i v_\alpha
    =
    e^{2\pi i\omega_{i,\alpha}}v_\alpha .
\end{equation}
The allowed internal momenta are then
\begin{equation}
    p_{i,n_i,\alpha}
    =
    \frac{n_i+\omega_{i,\alpha}}{R_i},
    \qquad
    n_i\in\mathbb Z ,
\label{eq:q4-multicycle-momenta}
\end{equation}
and for a rectangular flat metric
\begin{equation}
    m_{\vec n,\alpha}^2
    =
    m_{0}^2
    +
    \sum_{i=1}^{r}
    \frac{(n_i+\omega_{i,\alpha})^2}{R_i^2}.
\end{equation}

Geometrically, each common monodromy eigenspace sees a translated
momentum lattice. If $\Lambda^\ast$ denotes the dual lattice of the
torus and
\begin{equation}
    \boldsymbol{\delta}_\alpha
    =
    \left(
    \frac{\omega_{1,\alpha}}{R_1},
    \ldots,
    \frac{\omega_{r,\alpha}}{R_r}
    \right),
\end{equation}
then the allowed momenta lie in the affine lattice
\begin{equation}
    p_\alpha
    \in
    \Lambda^\ast+\boldsymbol{\delta}_\alpha .
\label{eq:q4-shifted-dual-lattice}
\end{equation}
Even on a flat torus there is therefore no unique shifted tower: the
spectrum and its degeneracies depend on the geometry, the field
representation, and the complete set of monodromy eigenvalues.

\narrativeheading{Beyond flat compactifications}

On a general internal space the Kaluza--Klein spectrum is determined by
the appropriate twisted internal operator,
\begin{equation}
    \operatorname{Spec}_{\rm KK}(\rho)
    =
    \operatorname{Spec}\!\left(\mathcal O_{K,\rho}\right),
\label{eq:q4-general-spectral-statement}
\end{equation}
whose eigenvalues need not be equally spaced. For scalars,
$\mathcal O_{K,\rho}$ is typically a covariant Laplacian; for spinors,
the corresponding twisted Dirac operator; and for gauge or higher-spin
fields, the appropriate gauge-fixed kinetic operator or system. A
monodromy may lift zero modes, split degeneracies, and change
multiplicities, but on a general compact manifold there is no universal
analogue of the rigid $(n+\omega)/R$ translation.

The discussion above concerns monodromies that act on fields transported
around ordinary one-cycles. This should be distinguished from another
type of global datum that will appear later in the review: holonomies of
higher-form gauge fields over higher-dimensional cycles and the associated
wrapped extended objects. These involve a different geometric structure
and will be treated separately when we discuss higher cycles.

\narrativeheading{Bibliographical remarks}

The shifted Kaluza--Klein spectra generated by non-trivial
higher-dimensional boundary conditions are already part of the original
Scherk--Schwarz construction
\cite{Scherk:1978ta,Scherk:1979zr}. Their formulation in terms of
sections of flat bundles and monodromy representations is the standard
geometric language reviewed, for example, in
Ref.~\cite{Husemoller:1994}. The equivalence between twisted boundary
conditions and a periodic description with a flat connection is one of
the recurring forms of the Scherk--Schwarz mechanism; its interpretation
in supergravity and in terms of Wilson lines will be developed in later
questions.

\begin{takehome}
\begin{takehomeitems}

\item
A monodromy changes the global spectral problem even when the local
differential equations are unchanged.

\item
The Kaluza--Klein masses are eigenvalues of the appropriate twisted
internal operator acting on sections of the bundle defined by the
monodromy.

\item
On a flat circle, a phase monodromy gives
$p_n=(n+\omega)/R$. Equivalently, one may use periodic fields and a
flat connection in the internal derivative.

\item
For matrix-valued monodromies, different field components may acquire
different shifts; on a flat torus each common eigenspace sees an affine
momentum lattice.

\item
On a general compact manifold there is no universal
$(n+\omega)/R$ formula.

\end{takehomeitems}
\end{takehome}

\nextquestion
{In what sense is Scherk--Schwarz supersymmetry breaking spontaneous?}

\end{reviewquestion}
%%%%%%%%%%%%%%%%%%%%%%%%%%%%%%%%%%%%%%%%%%%%%%%%%%%%%%%%%%%%%%%
\begin{reviewquestion}
{In what sense is Scherk--Schwarz supersymmetry breaking spontaneous?}
\label{sec:ss-spontaneous}

\begin{essentialidea}
Spontaneous supersymmetry breaking means that supersymmetry remains a
symmetry of the theory but is not preserved by the chosen background.
Explicit breaking means instead that the corresponding supersymmetry is
not a symmetry of the theory itself.

In a Scherk--Schwarz compactification, the absence of an invariant
supercharge establishes that supersymmetry is broken. It does not by
itself determine the nature of that breaking. In continuous
$R$-symmetry reductions of supergravity, the spontaneous interpretation
can be made explicit: the broken local supersymmetries remain part of
the gauge structure, and the Goldstino and super-Higgs mechanism provide
their dynamical realization. Other constructions must be examined
separately.
\end{essentialidea}

\narrativeheading{Symmetry of the theory and symmetry of the state}

For an ordinary global symmetry, the distinction between explicit and
spontaneous breaking is straightforward. In spontaneous breaking, the
symmetry remains a symmetry of the action but is not preserved by the
vacuum. For rigid supersymmetry one may write schematically
\begin{equation}
    \delta_\epsilon S=0,
    \qquad
    Q_A|0\rangle\neq0
\end{equation}
for at least one supercharge $Q_A$. Explicit breaking means instead that
the corresponding supersymmetry is not a symmetry of the theory in the
first place.

For local supersymmetry the language is slightly different because
supersymmetry is a gauge symmetry. A bosonic background is
supersymmetric when there exists a non-zero supersymmetry parameter for
which all fermionic variations vanish. If no such parameter exists, the
background breaks supersymmetry. When the broken local supersymmetry
nevertheless remains part of the gauge structure of the theory, the
breaking is spontaneous: a fermionic direction has an inhomogeneous
supersymmetry variation and provides the Goldstino, which is absorbed by
the gravitino through the super-Higgs mechanism.

Scherk--Schwarz compactification makes this distinction less immediate.
A non-trivial monodromy may eliminate every globally defined
supersymmetry parameter, so there is no ambiguity about whether
supersymmetry survives: it does not. The more subtle question is what
the word \emph{spontaneous} means in this situation. The answer depends
on the physical status of the compactification data. One must determine
whether the missing supersymmetries remain symmetries of the same
underlying dynamical theory and fail only to be preserved by the chosen
configuration, or whether the global definition of the compactified
theory has removed them from its symmetry structure. The spectrum alone
cannot decide between these possibilities.

\narrativeheading{The continuous Scherk--Schwarz case}

Consider a continuous $R$-symmetry monodromy
\begin{equation}
    U_R(\omega)
    =
    \exp\!\left(2\pi i\omega\,\mathcal R\right),
\end{equation}
where $\mathcal R$ denotes a generator of the $R$-symmetry group.
Its action on the supersymmetry-parameter space is generated by
$\mathbb R_\epsilon$, in the notation of
Question~\ref{sec:twist-supersymmetry}. Supersymmetry is restored whenever
the induced monodromy acts trivially on the supercharges. In the simplest
normalization, the branch connected to the untwisted point contains
\begin{equation}
    \omega=0,
\end{equation}
up to the identifications of the twist parameter.

On a flat circle the characteristic Scherk--Schwarz splittings behave as
\begin{equation}
    m_{\rm SS}
    \sim
    \frac{|\omega|}{R},
    \label{eq:q5-ss-scale}
\end{equation}
up to $R$-charges and normalization conventions. Supersymmetry is then
restored continuously as the monodromy becomes trivial on the
supercharges.

Continuity of the parameter, however, is not sufficient to establish
spontaneous breaking. A continuous boundary-condition parameter need not
be the expectation value of a dynamical field, and a continuous family
of theories with different explicit breaking parameters is not the same
thing as a continuous family of configurations of one supersymmetric
dynamical theory. This distinction is particularly important in deciding
how the Scherk--Schwarz parameter should be interpreted.

In supergravity, there are formulations in which the Scherk--Schwarz
twist is represented by an $R$-symmetry connection or holonomy belonging
to the locally supersymmetric theory itself. In appropriate off-shell
descriptions, the same information may be encoded in the radion sector.
In such realizations the monodromy is not introduced as an explicit
supersymmetry-breaking operator: different values correspond to
different configurations of a theory whose local supersymmetry remains
part of its gauge structure. The accompanying Goldstino and super-Higgs
mechanism then provide the dynamical content required for the
spontaneous interpretation.

\narrativeheading{Two ways of restoring supersymmetry}

Equation~\eqref{eq:q5-ss-scale} also shows that the Scherk--Schwarz
splitting vanishes in the decompactification limit,
\begin{equation}
    m_{\rm SS}\longrightarrow0
    \qquad
    (R\to\infty).
\end{equation}
There are therefore two conceptually different ways in which the
splitting may disappear: the monodromy can become trivial on the
supercharges, $\omega\to0$, or the compactification scale itself can
vanish, $R^{-1}\to0$, while the rigid monodromy remains non-trivial.
The two limits should not be identified. In the latter case the
effective dimensionality changes, and when $R$ is a modulus the
decompactification limit is typically at infinite distance in moduli
space. Supersymmetry restoration as $R\to\infty$ therefore shows that
the breaking scale is geometrical, but does not by itself establish that
the finite-radius breaking is spontaneous.

A more decisive diagnostic is the realization of the broken symmetry in
the low-energy theory. For global supersymmetry, spontaneous breaking
requires a Goldstino, equivalently a fermionic direction with an
inhomogeneous supersymmetry variation on the vacuum. For local
supersymmetry, the corresponding realization is the super-Higgs
mechanism. The next two questions will identify the Goldstino in the
Scherk--Schwarz compactification and explain the role of the full
Kaluza--Klein gravitino tower. In continuous supergravity reductions one
indeed finds the fermionic shifts and gravitino--Goldstino mixing
expected from spontaneous breaking.

\narrativeheading{What each diagnostic tells us}

Several frequently invoked properties answer different logical
questions and should not be used interchangeably.
\begin{center}
\begin{tabularx}{0.98\linewidth}
{@{}p{0.20\linewidth} X X@{}}
\hline
\textbf{Diagnostic}
&
\textbf{What it shows}
&
\textbf{What it does not show}
\\
\hline
\\[-2mm]

Supersymmetric parent theory
&
The theory is supersymmetric before compactification.
&
That the missing supercharges remain symmetries of the full theory
after compactification.
\\[22mm]

Vanishing splittings as $R\to\infty$
&
The Scherk--Schwarz mass scale is controlled by the compact geometry
and disappears upon decompactification.
&
That the monodromy becomes trivial or that the finite-radius breaking
is spontaneous.
\\[22mm]

Continuous path to trivial monodromy
&
The family of compactifications contains a supersymmetric point.
&
That the parameter labels configurations of one dynamical theory rather
than distinct theories or externally prescribed boundary conditions.
\\[22mm]

No invariant supercharge
&
No global supersymmetry survives the compactification.
&
How the broken supersymmetry is realized dynamically.
\\[22mm]

Dynamical or gauge realization of the monodromy
&
The twist can be represented by data belonging to a fixed
supersymmetric theory.
&
By itself, that the chosen configuration realizes spontaneous
supersymmetry breaking.
\\[22mm]

Goldstino / super-Higgs realization
&
The broken supersymmetry has the characteristic low-energy realization
of spontaneous breaking.
&
That every construction with the same spectrum or monodromy has the
same interpretation.
\\[22mm]
\hline
\end{tabularx}
\end{center}

\narrativeheading{Why discrete twists require separate treatment}

Discreteness by itself does not decide whether supersymmetry breaking is
spontaneous. An ordinary orbifold or orientifold projection retains only
the invariant physical states and supercharges. A supercharge genuinely
projected out of the physical Hilbert space is no longer a symmetry
acting on the projected theory, so its absence should not be interpreted
as spontaneous breaking of that theory. Discrete Scherk--Schwarz
constructions need not have this interpretation: freely acting
orbifolds, for example, can provide realizations of spontaneous
supersymmetry breaking despite the discreteness of the quotient. What
matters is not whether the twist parameter is continuous or discrete,
but how the quotient acts on the physical state space and whether the
missing supersymmetry remains realized in the underlying theory.

Nor does the reappearance of supersymmetry in a decompactification limit
settle the issue by itself. It may simply mean that the global
distinction responsible for the breaking becomes invisible when the
compact cycle disappears. Discrete twists, spin structures, freely
acting orbifolds, and related constructions must therefore be analysed
at the level of their physical state spaces and supercharges. Their
relation to $(-1)^{F_s}$ and to continuous $R$-symmetry twists will be
discussed later in the review: sharing the name Scherk--Schwarz does not
by itself imply sharing the same dynamical mechanism.

\narrativeheading{Working criterion}

In this review, we shall use the term \emph{spontaneous
Scherk--Schwarz breaking} when the supersymmetry not preserved by the
compactification remains a symmetry of the same underlying dynamical
theory rather than being removed by the global definition or projection,
and when its broken realization exhibits the corresponding Goldstino or
super-Higgs structure. Continuous $R$-symmetry Scherk--Schwarz
reductions of supergravity provide the canonical example: the monodromy
can be represented through gauge or dynamical data of the supersymmetric
theory, while the broken local supersymmetry remains part of its gauge
structure and is realized through the super-Higgs mechanism.

By contrast, a continuous parameter, a radius-dependent mass splitting,
the absence of a local explicit breaking term, or the restoration of
supersymmetry in a limiting regime are useful characteristics of
particular constructions, but none of them alone determines whether the
breaking is spontaneous. The criterion adopted here deliberately
separates these useful diagnostics from the dynamical statement that we
ultimately wish to make.

\narrativeheading{Bibliographical remarks}

The spontaneous interpretation of generalized dimensional reduction is
already explicit in the original papers of Scherk and Schwarz
\cite{Scherk:1978ta,Scherk:1979zr}, where local supersymmetry is retained
while gravitino masses are generated by the compactification. Explicit
higher-dimensional realizations based on $R$-symmetry, including the
resulting fermion and scalar mass splittings, were developed further by
Fayet \cite{Fayet:1985ua,Fayet:1985kt}. The distinction between
supersymmetry restoration by varying a compactification datum and
restoration in a decompactification limit becomes especially important
in string theory; we shall return to the corresponding constraints and
their early analyses when the microscopic string construction is
introduced. The Goldstino and super-Higgs aspects relevant to the
field-theory discussion will instead be developed directly in the next
questions.

\begin{takehome}
\begin{takehomeitems}

\item
The absence of an invariant supercharge establishes that supersymmetry
is broken, but does not by itself determine whether the breaking is
spontaneous or explicit.

\item
The central question is whether the missing supercharges remain
symmetries of the same underlying dynamical theory, or are removed by
the global definition or projection of the compactified theory.

\item
A continuous path to a supersymmetric point, or the disappearance of
mass splittings as $R\to\infty$, are useful diagnostics but are not by
themselves sufficient to establish spontaneous supersymmetry breaking.

\item
In continuous supergravity Scherk--Schwarz reductions, the monodromy
can be realized through gauge or dynamical data of the supersymmetric
theory, while the Goldstino and super-Higgs mechanism provide the
characteristic dynamical realization of spontaneous breaking.

\item
Discrete twists and projections require a separate analysis: what
matters is not discreteness itself, but whether the missing
supersymmetry remains realized in the underlying physical theory.

\end{takehomeitems}
\end{takehome}

\nextquestion
{Where is the Goldstino in a Scherk--Schwarz compactification?}

\end{reviewquestion}
%%%%%%%%%%%%%%%%%%%%%%%%%%%%%%%%%%%%%%%%%%%%%%%%%%%%%%%%%%%%%%%
\begin{reviewquestion}
{Where is the Goldstino in a Scherk--Schwarz compactification?}
\label{sec:ss-goldstino}

\begin{essentialidea}
The fermionic Stückelberg structure of a Kaluza--Klein gravitino tower
is already present before any Scherk--Schwarz twist is introduced. For
a five-dimensional gravitino, the internal component transforms at
non-zero Kaluza--Klein level as
\[
    \delta\psi_y^{(n)}
    \sim
    \frac{in}{R}\,\epsilon^{(n)} .
\]

A continuous Scherk--Schwarz twist does not create this structure. It
shifts the internal operator,
\[
    \frac{in}{R}
    \longrightarrow
    \frac{in}{R}+\mathcal M ,
\]
where $\mathcal M$ acts on the relevant $R$-symmetry indices. The
internal gravitino components selected by this shifted operator provide
the corresponding would-be Goldstino directions.

Their absorption by the gravitini, and the resulting spin-$\tfrac32$
mass matrix, are properties of the reduced Rarita--Schwinger
Lagrangian and will be established in the next question.
\end{essentialidea}

\narrativeheading{Fermion shifts and Goldstino directions}

On a bosonic background, lower-dimensional spin-$\tfrac12$ fermions
transform schematically as
\begin{equation}
    \delta_\epsilon\chi^a
    =
    \mathcal F^a{}_A\,\epsilon^A+\cdots ,
\label{eq:q6-general-fermion-shift}
\end{equation}
where $\mathcal F^a{}_A$ is the fermion-shift matrix. 
For a broken supersymmetry direction, the non-vanishing vector
$\mathcal F^a{}_A\epsilon^A$ identifies the corresponding would-be
Goldstino direction in fermion space. The Goldstino subspace is
therefore $\operatorname{Im}\mathcal F$. The physical normalized 
Goldstino combination also depends on the
fermionic kinetic terms, a point that becomes relevant when several
fermions contribute.

Parameters belonging to $\ker\mathcal F$ are candidates for preserved
supersymmetries, provided that the remaining Killing-spinor equations
are satisfied. For the flat-circle Scherk--Schwarz compactifications
considered below, this reduces to the monodromy criterion discussed in
the previous questions.

\narrativeheading{The Stückelberg structure of the gravitino KK tower}

Consider a five-dimensional gravitino compactified on a circle of
radius $R$, with $\Psi_M\to(\Psi_\mu,\Psi_y)$. Its local supersymmetry
transformation contains
\begin{equation}
    \delta_\epsilon\Psi_M=D_M\epsilon+\cdots ,
    \qquad
    \delta_\epsilon\Psi_y=D_y\epsilon+\cdots .
\label{eq:q6-gravitino-transformation}
\end{equation}
The internal gravitino is therefore directly sensitive to the operator
governing the dependence of the supersymmetry parameter on the compact
coordinate, which explains its special role in the Goldstino sector.

Before introducing a twist, expand
\begin{equation}
    \epsilon(x,y)
    =
    \sum_{n\in\mathbb Z}
    \epsilon^{(n)}(x)e^{iny/R},
    \qquad
    \Psi_y(x,y)
    =
    \sum_{n\in\mathbb Z}
    \psi_y^{(n)}(x)e^{iny/R}.
\end{equation}
The Fourier expansion then gives
\begin{equation}
    \delta\psi_y^{(n)}
    =
    \frac{in}{R}\epsilon^{(n)}+\cdots .
\label{eq:q6-untwisted-kk-shift}
\end{equation}
For $n\neq0$, this is the fermionic analogue of the ordinary
Kaluza--Klein Stückelberg transformation. The internal gravitino modes
supply the spin-$\tfrac12$ Stückelberg degrees of freedom associated
with the massive spin-$\tfrac32$ Kaluza--Klein tower.

Question~\ref{sec:twisted-kk-spectrum} showed that a Scherk--Schwarz
monodromy shifts the compact momentum. Its effect on the gravitino is
therefore naturally expected to deform this pre-existing Stückelberg
operator rather than to generate an entirely new structure.

\begin{figure}[t]
    \centering
    \includegraphics[width=\linewidth]{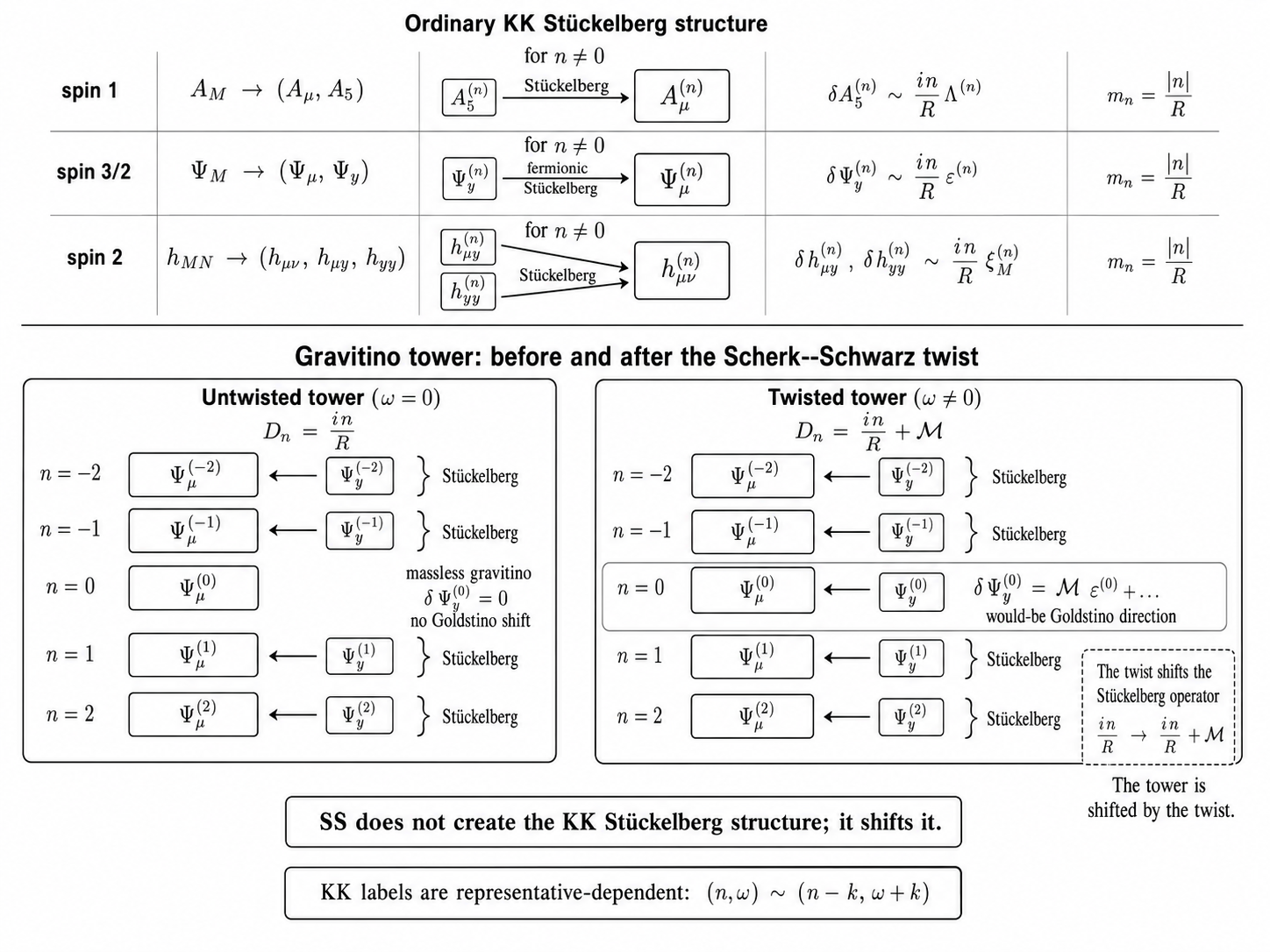}
    \caption{%
    Kaluza--Klein Stückelberg structure and its Scherk--Schwarz
    deformation. The upper panel recalls the common higher-dimensional
    origin of massive Kaluza--Klein fields of spin $1$, $3/2$, and $2$:
    at non-zero Kaluza--Klein level, internal field components provide
    the Stückelberg degrees of freedom required by the corresponding
    massive lower-dimensional fields. For the gravitino tower, the
    untwisted fermionic transformation is governed by $in/R$. A
    continuous Scherk--Schwarz twist shifts this operator to
    $in/R+\mathcal M$. The twist therefore deforms a Stückelberg
    structure already present in the higher-dimensional theory. The
    integer Kaluza--Klein label depends on the representative chosen for
    the twist, whereas the complete shifted tower is invariant under
    the corresponding relabelling.
    }
    \label{fig:q6-kk-stueckelberg}
\end{figure}

\narrativeheading{Adding the Scherk--Schwarz twist}

Write the continuous $R$-symmetry monodromy acting on the
supersymmetry parameters as
\begin{equation}
    U_{R,\epsilon}
    =
    \exp\!\left(2\pi R\,\mathcal M\right)
    =
    \exp\!\left(2\pi i\omega\,\mathbb R_\epsilon\right),
    \qquad
    \mathcal M
    =
    \frac{i\omega}{R}\,\mathbb R_\epsilon .
\label{eq:q6-Mcal-definition}
\end{equation}
Here $\mathbb R_\epsilon$ is the action of the $R$-symmetry generator
$\mathcal R$ on the supersymmetry-parameter space $\mathcal S$, in the
notation of Questions~\ref{sec:twist-supersymmetry}
and~\ref{sec:ss-spontaneous}.

Using the periodic-field description introduced in
Question~\ref{sec:twisted-kk-spectrum},
\begin{equation}
    \epsilon(x,y)
    =
    e^{y\mathcal M}\widetilde\epsilon(x,y),
    \qquad
    \Psi_M(x,y)
    =
    e^{y\mathcal M}\widetilde\Psi_M(x,y),
\label{eq:q6-periodic-fields}
\end{equation}
the internal derivative becomes
$\partial_y\to\partial_y+\mathcal M$. The internal gravitino
transformation is consequently
\begin{equation}
    \delta_{\widetilde\epsilon}\widetilde\Psi_y
    =
    \left(
        \partial_y+\mathcal M
    \right)
    \widetilde\epsilon+\cdots ,
\label{eq:q6-periodic-internal-variation}
\end{equation}
and at Kaluza--Klein level $n$,
\begin{equation}
    \delta\psi_y^{(n)}
    =
    \mathcal D_n\,\epsilon^{(n)}+\cdots ,
    \qquad
    \mathcal D_n
    \equiv
    \frac{in}{R}+\mathcal M .
\label{eq:q6-level-shift-operator}
\end{equation}

For the mode represented by $n=0$ in this periodic frame,
\begin{equation}
    \delta\psi_y^{(0)}
    =
    \mathcal M\,\epsilon^{(0)}
    +\cdots .
\label{eq:q6-goldstino-shift-zero-mode}
\end{equation}
The components of $\psi_y^{(0)}$ lying in
$\operatorname{Im}\mathcal M$ therefore provide the corresponding
would-be Goldstino directions. What distinguished the $n=0$ sector in
the untwisted theory was precisely the absence of the Stückelberg shift;
the Scherk--Schwarz deformation extends the same structure into that
sector.

\Needspace{5\baselineskip}
For extended supersymmetry, $\mathcal M$ is a matrix acting on
$R$-symmetry indices. In minimal five-dimensional supergravity, circle
compactification gives $\mathcal N=2$ supersymmetry in four dimensions,
and the supersymmetry parameters form an $SU(2)_R$ doublet. Thus
$\mathcal M$ acts as a $2\times2$ matrix. More supersymmetric parent
theories are treated analogously: only combinations selected by the
image of the shift operator transform inhomogeneously.

\narrativeheading{The Kaluza--Klein label is not invariant}

There is an important subtlety in calling the preceding field the
``zero-mode Goldstino.'' For an eigenstate of $R$-charge $q$, the
spectrum depends on the combination $n+\omega q$. An integer shift can
be compensated by relabelling the Kaluza--Klein tower,
\begin{equation}
    \omega q\longrightarrow\omega q+k,
    \qquad
    n\longrightarrow n-k,
    \qquad
    k\in\mathbb Z .
\label{eq:q6-kk-relabeling}
\end{equation}
Consequently, the statement that the Scherk--Schwarz super-Higgs
mechanism involves the $n=0$ gravitino is not invariant by itself. A
representative of the monodromy class may be chosen in which the light
gravitino carries that label, but another representative can assign a
different integer label to the same physical state.

The invariant datum is the complete shifted operator $\mathcal D_n$,
or equivalently the monodromy together with the full Kaluza--Klein
tower. The fundamental domain of the twist parameter and the
classification of inequivalent values of $\omega$ will be discussed
later.

\narrativeheading{From the Goldstino direction to the super-Higgs mechanism}

The Fourier components $\epsilon^{(n)}$ are all modes of the same
higher-dimensional local supersymmetry parameter; there is no change
of gauge status between $n=0$ and $n\neq0$. What changes is the action
of the internal operator. Without the twist it is $in/R$, whereas the
Scherk--Schwarz deformation replaces it by
\begin{equation}
    \mathcal D_n
    =
    \frac{in}{R}+\mathcal M .
\end{equation}
The same operator that produces the inhomogeneous transformation of
the internal gravitino will reappear in the reduced Rarita--Schwinger
Lagrangian, both in the spin-$\tfrac32$ mass matrix and in the
gravitino--Goldstino mixing.

The transformation law therefore identifies the would-be Goldstino
direction, but does not yet prove that it is absorbed or determine the
physical gravitino masses. These are dynamical statements about the
reduced action and are the subject of the next question.

For $|\omega|\ll1$, the gravitino that becomes massless at the
supersymmetric point may have
\begin{equation}
    m_{3/2}
    \sim
    \frac{|\omega|}{R}
    \ll
    \frac1R ,
\end{equation}
up to the relevant $R$-charge and normalization. At energies
$E\ll1/R$, the heavier Kaluza--Klein gravitini can then be integrated
out, leaving a four-dimensional supergravity description of the light
sector. This low-energy separation does not make the label $n=0$
physical: a change of representative can still relabel the complete
tower while leaving its spectrum unchanged.

\narrativeheading{Additional fermion shifts}

The internal gravitino need not exhaust the Goldstino sector. In a
general compactification, additional spin-$\tfrac12$ fermions may have
non-zero shifts,
\begin{equation}
    \delta\chi^a
    =
    \mathcal F^a{}_A\,\epsilon^A+\cdots .
\end{equation}
The physical Goldstino is then the linear combination selected by the
complete fermion-shift matrix together with the fermionic kinetic
terms. Modulini, gaugini, hyperini, or localized fermions may therefore
modify the physical combination without changing the basic role of the
twisted internal gravitino derivative.

The discussion here assumes a continuous supergravity realization in
which the broken supersymmetries remain part of the
higher-dimensional local gauge symmetry. It should not automatically
be applied to projections that remove the corresponding supercharges
from the physical theory. Orbifolds, localized breaking sectors, spin
structures, and $(-1)^{F_s}$ projections will be treated when those
ingredients are introduced.

\narrativeheading{Bibliographical remarks}

The appearance of the internal gravitino component as the fermionic
degree of freedom associated with Scherk--Schwarz supersymmetry breaking
is already visible in the original supergravity construction of Scherk
and Schwarz \cite{Scherk:1978ta,Scherk:1979zr}. Higher-dimensional
supersymmetric realizations and their fermionic mass spectra were
developed further by Fayet
\cite{Fayet:1985ua,Fayet:1985kt}. The relation between the fermion shift, 
the gravitino mass matrix, and
the super-Higgs mechanism will be established directly from the reduced
Rarita--Schwinger action in the next question.

\begin{takehome}
\begin{takehomeitems}

\item
The untwisted Kaluza--Klein gravitino tower already has a fermionic
Stückelberg structure,
$\delta\psi_y^{(n)}\sim(in/R)\epsilon^{(n)}$.

\item
A continuous Scherk--Schwarz twist replaces $in/R$ by
$\mathcal D_n=in/R+\mathcal M$. The internal gravitino components in
its image provide the corresponding would-be Goldstino directions.

\item
The designation of a particular gravitino as the $n=0$ mode depends
on the representative of the twist. The invariant datum is the
complete shifted tower, not an individual KK label.

\item
A fermion shift identifies the would-be Goldstino direction; its
absorption and the physical spin-$\tfrac32$ masses follow from the
reduced Rarita--Schwinger Lagrangian.

\item
If other fermions also carry shifts, the physical Goldstino is selected
by the complete fermion-shift and kinetic structure.

\end{takehomeitems}
\end{takehome}

\nextquestion
{How does the super-Higgs mechanism work in a Scherk--Schwarz compactification?}

\end{reviewquestion}
%%%%%%%%%%%%%%%%%%%%%%%%%%%%%%%%%%%%%%%%%%%%%%%%%%%%%%%%%%%%%%%
\begin{reviewquestion}
{How does the super-Higgs mechanism work in a Scherk--Schwarz
compactification?}
\label{sec:ss-superhiggs}

\begin{essentialidea}
For a five-dimensional gravitino, the ordinary Kaluza--Klein tower
already possesses the fermionic Stückelberg structure required by
massive spin-$\tfrac32$ fields. A Scherk--Schwarz twist does not create
this structure: it deforms the internal operator that controls it.

The same deformed operator appears in the gravitino--Goldstino mixing
and in the spin-$\tfrac32$ spectrum. In an $R_\xi$ gauge the mixing can
be cancelled explicitly; in the unitary-gauge limit the would-be
Goldstinos cease to propagate independently and provide the
longitudinal helicities of the massive gravitini.

This super-Higgs mechanism is a statement about the full fermionic
system and does not require the gravitino mass to be parametrically
below the Kaluza--Klein scale. The high-energy equivalence theorem is
a separate statement whose practical usefulness depends on the
hierarchy of scales.
\end{essentialidea}

\narrativeheading{Specializing to five dimensions}

The preceding questions were formulated, whenever possible, for a
compactification from $d+1$ to $d$ dimensions. We now specialize to
five-dimensional supergravity compactified to four dimensions.
Accordingly, we write the compact coordinate as $x^5\equiv y$ and
denote the internal gravitino components by $\psi_{5I}$. This
specialization will allow us to use explicitly four-dimensional Weyl,
Majorana, and helicity language.

Consider minimal five-dimensional supergravity, in which the
gravitini form an $SU(2)_R$ doublet. Their four-dimensional and
internal components will be denoted
\begin{equation}
    \psi_{\mu I},
    \qquad
    \psi_{5I},
    \qquad
    I=1,2 .
\end{equation}
As already anticipated in Question~\ref{sec:ss-goldstino}, the
internal components behave as spin-$\tfrac12$ fermions from the
four-dimensional point of view and provide the Stückelberg directions
associated with the massive gravitini.

\newpage
\narrativeheading{From Kaluza--Klein Stückelberg to super-Higgs}

After the field redefinitions required to put the spin-$\tfrac12$
kinetic terms in canonical form, the relevant bulk bilinears contain
schematically
\begin{align}
\mathcal L_{\rm bilinear}
\supset{}&
\psi_{\mu1}\sigma^{\mu\nu}\partial_5\psi_{\nu2}
-
\psi_{\mu2}\sigma^{\mu\nu}\partial_5\psi_{\nu1}
\nonumber\\
&-
\frac{\omega}{R}
\left(
\psi_{\mu1}\sigma^{\mu\nu}\psi_{\nu1}
+
\psi_{\mu2}\sigma^{\mu\nu}\psi_{\nu2}
\right)
\nonumber\\
&-
\frac{i\sqrt6}{2}
\left[
\partial_5\psi_{51}\sigma^\mu\psi_{\mu1}
+
\partial_5\psi_{52}\sigma^\mu\psi_{\mu2}
\right.
\nonumber\\
&\hspace{22mm}\left.
+
\frac{\omega}{R}
\left(
\psi_{52}\sigma^\mu\psi_{\mu1}
-
\psi_{51}\sigma^\mu\psi_{\mu2}
\right)
\right]
+\cdots .
\label{eq:q7-bilinear-system}
\end{align}
The numerical coefficients depend on normalization and spinor
conventions. The structural point is that the compact derivative and
the Scherk--Schwarz deformation always appear in the same
$SU(2)_R$-covariant combinations.

In the periodic-field representation these combinations are generated
by
\begin{equation}
    \mathcal D_5
    =
    \partial_5
    +
    \frac{\omega}{R}
    \begin{pmatrix}
        0 & 1\\
        -1 & 0
    \end{pmatrix},
\label{eq:q7-D5}
\end{equation}
in the normalization adopted here. Acting on the internal gravitino
doublet gives
\begin{align}
    g_1
    &=
    \partial_5\psi_{51}
    +
    \frac{\omega}{R}\psi_{52},
    \nonumber\\
    g_2
    &=
    \partial_5\psi_{52}
    -
    \frac{\omega}{R}\psi_{51}.
\label{eq:q7-goldstino-combinations}
\end{align}
These are precisely the would-be Goldstino combinations entering the
gravitino--Goldstino mixing. The same operator that generated the
inhomogeneous fermion shifts in Question~\ref{sec:ss-goldstino}
therefore appears directly in the reduced action.

\narrativeheading{The $R_\xi$ gauge}

The mixing between $\psi_{\mu I}$ and the Stückelberg fermions can be
removed without eliminating either field. Introduce
\begin{align}
    h_1
    &=
    \sigma^\mu\psi_{\mu1}
    -
    \frac{\sqrt6}{2}\,
    \xi\,
    \frac{\sigma^\mu\partial_\mu}{\partial^2}\,
    g_1,
    \nonumber\\
    h_2
    &=
    \sigma^\mu\psi_{\mu2}
    -
    \frac{\sqrt6}{2}\,
    \xi\,
    \frac{\sigma^\mu\partial_\mu}{\partial^2}\,
    g_2 ,
\end{align}
and add the fermionic gauge-fixing term
\begin{equation}
    \mathcal L_{\rm GF}
    =
    -\frac{i}{2\xi}
    \left(
        h_1\sigma^\mu\partial_\mu h_1
        +
        h_2\sigma^\mu\partial_\mu h_2
    \right)
    +\text{h.c.}
\label{eq:q7-Rxi}
\end{equation}
The cross terms generated by
Eq.~\eqref{eq:q7-Rxi} cancel the mixing in
Eq.~\eqref{eq:q7-bilinear-system}. The separation into a
spin-$\tfrac32$ field and a Stückelberg spinor is gauge dependent,
while physical poles and gauge-invariant observables are not.

This form of the gauge fixing also makes clear why the discussion is
not intrinsically tied to one Kaluza--Klein level. The functions
$g_I$ are five-dimensional objects. After Fourier expansion or, more
generally, after diagonalization of the complete fermionic mass
matrix, they provide the Goldstino combination associated with each
massive gravitino eigenstate.

\narrativeheading{Unitary gauge and the super-Higgs mechanism}

In the convention of Eq.~\eqref{eq:q7-Rxi}, unitary gauge is obtained
for $\xi\to\infty$. The would-be Goldstino combinations then obey
\begin{equation}
    g_1=g_2=0,
\end{equation}
or, for the pure bulk Scherk--Schwarz system,
\begin{align}
    \partial_5\psi_{51}
    +
    \frac{\omega}{R}\psi_{52}
    &=0,
    \nonumber\\
    \partial_5\psi_{52}
    -
    \frac{\omega}{R}\psi_{51}
    &=0 .
\label{eq:q7-unitary-gauge}
\end{align}
The would-be Goldstinos no longer propagate as independent fields.
Their degrees of freedom have become the helicity-$\tfrac12$
components of the massive gravitini.

This is the super-Higgs mechanism in the present compactification.
It should be distinguished from an approximation in which only a
light four-dimensional gravitino is retained. Even at $\omega=0$,
every non-zero Kaluza--Klein gravitino already undergoes its ordinary
fermionic Stückelberg mechanism. Turning on $\omega$ simply extends
the same structure to the direction that contained the massless
gravitino at the supersymmetric point.

\narrativeheading{The gravitino spectrum}

In unitary gauge, the bulk equations for a four-dimensional mass
eigenstate take the form
\begin{align}
    \partial_5\psi_{\mu2}
    +
    \left(
        m-\frac{\omega}{R}
    \right)\psi_{\mu1}
    &=0,
    \nonumber\\
    \partial_5\psi_{\mu1}
    -
    \left(
        m-\frac{\omega}{R}
    \right)\psi_{\mu2}
    &=0 .
\label{eq:q7-gravitino-eom}
\end{align}
Periodic solutions give the signed eigenvalues
\begin{equation}
    m_n^{\rm eigen}
    =
    \frac{n+\omega}{R},
    \qquad
    n\in\mathbb Z ,
\label{eq:q7-signed-spectrum}
\end{equation}
up to the sign convention for the $SU(2)_R$ generator, and hence the
physical masses
\begin{equation}
    m_n
    =
    \frac{|n+\omega|}{R}.
\label{eq:q7-gravitino-spectrum}
\end{equation}

The mass spectrum and the Goldstino shifts are therefore governed by
the same deformation. At $\omega=0$, the non-zero Kaluza--Klein
levels are already massive spin-$\tfrac32$ states with their
Stückelberg fermions, while the supersymmetric zero-mode gravitino is
massless. Switching on the twist produces
\begin{equation}
    \frac{n}{R}
    \longrightarrow
    \frac{n+\omega}{R},
\end{equation}
and the direction that was massless at the supersymmetric point now
participates in the same super-Higgs structure.

As emphasized in Question~\ref{sec:ss-goldstino}, the integer label is
not itself invariant. For an $R$-charge eigenstate, the physical
combination is $n+\omega q$, and a change of representative may shift
$\omega q$ by an integer while relabelling the entire tower. The
super-Higgs mechanism belongs to the shifted spectrum, not to a
privileged label called $n=0$.

\narrativeheading{What the equivalence theorem does, and does not, say}

The super-Higgs mechanism above is a statement about the complete
quadratic fermionic system. No high-energy limit was used. The
high-energy equivalence theorem is a different statement.

After diagonalizing the fermionic system, let
$\widehat\psi_{\mu a}$ be a gravitino mass eigenstate of physical mass
$m_a$, and let $\widehat\eta_a$ denote the corresponding Goldstino
eigenstate. For an on-shell amplitude containing external
helicity-$\tfrac12$ gravitini, the equivalence theorem gives
schematically
\begin{equation}
    \mathcal A
    \left(
        \widehat\psi_{a_1}^{\,\pm1/2},
        \ldots,
        \widehat\psi_{a_N}^{\,\pm1/2},
        X
    \right)
    =
    \mathcal A
    \left(
        \widehat\eta_{a_1},
        \ldots,
        \widehat\eta_{a_N},
        X
    \right)
    +
    \mathcal O\!\left(
        \max_i\frac{m_{a_i}}{E_i}
    \right),
\label{eq:q7-equivalence-theorem}
\end{equation}
up to conventional phases and normalizations. Here the hard
kinematic scales $E_i$ must be large compared with the masses of the
external longitudinal gravitini involved in the process. The theorem
does not require the energy to exceed the masses of the entire
Kaluza--Klein tower.

It is important not to read more into
Eq.~\eqref{eq:q7-equivalence-theorem} than it says. The theorem
replaces external longitudinal polarizations by the corresponding
Goldstino eigenstates. It does not integrate out the rest of the
Kaluza--Klein tower, nor does it turn a genuinely higher-dimensional
high-energy process into a theory containing only a finite number of
four-dimensional fields.

Its simplest use occurs when the gravitino is parametrically lighter
than the Kaluza--Klein scale. For a pure Scherk--Schwarz circle,
\begin{equation}
    m_{3/2}
    \simeq
    \frac{|\omega|}{R},
\end{equation}
so that for $|\omega|\ll1$ there is a parametrically large window
\begin{equation}
    m_{3/2}
    \ll
    E
    \ll
    \frac1R .
\label{eq:q7-4d-window}
\end{equation}
In this regime the excited Kaluza--Klein states are genuinely
decoupled and the equivalence theorem reduces to the familiar
four-dimensional supergravity statement. From the low-energy
gravitino and its eaten Goldstino alone, one need not be able to
reconstruct whether the microscopic supersymmetry breaking originated
from a bulk Scherk--Schwarz background or from a localized dynamical
sector: different higher-dimensional realizations can match onto the
same light four-dimensional super-Higgs system.

The situation changes as $|\omega|$ becomes of order one. The
gravitino mass then approaches the compactification scale and the
window \eqref{eq:q7-4d-window} disappears. The super-Higgs mechanism
itself remains perfectly well defined: it can still be followed
through the Kaluza--Klein gravitino and Goldstino system. What is lost
is the parametrically controlled interpretation in terms of one
isolated four-dimensional gravitino.

A finite truncation of the tower can still display the Stückelberg
and super-Higgs structure of the modes that have been retained. Such
a truncation should not, however, be used by itself to infer the
high-energy behaviour of the complete compactified theory. Once the
energy probes the Kaluza--Klein scale, additional states become
dynamically relevant, and cancellations or sum rules involving the
tower can matter for questions such as perturbative unitarity. The
equivalence theorem does not remove this requirement.

This point becomes particularly sharp near the half-twist
$\omega=1/2$. There is then no parametrically isolated light
gravitino: the two lightest Majorana states are already degenerate at
$1/(2R)$. Their individual super-Higgs structure may still be followed
in the Majorana description, while their recombination into a Dirac
gravitino is a separate symmetry statement that will be established
later in the review.

\narrativeheading{Localized breaking and additional Goldstino directions}

Boundaries can modify this picture in two logically distinct ways. A
localized gravitino bilinear may simply represent another distribution
of the same $R$-symmetry rotation between bulk and boundary data. In
the elementary interval problem, localized masses $M_0$ and $M_\pi$
give
\begin{equation}
    m_n^{\rm eigen}
    =
    \frac{1}{R}
    \left[
        n+\omega
        +
        \frac{1}{\pi}
        \left(
            \arctan M_0+\arctan M_\pi
        \right)
    \right],
\label{eq:q7-localized-spectrum}
\end{equation}
in the normalization adopted here. A cancellation among the bulk and
boundary contributions can therefore produce a gravitino much lighter
than $1/R$, again leaving a genuine four-dimensional window in which
the equivalence theorem is useful.

A genuine localized supersymmetry-breaking sector contains more
information than such a boundary rotation. Its fermions may themselves
have inhomogeneous supersymmetry transformations and therefore add
independent would-be Goldstino directions. The physical Goldstino
eaten by a given gravitino is then a combination of bulk and localized
fermions. If the number of independent would-be Goldstino directions
exceeds the number of broken local supersymmetries, orthogonal
combinations remain in the physical spectrum as pseudo-Goldstinos.

Consequently, two compactifications can have the same light gravitino
mass while differing in their low-energy fermionic content. In
particular, pseudo-Goldstinos can retain information about several
independent supersymmetry-breaking sectors that is invisible in the
gravitino mass alone. The distinction between localized gravitino
rotations and genuine localized supersymmetry breaking will be treated
systematically when boundary terms are discussed.

\narrativeheading{Beyond the flat circle}

The explicit formulae above exploit the simplicity of flat
five-dimensional supergravity on a circle or interval. On a more
general compact space there need be no Fourier label $n$ and no
universal shift $(n+\omega)/R$. The structural relation survives:
the internal supersymmetry operator determines the fermion shifts and
enters the spin-$\tfrac32$ spectral problem, while the reduced
Rarita--Schwinger action determines how the corresponding Stückelberg
directions are absorbed.

The flat circle is therefore not special because the super-Higgs
mechanism exists only there, but because the entire construction can
be followed analytically from the higher-dimensional gauge
transformation to the complete gravitino spectrum.

\narrativeheading{Bibliographical remarks}

The Goldstino associated with spontaneous supersymmetry breaking and
its absorption by the gravitino were developed in the early literature
on supersymmetry and supergravity
\cite{Fayet:1974jb,Volkov:1973jd,Deser:1977uq,
Cremmer:1978hn,Fayet:1977vd}. The high-energy equivalence theorem for
the longitudinal gravitino was formulated and established in
Refs.~\cite{Casalbuoni:1988qd,Casalbuoni:1988kv}.

The role of the complete Kaluza--Klein tower in the super-Higgs
mechanism, including the possibility that the physical gravitino and
Goldstino are mixtures of many Fourier modes, was analysed in
Ref.~\cite{Meissner:1999ja}. The equivalence theorem for
Kaluza--Klein gravitino mass eigenstates, together with the
corresponding four- and five-dimensional $R_\xi$ gauges, was developed
in Ref.~\cite{DeCurtis:2003hs}. An explicit treatment including both
bulk Scherk--Schwarz breaking and localized supersymmetry-breaking
$F$-terms, in general $R_\xi$ and unitary gauges, as well as the
resulting pseudo-Goldstinos, was given in
Ref.~\cite{Benakli:2007zza}.

Generalized Scherk--Schwarz boundary conditions and localized
gravitino masses in five-dimensional supergravity were developed in
Refs.~\cite{Bagger:2001qi,vonGersdorff:2002tj,
Gherghetta:2001sa,Rattazzi:2003rj}. We shall return to this literature
more systematically when discussing orbifolds, intervals, and
localized boundary terms.

\begin{takehome}
\begin{takehomeitems}

\item
The super-Higgs mechanism in a Scherk--Schwarz compactification is the
deformation of the fermionic Stückelberg structure already present in
the ordinary Kaluza--Klein gravitino tower.

\item
The same twisted internal operator controls the Goldstino shifts, the
gravitino--Goldstino mixing, and the spin-$\tfrac32$ spectrum. An
$R_\xi$ gauge separates the two sectors explicitly, while unitary
gauge removes the would-be Goldstinos.

\item
On a flat circle the gravitino spectrum is
$m_n=|n+\omega|/R$, up to $R$-charge normalization and Kaluza--Klein
relabelling.

\item
The super-Higgs mechanism does not require
$m_{3/2}\ll1/R$. The latter hierarchy is instead what permits a
controlled four-dimensional truncation to an isolated light gravitino.

\item
When $m_{3/2}\ll1/R$, the high-energy equivalence theorem can be used
inside the four-dimensional window
$m_{3/2}\ll E\ll1/R$. For an order-one twist this window disappears:
the theorem does not justify replacing the full Kaluza--Klein system
by a finite four-dimensional truncation.

\item
Localized gravitino rotations can modify the effective spectral shift,
whereas genuine localized supersymmetry-breaking sectors can introduce
additional Goldstino directions. Orthogonal uneaten combinations may
survive as pseudo-Goldstinos.

\end{takehomeitems}
\end{takehome}

\nextquestion
{How does Scherk--Schwarz compactification work on orbifolds and
intervals?}

\end{reviewquestion}
%%%%%%%%%%%%%%%%%%%%%%%%%%%%%%%%%%%%%%%%%%%%%%%%%%%%%%%%%%%%%%%
%%%%%%%%%%%%%%%%%%%%%

\begin{reviewquestion}
{How does Scherk--Schwarz compactification work on orbifolds and intervals?}
\label{sec:ss-orbifolds}

\begin{essentialidea}

The orbifold $S^1/\mathbb Z_2$ is most cleanly described from its
covering space $\mathds{R}$. Its orbifold group is generated by the
two reflections
\[
r_0:y\mapsto-y,
\qquad
r_\pi:y\mapsto2\pi R-y,
\]
whose product is the translation by one circumference,
\[
t=r_\pi r_0,
\qquad
t(y)=y+2\pi R.
\]

For a chosen equivariant lift of the orbifold action to a field
$\Phi$, let
\[
P_0^{(\Phi)}=\rho_\Phi(r_0),
\qquad
P_\pi^{(\Phi)}=\rho_\Phi(r_\pi).
\]
Then
\[
\bigl(P_0^{(\Phi)}\bigr)^2
=
\bigl(P_\pi^{(\Phi)}\bigr)^2
=
\mathbf1,
\qquad
U_{\rm cov}^{(\Phi)}
=
P_\pi^{(\Phi)}P_0^{(\Phi)},
\]
and therefore
\[
P_0^{(\Phi)}
U_{\rm cov}^{(\Phi)}
P_0^{(\Phi)}
=
\bigl(U_{\rm cov}^{(\Phi)}\bigr)^{-1}.
\]

These relations follow from the orbifold group. They constrain how a
Scherk--Schwarz deformation can be represented, but they do not fix
its value. A continuous twist is additional compactification data. It
may be represented by an $\omega$-dependent relative orientation of
the endpoint involutions or, in another trivialization, by a flat bulk
connection.

The quotient has the interval $0\leq y\leq\pi R$ as its underlying
topological space. In the interval description, the relevant global
information is carried jointly by the endpoint projections and by the
parallel transport between them. Supersymmetry survives only when
these data admit a common global Killing spinor.

\end{essentialidea}

\narrativeheading{Why orbifolds?}

The circle is the natural starting point for a pedagogical discussion
of Scherk--Schwarz monodromy. It is smooth, has a single
non-contractible direction, and makes the basic identification
particularly transparent. For phenomenological applications,
however, it is often too simple. Starting, for example, from
five-dimensional supergravity, compactification on a circle does not
by itself produce a chiral four-dimensional fermion spectrum. One
therefore wants a compactification that can project out part of the
higher-dimensional field content and, at the same time, reduce the
amount of supersymmetry seen in the lower-dimensional theory.

Orbifolds provide the simplest way of doing this. Their appearance in
high-energy physics came most prominently through string
compactification, although the mathematical notion is older. Satake
introduced what he called \(V\)-manifolds in the 1950s, namely spaces
that are locally quotients of smooth manifolds by finite groups.
Thurston later developed this point of view extensively and
popularized the term ``orbifold''. In string theory, the construction
acquired a much more concrete physical meaning in the work of Dixon,
Harvey, Vafa and Witten: strings were quantized on quotients of smooth
compactification spaces by discrete symmetries, with the quotient
supplemented by the twisted sectors required for a consistent
worldsheet theory.

This last point is worth keeping in mind. An orbifold in string theory
is not merely a singular quotient space on which one decides to put
fields. The quotient is part of a microscopic construction, and
worldsheet consistency supplies additional sectors that have no
analogue in a naive field theory obtained by simply deleting points or
imposing boundary conditions. In a low-energy field-theory or
supergravity treatment, by contrast, one can work directly with the
quotient or with a fundamental domain and specify the corresponding
projection, equivariance and localized data explicitly. The present
question adopts this latter viewpoint. Later, when we return to
string theory, the additional microscopic requirements will become
important again.

Our basic example is \(S^1/\mathbb Z_2\), whose underlying
topological space is an interval. Relative to the circle, however, a
bulk field must now be assigned not only a monodromy around the
covering circle but also transformation laws under the two reflections
whose fixed loci become the endpoints. These data cannot be chosen
independently: together they must define a representation of the full
orbifold group.

This is where Scherk--Schwarz compactification on an orbifold differs
from simply taking the circle formula and cutting the circle in half.
The twist, the two endpoint actions and the covering-space translation
are different descriptions of related group data, and their
compatibility must be imposed before discussing the spectrum or
supersymmetry.

We will first describe the orbifold as a quotient of its covering
line, then lift the group action to the fields, and finally ask how
local endpoint projections can coexist with, or obstruct, a globally
defined supersymmetry. Localized dynamics at the fixed points is a
separate issue and will be left to the next question.

\narrativeheading{The orbifold as a quotient of the covering line}

Begin with the real line $\mathds R$, with coordinate $y$. Define the
two reflections
\[
r_0(y)=-y,
\qquad
r_\pi(y)=2\pi R-y.
\]
Both are involutions, \(r_0^2=r_\pi^2=\mathrm{id}_{\mathds R}\), and
their product is the translation
\begin{equation}
t\equiv r_\pi r_0,
\qquad
t(y)=y+2\pi R.
\label{eq:q8-translation-from-reflections}
\end{equation}

The group generated by the two reflections is the infinite dihedral
group,
\begin{equation}
D_\infty
=
\left\langle
r_0,r_\pi
\ \middle|\
r_0^2=r_\pi^2=1
\right\rangle .
\label{eq:q8-dihedral-group}
\end{equation}
Using \(t=r_\pi r_0\), the same group can equivalently be presented as
\[
D_\infty
=
\left\langle
t,r_0
\ \middle|\
r_0^2=1,\quad r_0tr_0=t^{-1}
\right\rangle .
\]

The translation subgroup \(\langle t\rangle\simeq\mathbb Z\)
produces the circle,
\[
S^1=\mathds R/(2\pi R\mathbb Z),
\]
and quotienting in addition by a reflection gives
\begin{equation}
S^1/\mathbb Z_2
\simeq
\mathds R/D_\infty.
\label{eq:q8-orbifold-quotient}
\end{equation}

Its underlying topological space is the interval \([0,\pi R]\). What
distinguishes the orbifold description from a bare interval is not the
point-set topology, nor necessarily the field content, but the
isotropy and equivariance data inherited from the covering-space
action.

The endpoints \(y=0\) and \(y=\pi R\) are fixed loci with stabilizer
\(\mathbb Z_2\). Bulk fields carry compatible lifts of both
reflections. A degree of freedom intrinsically localized at one
endpoint, on the other hand, only has to transform consistently under
the stabilizer of that endpoint; it need not carry the complete bulk
monodromy around the covering circle.

An ordinary field theory formulated directly on an interval can
therefore reproduce the same physics if the corresponding endpoint
actions, bulk identifications and consistency conditions are supplied
by hand. The orbifold description packages these data into the group
action from which they originate.

\narrativeheading{Fields as equivariant sections}

Let \(V_\Phi\) denote the local representation space in which the
components of a field \(\Phi\) transform. For a scalar multiplet this
may simply be an internal multiplet space; for a spinor, \(V_\Phi\)
contains the corresponding spin representation, possibly together
with gauge or \(R\)-symmetry indices. Globally, \(\Phi\) is a section
of the appropriate associated bundle, but \(V_\Phi\) is sufficient
for describing the orbifold action locally.

An element \(g\in D_\infty\) sends \(y\) to \(g\!\cdot\! y\), while
the field components at the two points are related by an allowed
transformation. Thus one imposes
\begin{equation}
\Phi(g\!\cdot\!y)
=
\rho_\Phi(g)\,\Phi(y),
\label{eq:q8-equivariance}
\end{equation}
rather than requiring the field to be invariant under every
geometrical operation.

For a single real scalar, \(\rho_\Phi(g)\) may reduce to a sign,
giving the familiar even or odd parity. For a multiplet it may rotate,
reflect or permute components, for example through a gauge or
\(R\)-symmetry transformation. Since successive geometrical
operations must be represented consistently on the fields,
\[
\rho_\Phi:
D_\infty
\longrightarrow
GL(V_\Phi)
\]
is a representation whose image is built from transformations allowed
by the parent theory.

For the two reflections define
\[
P_0^{(\Phi)}=\rho_\Phi(r_0),
\qquad
P_\pi^{(\Phi)}=\rho_\Phi(r_\pi).
\]
Their involutive character gives
\[
\bigl(P_0^{(\Phi)}\bigr)^2
=
\bigl(P_\pi^{(\Phi)}\bigr)^2
=
\mathbf1.
\]
Since \(t=r_\pi r_0\), the lift of the translation is not independent:
\begin{equation}
U_{\rm cov}^{(\Phi)}
\equiv
\rho_\Phi(t)
=
P_\pi^{(\Phi)}P_0^{(\Phi)}.
\label{eq:q8-translation-lift}
\end{equation}
Hence
\[
\Phi(y+2\pi R)
=
U_{\rm cov}^{(\Phi)}\Phi(y).
\]

The two endpoint involutions therefore determine the monodromy seen
after one complete turn around the covering circle. In this
trivialization, the Scherk--Schwarz twist is encoded in their relative
action rather than introduced as an unrelated third operation.

\narrativeheading{Covering-space monodromy and equivalent trivializations}

The relation
\[
U_{\rm cov}^{(\Phi)}
=
P_\pi^{(\Phi)}P_0^{(\Phi)}
\]
does not mean that the geometry of \(S^1/\mathbb Z_2\) chooses a
Scherk--Schwarz parameter. The geometry fixes the orbifold group
relations; how those relations are represented on the fields is part
of the compactification data.

For a continuous Scherk--Schwarz deformation, one possible
description places the parameter in the relative orientation of the
two endpoint involutions, so that
\(U_{\rm cov}^{(\Phi)}=U_{\rm SS}^{(\Phi)}(\omega)\).

The same compactification may instead be described with fixed endpoint
involutions and with the Scherk--Schwarz dependence moved into the
compact derivative by a \(y\)-dependent change of field basis,
\begin{equation}
D_y
=
\partial_y+\mathcal A_y,
\qquad
\mathcal A_y
\propto
\frac{\omega}{R}\,\mathbb R_\Phi .
\label{eq:q8-flat-connection}
\end{equation}
At this level \(\mathcal A_y\) records the same monodromy in a
periodic field basis. When the twisting symmetry is realized locally
in supergravity, the same object can be the internal component of the
corresponding symmetry connection, and the description becomes a
Wilson-line background in the usual gauge-theory sense.

For the backgrounds considered here the connection is flat, so its
local curvature vanishes even though its parallel transport can carry
non-trivial global information. The two descriptions differ only in
the choice of field basis used to compare fibres at different points:
the same data can be moved between the relative endpoint orientation
and a flat bulk connection without changing the compactification.

On the interval this is particularly transparent. A flat connection
can be removed from the bulk by a \(y\)-dependent transformation, but
that transformation changes the relative field bases at
\(y=0\) and \(y=\pi R\), and hence changes the endpoint data. On the
covering circle the corresponding transformation need not be
single-valued; its failure to be single-valued retains the global
monodromy.

The orbifold group law also restricts the admissible endpoint data.
Suppressing the label \((\Phi)\), one has
\(P_\pi=U_{\rm cov}P_0\). Together with
\(P_0^2=P_\pi^2=\mathbf1\), this implies
\begin{equation}
P_0U_{\rm cov}P_0
=
U_{\rm cov}^{-1}.
\label{eq:q8-PUP}
\end{equation}
This is simply the representation on \(V_\Phi\) of
\(r_0tr_0=t^{-1}\).

The continuous Scherk--Schwarz freedom does not arise by relaxing
this relation. It arises because a fixed group relation can admit
continuous families of representations on the field space.

\narrativeheading{Continuous twists and the orbifold compatibility condition}

Take a continuous Scherk--Schwarz deformation generated by a
one-parameter subgroup of an allowed symmetry of the parent theory.
For a field \(\Phi\), choose the convention
\begin{equation}
U_{\rm SS}^{(\Phi)}(\omega)
=
\exp\!\left(
2\pi i\omega\,\mathbb R_\Phi
\right),
\label{eq:q8-one-parameter-U}
\end{equation}
where \(\mathbb R_\Phi\) represents the chosen generator on
\(V_\Phi\).

Work in the trivialization in which the flat bulk connection has been
gauged away. The Scherk--Schwarz data then reside in the relative
endpoint involutions:
\[
U_{\rm cov}^{(\Phi)}
=
U_{\rm SS}^{(\Phi)}(\omega),
\qquad
P_\pi^{(\Phi)}(\omega)
=
U_{\rm SS}^{(\Phi)}(\omega)P_0^{(\Phi)}.
\]

Suppose \(P_0^{(\Phi)}\) is fixed and this whole one-parameter family
satisfies the orbifold compatibility condition near the identity.
Equation~\eqref{eq:q8-PUP} gives
\[
P_0^{(\Phi)}
e^{2\pi i\omega\mathbb R_\Phi}
P_0^{(\Phi)}
=
e^{-2\pi i\omega\mathbb R_\Phi}.
\]
Differentiating at \(\omega=0\) yields
\begin{equation}
P_0^{(\Phi)}
\mathbb R_\Phi
P_0^{(\Phi)}
=
-\mathbb R_\Phi,
\qquad\text{or equivalently}\qquad
\{P_0^{(\Phi)},\mathbb R_\Phi\}=0.
\label{eq:q8-generator-anticommute}
\end{equation}

This statement concerns the generator of a continuous family, not an
isolated group element. A special value of \(U_{\rm SS}\) can satisfy
the orbifold relation even when an arbitrarily chosen logarithm of
that element does not obey
Eq.~\eqref{eq:q8-generator-anticommute}.

The orbifold geometry therefore constrains which continuous
deformations are compatible with a chosen endpoint involution, but it
does not determine \(\omega\). The latter is additional
compactification data.

\narrativeheading{A simple \(SU(2)\) example}

Consider an \(SU(2)\) doublet and choose \(P_0=\sigma_3\). Let the
continuous deformation lie along the \(\sigma_2\) direction,
\begin{equation}
U_{\rm SS}(\omega)
=
e^{2\pi i\omega\sigma_2}.
\label{eq:q8-SU2-U}
\end{equation}
With this normalization \(\omega\sim\omega+1\), since the eigenvalues
of \(\sigma_2\) are \(\pm1\).

Because
\(\sigma_3\sigma_2\sigma_3=-\sigma_2\), one immediately obtains
\(P_0U_{\rm SS}P_0=U_{\rm SS}^{-1}\). The orbifold relation is
therefore satisfied for every \(\omega\), and the second endpoint
involution is
\begin{equation}
P_\pi(\omega)
=
U_{\rm SS}(\omega)P_0
=
e^{2\pi i\omega\sigma_2}\sigma_3.
\label{eq:q8-SU2-Ppi}
\end{equation}
Indeed,
\[
P_\pi(\omega)^2
=
U_{\rm SS}P_0U_{\rm SS}P_0
=
\mathbf1.
\]

The same matrix can be written as
\begin{equation}
P_\pi(\omega)
=
e^{i\pi\omega\sigma_2}
P_0
e^{-i\pi\omega\sigma_2}.
\label{eq:q8-SU2-conjugate-projectors}
\end{equation}
The two endpoint involutions are therefore conjugate, and their
\(+1\) eigenspaces are related by the corresponding \(SU(2)\)
rotation. The parameter \(\omega\) measures their relative
orientation.

Nothing in the orbifold group changes as \(\omega\) varies; only its
representation on the doublet changes. At the half-twist,
\(\omega=\tfrac12\), one has
\(U_{\rm SS}(\tfrac12)=-\mathbf1\). We shall return to this special
point in Question~\ref{sec:ss-half-twist}.

\narrativeheading{Local projections and global supersymmetry}

The orbifold action must also be specified on the supersymmetry
parameter. Let \(\mathscr P_0\) and \(\mathscr P_\pi\) denote the full
involutions acting on the supersymmetry-parameter space
\(\mathcal S\):
\begin{align}
\epsilon(-y)
&=
\mathscr P_0\,\epsilon(y),
\nonumber\\
\epsilon(2\pi R-y)
&=
\mathscr P_\pi\,\epsilon(y).
\label{eq:q8-susy-orbifold-action}
\end{align}
They obey
\(\mathscr P_0^2=\mathscr P_\pi^2=\mathbf1\).

The notation \(\mathscr P_b\) distinguishes the complete action on the
supersymmetry parameter from the internal matrices that may act on
other fields. For a spinorial parameter, \(\mathscr P_b\) may combine
the spinorial action of the reflection with an internal
\(R\)-symmetry action, depending on the parent theory.

The supersymmetries locally compatible with the two fixed points are
the respective \(+1\) eigenspaces
\begin{equation}
\mathcal S_0
=
\ker(\mathscr P_0-\mathbf1),
\qquad
\mathcal S_\pi
=
\ker(\mathscr P_\pi-\mathbf1).
\label{eq:q8-endpoint-susy-spaces}
\end{equation}

First choose a trivialization in which the flat Scherk--Schwarz
connection has been removed from the bulk. The Killing-spinor
equation along the interval then reduces to
\(\partial_y\epsilon=0\), while the relative endpoint involutions
carry the Scherk--Schwarz information. A globally preserved
supersymmetry must satisfy both endpoint conditions, so
\begin{equation}
\mathcal S_{\rm global}
=
\mathcal S_0\cap\mathcal S_\pi.
\label{eq:q8-global-SUSY-simple}
\end{equation}
It is perfectly possible for both
\(\dim\mathcal S_0\) and \(\dim\mathcal S_\pi\) to be non-zero while
\(\mathcal S_0\cap\mathcal S_\pi=\{0\}\). Each endpoint then admits a
local supersymmetry subspace, but no common supersymmetry extends over
the whole interval.

In another trivialization the endpoint involutions may be kept fixed
while the same Scherk--Schwarz deformation is placed in a flat bulk
connection. The Killing-spinor equation becomes
\begin{equation}
D_5\epsilon
=
\left(
\partial_5+\mathcal A_5(y)
\right)\epsilon
=
0.
\label{eq:q8-killing-general}
\end{equation}
For an initial value \(\epsilon_0=\epsilon(0)\), its solution is
\begin{equation}
\epsilon(y)
=
\mathcal U(y,0)\epsilon_0,
\qquad
\mathcal U(y,0)
=
\mathcal P
\exp\!\left[
-\int_0^y ds\,\mathcal A_5(s)
\right].
\label{eq:q8-transport-operator}
\end{equation}

The endpoint conditions then require
\[
\epsilon_0\in\mathcal S_0,
\qquad
\mathcal U(\pi R,0)\epsilon_0\in\mathcal S_\pi.
\]
Subject to any remaining Killing-spinor equations imposed by the
background, global supersymmetry exists precisely when
\begin{equation}
\mathcal S_0
\cap
\mathcal U(\pi R,0)^{-1}\mathcal S_\pi
\neq
\{0\}.
\label{eq:q8-exact-killing-condition}
\end{equation}
Equivalently,
\(\mathcal U(\pi R,0)\mathcal S_0\cap\mathcal S_\pi\neq\{0\}\).

One may instead compare both endpoint conditions in the fibre over
\(y=0\). Transporting the second endpoint involution back to that
fibre gives
\[
\widehat{\mathscr P}_\pi
=
\mathcal U(\pi R,0)^{-1}
\mathscr P_\pi
\mathcal U(\pi R,0),
\]
so that
\begin{equation}
\mathcal S_{\rm global}
=
\ker(\mathscr P_0-\mathbf1)
\cap
\ker(\widehat{\mathscr P}_\pi-\mathbf1),
\label{eq:q8-global-susy-one-fibre}
\end{equation}
again subject to the other Killing-spinor equations of the
background.

The endpoint involutions and the bulk transport must always be written
in the same trivialization. A change of trivialization can move the
same Scherk--Schwarz datum between relative endpoint orientation and
flat bulk connection. Combining the two descriptions as independent
ingredients would count the twist twice.

\narrativeheading{From orbifold parity to boundary conditions}

For completeness, consider a smooth scalar near \(y=0\) satisfying
\[
\phi(-y)=p\,\phi(y),
\qquad
p=\pm1.
\]
For odd parity, evaluation at the fixed point gives \(\phi(0)=0\).
For even parity, differentiation of the parity relation gives
\(\partial_y\phi(0)=0\).

Thus, for a smooth scalar in the absence of localized interactions,
odd parity gives Dirichlet data and even parity gives Neumann data.
The same reasoning applies at \(y=\pi R\) in the local coordinate
\(y-\pi R\).

These conditions follow from smooth equivariance; they are not the
general definition of boundary conditions on an orbifold. Localized
operators modify the variational problem and can produce Robin, jump,
or more general matching conditions. This is the subject of the next
question.

\narrativeheading{Bibliographical remarks}

The mathematical notion underlying orbifolds can be traced to
Satake's \(V\)-manifolds and was developed further in the work of
Thurston. Orbifold compactifications in string theory are described
in Refs.~\cite{Dixon:1985jw,Dixon:1986jc}.

The ingredients needed for Scherk--Schwarz supersymmetry breaking
already appear in the original generalized reductions
Refs.~\cite{Scherk:1978ta,Scherk:1979zr}, where compact-coordinate
dependence generated by a symmetry is used to produce fermion masses.
The appearance of the compact component of a local internal-symmetry
connection in the fermionic mass matrix can be seen explicitly in
Ref.~\cite{Cremmer:1979uq}.

For \(S^1/\mathbb Z_2\) supergravity, descriptions in which the
Scherk--Schwarz data are distributed between endpoint actions,
localized terms and a flat \(R\)-symmetry connection can be found in
Refs.~\cite{Bagger:2001qi,Biggio:2002rb,vonGersdorff:2001ak,
Delgado:2002xf}.

\begin{takehome}
\begin{takehomeitems}

\item
The orbifold \(S^1/\mathbb Z_2\) is described on its covering line by
the infinite dihedral group generated by the reflections about
\(0\) and \(\pi R\).

\item
For a chosen equivariant lift, the two endpoint involutions determine
the induced covering-space translation,
\[
P_0^2=P_\pi^2=\mathbf1,
\qquad
U_{\rm cov}=P_\pi P_0,
\qquad
P_0U_{\rm cov}P_0=U_{\rm cov}^{-1}.
\]
These relations follow from the orbifold group and do not by
themselves fix a Scherk--Schwarz parameter.

\item
For a continuous family
\(U_{\rm SS}^{(\Phi)}(\omega)
=e^{2\pi i\omega\mathbb R_\Phi}\),
compatibility with a fixed \(P_0^{(\Phi)}\) near the identity requires
\[
P_0^{(\Phi)}
\mathbb R_\Phi
P_0^{(\Phi)}
=
-\mathbb R_\Phi.
\]

\item
The same Scherk--Schwarz deformation can be represented either through
the relative endpoint orientation or through a flat bulk connection.
These are different trivializations of the same compactification data,
not independent sources of twisting.

\item
Global supersymmetry requires compatibility between the endpoint
projections and the bulk transport,
\[
\mathcal S_0
\cap
\mathcal U(\pi R,0)^{-1}\mathcal S_\pi
\neq
\{0\},
\]
with all quantities expressed in one common trivialization.

\item
For smooth bulk scalars without localized interactions, odd and even
orbifold parities imply Dirichlet and Neumann data, respectively.
Localized terms can modify these dynamical boundary or matching
conditions.

\end{takehomeitems}
\end{takehome}

\nextquestion
{How do localized boundary terms modify Scherk--Schwarz breaking?}

\end{reviewquestion}
%%%%%%%%%%%%%%%%%%%%%%%%%%%%%%%%%%%%%%%%%%%%%%%%%%%%%%%%%%%%%%%
\begin{reviewquestion}
{How do localized boundary terms modify Scherk--Schwarz breaking?}
\label{sec:ss-localized-boundary-terms}

\begin{essentialidea}

Localized operators do not modify the orbifold group relations
established in Question~\ref{sec:ss-orbifolds}. They modify the
dynamical and spectral problem obeyed by the fields.

A localized potential, a boundary mass, a boundary kinetic term, or a
fermion bilinear changes the variational problem and can therefore
modify the boundary or matching conditions, the Kaluza--Klein
eigenvalues, the wave functions and their normalization, and the
couplings of bulk modes to localized sectors.

For a second-order bulk field, a localized mass produces a Robin
condition. A kinetic term containing derivatives tangent to the
boundary produces an eigenvalue-dependent boundary condition and
changes the inner product with respect to which the Kaluza--Klein
modes are normalized. The resulting problem is naturally a
generalized Sturm--Liouville problem.

For first-order fermionic systems, gravitini among them, localized
bilinears instead give matching conditions for the fields themselves.
A pure localized mass can rotate the allowed fermionic subspace at an
endpoint by an energy-independent angle, whereas a localized kinetic
term generically makes this rotation depend on the four-dimensional
eigenvalue.

This distinction limits the familiar statement that a boundary term
can be traded for a Scherk--Schwarz twist. A boundary interaction can
be represented by a constant effective twist only when its complete
effect is an allowed, energy-independent symmetry rotation. Generic
localized kinetic terms reshape the tower in a way that cannot be
summarized by one universal value of $\omega_{\rm eff}$.

Localized terms are also part of the quantum effective theory rather
than optional decorations. Fixed points and boundaries break
translation invariance in the compact direction, and ultraviolet
divergences can be supported on them. Local boundary
renormalization must, however, be distinguished from the finite
contribution that requires propagation over a non-zero compact
distance. The latter is non-local and has a natural Casimir
interpretation: the boundaries determine how fields are reflected,
while the compact geometry and Scherk--Schwarz data determine their
transport between successive reflections.

\end{essentialidea}

\narrativeheading{Localized operators deform the Kaluza--Klein problem}

Question~\ref{sec:ss-orbifolds} distinguished the orbifold
equivariance data from the dynamical boundary conditions following
from the action. Localized operators are where this distinction
becomes unavoidable.

Let $L=\pi R$ and consider a real scalar field on the interval
$0\leq y\leq L$. For this first part we return to a general
compactification from $d+1$ to $d$ dimensions, with
\[
x^M=(x^\mu,y),
\qquad
M,N=0,\ldots,d,
\qquad
\mu,\nu=0,\ldots,d-1,
\]
and use the mostly-plus metric
\[
\eta_{MN}=\operatorname{diag}(-,+,\ldots,+).
\]

Expand
\begin{equation}
    \phi(x,y)
    =
    \sum_n
    \varphi_n(x)f_n(y),
\end{equation}
where the lower-dimensional fields are defined by
\begin{equation}
    \Box_d\varphi_n
    =
    m_n^2\varphi_n,
    \qquad
    \Box_d
    \equiv
    \eta^{\mu\nu}\partial_\mu\partial_\nu.
\label{eq:q9-mass-convention}
\end{equation}

Consider the quadratic action
\begin{align}
S
={}&
-\frac12
\int d^dx
\int_0^Ldy\,
\left[
(\partial_M\phi)(\partial^M\phi)
+
M^2\phi^2
\right]
\nonumber\\
&
-\frac12
\sum_{b=0,L}
\int d^dx\,
\left[
r_b\,
\partial_\mu\phi\,\partial^\mu\phi
+
\lambda_b\,\phi^2
\right]_{y=b}.
\label{eq:q9-scalar-action}
\end{align}
The coefficient $\lambda_b$ is a localized quadratic potential,
whereas $r_b$ multiplies a kinetic term involving derivatives tangent
to the boundary. We restrict here to such tangential kinetic terms.
Localized operators containing normal derivatives require additional
care in the thin-boundary limit and will not be included in the
simple spectral problem below.

Away from the endpoints,
\begin{equation}
    \left(
        -\partial_y^2+M^2
    \right)f_n
    =
    m_n^2f_n.
\label{eq:q9-bulk-mode-equation}
\end{equation}

Let $n_0^y=-1$ and $n_L^y=+1$ be the outward normals at the two
endpoints. The boundary part of the variation is
\begin{equation}
\delta S\big|_{y_b}
=
\int d^dx\,
\left[
-\,n_b^y\partial_y\phi
+
r_b\Box_d\phi
-
\lambda_b\phi
\right]
\delta\phi\big|_{y_b}.
\end{equation}
Using Eq.~\eqref{eq:q9-mass-convention}, stationarity of the action
gives
\begin{equation}
    n_b^y\partial_y f_n(y_b)
    +
    \left(
        \lambda_b-r_bm_n^2
    \right)
    f_n(y_b)
    =
    0.
\label{eq:q9-general-boundary-condition}
\end{equation}
Thus
\begin{align}
    f_n'(0)
    &=
    \left(
        \lambda_0-r_0m_n^2
    \right)f_n(0),
\label{eq:q9-boundary-condition-zero}
\\
    f_n'(L)
    &=
    -
    \left(
        \lambda_L-r_Lm_n^2
    \right)f_n(L).
\label{eq:q9-boundary-condition-L}
\end{align}

The localized mass and the localized kinetic term therefore enter the
same boundary condition in physically different ways. The former
gives an eigenvalue-independent Robin coefficient. The latter
multiplies the spectral parameter itself.

A more general localized potential $V_b(\phi)$ can also change the
background profile. Its boundary equation is
\begin{equation}
    n_b^y\partial_y\phi
    +
    V_b'(\phi)
    =
    0.
\end{equation}
For fluctuations $\phi=\bar\phi+\delta\phi$ around a background
$\bar\phi$, the coefficient entering the quadratic boundary condition
is simply
\[
\lambda_b
=
V_b''\!\left(\bar\phi(y_b)\right).
\]
Localized potentials can therefore affect both the background and the
spectrum of fluctuations around it.

\narrativeheading{A generalized Sturm--Liouville problem}

It is useful first to recall the standard Sturm--Liouville problem,
\begin{equation}
-\frac{d}{dy}
\left(
p(y)\frac{df}{dy}
\right)
+
q(y)f
=
\lambda\,w(y)f,
\end{equation}
where $w(y)$ is the weight and $\lambda$ the spectral parameter. The
ordinary Kaluza--Klein equation is already of this form, with
$\lambda=m_n^2$. Localized masses add singular contributions to the
quadratic operator, while localized kinetic terms add singular
contributions to the weight. The latter also make the boundary
conditions depend explicitly on $m_n^2$.

There is a useful variational formulation that collects these facts
into a single generalized eigenvalue problem. Let $g(y)$ be an
arbitrary sufficiently regular test profile. More precisely, $g$
belongs to the form domain of the problem, for example
$H^1([0,L])$ in the scalar example considered here, subject only to
kinematic constraints that are independent of the eigenvalue.

Define
\begin{align}
\mathcal Q[f,g]
={}&
\int_0^Ldy\,
\left[
\partial_yf^*\partial_yg
+
M^2f^*g
\right]
+
\sum_{b=0,L}
\lambda_b f^*(y_b)g(y_b),
\label{eq:q9-quadratic-form-Q}
\\[1mm]
\mathcal N[f,g]
={}&
\int_0^Ldy\,
f^*g
+
\sum_{b=0,L}
r_b f^*(y_b)g(y_b).
\label{eq:q9-quadratic-form-N}
\end{align}
For real $M^2$, $\lambda_b$, and $r_b$, both forms are Hermitian.

Multiplying Eq.~\eqref{eq:q9-bulk-mode-equation} by $g^*$,
integrating by parts, and using the boundary conditions
\eqref{eq:q9-general-boundary-condition}, one obtains
\begin{equation}
    \mathcal Q[g,f_n]
    =
    m_n^2\,
    \mathcal N[g,f_n]
\label{eq:q9-generalized-eigenproblem}
\end{equation}
for every admissible test profile $g$. This is not an additional
assumption: it is simply the bulk equation and the boundary
conditions written in weak form.

If $\mathcal N$ is positive definite, setting $g=f_n$ gives the
Rayleigh quotient
\begin{equation}
    m_n^2
    =
    \frac{
        \mathcal Q[f_n,f_n]
    }{
        \mathcal N[f_n,f_n]
    }.
\label{eq:q9-Rayleigh-quotient}
\end{equation}
Both numerator and denominator are real, so the eigenvalues are real.
Hermiticity also implies that two modes with distinct eigenvalues are
orthogonal with respect to the kinetic form,
\begin{equation}
    \mathcal N[f_m,f_n]=0,
    \qquad
    m_m^2\neq m_n^2.
\label{eq:q9-generalized-orthogonality}
\end{equation}

This is the important point: localized kinetic terms do not merely
perturb the potential of an ordinary Sturm--Liouville problem. They
change the kinetic metric with respect to which the problem is
Hermitian and the modes are normalized.

For $r_b\geq0$, the localized pieces make non-negative contributions
to this norm. Negative coefficients require more care. If sufficiently
large in magnitude, they can make the kinetic form indefinite and
generate zero- or negative-norm states, signalling that the effective
description is not physically acceptable without additional
structure.

In distributional notation the same problem is often summarized as
\begin{equation}
\left[
-\partial_y^2
+
M^2
+
\sum_b\lambda_b\delta_b
\right]
f_n
=
m_n^2
\left[
1+
\sum_b r_b\delta_b
\right]
f_n,
\label{eq:q9-generalized-Sturm-Liouville}
\end{equation}
where the endpoint distributions are normalized so as to reproduce
Eqs.~\eqref{eq:q9-quadratic-form-Q} and
\eqref{eq:q9-quadratic-form-N}.

\narrativeheading{Attraction, repulsion, and normalization}

The Rayleigh quotient gives a precise version of the intuitive
statement that a localized interaction can push a wave function away
from a boundary.

For $\lambda_0>0$, a profile with a large value of $|f(0)|$ pays the
positive contribution $\lambda_0|f(0)|^2$ in the numerator of
Eq.~\eqref{eq:q9-Rayleigh-quotient}. Low-lying modes therefore tend to
reduce their support at $y=0$. In the strong-coupling limit
$\lambda_0\to+\infty$, the Robin condition approaches $f_n(0)=0$ and
the endpoint becomes effectively Dirichlet.

A negative localized mass acts in the opposite direction. It can
enhance a wave function near the boundary and can generate a
boundary-localized light state. If the attraction is sufficiently
strong, the lowest value of $m_n^2$ can become negative, signalling
an instability of the assumed background.

A positive localized kinetic term has a different interpretation.
Since $r_b$ appears in $\mathcal N$, it changes the normalization of
the mode rather than adding a static potential. It also enters the
boundary condition through $r_bm_n^2$, so its influence depends on
the eigenvalue.

For a massive mode, a large positive $r_bm_n^2$ can substantially
distort the profile and suppress its canonically normalized overlap
with the boundary. A zero mode behaves differently. If
$M=\lambda_0=\lambda_L=0$, the constant profile remains a solution,
\begin{equation}
    f_0(y)
    =
    \frac{1}{
        \sqrt{L+r_0+r_L}
    },
\label{eq:q9-zero-mode-profile}
\end{equation}
provided the norm is positive. Its shape remains flat: the localized
kinetic terms have changed its normalization, and hence its effective
couplings, rather than repelling it spatially.

Indeed, if a mode couples to an operator localized at $y_b$, its
effective coupling is proportional to its normalized boundary value,
\begin{equation}
    g_{n,b}^{\rm eff}
    \propto
    f_n(y_b).
\label{eq:q9-boundary-coupling}
\end{equation}
Boundary kinetic terms can consequently affect physical couplings
even when the light-mode profile itself looks almost unchanged.

\narrativeheading{Matching conditions on the covering space}

The same localized action can be described on the covering space by
delta-function sources. For a second-order scalar equation,
integration across a fixed point gives a jump in the first derivative.
Denoting by hatted quantities the coefficients in a chosen
covering-space delta-function convention,
\begin{equation}
    f_n'(0^+)-f_n'(0^-)
    =
    \left(
        \widehat\lambda_0
        -
        \widehat r_0m_n^2
    \right)
    f_n(0).
\label{eq:q9-scalar-jump}
\end{equation}
The relation between hatted and interval coefficients depends on the
normalization adopted for the orbifold delta distribution.

For an even profile, $f_n'(0^-)=-f_n'(0^+)$, and the jump condition
reduces to the interval Robin condition with the corresponding
normalization. Thus the interval boundary condition and the
covering-space jump condition are two descriptions of the same
variational problem.

The distinction becomes even more important for first-order systems.
There the field itself can have a finite jump. Localized fermion
bilinears are consequently more naturally described by matching
matrices, or equivalently by rotated allowed boundary subspaces, than
by the usual scalar terminology of Neumann and Dirichlet conditions.

\narrativeheading{Boundary terms are generated by renormalization}

Localized operators should not be regarded as optional decorations of
an orbifold effective theory. A fixed point breaks translation
invariance in the compact direction. Short-distance singularities can approach the fixed locus and generate ultraviolet
divergences supported there.

Bulk interactions may consequently renormalize boundary masses,
boundary kinetic terms, gauge kinetic terms, superpotentials, and
other localized operators allowed by the symmetries preserved at the
fixed point. An effective theory must contain the corresponding
counterterm basis. Setting one of these coefficients to zero is
generically a renormalization condition at a chosen scale, rather
than a prediction of the orbifold geometry. Particular coefficients
may of course remain zero when protected by an additional symmetry or
when the relevant contributions cancel.

This local renormalization must be distinguished from finite
dependence on the complete compactification. Schematically,
\begin{equation}
    \Gamma_{\rm 1\text{-}loop}
    =
    \Gamma_{\rm bulk}^{\rm local}
    +
    \Gamma_0^{\rm local}
    +
    \Gamma_L^{\rm local}
    +
    \Gamma_{\rm nonlocal}(L).
\label{eq:q9-local-nonlocal-decomposition}
\end{equation}
The first three terms contain local bulk and boundary counterterms,
together with scheme-dependent local finite pieces. The last term
requires propagation over a non-zero distance and can depend on the
relative boundary data, the compactification length, and a
Scherk--Schwarz holonomy.

We are not yet attempting a complete calculation of the
Scherk--Schwarz one-loop potential. The purpose of the next two
subsections is instead to make this local/non-local distinction
physically transparent and to identify how boundary data and bulk
transport enter the non-local part.

\newpage
\narrativeheading{The first Casimir interpretation}

Equation~\eqref{eq:q9-local-nonlocal-decomposition} is the first place
where the analogy with the Casimir effect becomes particularly useful.

A loop that explores only an arbitrarily small neighbourhood of one
boundary is local. It can renormalize operators already present there,
such as a boundary mass or kinetic term. These contributions are
ultraviolet sensitive, do not require the second boundary, and do not
know its distance.

A qualitatively different contribution is obtained when the field
propagates from one boundary to the other and back. Such a process
cannot be assigned to either endpoint independently. It knows the
distance $L$, the reflection law at both boundaries, and, when the
field has internal indices, any transformation accumulated while
propagating between them. This is the genuinely Casimir-like piece.

Consider first one real scalar with no internal indices. Wick rotate
the non-compact momentum to Euclidean space and define
\begin{equation}
    \kappa
    =
    \sqrt{k_E^2+M^2}.
\label{eq:q9-kappa}
\end{equation}
At fixed $k_E$, propagation through a distance $L$ gives a factor
$e^{-\kappa L}$; a complete trip from $0$ to $L$ and back therefore
carries $e^{-2\kappa L}$.

Let $\mathscr R_0(\kappa)$ and $\mathscr R_L(\kappa)$ denote the
corresponding Euclidean reflection amplitudes. A localized mass
changes the Robin coefficient and hence changes $\mathscr R_b$. A
localized kinetic term makes the boundary condition momentum
dependent and therefore generally makes $\mathscr R_b$ momentum
dependent as well.

One complete round trip consequently carries
\begin{equation}
    \mathscr R_0(\kappa)
    \mathscr R_L(\kappa)
    e^{-2\kappa L}.
\label{eq:q9-one-round-trip}
\end{equation}
Repeated reflections generate paths that cross the interval two,
four, six, and progressively more times. The functional determinant
sums all of them.

For one real bosonic degree of freedom with Euclidean quadratic
operator $\Delta$,
\begin{equation}
    \Gamma_{\rm 1\text{-}loop}
    =
    \frac12\Tr\log\Delta.
\end{equation}
Dividing by the volume
$\mathcal V_d=\int d^dx_E$ of the $d$ non-compact Euclidean
dimensions turns the corresponding part of the trace into
\[
    \frac1{\mathcal V_d}
    \Tr_{\mathbb R^d}
    \longrightarrow
    \int
    \frac{d^dk_E}{(2\pi)^d}.
\]

After the pieces that are local in the bulk or at either boundary
have been separated and renormalized, the part involving both
boundaries has the structural form
\begin{equation}
V_{\rm nonlocal}(L)
=
\frac12
\int
\frac{d^dk_E}{(2\pi)^d}
\log
\left[
1
-
\mathscr R_0(\kappa)
\mathscr R_L(\kappa)
e^{-2\kappa L}
\right].
\label{eq:q9-Casimir-scalar}
\end{equation}
The precise reflection coefficients, and possible extra isolated
contributions such as boundary bound states, depend on the spectral
problem under consideration. Equation~\eqref{eq:q9-Casimir-scalar}
is meant to display the universal round-trip structure.

Its dimensional content is already informative. The argument of the
logarithm is dimensionless, so
$[V_{\rm nonlocal}]=(\mathrm{mass})^d$. If $L$ is the only physical
length, defining $q_E=Lk_E$ makes the scaling explicit. For a
massless field,
\begin{equation}
V_{\rm nonlocal}(L)
=
\frac{1}{2L^d}
\int
\frac{d^dq_E}{(2\pi)^d}
\log
\left[
1
-
\mathscr R_0(q_E)
\mathscr R_L(q_E)
e^{-2|q_E|}
\right],
\end{equation}
with any further dimensionless dependence contained in the
reflection coefficients. Thus the scaling is $V_{\rm nonlocal}(L)\propto L^{-d}$.

With further scales present,
\begin{equation}
V_{\rm nonlocal}(L)
=
\frac1{L^d}
F\!\left(
ML,
\lambda_0L,
\lambda_LL,
\frac{r_0}{L},
\frac{r_L}{L},
\ldots
\right),
\end{equation}
where $F$ is dimensionless.

The logarithm gives an equally useful physical interpretation. With
\[
X(\kappa)
=
\mathscr R_0(\kappa)
\mathscr R_L(\kappa)
e^{-2\kappa L},
\]
one has
\begin{equation}
    \log(1-X)
    =
    -\sum_{\ell=1}^{\infty}
    \frac{X^\ell}{\ell},
\end{equation}
where
\[
X^\ell
=
\left[
\mathscr R_0(\kappa)
\mathscr R_L(\kappa)
\right]^\ell
e^{-2\ell\kappa L}.
\]
The integer $\ell$ counts complete round trips, while the factor
$1/\ell$ is the familiar combinatorial weight generated by the
logarithm of the determinant.

This expansion also makes the ultraviolet behaviour transparent.
Every contribution that probes both boundaries contains at least
$e^{-2\kappa L}$ and is therefore exponentially suppressed when
$\kappa\to\infty$. High-momentum fluctuations can renormalize the
bulk or an individual boundary locally, but they cannot probe two
separated boundaries without paying the finite propagation distance.
Once the allowed local counterterms have been included, the genuinely
non-local contribution is ultraviolet finite.

\narrativeheading{Internal indices and Scherk--Schwarz transport}

For a multiplet, reflections need not act diagonally on the internal
components. The reflection amplitudes become matrices
$\mathscr R_0(\kappa)$ and $\mathscr R_L(\kappa)$, and propagation
between the endpoints may itself rotate the internal state.

Let
\begin{equation}
\mathcal U(L,0)
=
\mathcal P
\exp
\left[
-\int_0^Ldy\,\mathcal A_y
\right]
\label{eq:q9-internal-transport}
\end{equation}
denote parallel transport in the relevant internal space. For a flat
Scherk--Schwarz connection, $\mathcal A_y$ has vanishing local
curvature but can nevertheless produce non-trivial transport between
the endpoints.

Combining ordinary Euclidean propagation with this internal
transport gives
\begin{align}
    \mathcal T_{0L}(\kappa)
    &=
    e^{-\kappa L}\,
    \mathcal U(L,0),
\label{eq:q9-T0L}
\\
    \mathcal T_{L0}(\kappa)
    &=
    e^{-\kappa L}\,
    \mathcal U(0,L).
\label{eq:q9-TL0}
\end{align}
For ordinary parallel transport by a non-singular connection,
$\mathcal U(0,L)=\mathcal U(L,0)^{-1}$.

Choose as reference state a wave immediately after reflection at
$y=0$, propagating toward $L$. One complete round trip is then
\[
0^+
\ \xrightarrow{\ \mathcal T_{0L}\ }\ 
L^-
\ \xrightarrow{\ \mathscr R_L\ }\ 
L^-
\ \xrightarrow{\ \mathcal T_{L0}\ }\ 
0^+
\ \xrightarrow{\ \mathscr R_0\ }\ 
0^+ .
\]
The last reflection is essential: arrival back at $y=0$ has reversed
the propagation direction, so the state has not yet returned to the
chosen reference state.

The corresponding round-trip operator is
\begin{equation}
    \mathcal T_{\rm rt}(\kappa)
    =
    \mathscr R_0(\kappa)
    \mathcal T_{L0}(\kappa)
    \mathscr R_L(\kappa)
    \mathcal T_{0L}(\kappa).
\label{eq:q9-round-trip-operator}
\end{equation}
The order of these matrices matters. In particular,
$\mathcal U(0,L)=\mathcal U(L,0)^{-1}$ does not make the
Scherk--Schwarz information disappear because the two transport
operators are separated by reflections that need not commute with
them.

The scalar expression \eqref{eq:q9-Casimir-scalar} is accordingly
replaced by the structural matrix formula
\begin{equation}
V_{\rm nonlocal}(L)
=
\frac{\eta_\Phi}{2}
\int
\frac{d^dk_E}{(2\pi)^d}
\log\det
\left[
\mathbf1
-
\mathcal T_{\rm rt}(\kappa)
\right].
\label{eq:q9-Casimir-scattering}
\end{equation}
Here the determinant acts on the internal space. The overall sign,
multiplicity, and precise power of the determinant depend on the
statistics, reality properties, spin and gauge fixing of the field;
$\eta_\Phi$ is only a compact reminder of this bookkeeping.

The physical organization is nevertheless universal:
\[
\begin{array}{ccl}
\text{localized boundary dynamics}
&\longrightarrow&
\text{reflection matrices},
\\[1mm]
\text{bulk propagation and Scherk--Schwarz data}
&\longrightarrow&
\text{transport matrices},
\\[1mm]
\text{complete round trips}
&\longrightarrow&
\text{non-local one-loop contribution}.
\end{array}
\]

A localized mass modifies the reflection law. A localized kinetic term
makes it momentum dependent. A Scherk--Schwarz deformation enters the
ordered transport between the endpoints. The non-local determinant
depends on all of these data together.

\narrativeheading{Where the Casimir analogy stops}

The analogy is useful precisely because it separates local from
non-local physics; it should not be pushed beyond that point.

An orbifold fixed point is not a material plate. Localized masses,
kinetic terms, superpotentials and other couplings are parameters of
the effective field theory. Their coefficients are renormalized by
local ultraviolet physics and are not predicted by a Casimir
calculation.

The Casimir-like quantity is instead the piece that cannot be
assigned to either endpoint separately because it requires
propagation across a non-zero distance and back. Local boundary terms
enter that calculation as input through their reflection matrices.

This distinction remains true when boundaries are idealized as
infinitely thin defects. The idealization can itself produce surface
divergences, but those divergences are localized and are absorbed by
local counterterms. They do not contradict the ultraviolet finiteness
of the part that genuinely probes the finite separation $L$.

The same logic will reappear later after Poisson resummation of
Kaluza--Klein sums. The zero-winding sector can contain contributions
local in the higher-dimensional theory, whereas non-zero winding
sectors necessarily probe the compact geometry globally. The
scattering and winding descriptions are two ways of recognizing the
same separation.

\narrativeheading{Localized terms for a fermionic doublet}

We now turn from second-order fields to first-order fermionic systems.
Consider a two-component fermion
\[
\Psi
=
\begin{pmatrix}
\psi_1\\
\psi_2
\end{pmatrix},
\]
with components of opposite orbifold parity.

Without localized interactions, the orbifold projection selects a
one-dimensional subspace of this doublet at each endpoint. With
localized interactions, it is useful to formulate the resulting
conditions in terms of one-sided limits from within the physical
interval:
\begin{align}
    \psi_2(0^+)
    &=
    \beta_0(m_n)\,
    \psi_1(0^+),
\label{eq:q9-fermion-bc-zero}
\\
    \psi_2(L^-)
    &=
    -\beta_L(m_n)\,
    \psi_1(L^-).
\label{eq:q9-fermion-bc-L}
\end{align}
The relative sign at the second endpoint depends on the orientation
and parity conventions. The values on the opposite sides of the
fixed points are related by orbifold equivariance and are not
independent boundary data.

For a pure localized mass, the matching coefficient is independent of
the KK eigenvalue, $\beta_b(m_n)=M_b$. The first endpoint then selects
the line proportional to $(1,M_0)^T$, while the second selects, with
the convention above, the line proportional to $(1,-M_L)^T$.

Introducing the angle $\Theta_b=\arctan M_b$ makes the geometrical
meaning transparent: a pure localized mass rotates the
one-dimensional fermionic subspace allowed at the boundary by a fixed
angle.

A localized kinetic term changes this conclusion. Since it contains
derivatives tangent to the boundary, its contribution to the matching
condition depends on the lower-dimensional momentum and hence, on
shell, on the eigenvalue. Schematically,
\begin{equation}
    \beta_b(m_n)
    =
    M_b
    +
    r_b^{(\psi)}m_n
    +
    \cdots ,
\label{eq:q9-energy-dependent-fermion-matching}
\end{equation}
where the precise dependence depends on the fermion system and the
localized operator.

Different KK levels can see different allowed boundary
directions. This simple observation will be crucial below: a fixed
boundary rotation may be equivalent to Scherk--Schwarz data, whereas
an eigenvalue-dependent one generally is not.

\narrativeheading{Localized gravitino masses and the supersymmetry parameter}

We now specialize again to minimal five-dimensional supergravity
compactified on $S^1/\mathbb Z_2$. The four-dimensional gravitini
form an $SU(2)_R$ doublet $\psi_{\mu I}$, $I=1,2$, and the
supersymmetry parameter carries the corresponding doublet index.

At $y=0$, take $\psi_{\mu1}$ to be even and $\psi_{\mu2}$ odd. At
$y=L=\pi R$, introduce
\begin{align}
    \psi_{\mu+}
    &=
    \cos(\pi\omega)\,\psi_{\mu1}
    -
    \sin(\pi\omega)\,\psi_{\mu2},
\nonumber\\
    \psi_{\mu-}
    &=
    \sin(\pi\omega)\,\psi_{\mu1}
    +
    \cos(\pi\omega)\,\psi_{\mu2},
\label{eq:q9-gravitino-plus-minus}
\end{align}
and analogously
\begin{align}
    \epsilon_+
    &=
    \cos(\pi\omega)\,\epsilon_1
    -
    \sin(\pi\omega)\,\epsilon_2,
\nonumber\\
    \epsilon_-
    &=
    \sin(\pi\omega)\,\epsilon_1
    +
    \cos(\pi\omega)\,\epsilon_2.
\label{eq:q9-epsilon-plus-minus}
\end{align}

Without localized interactions, $\psi_{\mu1}$ and $\psi_{\mu+}$ are
the even gravitino components at the two fixed points. A localized
gravitino bilinear changes the matching of the odd component. In the
normalization used here,
\begin{align}
    \psi_{\mu2}(0^+)
    &=
    M_0\,\psi_{\mu1}(0^+),
\nonumber\\
    \psi_{\mu-}(L^-)
    &=
    -\,M_\pi\,\psi_{\mu+}(L^-).
\label{eq:q9-gravitino-boundary-relations}
\end{align}

The corresponding restrictions on the supersymmetry parameter are
not independent assumptions. Local supersymmetry relates
$\psi_{\mu I}$ to $\epsilon_I$, schematically through
\begin{equation}
    \delta\psi_{\mu I}
    =
    2D_\mu\epsilon_I+\cdots .
\label{eq:q9-gravitino-susy-variation}
\end{equation}
For a supersymmetric boundary coupling, the space of allowed gravitino
boundary values must be stable under the allowed supersymmetry
transformations. The same matching therefore appears for the
parameter,
\begin{align}
    \epsilon_2(0^+)
    &=
    M_0\,\epsilon_1(0^+),
\nonumber\\
    \epsilon_-(L^-)
    &=
    -\,M_\pi\,\epsilon_+(L^-).
\label{eq:q9-susy-boundary-relations}
\end{align}
These conditions also ensure the cancellation of the distributional
terms generated when derivatives act on discontinuous odd
supersymmetry parameters.

The two localized masses define the angles
$\Theta_0=\arctan M_0$ and $\Theta_\pi=\arctan M_\pi$. Thus the
spin-$\tfrac32$ boundary problem is geometrically a problem of
comparing two one-dimensional directions in the $SU(2)_R$ doublet
space.

These directions describe only the restrictions coming from the
gravitino sector. If localized matter fermions are present, their
supersymmetry variations can impose additional conditions.

\narrativeheading{What the lower-dimensional theory sees}

The physical four-dimensional gravitini are obtained by solving the
complete compact-direction eigenvalue problem. Write
\begin{equation}
    \psi_{\mu I}(x,y)
    =
    \sum_n
    f_{nI}(y)\,
    \psi_{\mu,n}(x).
\label{eq:q9-gravitino-KK-expansion}
\end{equation}
The two-component object $f_{nI}(y)$ is the internal profile, while
$\psi_{\mu,n}(x)$ is the corresponding four-dimensional gravitino
field.

A massless gravitino exists only when the global spectral problem
admits a non-zero solution with $m_n=0$ satisfying the bulk equations
and both endpoint matching conditions.

If localized kinetic terms are present, its normalization is also
defined by the kinetic metric obtained from the reduced action,
schematically
\begin{equation}
    \int_0^Ldy\,
    f_m^\dagger(y)\,
    W(y)\,
    f_n(y)
    =
    \delta_{mn},
\label{eq:q9-gravitino-mode-normalization}
\end{equation}
where $W$ includes any localized kinetic contributions.

A profile whose direction in $SU(2)_R$ space changes along the
interval can therefore represent one perfectly ordinary
four-dimensional gravitino. There is no requirement that its separate
higher-dimensional components be constant or have independent
four-dimensional interpretations.

For the flat locally supersymmetric system considered below, the
massless-gravitino equation and the gravitino Killing-spinor equation
are closely related. The existence of a zero-mode gravitino is an important test of unbroken supersymmetry, although, as
we shall see, it is not the complete test when localized matter
fermions also have non-zero shifts.

\narrativeheading{The zero-mode condition in twisted variables}

First use the description in which the Scherk--Schwarz datum resides
in the twisted identification and in the relative endpoint actions.
There is then no explicit flat connection in the open interval.

For a candidate massless gravitino, the bulk profile is constant,
\[
    f_0(y)
    =
    \begin{pmatrix}
        v_1\\
        v_2
    \end{pmatrix}.
\]
The first boundary condition gives
\begin{equation}
    v_2=M_0v_1.
\label{eq:q9-zero-mode-first-boundary}
\end{equation}

At the second endpoint, define
\[
v_+
=
\cos(\pi\omega)v_1
-
\sin(\pi\omega)v_2,
\qquad
v_-
=
\sin(\pi\omega)v_1
+
\cos(\pi\omega)v_2.
\]
The matching condition requires
\begin{equation}
    v_-=-M_\pi v_+.
\label{eq:q9-zero-mode-second-boundary}
\end{equation}

Substituting Eq.~\eqref{eq:q9-zero-mode-first-boundary} gives
\begin{equation}
\left[
\left(
1-M_0M_\pi
\right)\sin(\pi\omega)
+
\left(
M_0+M_\pi
\right)\cos(\pi\omega)
\right]
v_1
=
0.
\label{eq:q9-zero-mode-algebraic-condition}
\end{equation}
For a non-zero profile this is equivalently
\[
\sin\!\left(
\pi\omega+\Theta_0+\Theta_\pi
\right)=0,
\]
or
\begin{equation}
    \pi\omega+\Theta_0+\Theta_\pi
    =
    n\pi,
    \qquad
    n\in\mathbb Z.
\label{eq:q9-zero-mode-angle-condition}
\end{equation}

The half-twist gives a particularly transparent example. For
$\omega=1/2$ and $M_0=M_\pi=0$, the first endpoint selects
$\operatorname{span}(1,0)^T$, whereas the second selects the
orthogonal line $\operatorname{span}(0,1)^T$. No non-zero constant
profile belongs to both, so no massless four-dimensional gravitino
exists.

Localized masses can compensate the misalignment. At the half-twist,
a zero mode is restored when
\[
\Theta_0+\Theta_\pi
=
-\frac{\pi}{2}
\pmod\pi.
\]

\narrativeheading{The same compactification in periodic variables}

The same physics can be described with periodic fields and a flat bulk
connection. Let
\[
R[\alpha]
=
\begin{pmatrix}
\cos\alpha&\sin\alpha\\
-\sin\alpha&\cos\alpha
\end{pmatrix}
\]
and write
\begin{equation}
\begin{pmatrix}
\psi_{\mu1}\\
\psi_{\mu2}
\end{pmatrix}
=
R[\alpha(y)]
\begin{pmatrix}
\widetilde\psi_{\mu1}\\
\widetilde\psi_{\mu2}
\end{pmatrix},
\qquad
\alpha(y+2\pi R)
=
\alpha(y)+2\pi\omega.
\label{eq:q9-periodic-field-rotation}
\end{equation}
The periodic fields acquire the connection
\begin{equation}
\partial_y\psi_\mu
=
R[\alpha(y)]
\left[
\partial_y
+
\alpha'(y)J
\right]
\widetilde\psi_\mu,
\qquad
J=
\begin{pmatrix}
0&1\\
-1&0
\end{pmatrix}.
\label{eq:q9-periodic-bulk-connection}
\end{equation}

The supersymmetry parameter, endpoint projectors and localized action
must all be transformed by the same change of variables.

The pure half-twist makes the equivalence especially easy to see. In
the twisted-field description,
\[
\mathcal A_y=0,
\qquad
\mathcal L_0=\operatorname{span}(Q_1),
\qquad
\mathcal L_\pi=\operatorname{span}(Q_2),
\]
so a constant zero-mode profile cannot satisfy both endpoint
conditions.

In periodic variables the endpoint lines may instead be aligned,
\[
\widetilde{\mathcal L}_0
=
\widetilde{\mathcal L}_\pi
=
\operatorname{span}(Q_1),
\]
but the flat connection now transports the initial direction according
to $\mathcal U(L,0)Q_1\propto Q_2$. The transported zero-mode
candidate again fails the second endpoint condition.

The two descriptions therefore agree. A spurious zero mode appears
only if one combines the endpoint line from the twisted description
with the non-trivial transport from the periodic description. That
would insert the same Scherk--Schwarz datum twice. Endpoint data and
bulk transport must always belong to one common trivialization.

\narrativeheading{The Killing-spinor equations give the same condition}

The equivalence can also be seen directly from the Killing-spinor
equations. In periodic variables let $\alpha(y)$ denote the rotation
angle introduced above. If it contains localized jumps, one may
understand it as the thin-boundary limit of a smooth regularization.

For the flat background and the conventions used here, the relevant
equations take the form
\begin{align}
    \partial_y\widetilde\epsilon_1
    +
    \alpha'(y)\widetilde\epsilon_2
    &=
    0,
\nonumber\\
    \partial_y\widetilde\epsilon_2
    -
    \alpha'(y)\widetilde\epsilon_1
    &=
    2M_0\widetilde\epsilon_1\delta(y)
    +
    2M_\pi\widetilde\epsilon_1\delta(y-L).
\label{eq:q9-periodic-killing-equations}
\end{align}
Integration across the two fixed points reproduces
\begin{equation}
    \widetilde\epsilon_2(0^+)
    =
    M_0\widetilde\epsilon_1(0),
    \qquad
    \widetilde\epsilon_2(L^-)
    =
    -M_\pi\widetilde\epsilon_1(L).
\label{eq:q9-periodic-killing-boundaries}
\end{equation}
Solving between the boundaries then gives once again
\begin{equation}
    \pi\omega+\Theta_0+\Theta_\pi
    =
    n\pi.
\label{eq:q9-killing-angle-condition}
\end{equation}

This is an important consistency check, but at this stage it concerns
only the gravitino/Killing-spinor sector. Variations of additional
localized fermions have not yet been imposed.

\narrativeheading{The gravitino spectrum gives the same answer}

The same condition follows directly from the spin-$\tfrac32$
spectrum. Solving the first-order bulk equations for a signed
four-dimensional mass $m$ and imposing the endpoint matching
conditions gives
\begin{equation}
m_n^{\rm signed}
=
\frac1R
\left[
n+\omega
+
\frac{\Theta_0+\Theta_\pi}{\pi}
\right],
\qquad
n\in\mathbb Z.
\label{eq:q9-combined-spectrum}
\end{equation}
The physical masses are simply
$m_n^{\rm phys}=|m_n^{\rm signed}|$.

A zero appears precisely when
\[
n+\omega
+
\frac{\Theta_0+\Theta_\pi}{\pi}
=
0
\]
for some $n\in\mathbb Z$, which is equivalent to
Eq.~\eqref{eq:q9-killing-angle-condition}.
For this restricted class of energy-independent boundary rotations,
one may therefore define
\begin{equation}
    \omega_{\rm eff}
    =
    \omega
    +
    \frac{\Theta_0+\Theta_\pi}{\pi}
    \pmod1.
\label{eq:q9-effective-twist}
\end{equation}
The spin-$\tfrac32$ tower remains equally spaced and depends only on
$\omega_{\rm eff}$. The agreement between the spectral zero and the Killing-spinor
condition is a consequence of the locally supersymmetric flat
bulk-plus-boundary gravitino system. It should not yet be promoted to
a statement about every fermion in a more general boundary sector.

\narrativeheading{The invariant geometric statement}

The calculation can now be summarized without referring to a
particular field basis.

In one fixed trivialization, let
$\mathcal B_0^{(3/2)}$ and $\mathcal B_\pi^{(3/2)}$ denote the
one-dimensional subspaces of allowed gravitino-doublet boundary
values at $y=0$ and $y=L$. Let $\mathcal U(L,0)$ denote the transport
operator obtained from the bulk zero-mode equation, or equivalently
the parallel transport generated by the connection $\mathcal A_y$ in
a trivialization where that connection is explicit.

A gravitino zero mode exists precisely when
\begin{equation}
    \mathcal U(L,0)
    \mathcal B_0^{(3/2)}
    \cap
    \mathcal B_\pi^{(3/2)}
    \neq
    \{0\}.
\label{eq:q9-geometric-zero-mode}
\end{equation}
This merely states that an allowed boundary value at one end can be
transported by the bulk equation into an allowed boundary value at
the other.

Under a change of basis $\psi'(y)=G(y)^{-1}\psi(y)$, the three
ingredients transform together,
\begin{align}
    \mathcal B_0^{\prime\,(3/2)}
    &=
    G(0)^{-1}\mathcal B_0^{(3/2)},
\nonumber\\
    \mathcal B_\pi^{\prime\,(3/2)}
    &=
    G(L)^{-1}\mathcal B_\pi^{(3/2)},
\nonumber\\
    \mathcal U'(L,0)
    &=
    G(L)^{-1}
    \mathcal U(L,0)
    G(0).
\label{eq:q9-trivialization-transform}
\end{align}
The dimension of the intersection in
Eq.~\eqref{eq:q9-geometric-zero-mode} is therefore invariant.

This is the cleanest way of stating the double-counting warning:
changing trivialization moves information between transport and
endpoint data, but cannot create or remove a zero mode.

\narrativeheading{Localized kinetic terms do not define one universal twist}

The simple quantity $\omega_{\rm eff}$ exists because the two
boundary rotations above are independent of the KK eigenvalue.

With a localized fermion kinetic term, the matching coefficient
becomes a function $\beta_b(m)$ and the associated angle
$\Theta_b(m)=\arctan\beta_b(m)$ depends on the state being quantized.
The spectral condition then has the implicit form
\begin{equation}
    m_nR
    =
    n+\omega
    +
    \frac{
        \Theta_0(m_n)+\Theta_\pi(m_n)
    }{\pi},
\label{eq:q9-energy-dependent-quantization}
\end{equation}
up to the conventions used for the signed tower.

The eigenvalue now appears on both sides. Different KK states
generically see different boundary directions, their spacing need not
remain uniform, and no single number $\omega_{\rm eff}$ can encode
the whole spectrum.

A field redefinition can redistribute an energy-independent symmetry
rotation between the bulk and the endpoints. It cannot eliminate the
eigenvalue-dependent dynamics of a genuine localized kinetic term.

\narrativeheading{What does the gravitino zero-mode condition test?}

The condition derived above refers to the gravitino boundary-value
problem defined by the bulk equations and by the localized gravitino
bilinears included in that problem. For the flat locally
supersymmetric system considered here, a zero of the spin-$\tfrac32$
spectrum coincides with a solution of the corresponding gravitino
Killing-spinor equations.

This is not yet a statement about a more general localized sector.
If additional boundary fermions are present, their supersymmetry
variations must be imposed as well. In particular, a localized
fermion may have an inhomogeneous transformation of the form
\begin{equation}
    \delta\chi_b^{\,i}
    =
    F_b^{\,i}\epsilon_b+\cdots .
\label{eq:q9-boundary-goldstino-shift}
\end{equation}
A supersymmetry parameter preserved by the complete configuration must
satisfy these conditions together with the bulk and gravitino
Killing-spinor equations.

Thus Eq.~\eqref{eq:q9-zero-mode-angle-condition} is complete for the
gravitino system from which it was derived. Adding an independent
localized breaking sector enlarges the fermionic problem.

\narrativeheading{When is a localized gravitino mass just a
Scherk--Schwarz rotation?}

The spectrum
Eq.~\eqref{eq:q9-combined-spectrum} shows that an
energy-independent gravitino boundary mass enters through the angles
$\Theta_0$ and $\Theta_\pi$ in exactly the same way as a shift of the
Scherk--Schwarz parameter. This is not accidental.

A $y$-dependent $SU(2)_R$ change of variables can redistribute a
fixed total monodromy between a smooth bulk connection and localized
rotations at the two endpoints. If the corresponding endpoint angles
are $\Omega_0$ and $\Omega_\pi$, one may write schematically
\begin{equation}
    \pi\omega
    =
    \pi\omega_B
    +
    \Omega_0
    +
    \Omega_\pi .
\label{eq:q9-generalized-SS-splitting}
\end{equation}
The bulk and boundary descriptions then differ only by the
trivialization used to represent the same global Scherk--Schwarz
datum.

This equivalence requires more than agreement of the gravitino mass
spectrum: the boundary operation must represent the same allowed
symmetry transformation on the complete system.

A genuine localized supersymmetry-breaking sector can instead contain
additional fermions with independent shifts and hence additional
would-be Goldstino directions. This information is not encoded in the
shifted spin-$\tfrac32$ spectrum.

\narrativeheading{Discontinuous representatives and one-sided limits}

The gravitino matching condition derived above provides a simple
example of how an orbifold field can become discontinuous at a fixed
point. Consider $y=0$, with $\psi_{\mu1}$ even and $\psi_{\mu2}$ odd
under $y\to-y$.

For a four-dimensional gravitino eigenstate of signed mass $m$, the
bulk equations are first order in the compact coordinate. In the
presence of a localized gravitino mass, one of them has, in the
conventions used here, the schematic form
\begin{equation}
    \partial_y\psi_{\mu2}
    +
    m\,\psi_{\mu1}
    =
    2M_0\,\delta(y)\,\psi_{\mu1},
\label{eq:q9-localized-gravitino-eom}
\end{equation}
while the companion equation contains no delta-function source at
$y=0$. The precise normalization of $M_0$ depends on the convention
for the orbifold delta function; the argument below is independent of
this choice once the convention is fixed.

Integrating Eq.~\eqref{eq:q9-localized-gravitino-eom} across an
infinitesimal neighbourhood of the fixed point gives
\begin{align}
    \psi_{\mu2}(0^+)-\psi_{\mu2}(0^-)
    &+
    m\int_{-\varepsilon}^{+\varepsilon}
    dy\,\psi_{\mu1}(y)
    \nonumber\\
    &=
    2M_0\,\psi_{\mu1}(0).
\end{align}
The second term vanishes as $\varepsilon\to0$ because it is the
integral of a regular bulk contribution. The localized source
therefore fixes the finite jump,
\begin{equation}
    \psi_{\mu2}(0^+)-\psi_{\mu2}(0^-)
    =
    2M_0\,\psi_{\mu1}(0).
\label{eq:q9-gravitino-jump}
\end{equation}

Orbifold equivariance relates the two one-sided limits. Since
$\psi_{\mu1}$ is even and $\psi_{\mu2}$ odd,
\begin{equation}
    \psi_{\mu1}(0^-)=\psi_{\mu1}(0^+),
    \qquad
    \psi_{\mu2}(0^-)=-\psi_{\mu2}(0^+).
\label{eq:q9-discontinuous-parity}
\end{equation}
The even component is continuous, so we may identify
$\psi_{\mu1}(0)$ with either one-sided limit. Substituting the parity
relations into Eq.~\eqref{eq:q9-gravitino-jump} then gives
\begin{equation}
    \psi_{\mu2}(0^+)
    =
    M_0\,\psi_{\mu1}(0^+),
\label{eq:q9-discontinuous-example-plus}
\end{equation}
which is precisely the matching condition used above.

This calculation also explains an apparent paradox of orbifold
language. An odd field is often said to vanish at a fixed point. That
statement assumes continuity. Here odd parity instead requires
\[
    \psi_{\mu2}(0^-)
    =
    -\psi_{\mu2}(0^+),
\]
and the two limits need not vanish separately. The localized
first-order interaction has replaced continuity by the matching
condition~\eqref{eq:q9-discontinuous-example-plus}.

The discontinuity itself is representation dependent. To see this,
collect the two components into
\[
    \Psi_\mu
    =
    \begin{pmatrix}
        \psi_{\mu1}\\
        \psi_{\mu2}
    \end{pmatrix}
\]
and perform the piecewise rotation
\begin{equation}
    \Psi_\mu(y)
    =
    R[-\Theta_0\,\varepsilon(y)]\,
    \widetilde\Psi_\mu(y),
    \qquad
    \tan\Theta_0=M_0,
\label{eq:q9-discontinuous-rotation}
\end{equation}
where
\[
R[\alpha]
=
\begin{pmatrix}
\cos\alpha&\sin\alpha\\
-\sin\alpha&\cos\alpha
\end{pmatrix},
\qquad
\varepsilon(y)
=
\begin{cases}
+1,&y>0,\\
-1,&y<0.
\end{cases}
\]
Take $\widetilde\psi_{\mu2}$ to be continuous and odd, so that its
one-sided limits vanish at the fixed point. Equation
\eqref{eq:q9-discontinuous-rotation} then gives
\[
\frac{\psi_{\mu2}(0^+)}{\psi_{\mu1}(0^+)}
=
\tan\Theta_0=M_0,
\qquad
\frac{\psi_{\mu2}(0^-)}{\psi_{\mu1}(0^-)}
=
-\tan\Theta_0,
\]
and reproduces exactly the discontinuous representative obtained from
the matching condition.

The information has not disappeared: it has moved into the
differential operator. Since
$\partial_y\varepsilon(y)=2\delta(y)$, differentiating the
$y$-dependent rotation generates a localized contribution,
schematically
\begin{equation}
    R^{-1}\partial_y R
    \propto
    \Theta_0\,\delta(y).
\end{equation}
Thus a discontinuous field with a simple bulk operator and a
continuous field with a singular localized operator can describe the
same matching problem.

What is invariant is neither the value assigned to an odd
field exactly at the fixed point nor the continuity of a chosen field
representative. The physical information is contained in the
one-sided limits and in the matching relation that connects them.

\narrativeheading{Putting the pieces together}

Localized boundary operators affect Scherk--Schwarz compactifications
at several logically distinct levels.

At the classical level, they modify the global spectral problem.
For second-order fields this appears through Robin or
eigenvalue-dependent boundary conditions and through a modified
kinetic inner product. For first-order fermionic systems it appears as
matching conditions between components.

For energy-independent gravitino boundary rotations, the effect on
the spin-$\tfrac32$ spectrum can be summarized by
\begin{equation}
    \omega_{\rm eff}
    =
    \omega
    +
    \frac{\arctan M_0+\arctan M_\pi}{\pi}
    \pmod1.
\label{eq:q9-effective-twist-summary}
\end{equation}
This description ceases to be universal when the boundary conditions
depend on the eigenvalue, as happens generically in the presence of
localized kinetic terms.

At the supersymmetric level, one must distinguish a redistribution of
the same Scherk--Schwarz rotation from an independent localized
breaking sector. The latter can introduce additional fermion shifts
and Goldstino directions and therefore requires analysis of the
complete fermionic system.

At the quantum level, bulk and boundary operators require local
renormalization. This local ultraviolet dependence is distinct from
the finite part that requires propagation over the compact interval.
The latter depends on the relative boundary data and on the
Scherk--Schwarz transport and admits the Casimir interpretation
developed above.

\narrativeheading{Bibliographical remarks}

Localized gravitino masses, discontinuous orbifold representatives,
and their relation to generalized Scherk--Schwarz descriptions were
studied in Refs.~\cite{Bagger:2001qi,Biggio:2002rb}. The Wilson-line
description of Scherk--Schwarz breaking on orbifolds was discussed in
Ref.~\cite{vonGersdorff:2001ak}, and related brane-assisted
constructions in Ref.~\cite{Delgado:2002xf}.
The interplay between bulk Scherk--Schwarz breaking, localized
supersymmetry-breaking sectors, the gravitino tower, and
pseudo-Goldstinos was analysed explicitly in
Ref.~\cite{Benakli:2007zza}. Related treatments of localized
supersymmetry breaking in five-dimensional supergravity include
Refs.~\cite{Rattazzi:2003rj,Bagger:2004rr}.
Boundary-localized counterterms and kinetic operators are generic
ingredients of orbifold effective field theories. Their
renormalization and the complementary finite non-local contributions
will reappear when we discuss quantum effects.

\begin{takehome}
\begin{takehomeitems}

\item
Localized operators modify the boundary-value problem, not the
orbifold group relations. They can change the Kaluza--Klein spectrum,
wave functions, normalization, and localized couplings.

\item
Localized masses and localized kinetic terms play different roles.
The former modify the boundary potential; the latter also modify the
kinetic norm and generally make the boundary conditions
eigenvalue dependent.

\item
For first-order fermionic systems, localized terms are naturally
described by matching conditions or rotated boundary subspaces.

\item
Energy-independent gravitino boundary rotations can be absorbed into
an effective Scherk--Schwarz angle. Eigenvalue-dependent boundary
conditions cannot in general be represented by one universal
$\omega_{\rm eff}$.

\item
A genuine localized supersymmetry-breaking sector can contain
additional fermion shifts and Goldstino directions. Agreement of the
gravitino spectrum alone does not establish equivalence to a
Scherk--Schwarz rotation.

\item
Boundary ultraviolet divergences are local and renormalize boundary
operators. The finite contribution requiring propagation between the
two boundaries is non-local and has the Casimir interpretation
described above.

\end{takehomeitems}
\end{takehome}

\nextquestion
{How is a Scherk--Schwarz twist related to an \(R\)-symmetry Wilson line and to the radion \(F\)-term?}

\end{reviewquestion}
%%%%%%%%%%%%%%%%%%%%%%%%%%%%%%%%%%%%%%%%%%%%%%%%%

\begin{reviewquestion}
{How is a Scherk--Schwarz twist related to an \texorpdfstring{\(R\)}{R}-symmetry Wilson line
and to the radion \texorpdfstring{\(F\)}{F}-term?}
\label{sec:ss-wilson-radion}

\begin{essentialidea}

A continuous Scherk--Schwarz twist can be moved from the boundary
condition into the compact derivative by a \(y\)-dependent change of
field basis. Periodic fields then see a flat connection whose
holonomy contains the same global information as the original
monodromy.

When the twisting symmetry acts on the supercharges, it is an
\(R\)-symmetry. In supergravity the corresponding connection enters
the transport of the supersymmetry parameter and the gravitino
sector.

In a lower-dimensional effective supergravity obtained from a more
complete compactification, the same breaking may appear through
auxiliary fields of several compactification moduli. In a
five-dimensional description adapted to a single radial modulus,
the relevant auxiliary direction is packaged into the radion
\(F\)-term. The relation between that \(F\)-term and the
Scherk--Schwarz parameter depends on the choice of fields and
normalization.

\end{essentialidea}

\narrativeheading{From the Scherk--Schwarz ansatz to a flat connection}

Consider a field \(\Phi\) on a circle of radius \(R\), with
\begin{equation}
\Phi_{\rm tw}(y+2\pi R)
=
U_{\rm SS}^{(\Phi)}(\omega)\,
\Phi_{\rm tw}(y),
\qquad
U_{\rm SS}^{(\Phi)}(\omega)
=
\exp\!\left(
2\pi i\omega\mathbb R_\Phi
\right).
\label{eq:q10-monodromy}
\end{equation}
Here \(\mathbb R_\Phi\) denotes the generator of the twisting symmetry
in the representation carried by \(\Phi\). This equation fixes our
normalization of \(\omega\).

Introduce the periodic representative
\begin{equation}
\Phi_{\rm per}(y)
=
\exp\!\left(
-i\frac{\omega y}{R}\mathbb R_\Phi
\right)
\Phi_{\rm tw}(y).
\label{eq:q10-periodic-representative}
\end{equation}
It satisfies
\[
\Phi_{\rm per}(y+2\pi R)=\Phi_{\rm per}(y).
\]
Using the inverse transformation gives
\begin{equation}
\partial_y\Phi_{\rm tw}
=
e^{i\omega y\mathbb R_\Phi/R}
\left(
\partial_y
+
i\frac{\omega}{R}\mathbb R_\Phi
\right)
\Phi_{\rm per}.
\label{eq:q10-derivative-redefinition}
\end{equation}
The compact derivative acting on periodic fields therefore takes the
form
\begin{equation}
D_y
=
\partial_y+\mathcal A_y^{(\Phi)},
\qquad
\mathcal A_y^{(\Phi)}
=
i\frac{\omega}{R}\mathbb R_\Phi .
\label{eq:q10-flat-connection}
\end{equation}
For a constant twist this connection is flat.

The matrix obtained by going once around the positively oriented
circle is
\begin{equation}
\mathcal H_\Phi
=
\mathcal P
\exp\!\left[
\oint dy\,\mathcal A_y^{(\Phi)}
\right]
=
e^{2\pi i\omega\mathbb R_\Phi}
=
U_{\rm SS}^{(\Phi)}(\omega).
\label{eq:q10-holonomy}
\end{equation}
We denote the holonomy by \(\mathcal H_\Phi\), reserving \(W\) for
the four-dimensional superpotential.

The transport convention used in
Question~\ref{sec:ss-localized-boundary-terms} was
\[
\mathcal U(y_2,y_1)
=
\mathcal P
\exp\!\left[
-\int_{y_1}^{y_2}dy\,\mathcal A_y
\right].
\]
For the constant connection considered here, the transport operator
along the same oriented path is therefore the inverse of the
holonomy matrix,
\[
\mathcal U(2\pi R,0)
=
\mathcal H_\Phi^{-1}.
\]
For a general non-Abelian, \(y\)-dependent connection, the relation
between a path-ordered exponential and its inverse also involves
reversing the path ordering.

\narrativeheading{When the twisting symmetry acts on supersymmetry}

The preceding construction applies to any continuous symmetry
admitted by the higher-dimensional theory. If the symmetry acts on
the supersymmetry generators, its action on the supersymmetry
parameter is described by
\begin{equation}
D_y\epsilon
=
\left(
\partial_y
+
i\frac{\omega}{R}\mathbb R_\epsilon
\right)\epsilon .
\label{eq:q10-susy-connection}
\end{equation}
The corresponding holonomy is
\[
\mathcal H_\epsilon
=
\exp\!\left(
2\pi i\omega\mathbb R_\epsilon
\right).
\]
Subject to the remaining Killing-spinor equations, the supersymmetries
preserved on the flat circle obey
\begin{equation}
\mathcal S_{\rm preserved}
=
\ker\!\left(
\mathcal H_\epsilon-\mathbf 1
\right).
\label{eq:q10-holonomy-preserved-susy}
\end{equation}
This is the invariant-subspace condition encountered earlier, now
written in the connection basis.

The ingredients needed for this interpretation are already visible
in the original generalized reductions
\cite{Scherk:1978ta,Scherk:1979zr}. In the supergravity example of
Ref.~\cite{Scherk:1978ta}, the compact-coordinate dependence acts
chirally on the gravitino while leaving the vielbein untwisted. A
formulation in which the compact component of a local internal
connection enters the fermionic mass terms can be found in
Ref.~\cite{Cremmer:1979uq}.

When the symmetry is realized locally in the supergravity
formulation, \(\mathcal A_y^{(R)}\) is an \(R\)-symmetry connection
and the flat background may be described as an \(R\)-symmetry Wilson
line. A detailed off-shell realization on an orbifold can be found
in Ref.~\cite{vonGersdorff:2001ak}.

The same connection appears in the local supersymmetry transformation
of the internal gravitino,
\begin{equation}
\delta\psi_y
=
2D_y\epsilon+\cdots .
\label{eq:q10-internal-gravitino-variation}
\end{equation}
With the notation introduced in
Question~\ref{sec:ss-goldstino}, the constant part of the internal
operator is
\begin{equation}
\mathcal M
=
i\frac{\omega}{R}\mathbb R_\epsilon .
\label{eq:q10-M-as-R-connection}
\end{equation}
This identifies explicitly the matrix \(\mathcal M\) used there with
the action of the Scherk--Schwarz generator on the supersymmetry
parameters.

\narrativeheading{A brief comment on the interval}

On \(S^1/\mathbb Z_2\), the underlying topological space is an
interval and contains no non-contractible closed loop. The circle
holonomy should therefore be understood on the covering space. In an
interval description the same compactification data are distributed
between the endpoint actions and the open transport between them.

Using the convention of
Question~\ref{sec:ss-localized-boundary-terms}, the latter is simply
\begin{equation}
\mathcal U(\pi R,0)
=
\mathcal P
\exp\!\left[
-\int_0^{\pi R}dy\,\mathcal A_y
\right].
\label{eq:q10-interval-transport}
\end{equation}
Under a change of basis \(G(y)\),
\begin{align}
\mathcal U(\pi R,0)
&\longrightarrow
G(\pi R)^{-1}
\mathcal U(\pi R,0)
G(0),
\nonumber\\
P_b
&\longrightarrow
G(y_b)^{-1}P_bG(y_b),
\qquad y_b=0,\pi R.
\label{eq:q10-interval-transformations}
\end{align}
Neither the open transport nor the endpoint matrices should be
interpreted separately. This is the change-of-trivialization freedom
already used in Questions~\ref{sec:ss-orbifolds} and
\ref{sec:ss-localized-boundary-terms}.

\narrativeheading{Compactification moduli and the radial direction}

The effective theory obtained from a complete compactification need
not contain a field whose name and normalization coincide with the
radion of a simple five-dimensional truncation. In four-dimensional
\(N=1\) supergravity, let
\[
G=K+\ln|W|^2 .
\]
On a bosonic background the chiral auxiliary fields are proportional,
up to standard convention-dependent factors, to
\[
F^I\propto e^{G/2}G^I,
\qquad
G^I=G^{I\bar J}G_{\bar J}.
\]
Supersymmetry breaking can therefore point along several coordinates
of the compactification-moduli space.

A string-derived effective-supergravity description of
coordinate-dependent compactification can be found in
Ref.~\cite{Ferrara:1994kg}. In one of the examples discussed there,
the supersymmetry-breaking direction satisfies
\begin{equation}
G^S G_S
=
G^{T'}G_{T'}
=
G^{U'}G_{U'}
=
1,
\label{eq:q10-moduli-goldstino}
\end{equation}
while
\begin{equation}
m_{3/2}^2
=
\frac{k^2}
{(S+\bar S)(T'+\bar T')(U'+\bar U')}
=
\frac{(e_1+e_2)^2g_U^2}{2R^2},
\label{eq:q10-string-gravitino-mass}
\end{equation}
with
\begin{equation}
R
=
2\sqrt{
(T'+\bar T')(U'+\bar U')
}.
\label{eq:q10-string-radius}
\end{equation}
In the conventions of Ref.~\cite{Ferrara:1994kg},
\(k=(e_1+e_2)/2\) is the constant term of the superpotential
in Planck units, \(e_1,e_2\) are quantized charges, and
\(g_U\) is the unified gauge coupling. These quantities belong
to that string-derived example, rather than to the general
Scherk--Schwarz circle.
The effective supergravity describes the Scherk--Schwarz
background through non-vanishing auxiliary directions in the
compactification-moduli sector and gives the characteristic scaling
\(m_{3/2}\propto R^{-1}\).

The variables \(S,T',U'\) are coordinates adapted to that
four-dimensional effective description. They need not coincide with
fields adapted to a higher-dimensional Einstein-frame
parametrization. The physical radius in
Eq.~\eqref{eq:q10-string-radius} is already a combination of the
\(T'\)- and \(U'\)-type moduli, and the relation between the
higher-dimensional metric and the four-dimensional Einstein frame
also involves the dilaton sector.

If the radial degree of freedom can be isolated locally as a chiral
coordinate
\[
T_{\rm rad}=T_{\rm rad}(Z^I),
\]
its auxiliary field transforms under a holomorphic change of
coordinates as
\begin{equation}
F^{T_{\rm rad}}
=
\frac{\partial T_{\rm rad}}{\partial Z^I}\,F^I .
\label{eq:q10-radion-auxiliary-combination}
\end{equation}
Several non-zero \(F^I\) in one parametrization can therefore
contribute to a single auxiliary direction in another. Determining
the precise combination requires the map between the
four-dimensional moduli and the higher-dimensional radion. We shall
not identify any one of the auxiliary fields appearing in
Eq.~\eqref{eq:q10-moduli-goldstino} with \(F_T\) by itself.

\narrativeheading{An adapted five-dimensional radion description}

For a flat five-dimensional theory in which the radial degree of
freedom has already been isolated, the relation becomes particularly
simple. A four-dimensional \(N=1\) superfield formulation with
dimensionless coordinate
\(\hat y\sim\hat y+2\pi\) and
\[
ds^2
=
\eta_{\mu\nu}dx^\mu dx^\nu+R^2d\hat y^2
\]
can be written in terms of the radion multiplet
\begin{equation}
T_{\rm rad}
=
R+iB_5+\theta\Psi_R^5+\theta^2F_T .
\label{eq:q10-radion-superfield}
\end{equation}
A realization of this form can be found in
Ref.~\cite{Marti:2001iw}.

For a real constant \(F_T\), the non-zero Kaluza--Klein levels of the
five-dimensional gaugino doublet have mass matrix
\begin{equation}
\mathcal M_n
=
\frac{1}{R}
\begin{pmatrix}
F_T/2 & n\\
n & F_T/2
\end{pmatrix},
\qquad n\geq1.
\label{eq:q10-radion-gaugino-matrix}
\end{equation}
Its eigenvalues give
\begin{equation}
m_{n,\pm}
=
\frac{|n\pm F_T/2|}{R},
\label{eq:q10-radion-gaugino-spectrum}
\end{equation}
while the zero-mode gaugino has mass
\[
m_{\lambda,0}
=
\frac{|F_T|}{2R}.
\]
In the normalization used in that formulation, the
Scherk--Schwarz shift is
\begin{equation}
q_R
=
\frac{F_T}{2}.
\label{eq:q10-FT-qR}
\end{equation}

The relation can also be displayed directly by a field redefinition.
For an \(R\)-charge eigenstate \(q=\pm1\), consider
\begin{equation}
\mathscr D_q
=
\frac{1}{R}
\left(
-i\partial_{\hat y}
+
q\frac{F_T}{2}
\right).
\label{eq:q10-radion-compact-operator}
\end{equation}
Define
\begin{equation}
\lambda_q^{\rm per}(\hat y)
=
e^{-iqF_T\hat y/2}
\lambda_q^{\rm tw}(\hat y).
\label{eq:q10-radion-field-redefinition}
\end{equation}
The constant term in
Eq.~\eqref{eq:q10-radion-compact-operator} is removed, while the
periodicity of \(\lambda_q^{\rm per}\) gives
\begin{equation}
\lambda_q^{\rm tw}(\hat y+2\pi)
=
e^{i\pi qF_T}
\lambda_q^{\rm tw}(\hat y).
\label{eq:q10-radion-induced-monodromy}
\end{equation}
Comparing with
Eq.~\eqref{eq:q10-monodromy}, for a generator with eigenvalues
\(q=\pm1\),
\begin{equation}
\omega
=
\frac{F_T}{2}
\label{eq:q10-omega-FT}
\end{equation}
in these conventions. A component treatment of the corresponding
field redefinition can be found in
Ref.~\cite{Kaplan:2001cg}.

The numerical relation in
Eq.~\eqref{eq:q10-omega-FT} is not formulation independent.
Rescaling the radion superfield or the \(R\)-symmetry generator changes
it. The invariant statement is the agreement of the monodromy, or of
the corresponding shifted compact operator, after the fields and
generators have been normalized consistently.

\begin{figure}[t]
    \centering
    \includegraphics[width=0.88\textwidth]
    {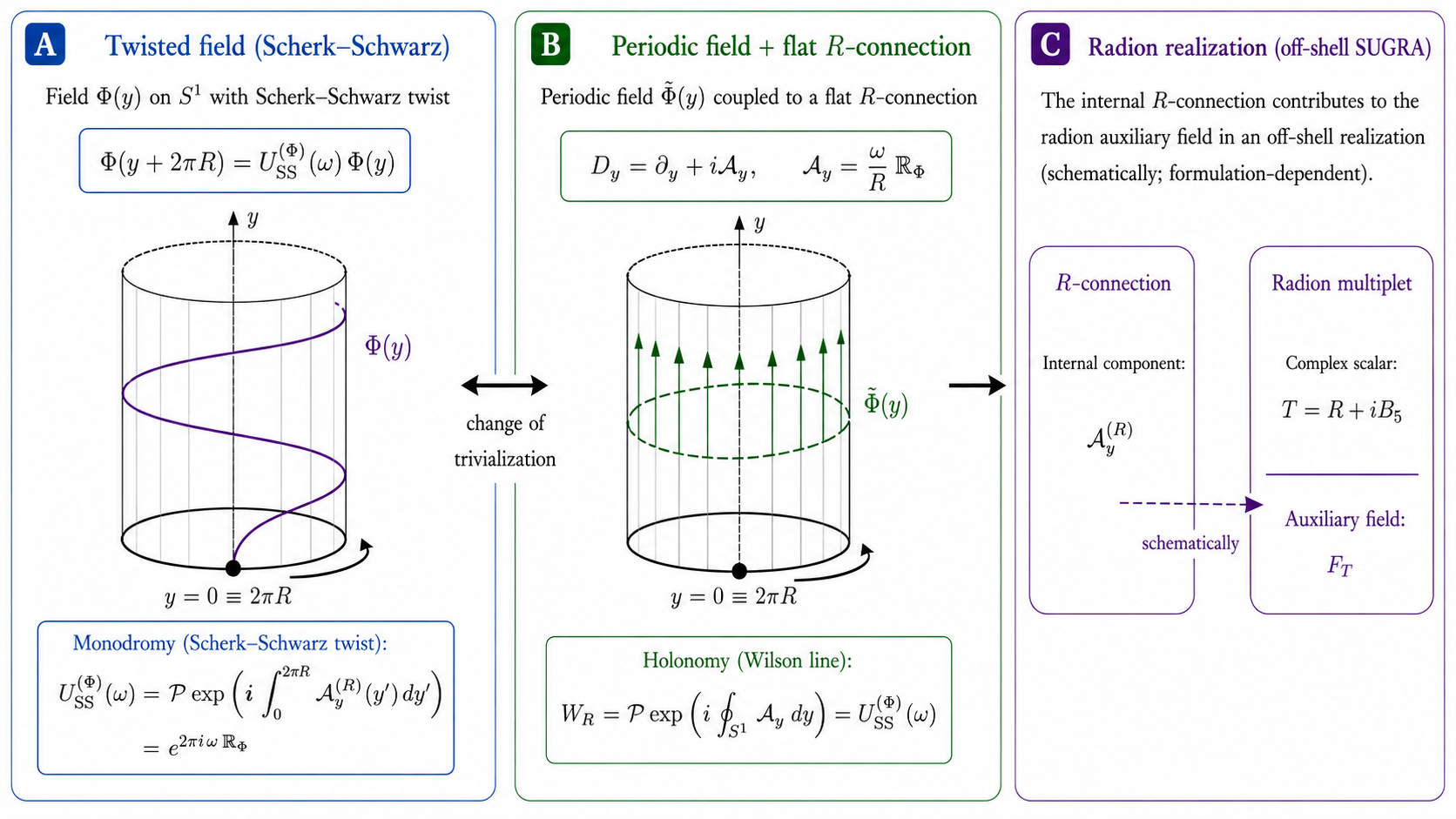}
    \caption{
    Three descriptions used in this question. In the twisted basis,
    the compactification is specified by the monodromy
    \(U_{\rm SS}\). A \(y\)-dependent change of field basis gives
    periodic fields with a flat \(R\)-symmetry connection. In a
    formulation adapted to the radial multiplet, the same
    Scherk--Schwarz shift can be encoded in a non-zero auxiliary
    field \(F_T\). The numerical relation between \(F_T\) and the
    twist parameter depends on the normalization of the radion and of
    the \(R\)-symmetry generator.
    }
    \label{fig:ss-wilson-radion}
\end{figure}

\narrativeheading{Bibliographical remarks}

The essential ingredients of generalized dimensional reduction,
including a compact-coordinate dependence generated by a symmetry
acting on the gravitino, can already be found in
Refs.~\cite{Scherk:1978ta,Scherk:1979zr}. A description in which the
compact component of a local internal-symmetry connection contributes
to the fermion mass matrix can be found in
Ref.~\cite{Cremmer:1979uq}.

Descriptions of coordinate-dependent compactifications in
string-derived four-dimensional effective supergravities, with
supersymmetry breaking along compactification-modulus directions, can
be found in
Refs.~\cite{Kounnas:1988ye,Ferrara:1988jx,Ferrara:1994kg}.
A five-dimensional off-shell formulation making the
\(R\)-symmetry-connection description explicit on orbifolds can be
found in Ref.~\cite{vonGersdorff:2001ak}. A superfield formulation in
which the radial multiplet is isolated and its auxiliary component
produces the shifted Kaluza--Klein spectrum can be found in
Ref.~\cite{Marti:2001iw}. The relation to twisted boundary conditions
through a field redefinition is discussed in
Ref.~\cite{Kaplan:2001cg}.

% Keep the preceding figure clear of the take-home box.
\FloatBarrier
\begin{takehome}
\begin{takehomeitems}

\item
A continuous Scherk--Schwarz monodromy can be represented by periodic
fields coupled to a flat connection. The two forms are related by a
\(y\)-dependent change of field basis.

\item
When the twisting symmetry acts on the supercharges, the corresponding
connection is an \(R\)-symmetry connection. The same operator enters
the transport of the supersymmetry parameter and the internal
gravitino transformation.

\item
In a complete effective supergravity, the radial degree of freedom
need not coincide with one of the conventional moduli coordinates.
Its auxiliary field can be a combination of the \(F\)-terms appearing
separately in another parametrization.

\item
In a flat five-dimensional formulation adapted to a single radion
multiplet, a constant \(F_T\) reproduces the Scherk--Schwarz shifted
spectrum. The relation \(\omega=F_T/2\) holds in the normalization
used above and is not a universal component identity.

\end{takehomeitems}
\end{takehome}

\nextquestion
{What is special about the half-twist, and why does a Dirac gravitino
appear?}

\end{reviewquestion}
%%%%%%%%%%%%%%%%%%%%%%%%%%%%%%%%%%%%%%%%%%%%%%%%%
% Question 11 v10. Physical half-twist and the Dirac gravitino.
% Split from Question11v8: the detailed superspace construction is
% in Question12v2. Original scientific equation labels are retained.
%%%%%%%%%%%%%%%%%%%%%%%%%%%%%%%%%%%%%%%%%%%%%%%%%

\begin{reviewquestion}
{What is special about the half-twist, and why does a Dirac gravitino
appear?}
\label{sec:ss-half-twist}

\begin{essentialidea}
At the half-twist, $\omega=1/2$, the Kaluza--Klein masses of the
five-dimensional gravitino occur in equal-mass pairs. The two ends of
the orbifold preserve different supersymmetries, so none survives on
the whole interval. Yet a $U(1)_R$ symmetry remains. It allows the two
gravitini to form a Dirac field, with mass $1/(2R)$ for the lightest
pair, and forbids separate Majorana masses.

The Dirac mass is already present in the local five-dimensional
Rarita--Schwinger action. Its effective description in four-dimensional
superspace is a separate question, to which we turn next.
\end{essentialidea}

\narrativeheading{The physical picture}

There are two ways of thinking about the half-twist. The first is to
follow a field around the circle. With ordinary periodic boundary
conditions its compact momentum is an integer multiple of $1/R$.
An anti-periodic fermion picks up a minus sign on a complete turn;
its momenta are instead half-integer multiples of $1/R$. The first
possible mass is $1/(2R)$, followed by $3/(2R)$, $5/(2R)$, and so on.
There is no massless gravitino.

The second picture uses the interval $0\leq y\leq\pi R$. Each end
admits one direction in the doublet of supersymmetry parameters.
At the ordinary untwisted point the directions agree. At the
half-twist they are orthogonal. A spinor that is constant along the
interval cannot belong to both. This explains why the compactified
background has no unbroken supersymmetry, although the original
five-dimensional theory is locally supersymmetric.

A minus sign around the circle might seem to destroy every
continuous $R$-symmetry. It does not. On the gravitino doublet,
the half-twist is the central matrix $-\mathbf 1$, so it commutes
with every $SU(2)_R$ rotation. The orbifold projections restrict
these rotations, but leave a $U(1)_R$. The two gravitini carry
opposite charges under this $U(1)_R$. Their mixed mass has zero
charge, while either Majorana mass has a non-zero charge.
This is the physical distinction between an accidental degeneracy
and a Dirac gravitino.

The twist acts on the supersymmetry doublet, not on the four-dimensional graviton.
The graviton retains its massless zero mode. Its partner gravitino
does not. When the two gravitini are written as $N=1$ superfields,
their different multiplet assignments make the Dirac mass operator
less familiar than the spectrum itself.

The pairing of the signed Kaluza--Klein masses and the $U(1)_R$
selection rule can both be derived without four-dimensional superspace.

\narrativeheading{A symmetry of the signed spectrum}

We saw the shifted spectrum in Question~\ref{sec:twisted-kk-spectrum}:
for a phase twist on the circle, the signed masses are
\[
 m_n(\omega)=\frac{n+\omega}{R},\qquad n\in\mathbb Z.
\]
The sign is useful here, even though the physical mass of a given mode
is $M_n=|m_n|$. Let $\mathcal D_\omega$ act on the Fourier labels by
$\mathcal D_\omega|n\rangle=m_n(\omega)|n\rangle$, and introduce the
reflection $\mathcal J|n\rangle=|-n-1\rangle$. A short calculation gives
\begin{equation}
 \mathcal J\mathcal D_\omega\mathcal J^{-1}
 =-\mathcal D_{1-\omega}.
 \label{eq:q11-affine-spectrum-relation}
\end{equation}
In particular,
\begin{equation}
 \{\mathcal J,\mathcal D_{1/2}\}=0,\qquad
 m_{-n-1}\!\left(\tfrac12\right)
 =-m_n\!\left(\tfrac12\right).
 \label{eq:q11-half-spectral-pairing}
\end{equation}
There is no integer fixed by $n\mapsto-n-1$. The lowest pair has
signed masses $+1/(2R)$ and $-1/(2R)$, followed by
$\pm3/(2R)$, $\pm5/(2R)$, and so forth.

The equality of masses by itself does not guarantee a Dirac particle.
Two Majorana fermions may be degenerate without carrying a conserved
charge.
The labels $n$ and $-n-1$ can be used for the two Weyl
vector-spinors $\psi_{\mu k}$ and $\rho_{\mu k}$ of one orbifold level
$k\geq0$. This is one Dirac gravitino, not two copies of it. Pairing
alone does not protect it: that requires a symmetry of the full
boundary problem.

\narrativeheading{What the two boundaries preserve}

Take the gravitino $SU(2)_R$ doublet in the orbifold basis and set
\[
 P_0=\tau_3,\qquad
 U(\omega)=e^{2\pi i\omega\tau_2},\qquad
 P_\pi(\omega)=U(\omega)P_0.
\]
The $\tau_a$ act on the internal doublet, not on the Lorentz spinor
indices. The consistency condition $P_0U P_0=U^{-1}$ is automatic.
At the half-twist,
\begin{equation}
 U\!\left(\tfrac12\right)=-\mathbf1,
 \qquad P_\pi=-P_0=-\tau_3.
 \label{eq:q11-half-endpoints}
\end{equation}
We use the twisted-boundary frame: the bulk derivative is the
ordinary $\partial_y$, while the doublet obeys
$\epsilon(y+2\pi R)=U(\omega)\epsilon(y)$. A massless global
supersymmetry parameter would satisfy $\partial_y\epsilon=0$ and
both endpoint projections. At $\omega=1/2$ these projectors select
$(1,0)^T$ at $y=0$ and $(0,1)^T$ at $y=\pi R$. Thus
\begin{equation}
 \mathcal S_0\cap\mathcal S_\pi=\{0\}.
 \label{eq:q11-no-global-susy}
\end{equation}
Each endpoint still admits a local supersymmetry direction, but no
constant spinor belongs to both. The Goldstino and super-Higgs aspects
were discussed in Questions~\ref{sec:ss-spontaneous},
\ref{sec:ss-goldstino} and \ref{sec:ss-superhiggs}.

On the gravitino doublet, the central holonomy
$U(1/2)=-\mathbf1$ has the same action as $(-1)^F$.
The orbifold
projections select a smaller surviving group, namely
\begin{equation}
 G_{\rm bc}(\omega)=
 Z_{SU(2)_R}(P_0)\cap Z_{SU(2)_R}(P_\pi(\omega)).
 \label{eq:q11-common-centralizer}
\end{equation}
For generic $\omega$, the two projection axes are different, and
$G_{\rm bc}$ contains only $\{\pm\mathbf1\}$. At $\omega=1/2$
the axes coincide, up to a sign; the common centralizer is therefore
\begin{equation}
 G_{\rm bc}\bigl(\tfrac12\bigr)=U(1)_{\tau_3}.
 \label{eq:q11-centralizer-enhancement}
\end{equation}
At $\omega=0$ this group is also $U(1)$, but there
$P_\pi=P_0$ and one supersymmetry survives. The sign of the
second projection changes the physics.

The mass matrix takes different forms in the two familiar bases.
In the orbifold basis $Q_R=\tau_3$, and an allowed mass matrix is
proportional to $\tau_1$. With the change of basis
\[
 \psi_{\rm Maj}=V\psi_{\rm orb},\qquad
 V=\frac1{\sqrt2}
 \begin{pmatrix}1&1\\-i&i\end{pmatrix},\qquad
 V\tau_3V^{-1}=-\sigma_2,\qquad
 (V^{-1})^T\tau_1V^{-1}=\mathbf1.
\]
the mass matrix becomes proportional to $\mathbf1$ and the charge
generator becomes $-\sigma_2$. The two equal Majorana masses are
therefore invariant under real $SO(2)$ rotations: for
$O(\alpha)=e^{-i\alpha\sigma_2}$, one has $O^TO=\mathbf1$.
In the charged basis the same invariance is the off-diagonal Dirac
selection rule.

Indeed, for two Weyl gravitini with charges $+1$ and $-1$, a symmetric
mass matrix obeys
\begin{equation}
 Q_R^T\mathcal M_{3/2}+\mathcal M_{3/2}Q_R=0,
 \qquad Q_R=\begin{pmatrix}1&0\\0&-1\end{pmatrix}.
 \label{eq:q11-mass-selection-rule}
\end{equation}
The diagonal terms must vanish, leaving
\begin{equation}
 \mathcal M_{3/2}=\begin{pmatrix}0&m_D\\m_D&0\end{pmatrix}.
 \label{eq:q11-dirac-mass-matrix}
\end{equation}
The eigenvalues of the mass matrix in
Eq.~\eqref{eq:q11-dirac-mass-matrix} are $+m_D$ and $-m_D$.
They are the signed masses of the lowest pair in
Eq.~\eqref{eq:q11-half-spectral-pairing}, with $m_D=1/(2R)$.
For canonical Rarita--Schwinger kinetic terms, the mass bilinear
in the charged Weyl basis is
\begin{equation}
 \mathcal L_{\rm mass}
 =-2m_D\psi_{+\mu}\sigma^{\mu\nu}\psi_{-\nu}
 +\mathrm{h.c.}
 \label{eq:q11-RS-two-component-mass}
\end{equation}
The two off-diagonal entries carry one mass $m_D$ and give the
coefficient $2m_D$. The canonical normalization is checked in
Question~\ref{sec:ss-dirac-effective-operator}.
The action of the lowest pair is
\begin{equation}
 \mathcal L_{\rm phys}=
 \mathcal L_{\rm RS}[\psi_+]
 +\mathcal L_{\rm RS}[\psi_-]
 -2m_D\psi_{+\mu}\sigma^{\mu\nu}\psi_{-\nu}
 +\mathrm{h.c.},\qquad m_D=\frac1{2R}.
 \label{eq:q11-local-component-action}
\end{equation}
In this equation $\mathcal L_{\rm RS}[\psi]$ is the canonically
normalized Weyl Rarita--Schwinger kinetic term.

The $U(1)_R$ is a symmetry of the pure half-twisted bulk and its
boundary conditions. If localized masses or matter interactions
break it, the diagonal Majorana terms are no longer forbidden.

\narrativeheading{From the fifth derivative to a Dirac mass}

The mode profiles give a useful check that does not require any
four-dimensional superspace. Write $\ell=\pi R$ and
$m_k=(k+\tfrac12)/R$. In a twisted-boundary frame, the two components
of the five-dimensional gravitino have profiles proportional to
\begin{equation}
 \begin{split}
 \Psi^1_\mu(x,y)&=\sqrt{\frac2\ell}
  \sum_{k\geq0}\cos(m_ky)\,\psi_{\mu k}(x),\\
 \Psi^2_\mu(x,y)&=-\sqrt{\frac2\ell}
  \sum_{k\geq0}\sin(m_ky)\,\rho_{\mu k}(x).
 \end{split}
 \label{eq:q11-kk-profiles}
\end{equation}
The fields $\psi_{\mu k,\alpha}$ and $\rho_{\mu k,\alpha}$ are the
left-handed four-dimensional Weyl components of the
five-dimensional symplectic-Majorana doublet. Their dotted
conjugates give the corresponding right-handed components; the
symplectic-Majorana condition avoids an independent doubling of
the states.

The profiles in Eq.~\eqref{eq:q11-kk-profiles} obey the endpoint
projections directly:
\[
 \Psi^2_\mu(x,0)=0,\qquad
 \Psi^1_\mu(x,\ell)=0,
 \qquad
 \sin(m_k\ell)=(-1)^k.
\]
Thus $P_0=\tau_3$ selects the first component at $y=0$,
whereas $P_\pi=-\tau_3$ selects the second at $y=\ell$.

The fifth-dimensional Rarita--Schwinger term contains an internal
first-derivative coupling between the two Weyl components.
Its relevant overlap is
\begin{equation}
 \frac{2}{\ell}\int_0^\ell dy\,\cos(m_ky)
 \partial_y\sin(m_{k'}y)
 =m_{k'}\delta_{kk'}.
 \label{eq:q11-kk-overlap}
\end{equation}
The term with the fifth derivative acting on the other component
gives the same overlap after integration by parts. With the
canonical Rarita--Schwinger normalization and the sine-profile sign
chosen in Eq.~\eqref{eq:q11-kk-profiles}, the two terms give
\begin{equation}
 \mathcal L_{\partial_y}\supset
 -2\sum_{k\geq0}m_k\,
 \psi_{\mu k}\sigma^{\mu\nu}\rho_{\nu k}
 +\mathrm{h.c.}
 \label{eq:q11-kk-dirac-bilinear}
\end{equation}
The two Weyl vector-spinors of each level form one Dirac gravitino.
The first one has mass $1/(2R)$; there is no massless gravitino at the
half-twist. The spectrum, boundary projections and $U(1)_R$ selection rule
account for the Dirac gravitino without introducing auxiliary
fields or four-dimensional superspace.

\narrativeheading{Moving away from the half-twist}

Finally let $\omega=\tfrac12-\varepsilon$, with
$|\varepsilon|\ll1$. The two lightest physical masses become
\[
 M_0=\frac{\tfrac12-\varepsilon}{R},\qquad
 M_{-1}=\frac{\tfrac12+\varepsilon}{R},\qquad
 \Delta M=\frac{2|\varepsilon|}{R},
\]
with the labels exchanged when $\varepsilon<0$. The endpoint
projections no longer have the same axis. Their common centralizer is
reduced to $\mathbb Z_2$, and diagonal Majorana terms are no longer
forbidden by the half-twist $U(1)_R$. The two nearly degenerate
states then form a pseudo-Dirac pair.

A localized rotation can give the same leading spectral splitting.
For energy-independent rotations of the gravitino boundary data,
Question~\ref{sec:ss-localized-boundary-terms},
Eq.~\eqref{eq:q9-combined-spectrum}, gives the shifted masses in
terms of $\omega+(\Theta_0+\Theta_\pi)/\pi$.
This is an equality of the gravitino spectra for that restricted
boundary problem. It says nothing by itself about localized fermions
or their Goldstino directions.

To describe the mass in $N=1$ superspace, we must now do more than
identify its eigenvalues. We need a gauge-invariant operator with the
correct kinetic normalization, together with the momentum dependence
of the full Kaluza--Klein tower. This is the subject of the next question.

\narrativeheading{Bibliographical remarks}

The pairing of the gravitini at the half-twist and its $U(1)_R$
selection rule are discussed in
\emph{(Pseudo-)Dirac Gravitinos}~\cite{Benakli:2026zwx}.
The mass spectrum follows from the five-dimensional Scherk--Schwarz
reduction, while the relation between boundary rotations and the
spectral shift was derived in
Question~\ref{sec:ss-localized-boundary-terms}.
The four-dimensional superspace operator is treated separately in
the next question.

\begin{takehome}
\begin{takehomeitems}

\item At $\omega=1/2$ the signed KK masses are paired by
$n\leftrightarrow -n-1$. Each pair gives one Dirac gravitino.
The first has mass $m_D=1/(2R)$.

\item The two orbifold boundaries select orthogonal supersymmetry
directions, $P_\pi=-P_0$. No global Killing spinor survives,
but their common $U(1)_R$ protects the Dirac mass.

\item Two degenerate Majorana gravitini do not automatically form
a symmetry-protected Dirac field. The residual $U(1)_R$ permits
the off-diagonal mass and excludes independent Majorana masses.

\item The mass follows directly from the fifth derivative in the
local five-dimensional theory. Rotating the boundary data or moving
away from $\omega=1/2$ can split the pair and produce a pseudo-Dirac
gravitino.

\end{takehomeitems}
\end{takehome}

\nextquestion
{How can the Dirac gravitino mass be described in effective superspace?}

\end{reviewquestion}

%%%%%%%%%%%%%%%%%%%%%%%%%%%%%%%%%%%%%%%%%%%%%%%%%%%%%%%%%%%%%%%
% Question 12 v3. Detailed construction of the effective superspace
% Dirac-gravitino operator, split from Question11v8.
% Equation labels inherited from Question11v8 are kept for compatibility.
%%%%%%%%%%%%%%%%%%%%%%%%%%%%%%%%%%%%%%%%%%%%%%%%%%%%%%%%%%%%%%%

\begin{reviewquestion}
{How can the Dirac gravitino mass be described in effective superspace?}
\label{sec:ss-dirac-effective-operator}

\begin{essentialidea}
At the half-twist, the fifth derivative gives a Dirac mass to the
two four-dimensional gravitini. The first belongs to the gravity
multiplet, together with the graviton. The second belongs to a
companion spin-$\tfrac32$ multiplet containing a vector. We would
like to see their mass in $N=1$ superspace, without choosing a
gravitino gauge.

A direct coupling between the two prepotentials fails. Its unprojected
form is not gauge invariant, while its transverse projection vanishes
identically. The invariant curvatures allow a non-zero transverse
mass, represented by an operator containing an inverse d'Alembertian.
The Kaluza--Klein tower fixes its momentum dependence.

The five-dimensional calculation also gives the diagonal kinetic
response. Two superspace invariants, one Hermitian and one chiral,
reproduce the fermionic curvature kernel after the boundary sources
have been properly identified. This is a quadratic, fixed-background
result. The full bosonic and auxiliary completion is a separate task.
\end{essentialidea}

\narrativeheading{The question in four dimensions}

Question~\ref{sec:ss-half-twist} gave the physical mass
$m_D=1/(2R)$ and its $U(1)_R$ protection. That answer follows from a
local five-dimensional action. In four-dimensional $N=1$ language,
however, the two gravitini cannot be treated as two copies of the
ordinary gravitational prepotential. We must respect the independent
gauge transformations of the two multiplets. We shall first test the
obvious bilinear, then construct the curvature operator and match its
fermionic part to the full KK response. The underlying five-dimensional
calculation and its component reductions are developed in
Ref.~\cite{Benakli:2026operator}.

\narrativeheading{Why the direct $H\Psi$ term fails}

Let $H_{\alpha\dot\alpha}$ be the real gravitational prepotential and
$\Psi_\alpha$ the spinor prepotential of the companion gravitino
multiplet. At the linearized level,
\begin{equation}
 \begin{split}
 \delta H_{\alpha\dot\alpha}
 &=\bar D_{\dot\alpha}L_\alpha-D_\alpha\bar L_{\dot\alpha},\\
 \delta\Psi_\alpha&=D_\alpha\Omega+\eta_\alpha,
 \qquad \bar D_{\dot\alpha}\eta_\beta=0.
 \end{split}
 \label{eq:q11-H-gauge-transformation}
\end{equation}
The parameters $L_\alpha$ and $\Omega$ are unconstrained, and
$\eta_\alpha$ is chiral. The complete off-shell multiplets also
contain compensators. First consider a bilinear made only from these
two prepotentials.

An obvious current is
\begin{equation}
 \begin{split}
 J_{\alpha\dot\alpha}(\Psi)
 &=\bar D_{\dot\alpha}\Psi_\alpha
   -D_\alpha\bar\Psi_{\dot\alpha},\\
 S_{H\Psi}&=\mu\int d^8z\,
 H^{\alpha\dot\alpha}J_{\alpha\dot\alpha}(\Psi),
 \qquad d^8z=d^4x\,d^2\theta\,d^2\bar\theta .
 \end{split}
 \label{eq:q11-naive-Hpsi-current}
\end{equation}
Vary $H$ while keeping $\Psi$ fixed.
The spinor gauge parameters are Grassmann odd. Thus integration by
parts uses $\int d^8z\,(D_\alpha L)J
=+\int d^8z\,L(D_\alpha J)$, and similarly for $\bar D$.
With this ordering and no surface terms,
\begin{equation}
 \delta_H S_{H\Psi}
 =\mu\int d^8z\,
 \left(L_\alpha\bar D_{\dot\alpha}J^{\alpha\dot\alpha}
       -\bar L_{\dot\alpha}D_\alpha J^{\alpha\dot\alpha}\right).
 \label{eq:q11-naive-Hpsi-variation}
\end{equation}
For arbitrary $L_\alpha$ and $\bar L_{\dot\alpha}$, we would need
\begin{equation}
 \bar D^{\dot\alpha}J_{\alpha\dot\alpha}=0,
 \qquad
 D^\alpha J_{\alpha\dot\alpha}=0.
 \label{eq:q11-transverse-current}
\end{equation}
The current above satisfies neither condition for an unconstrained
$\Psi$. It also changes under the companion gauge transformation:
$\delta_\Omega J_{\alpha\dot\alpha}
=\bar D_{\dot\alpha}D_\alpha\Omega
 -D_\alpha\bar D_{\dot\alpha}\bar\Omega$, which is generally
non-zero. The unprojected coupling is therefore not a gauge-invariant
mass term.

One may instead try the transverse
superspin-$\tfrac32$ projector $\Pi^T_{3/2}$. On unconstrained
prepotentials it is non-local, with inverse powers of $\Box$
(up to $\Box^{-2}$), and would give
\begin{equation}
 S_{\rm mix}^{T}=\mu\int d^8z\,
 H^{\alpha\dot\alpha}
 \bigl(\Pi^T_{3/2}J\bigr)_{\alpha\dot\alpha}.
 \label{eq:q11-projector-mixing}
\end{equation}
But the current is already a gauge variation of $H$:
\begin{equation}
 J_{\alpha\dot\alpha}(\Psi)
 =\left.\delta H_{\alpha\dot\alpha}\right|_{L=\Psi}.
 \label{eq:q11-current-is-gauge}
\end{equation}
The projector removes gauge directions. In particular,
\begin{equation}
 \Pi^T_{3/2}
 \bigl(\bar D_{\dot\alpha}L_\alpha
       -D_\alpha\bar L_{\dot\alpha}\bigr)=0.
 \label{eq:q11-projector-annihilates-gauge}
\end{equation}
Consequently,
\begin{equation}
 \boxed{\Pi^T_{3/2}J(\Psi)=0,
 \qquad S_{\rm mix}^{T}=0.}
 \label{eq:q11-Hpsi-projected-zero}
\end{equation}
This is an algebraic identity, without a field equation or gauge fixing.

The cancellation can also be checked without knowing the projector.
The two chiral curvatures used below are
\begin{equation}
 \begin{split}
 \mathcal W_{\alpha\beta\gamma}(V)
 &=-\frac14\bar D^2D_{(\alpha}
       \partial_{\beta}{}^{\dot\delta}
       V_{\gamma)\dot\delta},\\
 \mathcal F_{\alpha\beta}(\Psi)
 &=-\frac14\bar D^2D_{(\alpha}\Psi_{\beta)}.
 \end{split}
 \label{eq:q11-chiral-curvatures}
\end{equation}
Apply the first one to $V=J(\Psi)$, using
$\{D_\alpha,\bar D_{\dot\alpha}\}
=-2i\partial_{\alpha\dot\alpha}$. The two pieces give
\begin{align}
 \bar D^2D_{(\alpha}\partial_{\beta}{}^{\dot\delta}
       \bar D_{\dot\delta}\Psi_{\gamma)}
 &=-2i\,\partial_{(\beta}{}^{\dot\delta}
    \partial_{\alpha\dot\delta}\bar D^2\Psi_{\gamma)}=0,
 \label{eq:q11-direct-WJ-first}\\
 \bar D^2D_{(\alpha}\partial_{\beta}{}^{\dot\delta}
       D_{\gamma)}\bar\Psi_{\dot\delta}&=0.
 \label{eq:q11-direct-WJ-second}
\end{align}
The first zero follows because
$\partial_{\beta}{}^{\dot\delta}
\partial_{\alpha\dot\delta}\propto
\varepsilon_{\alpha\beta}\Box$ is antisymmetric in
$\alpha,\beta$. The second follows from
$D_{(\alpha}D_{\gamma)}=0$. We have used
$\bar D^2\bar D_{\dot\delta}=0$ in the first line. Equations~\eqref{eq:q11-direct-WJ-first} and
\eqref{eq:q11-direct-WJ-second} therefore give
$\mathcal W_{\alpha\beta\gamma}(J)=0$.

The same argument covers the real current with an arbitrary complex
phase,
\begin{equation}
 J^{(a)}_{\alpha\dot\alpha}
 =a\bar D_{\dot\alpha}\Psi_\alpha
  -\bar a D_\alpha\bar\Psi_{\dot\alpha}
 =\left.\delta H_{\alpha\dot\alpha}\right|_{L=a\Psi},
 \qquad a\in\mathbb C.
 \label{eq:q11-phase-generalization}
\end{equation}
Every gauge-invariant observable of the gravitational prepotential
that vanishes at $H=0$ vanishes on $H=J^{(a)}$ as well.

The conclusion concerns the specific current $J(\Psi)$.
It is not a no-go theorem for massive gravitini: the local five-dimensional
action already supplies such a mass, and a different four-dimensional
field content may include additional St\"uckelberg or compensator
superfields.

The curvatures $\mathcal W(H)$ and $\mathcal F(\Psi)$ in
Eq.~\eqref{eq:q11-chiral-curvatures} are both chiral and invariant
under the transformations in Eq.~\eqref{eq:q11-H-gauge-transformation}.
The lowest component of $\mathcal W(H)$ contains the first gravitino
curvature, and a spinor derivative of $\mathcal F(\Psi)$ contains the
second. Their coupling need not vanish.

\narrativeheading{The curvature operator and its component mass}

Take a real Abelian prepotential $V'$ and its chiral
field strength
\[
 \mathcal W'_\alpha=-\frac14\bar D^2D_\alpha V'.
\]
Its homogeneous auxiliary background is
\[
 \langle\mathcal W'^\alpha\rangle=\theta^\alpha D_0,
 \qquad \langle F'_{\mu\nu}\rangle=0.
\]
The auxiliary value $D_0$ supplies one spinor derivative in the
chiral integral and allows a fermionic curvature to couple to the
spinor descendant of the other. With a scalar momentum kernel
$\mathcal K$ we may write
\begin{equation}
 S_{D'}[\mathcal K]=\frac{ic_V}{2\Lambda_{\rm med}}
 \int d^4x\,d^2\theta\;
 \mathcal W'^\alpha
 \mathcal W_{\alpha\beta\gamma}(H)
 \mathcal K(\mathcal Q)
 \mathcal F^{\beta\gamma}(\Psi)
 +\mathrm{h.c.}
 \label{eq:q11-vector-spurion-operator}
\end{equation}
The coefficient $c_V$ is dimensionless, while $\Lambda_{\rm med}$ is the suppression
scale, and $\mathcal Q$ is the spectral operator with Euclidean
eigenvalue $q=p_E^2$. Throughout this question we use
$\eta_{\mu\nu}=\operatorname{diag}(-,+,+,+)$ and the
Wess--Bagger two-component spinor conventions. Thus, for a Fourier
mode $e^{ip_M\cdot x}$, $\Box\mapsto-p_M^2$; we take
$\mathcal Q=-\Box$ and write its Euclidean eigenvalue as
$q=p_E^2$. The continuation back to the mostly-plus Minkowski metric
is $q=p_M^2$, so physical mass poles lie at $q=-M^2$.
The kernel $\mathcal K=1/\mathcal Q$ has mass dimension $-2$.
With $R(H)=0$, $R(\Psi)=-1$ and
$R(\mathcal W')=+1$, the chiral integrand has the required $R$-charge
$+2$.

Why the inverse d'Alembertian? The answer is already contained in the
components of the curvatures. With the conventions of the spinor
contractions, we have
\begin{equation}
 \mathcal W_{\alpha\beta\gamma}(H)|=-C_{\alpha\beta\gamma}(\psi),
 \qquad
 D_{(\alpha}\mathcal F_{\beta\gamma)}(\Psi)|
 =iC_{\alpha\beta\gamma}(\rho),
 \label{eq:q11-curvature-descendants}
\end{equation}
where
$C_{\alpha\beta\gamma}(\psi)
=-2(\sigma^{\mu\nu})_{(\alpha\beta}
\partial_\mu\psi_{\nu\gamma)}$.
Thus each curvature carries one spacetime derivative. For opposite Fourier momenta $(-p_M,+p_M)$, the Wess--Bagger
raising convention used here gives, on the transverse and
gamma-traceless subspace,
\begin{equation}
 C_{\alpha\beta\gamma}(\psi;-p_M)
 C^{\alpha\beta\gamma}(\rho;+p_M)
 =-4p_M^2\,
 \psi^T_\mu(-p_M)\sigma^{\mu\nu}\rho^T_\nu(+p_M),
 \qquad q=p_M^2.
 \label{eq:q11-curvature-RS-identity}
\end{equation}
The minus sign belongs to the displayed lower--upper contraction,
with Fourier momenta $(-p_M,+p_M)$ kept in that order. Raising the
first spinor triple instead of the second reverses the sign:
\[
 C^{\alpha\beta\gamma}(\psi;-p_M)
 C_{\alpha\beta\gamma}(\rho;+p_M)
 =+4p_M^2\,\psi^T_\mu(-p_M)
   \sigma^{\mu\nu}\rho^T_\nu(+p_M).
\]
The calculation in Ref.~\cite{Benakli:2026operator} uses the opposite
Lorentzian metric signature, for which $q=-p_{M,\mathrm{there}}^2$.
With the mostly-plus metric used here, $q=p_M^2$. In both conventions
the lower--upper contraction is $-4q\,\psi\sigma\rho$ and
the upper--lower contraction is $+4q\,\psi\sigma\rho$.
Thus the signature change does not alter the $q$-dependent component
kernel. These identities hold on the transverse, gamma-traceless
sector at non-zero momentum, without imposing the massive
Rarita--Schwinger equation.

With $\mathcal K=1/\mathcal Q$, the two derivatives cancel the
factor $1/q$. Rephase the companion field, if necessary, so that the
mass bilinear has a positive coefficient. The operator then gives
\begin{equation}
 \left.\mathcal L_{D'}\right|_{3/2}
 =\frac{c_VD_0}{\Lambda_{\rm med}}
 \psi^T_\mu\sigma^{\mu\nu}\rho^T_\nu
 +\mathrm{h.c.}
 \label{eq:q11-vector-spurion-mass}
\end{equation}
For two canonically normalized Weyl gravitini the mass term is
$-2m_D\psi_\mu\sigma^{\mu\nu}\rho_\nu+\mathrm{h.c.}$.
Thus $m_D=|c_VD_0|/(2\Lambda_{\rm med})$. For diagonal kinetic
coefficients $Z_\psi$ and $Z_\rho$, canonical normalization gives
\[
 m_D=\frac{|c_VD_0|}
 {2\Lambda_{\rm med}\sqrt{Z_\psi Z_\rho}}.
\]
The factor $2$ in the definition of $m_D$ is essential. Later we
shall meet a crossed coefficient for collective boundary fields. It
is a coefficient in a non-local quadratic functional, not the pole
mass of a single gravitino.

\narrativeheading{The bosonic interaction}

The same superspace operator also contains a bosonic term. Define
\[
 f_{\beta\gamma}=\mathcal F_{\beta\gamma}(\Psi)|,
 \qquad
 t_{\beta\gamma}=D^\alpha
 \mathcal W_{\alpha\beta\gamma}(H)|.
\]
On the homogeneous $D_0$ background, the component expansion gives
\begin{equation}
 \left.\mathcal L_{D'}\right|_{\rm bos}
 =-\frac{ic_VD_0}{4\Lambda_{\rm med}}
 t_{\beta\gamma}\mathcal K(\mathcal Q)
 f^{\beta\gamma}+\mathrm{h.c.}
 \label{eq:q11-bosonic-curvature-coupling}
\end{equation}
This is a mixing between the trace descendant of the gravitational
curvature and the vector or auxiliary part of the companion multiplet.
In a conformal-supergravity component description,
$t_{\beta\gamma}$ contains the curl of an auxiliary $U(1)_R$
connection. Its coefficient depends on the normalization of that
auxiliary field; no numerical identification is needed to establish
the mixing in Eq.~\eqref{eq:q11-bosonic-curvature-coupling}.

One can see why the gravitational Weyl tensor does not enter.
Decompose
\begin{equation}
 D_\delta\mathcal W_{\alpha\beta\gamma}(H)|
 =C_{\delta\alpha\beta\gamma}
  +\frac34\varepsilon_{\delta(\alpha}t_{\beta\gamma)},
 \qquad
 C_{\delta\alpha\beta\gamma}
 =D_{(\delta}\mathcal W_{\alpha\beta\gamma)}(H)|.
 \label{eq:q11-weyl-trace-decomposition}
\end{equation}
The factor $3/4$ makes the epsilon trace equal to $t$. The completely
symmetric spinor $C_{\delta\alpha\beta\gamma}$ contains the
linearized graviton Weyl curvature. The $D$-spurion supplies one
$\theta$, and the gravitational supercurvature supplies another.
Since the two Grassmann coordinates are antisymmetric,
\begin{equation}
 \theta^\alpha\theta^\delta
 C_{\delta\alpha\beta\gamma}=0.
 \label{eq:q11-no-graviton-weyl-term}
\end{equation}
The trace term in Eq.~\eqref{eq:q11-weyl-trace-decomposition}
is the contribution to Eq.~\eqref{eq:q11-bosonic-curvature-coupling}.

The operator is bilinear in $H$ and $\Psi$, so it has no $hh$ term in
the first place. The useful result here is the identity of the
\emph{bosonic mixing}: it involves the auxiliary $U(1)_R$ connection,
not the graviton Weyl curvature. The masses and kinetic terms of the
remaining bosons require the complete quadratic action and elimination
of its auxiliary fields.

The first result can be read on its own. The prepotential coupling
is not gauge invariant and its transverse projection is zero,
whereas the curvature operator gives a non-zero Dirac mass.
The remaining calculations establish its normalization in the
five-dimensional theory. They are not needed for a first reading.

\narrativeheading{The complete KK matching}

The endpoint response contains the diagonal kinetic term as well
as the crossed bilinear. Their common normalization is needed to
compare the collective boundary variables with individual KK
mass eigenstates.

We start with one canonically normalized spin-$\tfrac32$ pair.
For four-component Majorana vector-spinors $\chi_{\mu A}$, we use
\begin{equation}
 \begin{split}
 \mathcal L_{\rm RS}[\chi_A]
 &=-\frac12\bar\chi_{\mu A}\gamma^{\mu\nu\rho}
    \partial_\nu\chi_{\rho A},\\
 \mathcal L_{\rm mass}
 &=-\frac12\mathcal M_{AB}\bar\chi_{\mu A}
    \gamma^{\mu\nu}\chi_{\nu B},\qquad
 \gamma^{\mu\nu}=\frac12[\gamma^\mu,\gamma^\nu].
 \end{split}
 \label{eq:q11-RS-four-component-normalization}
\end{equation}
The indices $A,B=1,2$ are summed, and
$\gamma^{\mu\nu\rho}$ denotes the normalized antisymmetrized
product of three gamma matrices. We use
$\{\gamma^\mu,\gamma^\nu\}=2\eta^{\mu\nu}$ and
$\eta=\operatorname{diag}(-,+,+,+)$. The factor $1/2$ is the
Majorana normalization. We follow the standard free-field
normalization discussed by Freedman and Van Proeyen.\footnote{D.~Z.~Freedman and A.~Van~Proeyen, \emph{Supergravity},
Cambridge University Press (2012), Chapter~5.}
For the two-component fields we use
$\sigma^{\mu\nu}=(\sigma^\mu\bar\sigma^\nu
-\sigma^\nu\bar\sigma^\mu)/4$ with Wess--Bagger index conventions.

The real Majorana basis has the mass matrix $m_D\mathbf1$.
The change of basis of Question~\ref{sec:ss-half-twist}
turns it into the off-diagonal matrix
of Eq.~\eqref{eq:q11-dirac-mass-matrix}. Translating to Weyl spinors
gives two equal crossed terms, hence the coefficient $2m_D$ in
Eq.~\eqref{eq:q11-RS-two-component-mass}.

On a transverse, gamma-traceless vector-spinor,
$\gamma^{\mu\nu\rho}p_\nu\chi_\rho=\slashed p\,\chi^\mu$
and $\gamma^{\mu\nu}\chi_\nu=-\chi^\mu$.
The first-order Euclidean operator for a single pair is
\begin{equation}
 \mathcal O_1(p_E)=i\slashed p_E\mathbf1_2+m\sigma_1,
 \qquad
 \mathcal O_1^{-1}(p_E)
 =\frac{-i\slashed p_E\mathbf1_2+m\sigma_1}{p_E^2+m^2}.
 \label{eq:q11-one-level-RS-test}
\end{equation}
The pole is at $p_E^2=-m^2$. The off-diagonal entry of this
first-order operator is $m$, although the associated crossed Weyl
bilinear has coefficient $2m$. The signed eigenvalues of $m\sigma_1$
are $\pm m$; for the lightest half-twisted level they agree with
Eq.~\eqref{eq:q11-half-spectral-pairing}.

The complete tower gives more information: the response of the bulk
field at the two ends of the interval.

Let $p=|p_E|$, $m_n=(n+\tfrac12)/R$ and $\ell=\pi R$.
At the half-twist, two sums are
\begin{equation}
 \sum_{n\in\mathbb Z}\frac1{p^2+m_n^2}
 =\frac\ell p\tanh(\ell p),\qquad
 \sum_{n\in\mathbb Z}\frac{(-1)^n m_n}{p^2+m_n^2}
 =\ell\,\operatorname{sech}(\ell p).
 \label{eq:q11-half-KK-sums}
\end{equation}
The first sum is absolutely convergent. The second is conditionally
convergent and is understood by pairing $n$ with $-n-1$, or by the
corresponding Abel limit. The unweighted mass-odd diagonal sum
$\sum_n m_n/(p^2+m_n^2)$ vanishes under symmetric principal-value
summation. The factor $(-1)^n$ in the crossed sum comes from the
relative endpoint wave-function signs.

For the transverse fermionic channel the unit-weight endpoint Green
matrix is
\begin{equation}
 G_\partial(p_E)=
 \begin{pmatrix}S&A\\A&S\end{pmatrix},\qquad
 S=-i\ell\frac{\not p_E}{p}\tanh(\ell p),\quad
 A=\ell\operatorname{sech}(\ell p),
 \label{eq:q11-endpoint-green}
\end{equation}
where Euclidean gamma matrices obey $\not p_E^2=p^2$.
It follows that $S^2-A^2=-\ell^2$, and hence
\begin{equation}
 K_\partial=G_\partial^{-1}
 =\frac1\ell\left[
 i\not p_E\frac{\tanh(\ell p)}p\,\mathbf1_2
 +\operatorname{sech}(\ell p)\,\sigma_1^{(\partial)}
 \right],
 \label{eq:q11-exact-endpoint-kernel}
\end{equation}
with $\sigma_1^{(\partial)}$ interchanging the two endpoints.
The crossed entry is
\begin{equation}
 K_{0\ell}(p)=\frac{\operatorname{sech}(\ell p)}\ell.
 \label{eq:q11-exact-crossed-kernel}
\end{equation}

The diagonal and crossed terms must be kept together when
interpreting this answer. In the first-order endpoint convention,
write
\begin{equation}
 K_\partial(p_E)
 =i\slashed p_E Z_\partial(q)\,\mathbf1_2
  +\mu_\partial(q)\sigma_1^{(\partial)},\qquad
 Z_\partial(q)=\frac{\tanh(\ell p)}{\ell p},\quad
 \mu_\partial(q)=\frac{\operatorname{sech}(\ell p)}\ell.
 \label{eq:q11-endpoint-diagonal-normalization}
\end{equation}
In particular $\mu_\partial=K_{0\ell}$, while
$Z_\partial(0)=1$. For one canonically normalized
Rarita--Schwinger pair, Eq.~\eqref{eq:q11-one-level-RS-test}
identified the off-diagonal first-order entry as the mass
parameter. Applying exactly that convention to the two entries
in Eq.~\eqref{eq:q11-endpoint-diagonal-normalization}
gives the momentum-dependent ratio
\begin{equation}
 m_\partial(q):=\frac{\mu_\partial(q)}{Z_\partial(q)}
 =\frac{p}{\sinh(\ell p)}.
 \label{eq:q11-endpoint-mass-ratio}
\end{equation}
At $q=0$, $m_\partial(0)=1/\ell$. This is the leading
coefficient of a derivative expansion of a collective response.
It is not a pole mass. The ratio has a finite analytic continuation at a KK pole,
$p=iM_k$ with $M_k=(k+\tfrac12)/R$, and gives
\begin{equation}
 m_\partial(-M_k^2)
 =\frac{M_k}{\sin(\ell M_k)}=(-1)^k M_k.
 \label{eq:q11-endpoint-on-pole-test}
\end{equation}
The sign alternates with the KK level and may be rephased.
The magnitude is the correct canonical mass. This tests the
normalization against the diagonal kinetic term, rather than
against the crossed coefficient alone.

The ratio in Eq.~\eqref{eq:q11-endpoint-on-pole-test} has a simple
interpretation: in the endpoint variables, the kinetic and mixed
terms share the same poles. Their ratio agrees with the signed mass
at each pole. The superfield calculation must reproduce both terms,
not just the crossed one.

For $|\ell p|<\pi/2$ one may expand
\[
 K_{0\ell}(p)=\frac1\ell\left[
 1-\frac{(\ell p)^2}{2}
 +\frac{5(\ell p)^4}{24}+O((\ell p)^6)\right].
\]
After $p\to iM$ the exact function has poles at
$M=(k+\tfrac12)/R$ for all $k\geq0$, with non-zero residues. A
finite rational KK sum cannot reproduce it exactly.

\narrativeheading{The curvature form of the boundary response}

The two boundary values originate from one five-dimensional
field. Two different linear spaces enter the calculation. The
fermionic component Hessian for one generic massive KK harmonic
has $76=32+8+32+4$ coefficients, of which $36$ are gauge
directions and $40$ are non-gauge coefficients. This is the
space on which the component operator and its generalized inverse
are constructed.

The boundary polarization uses a separate, redundant set of $80$
\emph{candidate source values} at the two ends of the interval.
The source count is not obtained by equating these values with the
$76$ bulk-field coefficients. On the left, the Ward condition is
$L(p)^Tj_0=0$ and its kernel has dimension $12$. On the right,
$B_\Omega(p)^Tj_\ell=0$ leaves $36$ directions. The unequal
numbers reflect the different endpoint source maps in this
polarization, including the larger, initially non-chiral right
curvature representative; they do not count physical gravitino
states. Four directions at each end have zero pairing with all
admissible sources. Consequently,
\begin{equation}
 \dim\mathcal J_{\rm adm}=12+36=48,\qquad
 \dim\operatorname{rad}\mathcal G=8,\qquad
 48-8=40=16+24.
 \label{eq:q11-source-quotient-count}
\end{equation}
The $16$ curvature coordinates consist of eight at each end;
the remaining $24$ are local-contact directions. The calculation
first uses the generalized inverse of the $76$-component bulk
Hessian, then imposes the endpoint Ward constraints and takes the
source quotient. This order is what removes the gauge-null
response without discarding the contact terms.

We already met the two functions generated by the KK tower. It is
convenient to name them
\begin{equation}
 f(q)=\frac{\tanh(\ell\sqrt q)}{\sqrt q},\qquad
 h(q)=\operatorname{sech}(\ell\sqrt q),\qquad
 qf(q)^2+h(q)^2=1.
 \label{eq:q11-fh-identity}
\end{equation}
The last relation is simply $\tanh^2x+\operatorname{sech}^2x=1$.
The diagonal and crossed curvature pairings are matrices
$\widehat C_{\rm d}(p)$ and $\widehat C_{\rm x}(p)$, acting on the
eight curvature components at an endpoint. With the source phases
and right-boundary orientation fixed as in the component reduction,
their two-point matrix is
\begin{equation}
 \mathcal C_{\rm curv}(p)=
 \begin{pmatrix}
 f(q)\widehat C_{\rm d}(p)&h(q)\widehat C_{\rm x}(p)\\
 h(q)\widehat C_{\rm x}(p)&f(q)\widehat C_{\rm d}(p)
 \end{pmatrix}.
 \label{eq:q11-curvature-covariance-16}
\end{equation}
The contact-source contributions cancel out of these four blocks.
At generic $q\ne0$ the component curvature matrices satisfy
\begin{equation}
 \widehat C_{\rm x}^{-1}\widehat C_{\rm d}
 =-q\widehat C_{\rm d}^{-1}\widehat C_{\rm x}.
 \label{eq:q11-curvature-matrix-relation}
\end{equation}
Combining this identity with Eq.~\eqref{eq:q11-fh-identity},
we obtain
\begin{equation}
 \mathcal K_{\rm curv}(p):=\mathcal C_{\rm curv}(p)^{-1}
 =\begin{pmatrix}
 qf(q)\widehat C_{\rm d}^{-1}&h(q)\widehat C_{\rm x}^{-1}\\
 h(q)\widehat C_{\rm x}^{-1}&qf(q)\widehat C_{\rm d}^{-1}
 \end{pmatrix}.
 \label{eq:q11-curvature-kernel-16}
\end{equation}
This is an equality of $16\times16$ matrices, not a replacement of
the full off-shell Hessian. The inverse is taken on the curvature
source quotient; the physical poles are found by continuation of
its two-point function.

\narrativeheading{The two superspace invariants}

We now look for $N=1$ expressions whose fermionic components give
Eq.~\eqref{eq:q11-curvature-kernel-16}. Both the diagonal term
and the crossed term must be present. The diagonal one is Hermitian
and involves a curvature and its conjugate. The crossed one is
holomorphic.

The left curvature is the ordinary chiral super-Weyl field. The
right curvature comes from a superfield $\mathcal R_\ell$ of the
five-dimensional theory that is not, in general, chiral in this
$N=1$ parametrization. On the fixed radion background let
\begin{equation}
 X_{\rm SS}=\theta^2F_E,\qquad
 \Xi=\frac{X_{\rm SS}}{F_E},\qquad
 \langle\Xi\rangle=\theta^2,\qquad F_E\ne0.
 \label{eq:q11-radion-spurion}
\end{equation}
For a curvature representative $\mathcal R_\ell^{(g)}$ in the
geometric normalization we define the chiral projection
\begin{equation}
 \mathcal W^{(g)}_{\ell\,\alpha\beta\gamma}
 =-\frac14\bar D^2
   (\bar\Xi\,\mathcal R^{(g)}_{\ell\,\alpha\beta\gamma}),
 \qquad
 \bar D_{\dot\alpha}\mathcal W^{(g)}_\ell=0,
 \qquad
 \mathcal W^{(g)}_\ell|=\mathcal R^{(g)}_\ell|.
 \label{eq:q11-right-chiral-projection}
\end{equation}
The final equality uses $-\bar D^2\bar\theta^2/4=1$ and holds
on this fixed background. It does not turn the original right
superfield into an unconstrained second chiral curvature.

We distinguish the geometric curvature variables from the variables
used to couple boundary sources. Let
\begin{equation}
 \begin{split}
 \mathcal R_0^{(s)}&=\mathcal W_0^{(s)}
   =c_0\mathcal W_0^{(g)},\\
 \mathcal R_\ell^{(s)}&=-c_\ell\mathcal R_\ell^{(g)},\qquad
 \widehat{\mathcal W}_\ell^{(s)}
 =-\frac14\bar D^2(\bar\Xi\mathcal R_\ell^{(s)})
 =-c_\ell\mathcal W_\ell^{(g)}.
 \end{split}
 \label{eq:q11-curvature-source-rescalings}
\end{equation}
The superscripts $(g)$ and $(s)$ mean geometric and source
normalization. In particular, the unnormalized chiral projection
$\mathcal W_\ell^{(g)}$ and the normalized
$\widehat{\mathcal W}_\ell^{(s)}$ are different objects. The real
minus sign is the chosen right-end source orientation: the odd
sine profiles give
$\Psi^2_\mu(x,\ell)=-\sqrt{2/\ell}\sum_k(-1)^k\rho_{\mu k}(x)$.
The component-to-superfield phase is fixed separately by their
kinetic and fifth-derivative terms.

Write $d^8z=d^4x\,d^2\theta\,d^2\bar\theta$ and
$d^6z=d^4x\,d^2\theta$. The two invariants are
\begin{equation}
 \begin{split}
 S_A^{\rm match}=2i\sum_{b\in\{0,\ell\}}\int d^8z\,
 \bar\Xi\Xi\,
 \mathcal R_{b\,\alpha\beta\gamma}^{(s)}
 \frac{f(q)}{q^2}\,
 \partial^{\alpha\dot\alpha}
 \partial^{\beta\dot\beta}
 \partial^{\gamma\dot\gamma}
 \overline{\mathcal R}_{b\,\dot\alpha\dot\beta\dot\gamma}^{(s)} .
 \end{split}
 \label{eq:q11-SA-diagonal}
\end{equation}
\begin{equation}
 S_B^{\rm match}=2\int d^6z\,\Xi\,
 \mathcal W_{0\,\alpha\beta\gamma}^{(s)}
 \frac{h(q)}q\,
 \widehat{\mathcal W}_\ell^{(s)\,\alpha\beta\gamma}
 +\mathrm{h.c.}
 \label{eq:q11-SB-crossed}
\end{equation}
In $S_A$ the three derivatives act on the conjugate curvature;
this is the Hermitian same-boundary kinetic term. The spurion
$\bar\Xi\Xi$ selects its lowest curvature components. The term
$S_B$ connects the two ends. Its spinor triples have the indicated
lowered and raised indices, in that order. The two normalizations
are fixed by the component calculation, not by choosing an
arbitrary coefficient for the mass term.

The spinor algebra behind this matching is worth recording. For
symmetric triples ordered as $(111,112,122,222)$, the contraction
of three antisymmetric epsilon tensors gives
\begin{equation}
 \mathsf E_3=
 \begin{pmatrix}
 0&0&0&1\\
 0&0&-3&0\\
 0&3&0&0\\
 -1&0&0&0
 \end{pmatrix}.
 \label{eq:q11-symmetric-triple-E3}
\end{equation}
The factors of $3$ count the ordered occurrences of the symmetric
mixed-index components. Replacing the epsilon contractions by the
appropriate momentum contractions produces a cubic matrix
$\mathsf P_3(p)$ for the diagonal term. After retaining the
opposite Fourier momenta, the Grassmann signs, and the factors
$2i$ and $2$ in Eqs.~\eqref{eq:q11-SA-diagonal} and
\eqref{eq:q11-SB-crossed}, their fermionic numerators are
\begin{equation}
 \begin{split}
 \mathsf D_A(p)&=q^3\operatorname{diag}
   (\widehat C_{\rm d}^{-1},\widehat C_{\rm d}^{-1}),\\
 \mathsf D_B(p)&=q\begin{pmatrix}
 0&\widehat C_{\rm x}^{-1}\\
 \widehat C_{\rm x}^{-1}&0
 \end{pmatrix}.
 \end{split}
 \label{eq:q11-graded-component-numerators}
\end{equation}
These are the polynomial numerators obtained from the graded
spinor contractions. In the diagonal term, the three displayed
spacetime derivatives give the symmetric-cube momentum map
$\mathsf P_3(p)$, which is cubic in $p_{\alpha\dot\alpha}$.
In the curvature basis $\widehat C_{\rm d}$ is itself cubic in
momentum. Thus $\widehat C_{\rm d}^{-1}$ has degree $-3$ and
$q^3\widehat C_{\rm d}^{-1}$ has degree $+3$, exactly that of
$\mathsf P_3$. The same source-matrix conventions give
$\widehat C_{\rm x}\sim q$ and hence
$q\widehat C_{\rm x}^{-1}$ of degree zero for the crossed
contraction. The component reduction fixes the coefficients as
well as these momentum degrees. The underlying matrices and
Grassmann contractions are given in the subsection
\emph{Component reduction of the superspace invariants} in the
appendix \emph{Endpoint Polarization and Fermionic Superspace
Matching} of Ref.~\cite{Benakli:2026operator}. Multiplying by the kernels in the two invariants gives
\begin{equation}
 \begin{split}
 \mathcal K^F_{A+B}(p)
 &=\frac{f(q)}{q^2}\mathsf D_A(p)
   +\frac{h(q)}q\mathsf D_B(p)\\
 &=\begin{pmatrix}
 qf(q)\widehat C_{\rm d}^{-1}&h(q)\widehat C_{\rm x}^{-1}\\
 h(q)\widehat C_{\rm x}^{-1}&qf(q)\widehat C_{\rm d}^{-1}
 \end{pmatrix}
 =\mathcal K_{\rm curv}(p).
 \end{split}
 \label{eq:q11-fermionic-16-match}
\end{equation}
This is the fermionic $16\times16$ matching obtained from the
local five-dimensional component system. The graded contractions
leading to Eq.~\eqref{eq:q11-graded-component-numerators} are
lengthy; Eq.~\eqref{eq:q11-fermionic-16-match} displays their
result and its comparison with the independently calculated
curvature kernel. Changing $2i$ in $S_A$ to $4i$, for example,
spoils the diagonal blocks, and reversing the sign of $S_B$
spoils the crossed blocks. The relative coefficient and sign are
therefore fixed within this two-invariant ansatz.

\narrativeheading{From curvature sources to canonical gravitini}

The endpoint matrix in Eq.~\eqref{eq:q11-exact-endpoint-kernel}
was written with unit weights for the signed KK modes. The
curvature-source action uses $c_0$ and $c_\ell$. We now compare
them with the canonical Rarita--Schwinger normalization.

Let $a_{0\ell}(q)$ be the coefficient of the crossed Weyl bilinear,
and $Z_b(q)$ the coefficient of the diagonal kinetic action for
$b\in\{0,\ell\}$, in the \emph{same} quadratic functional.
Eq.~\eqref{eq:q11-RS-two-component-mass} then gives
\begin{equation}
 m_{\rm curv}(q)
 =\frac{|a_{0\ell}(q)|}
 {2\sqrt{Z_0(q)Z_\ell(q)}}.
 \label{eq:q11-Weyl-mass-normalization}
\end{equation}
For real Euclidean $q\geq0$, and in the source conventions of
Eqs.~\eqref{eq:q11-SA-diagonal} and
\eqref{eq:q11-SB-crossed}, the component reductions
in Eq.~\eqref{eq:q11-fermionic-16-match} give the transverse
quadratic action in the form
\[
 \mathcal L_{\partial,\mathrm{TT}}^{(2)}
 =\sum_{b=0,\ell}Z_b(q)\mathcal L_{\rm RS}^{(0)}[\psi_b]
 +\bigl[a_{0\ell}(q)\psi_0^\mu
 \sigma_{\mu\nu}\psi_\ell^\nu+\mathrm{h.c.}\bigr],
\]
where $\mathcal L_{\rm RS}^{(0)}$ denotes the unit-coefficient
kinetic term. The computed coefficients are
\begin{equation}
 Z_b(q)=4|c_b|^2f(q),\quad b\in\{0,\ell\},
 \qquad |a_{0\ell}(q)|=8|c_0c_\ell|h(q),
 \qquad
 m_{\rm curv}(q)=\frac{h(q)}{f(q)}=m_\partial(q).
 \label{eq:q11-matched-curvature-kinetic-ratio}
\end{equation}
The factor $4$ comes from the curvature-to-Rarita--Schwinger
spinor contraction in the chosen normalization; the crossed
Weyl bilinear includes both symmetric off-diagonal entries, giving
the factor $8$. Both coefficients use the same
curvature-source maps. The absolute values in
Eq.~\eqref{eq:q11-matched-curvature-kinetic-ratio} refer to the
Euclidean normalization; the alternating signs at the KK poles
follow from analytic continuation of the signed ratio in
Eq.~\eqref{eq:q11-endpoint-mass-ratio}. They are not obtained by declaring the
crossed entry of the Green-function inverse to be a physical mass. The matrix calculation establishes the two fermionic blocks;
the conversion to the Weyl fields uses the same source maps and
the canonical Rarita--Schwinger normalization. This is a
fermionic result on the specified radion background, rather than
an off-shell reconstruction of the full multiplets.

For the geometric operator below the source normalization has
\begin{equation}
 |c_0c_\ell|=\frac{F_E}{8\pi}=\frac1{8\ell},
 \qquad F_E=\frac1R.
 \label{eq:q11-source-product}
\end{equation}
At $q=0$, since $f(0)=\ell$ and $h(0)=1$,
Eq.~\eqref{eq:q11-matched-curvature-kinetic-ratio} gives
\begin{equation}
 |a_{0\ell}(0)|=\frac1\ell,\qquad
 \sqrt{Z_0(0)Z_\ell(0)}=\frac12,\qquad
 |a^{\rm can}_{0\ell}(0)|=\frac2\ell.
 \label{eq:q11-boundary-bilinear-dictionary}
\end{equation}
The geometric mean depends only on $|c_0c_\ell|$. The
individual equalities $Z_0(0)=Z_\ell(0)=1/2$ would also require
$|c_0|=|c_\ell|$. The coefficient $2/\ell$ is the crossed
Weyl coefficient in canonical endpoint kinetic variables at
zero momentum. Its corresponding first-order entry is $1/\ell$.
Neither is the mass of one isolated mode. The quantities
\[
 K_{0\ell}(0)=\frac1{\pi R},\qquad
 m_{\rm light}=\frac1{2R}
\]
refer, respectively, to the zero-momentum endpoint response and
to the first KK pole. The relation among them is checked
independently by Eq.~\eqref{eq:q11-endpoint-on-pole-test}.

\narrativeheading{The Scherk--Schwarz curvature operator}

The geometric operator is more compact than the pair of matched
invariants. The normalized background is
$X_{\rm SS}=\theta^2F_E$ with $F_E=1/R$ at the half-twist.
This normalization is fixed by the component coupling to the
five-dimensional fifth derivative; it is not a new mass parameter.
With the geometric curvatures defined above, the transverse crossed
term is represented by
\begin{equation}
 S_{\rm SS}^{(3/2)}=-\frac1{4\pi}
 \int d^4x\,d^2\theta\;
 X_{\rm SS}\,
 \mathcal W_0^{(g)\,\alpha\beta\gamma}
 \frac{\operatorname{sech}(\ell\sqrt{\mathcal Q})}{\mathcal Q}
 \mathcal W^{(g)}_{\ell\,\alpha\beta\gamma}
 +\mathrm{h.c.}
 \label{eq:q11-SS-curvature-operator}
\end{equation}
The $1/\mathcal Q$ cancels the two derivatives in the
transverse gravitino curvatures, while $\operatorname{sech}$
is fixed by the entire KK tower. The sign depends on the sine-profile orientation and on the
position of the three raised spinor indices. In the conventions
used here, the latter gives the exact identity
\begin{equation}
 \mathcal W_{0\,\alpha\beta\gamma}^{(g)}
 \mathcal W_\ell^{(g)\,\alpha\beta\gamma}
 =(-1)^3\mathcal W_0^{(g)\,\alpha\beta\gamma}
 \mathcal W_{\ell\,\alpha\beta\gamma}^{(g)}
 =-\mathcal W_0^{(g)\,\alpha\beta\gamma}
 \mathcal W_{\ell\,\alpha\beta\gamma}^{(g)}.
 \label{eq:q11-triple-index-sign}
\end{equation}
Consequently, the source coefficient $-2c_0c_\ell=+F_E/(4\pi)$
in the first contraction becomes $-F_E/(4\pi)$ in the
geometric contraction of
Eq.~\eqref{eq:q11-SS-curvature-operator}. The matching condition is
\begin{equation}
 c_0c_\ell=-\frac{F_E}{8\pi}.
 \label{eq:q11-source-phase-product}
\end{equation}
The sign of $c_0c_\ell$ follows from the source rescaling and
the odd number of raised spinor indices in
Eq.~\eqref{eq:q11-triple-index-sign}; it is independent of the
choice of Lorentzian signature. To check the component sign in
our mostly-plus convention, the upper--lower contraction of
Eq.~\eqref{eq:q11-curvature-RS-identity} is $+4q$.
The factor $1/q$ in Eq.~\eqref{eq:q11-SS-curvature-operator}
therefore gives a crossed Weyl coefficient
$-F_E h(q)/\pi$ for the geometric right-end field
$\psi_{\ell,\mathrm{curv}}$. The endpoint-source phase is
$\psi_{\ell,\mathrm{bdry}}=-\psi_{\ell,\mathrm{curv}}$.
In that basis the coefficient is $+F_Eh(q)/\pi=h(q)/\ell$,
as required by Eq.~\eqref{eq:q11-exact-crossed-kernel}.
A rephasing changes the sign of an off-diagonal source correlator,
but not the pole positions or the absolute values of its residues.

The transverse component contraction of
Eq.~\eqref{eq:q11-SS-curvature-operator} gives
\begin{equation}
 |a_{0\ell}(q)|=\frac{F_E}{\pi}
 \operatorname{sech}(\ell\sqrt q)
 =\frac{h(q)}\ell
 =K_{0\ell}(q),\qquad F_E=\frac1R.
 \label{eq:q11-SS-coefficient-dictionary}
\end{equation}
It agrees with the crossed endpoint coefficient in the
chosen source convention. Equation~\eqref{eq:q11-SA-diagonal}
provides the diagonal invariant needed to compare this with
the canonical kinetic normalization. The full curvature
matching is Eq.~\eqref{eq:q11-fermionic-16-match}.

There is also a check relating the vector-background operator
of Eq.~\eqref{eq:q11-vector-spurion-operator} to the geometric
curvature operator. On fixed backgrounds
$\mathcal W'^\alpha=\theta^\alpha D_0$ and
$X=\theta^2F_E$, let $I_D[\mathcal K]$ denote the chiral
integral in Eq.~\eqref{eq:q11-vector-spurion-operator} without
its prefactor. Define
\[
 \begin{split}
 I_X[\mathcal K]&=\int d^4x\,d^2\theta\,X\,
  \mathcal W_0^{(g)\,\alpha\beta\gamma}\mathcal K
  \mathcal W^{(g)}_{\ell\,\alpha\beta\gamma},\\
 B[\mathcal K]&=\int d^4x\,
 (D^\alpha\mathcal W_{0\,\alpha\beta\gamma}^{(g)})|
 \mathcal K\mathcal F^{\beta\gamma}|.
 \end{split}
\]
The first operation is a graded integration by parts:
\[
 I_D[\mathcal K]=\frac{D_0}{2}\int d^4x\,
 \left[\mathcal W_{0\,\alpha\beta\gamma}^{(g)}|
 \mathcal K D^\alpha\mathcal F^{\beta\gamma}|
 -(D^\alpha\mathcal W_{0\,\alpha\beta\gamma}^{(g)})|
 \mathcal K\mathcal F^{\beta\gamma}|\right].
\]
The second uses the component identification, including the
right-end phase,
\[
 D_{(\alpha}\mathcal F_{\beta\gamma)}|
 =iC_{\alpha\beta\gamma}(\rho)
 =-i\mathcal W^{(g)}_{\ell\,\alpha\beta\gamma}|.
\]
Only after this identification do we obtain
\begin{equation}
 I_D[\mathcal K]
 =\frac{iD_0}{2F_E}I_X[\mathcal K]
   -\frac{D_0}{2}B[\mathcal K].
 \label{eq:q11-two-curvature-identity}
\end{equation}
The remainder $B$ has bosonic lowest components. Therefore
the two routes yield the same quadratic fermionic bilinear
when their coefficients and phases are matched. The equation
is a fixed-background component identity, not an equality of
unconstrained off-shell superspace functionals.

\narrativeheading{What the fermionic matching leaves undetermined}

In the radion-spurion convention of this five-dimensional
reduction, the fifth-derivative coupling is parametrized by
$F_E=2\omega/R$. At $\omega=\tfrac12$ this gives
$F_E=1/R=2m_{\rm light}$. Once the radion-spurion convention is
fixed, this value is set by the KK mass spectrum; the separate
normalizations $c_0$ and $c_\ell$ of the boundary curvature
sources were fixed by the matching above. Determining the
dynamical radion kinetic term requires the independent
radion--compensator reduction and is not part of the transverse
fermionic curvature matching.

It is possible to change the bosonic action without changing the
$16\times16$ fermionic curvature kernel. For example, the bosonic
supercurvature
\[
 \mathcal B_{\alpha\beta\gamma\delta}
 =D_{(\alpha}\mathcal W_{\beta\gamma\delta)}(H)
\]
allows the invariant
\begin{equation}
 \Delta S=\lambda\int d^8z\,\bar\Xi\Xi\,
 \mathcal B_{\alpha\beta\gamma\delta}
 \mathcal B^{\alpha\beta\gamma\delta}+\mathrm{h.c.}
 \label{eq:q11-bosonic-ambiguity}
\end{equation}
Its selected lowest component is bosonic. On the fixed background,
its component projection leaves Eq.~\eqref{eq:q11-fermionic-16-match}
unchanged while changing the bosonic Hessian. Thus the fermionic
matching is exact on its stated curvature sector, but it does not
uniquely specify the bosonic and auxiliary completion.

The $24$ contact directions in Eq.~\eqref{eq:q11-source-quotient-count}
provide another part of the reconstruction problem. The complete
quadratic boundary functional separates into curvature and contact
coordinates only after the appropriate mixed terms are included.
Its local contact subtraction depends on the chosen prescription;
the nonlocal curvature response does not. An off-shell completion
must reproduce the full bosonic, auxiliary and contact responses,
including the longitudinal constraints of the parent theory.

\narrativeheading{Bibliographical remarks}

The spectral pairing and the $U(1)_R$ protection of the Dirac
mass are discussed by Benakli and Goudeau in
\emph{(Pseudo-)Dirac Gravitinos}~\cite{Benakli:2026zwx}.
The local five-dimensional reduction, the curvature-source
matrices and the superspace matching discussed here are developed
in Ref.~\cite{Benakli:2026operator},
\emph{Anti-periodic Gravitino and Dirac Effective Operators}
(arXiv:2611.16047).
The two invariants $S_A^{\rm match}$ and $S_B^{\rm match}$ reproduce
its complete quadratic fermionic curvature kernel on the fixed
radion background. Their fermionic matching does not determine
all bosonic, auxiliary and contact couplings.
The linearized supergravity superfield formulation and its
five-dimensional extensions were developed in
Refs.~\cite{Linch:2002wg,Becker:2019ywd}. These provide formal
background for the superspace construction considered here.

\begin{takehome}
\begin{takehomeitems}

\item The current $J_{\alpha\dot\alpha}=\bar D_{\dot\alpha}\Psi_\alpha
-D_\alpha\bar\Psi_{\dot\alpha}$ is a pure gauge variation of $H$.
The direct $H\Psi$ term is not gauge invariant before projection,
and its transverse part vanishes: $\Pi^T_{3/2}J=0$.

\item A non-zero transverse Dirac mass is represented by a bilinear
of the gauge-invariant curvatures of the gravity and companion
spin-$\tfrac32$ multiplets. The inverse d'Alembertian removes
the extra derivatives carried by these curvatures.

\item The complete half-twisted KK tower gives
$K_{0\ell}(p)=\operatorname{sech}(\ell p)/\ell$. Its ratio to the
diagonal kinetic coefficient is $p/\sinh(\ell p)$ and reproduces
the mass $(-1)^k(k+\tfrac12)/R$ at each KK pole.

\item The Hermitian invariant $S_A^{\rm match}$ and the chiral
crossed invariant $S_B^{\rm match}$ reproduce the full
$16\times16$ fermionic curvature kernel on the fixed radion
background, with the normalization of sources and both kinetic
and mass terms included.

\item The fermionic $16\times16$ matching is exact on the fixed
radion background, but it does not uniquely fix the bosonic,
auxiliary and local contact terms of a complete off-shell action.

\end{takehomeitems}
\end{takehome}

\nextquestion
{How does Scherk--Schwarz compactification appear as a gauging in the
lower-dimensional supergravity?}

\end{reviewquestion}

% Question 13 v3, renumbered from Question12v3.tex.
% Original section/equation labels retained for cross-reference stability.
%%%%%%%%%%%%%%%%%%%%%%%%%%%%%%%%%%%%%%%%%%%%%%%%%%%%%%%%%%%%%%%

\begin{reviewquestion}
{How does Scherk--Schwarz compactification appear as a gauging in the
lower-dimensional supergravity?}
\label{sec:ss-gauging}

\begin{essentialidea}

A Scherk--Schwarz reduction does more than shift Kaluza--Klein masses.
It tells a gauge symmetry inherited from the compactification how to
act on the lower-dimensional fields.

For an ordinary circle reduction, local translations of the compact
coordinate are higher-dimensional diffeomorphisms. In the
lower-dimensional theory they are gauged by the Kaluza--Klein vector
\(A_\mu^0\). If the Scherk--Schwarz ansatz contains a generator
\(\mathcal M\) of a global symmetry of the parent theory, the same
Kaluza--Klein gauge transformation acts on the reduced fields through
\(\mathcal M\). This is the simplest rank-one gauging.

More generally, the embedding tensor specifies which global-symmetry
generator is gauged by each vector field,
\[
\Theta:
\mathcal E_{\rm vec}
\longrightarrow
\mathfrak g .
\]
Its image is initially only a vector subspace of the global symmetry
algebra \(\mathfrak g\). A quadratic constraint requires the chosen
generators to close consistently under gauge transformations, while a
separate linear constraint keeps only the components of \(\Theta\)
allowed by supersymmetry.

The constant embedding tensor is converted into the
scalar-dependent quantities entering the supergravity transformations
by dressing it with the scalar fields. The resulting \(T\)-tensor
determines the fermion shifts, mass matrices and, through supersymmetry
Ward identities, the scalar potential.

The same language extends beyond the elementary circle. A geometric
Scherk--Schwarz reduction can be organized by a constant Lie algebra
of internal frames. In double and exceptional field theory the
ordinary frame is replaced by a generalized frame, and the resulting
constant algebra is naturally Leibniz rather than Lie. Its
coefficients encode the lower-dimensional gauging.

A consistent gauging does not explicitly destroy local supersymmetry.
Supersymmetry breaking is a property of a particular bosonic
background and is decided by the simultaneous vanishing of all
fermionic supersymmetry variations.

\end{essentialidea}

\narrativeheading{The circle reduction as a rank-one gauging}

The simplest circle reduction already contains the essential
mechanism. The compact coordinate can be shifted locally in the
non-compact spacetime, and the mixed metric component becomes the
corresponding gauge field. A Scherk--Schwarz ansatz then determines
how this gauge transformation acts on the other reduced fields.

Let the compact coordinate be dimensionless,
\(\theta\sim\theta+2\pi\), and let \(G\) be a continuous symmetry
group of the higher-dimensional theory, with
\(\mathfrak g=\operatorname{Lie}(G)\). Choose
\(\mathcal M\in\mathfrak g\) and take
\begin{equation}
\Phi(x,\theta)
=
e^{\theta\mathcal M}\phi(x).
\label{eq:q12-SS-ansatz}
\end{equation}
Here \(\phi(x)\) is independent of the compact coordinate. We retain
only this field for clarity; the complete Kaluza--Klein tower is
obtained by multiplying the ansatz by the periodic factors
\(e^{in\theta}\), \(n\in\mathbb Z\), without changing the gauging
mechanism.

Going once around the circle gives
\begin{equation}
\Phi(x,\theta+2\pi)
=
U_{\rm SS}\Phi(x,\theta),
\qquad
U_{\rm SS}
=
e^{2\pi\mathcal M}.
\label{eq:q12-circle-monodromy}
\end{equation}
For the one-parameter convention used in the preceding questions,
\(\mathcal M=i\omega\mathbb R_\Phi\), so
\(U_{\rm SS}=e^{2\pi i\omega\mathbb R_\Phi}\).

The gauging itself comes from the gravitational sector. Ignoring the
radion-dependent Weyl factors, write
\begin{equation}
ds^2
=
g_{\mu\nu}(x)\,dx^\mu dx^\nu
+
R^2
\left(
d\theta+A_\mu^0(x)dx^\mu
\right)^2.
\label{eq:q12-KK-metric}
\end{equation}
The mixed metric component gives the Kaluza--Klein vector
\(A_\mu^0\). Under the higher-dimensional diffeomorphism
\(\theta\to\theta+\lambda(x)\), one has
\(d\theta\to d\theta+d\lambda\), so the metric keeps the form
Eq.~\eqref{eq:q12-KK-metric} provided
\begin{equation}
A_\mu^0
\longrightarrow
A_\mu^0-\partial_\mu\lambda.
\label{eq:q12-KK-vector-transformation}
\end{equation}

The Scherk--Schwarz ansatz now tells us how the same transformation
acts on \(\phi\). Since
\[
e^{\theta\mathcal M}\phi(x)
=
e^{(\theta+\lambda)\mathcal M}
e^{-\lambda\mathcal M}\phi(x),
\]
the lower-dimensional field transforms as
\begin{equation}
\phi(x)
\longrightarrow
e^{-\lambda(x)\mathcal M}\phi(x).
\label{eq:q12-lower-field-transformation}
\end{equation}
Its covariant derivative is therefore
\begin{equation}
D_\mu\phi
=
\left(
\partial_\mu-A_\mu^0\mathcal M
\right)\phi,
\label{eq:q12-simple-covariant-derivative}
\end{equation}
which indeed transforms in the same way as \(\phi\).

If \(t_{\rm KK}\) denotes the Abelian generator associated with circle
translations, the reduction has therefore defined the map
\begin{equation}
t_{\rm KK}
\longmapsto
\mathcal M\in\mathfrak g.
\label{eq:q12-rank-one-embedding}
\end{equation}
Only one independent direction of \(\mathfrak g\) participates. This
is the simplest rank-one Scherk--Schwarz gauging.

\narrativeheading{The embedding tensor as a linear map}

The circle example suggests the general question. A
lower-dimensional supergravity can contain several vector fields, and
each of them may gauge a different generator, or a different linear
combination of generators, of the global symmetry group. The
embedding tensor is the object that records this assignment.

Let
\[
A_\mu^M,
\qquad
M=1,\ldots,n_{\rm v},
\]
be the vector gauge potentials of the ungauged theory. Depending on
the compactification, they can include Kaluza--Klein vectors, vectors
descending from higher-dimensional gauge fields, and vectors already
belonging to lower-dimensional supergravity multiplets. The label \(M\) enumerates these fields. Associate with them the
finite-dimensional vector space
\[
\mathcal E_{\rm vec}
=
\operatorname{span}\{e_M\},
\]
where the abstract basis element \(e_M\) corresponds to \(A_\mu^M\).

Let \(G\) now denote the global symmetry group of the ungauged
lower-dimensional theory and
\(\mathfrak g=\operatorname{Lie}(G)\). Choose a basis \(t_\alpha\) of
\(\mathfrak g\), with
\begin{equation}
[t_\alpha,t_\beta]
=
f_{\alpha\beta}{}^\gamma t_\gamma .
\label{eq:q12-global-algebra}
\end{equation}
A gauging specifies which element of \(\mathfrak g\) is gauged by
each vector \(A_\mu^M\). This information is collected in the linear
map
\begin{equation}
\Theta:
\mathcal E_{\rm vec}
\longrightarrow
\mathfrak g,
\label{eq:q12-theta-map}
\end{equation}
defined by
\begin{equation}
\Theta(e_M)
=
X_M
=
\Theta_M{}^\alpha t_\alpha.
\label{eq:q12-XM}
\end{equation}
Thus \(X_M\) is the global-symmetry generator gauged by
\(A_\mu^M\), while \(\Theta_M{}^\alpha\) are the components of the
embedding tensor.

Before any consistency condition is imposed,
\[
\operatorname{Im}\Theta
=
\operatorname{span}\{X_M\}
\subseteq
\mathfrak g
\]
is only a vector subspace. There is no reason yet for it to be closed
under commutators, nor for an arbitrary choice of \(\Theta\) to be
compatible with supersymmetry.

When the gauging is consistent, the gauge-covariant derivative takes
the form
\begin{equation}
D_\mu
=
\partial_\mu
-
g_{\rm gau}A_\mu^M X_M,
\label{eq:q12-general-covariant-derivative}
\end{equation}
where \(g_{\rm gau}\) may alternatively be absorbed into the
normalization of \(\Theta\). For the circle reduction above, only the
Kaluza--Klein vector participates. If it is labelled by \(M=0\), one
can choose the normalization so that
\[
g_{\rm gau}X_0=\mathcal M,
\qquad
X_{\widehat M}=0
\]
for all remaining vector directions \(\widehat M\). Hence
\(\operatorname{rank}\Theta=1\).

\narrativeheading{The quadratic constraint: closure of the gauged generators}

The basic consistency question is simple. Once a set of generators
has been chosen to be gauged, a gauge transformation generated by one
of them must not take us outside that same set. The embedding tensor
must therefore be invariant under the gauge transformations that it
itself defines.

The global symmetry group can act on the vector fields. Let
\((t_\alpha)_N{}^P\) denote the matrix representing \(t_\alpha\) on
\(\mathcal E_{\rm vec}\),
\begin{equation}
t_\alpha:
e_N
\longmapsto
(t_\alpha)_N{}^P e_P.
\label{eq:q12-vector-representation}
\end{equation}
The generator \(X_M=\Theta_M{}^\alpha t_\alpha\) then acts on the
vector representation through
\begin{equation}
X_{MN}{}^P
\equiv
\Theta_M{}^\alpha
(t_\alpha)_N{}^P.
\label{eq:q12-XMN}
\end{equation}
The coefficient \(X_{MN}{}^P\) tells us how the gauge transformation
associated with \(A_\mu^M\) acts on the vector direction labelled by
\(N\).

To impose gauge invariance, temporarily regard \(\Theta\) as a tensor
transforming under \(G\). Since \(\Theta_N{}^\alpha\) carries both a
vector index \(N\) and an adjoint index \(\alpha\), the generator
\(X_M\) acts on both. With the conventions used here,
\begin{equation}
\delta_M\Theta_N{}^\alpha
=
X_{MN}{}^P\Theta_P{}^\alpha
+
\Theta_M{}^\beta
\Theta_N{}^\gamma
f_{\beta\gamma}{}^\alpha .
\label{eq:q12-delta-theta}
\end{equation}
The first term is the action on the vector index, while the second is
the adjoint action.
Gauge invariance requires
\(\delta_M\Theta_N{}^\alpha=0\), hence
\begin{equation}
\Theta_M{}^\beta
\Theta_N{}^\gamma
f_{\beta\gamma}{}^\alpha
+
X_{MN}{}^P
\Theta_P{}^\alpha
=
0.
\label{eq:q12-quadratic-components}
\end{equation}
This is the quadratic constraint. Both terms are quadratic in
\(\Theta\), since \(X_{MN}{}^P\) is itself linear in the embedding
tensor. The relative signs depend on the conventions chosen for the
generators and the covariant derivative; here they match
Eq.~\eqref{eq:q12-general-covariant-derivative}. Multiplying Eq.~\eqref{eq:q12-quadratic-components} by \(t_\alpha\)
gives
\begin{equation}
[X_M,X_N]
=
-
X_{MN}{}^P X_P.
\label{eq:q12-gauge-closure}
\end{equation}
Thus
\begin{equation}
[
\operatorname{Im}\Theta,
\operatorname{Im}\Theta
]
\subseteq
\operatorname{Im}\Theta.
\label{eq:q12-image-closed}
\end{equation}
Only after this condition is imposed can one identify
\[
\mathfrak g_{\rm gauge}
=
\operatorname{Im}\Theta
\subseteq\mathfrak g
\]
as the gauge subalgebra.

There is one subtlety. The coefficients \(X_{MN}{}^P\) need not be
antisymmetric in \(M,N\), even though
Eq.~\eqref{eq:q12-gauge-closure} is a commutator. Symmetrizing gives
\begin{equation}
X_{(MN)}{}^P X_P=0,
\label{eq:q12-symmetric-trivial}
\end{equation}
where
\(X_{(MN)}{}^P=\tfrac12(X_{MN}{}^P+X_{NM}{}^P)\). This does not imply
\(X_{(MN)}{}^P=0\), because the generators \(X_P\) need not be
linearly independent. Thus \(X_{MN}{}^P\) should not in general be
identified directly with the structure constants of the gauge
algebra. The symmetric part will reappear naturally in the
generalized Scherk--Schwarz description below.

\narrativeheading{The linear constraint: compatibility with supersymmetry}

Closure is not enough in a supergravity theory. Even a perfectly good
Lie subalgebra may fail to define an admissible supersymmetric gauging
because the corresponding couplings do not fit into the allowed
supergravity multiplets. The linear constraint selects precisely the
representation content that supersymmetry permits.

The embedding tensor is a linear map
\[
\Theta:
\mathcal E_{\rm vec}
\longrightarrow
\mathfrak g .
\]
It therefore belongs to the vector space of all such linear maps,
\[
\Theta
\in
\operatorname{Hom}
\left(
\mathcal E_{\rm vec},
\mathfrak g
\right),
\]
where \(\operatorname{Hom}(V,W)\) denotes the vector space of linear
maps from \(V\) to \(W\). For finite-dimensional vector spaces,
\[
\operatorname{Hom}
\left(
\mathcal E_{\rm vec},
\mathfrak g
\right)
\simeq
\mathcal E_{\rm vec}^{\,*}\otimes\mathfrak g.
\]
Hence
\begin{equation}
\Theta
\in
\mathcal E_{\rm vec}^{\,*}\otimes\mathfrak g.
\label{eq:q12-unconstrained-theta}
\end{equation}
In components, this is simply the statement that
\(\Theta_M{}^\alpha\) carries one index in the dual vector
representation and one in the adjoint representation of
\(\mathfrak g\).

The tensor product
\(\mathcal E_{\rm vec}^{\,*}\otimes\mathfrak g\) decomposes into
irreducible representations of \(G\). Supersymmetry allows only
particular components of this decomposition. Denote their sum by
\[
\mathcal R_\Theta
\subset
\mathcal E_{\rm vec}^{\,*}\otimes\mathfrak g.
\]
The linear constraint is
\begin{equation}
\Theta
\in
\mathcal R_\Theta,
\label{eq:q12-linear-representation}
\end{equation}
or, if \(\mathbb P_{\rm forbidden}\) projects onto the excluded
representations,
\begin{equation}
\mathbb P_{\rm forbidden}\Theta=0.
\label{eq:q12-linear-constraint}
\end{equation}

A standard example is maximal supergravity in four dimensions. Its
global symmetry group is \(E_{7(7)}\), the vectors transform in the
\(\mathbf{56}\), and the generators lie in the adjoint
\(\mathbf{133}\). Before imposing supersymmetry,
\[
\mathbf{56}\otimes\mathbf{133}
=
\mathbf{56}
\oplus
\mathbf{912}
\oplus
\mathbf{6480}.
\]
The linear constraint removes the \(\mathbf{56}\) and
\(\mathbf{6480}\), leaving
\[
\Theta\in\mathbf{912}.
\]
This example makes the distinction between the two constraints
explicit: the quadratic constraint asks whether the selected
generators form a consistent gauging, while the linear constraint
asks whether that gauging has the representation content permitted by
the supersymmetric theory. In formulations containing both electric
and magnetic potentials, additional locality conditions can also
arise.

\narrativeheading{Dressing the gauging: the \(T\)-tensor}

The embedding tensor specifies the gauging in a \(G\)-covariant way
and is independent of the scalar fields. The fermion shifts, however,
depend on the scalar configuration and are naturally written in
representations of the local isotropy group. The \(T\)-tensor is the
same gauging expressed in this scalar-dependent local frame.

Suppose, for definiteness, that the scalar manifold is locally a
coset
\[
\mathcal M_{\rm scalar}=G/H,
\]
where \(H\) is the isotropy subgroup of \(G\), and let
\(\phi^i\), \(i=1,\ldots,n_{\rm s}\), be local scalar coordinates. A
configuration \(\phi\) specifies a point of \(G/H\). To represent this
point by an element transforming under \(G\), choose a coset
representative
\[
\mathcal V(\phi)\in G.
\]
This is not an additional scalar degree of freedom. It is a choice of
element of \(G\) projecting to the point \(\phi\in G/H\).

The representative is not unique. With our convention,
\begin{equation}
\mathcal V(\phi)
\longrightarrow
\mathcal V(\phi)\,h(\phi),
\qquad
h(\phi)\in H,
\label{eq:q12-coset-representative}
\end{equation}
describes the same scalar point. This is the local \(H\) redundancy
of the coset description.

The usefulness of \(\mathcal V(\phi)\) is that it converts between two
natural descriptions of the theory. The embedding tensor carries
indices transforming under the global group \(G\), whereas the
fermions and their supersymmetry transformations are organized in
representations of the local group \(H\). Dressing the \(G\)-indices
of \(\Theta\) with the coset representative gives the conventional
\(T\)-tensor. To avoid confusion with the radion superfield \(T\) of
Question~\ref{sec:ss-wilson-radion}, we denote it by
\begin{equation}
\mathscr T(\phi)
=
\mathcal V(\phi)^{-1}\cdot\Theta,
\label{eq:q12-T-tensor}
\end{equation}
where the dot denotes the induced action on all relevant
\(G\)-indices.
No new gauging data are introduced. The tensor \(\Theta\) is
constant, while \(\mathscr T(\phi)\) is the same gauging expressed in
the \(H\)-covariant frame appropriate to the scalar point \(\phi\).
Its scalar dependence comes entirely from \(\mathcal V(\phi)\).

The \(T\)-tensor can be decomposed into irreducible representations of
\(H\). Particular components give the fermionic shift matrices,
schematically
\begin{equation}
S_{AB}(\phi)
=
\mathbb P_S\,\mathscr T(\phi),
\qquad
N^I{}_A(\phi)
=
\mathbb P_N\,\mathscr T(\phi),
\label{eq:q12-T-projections}
\end{equation}
where \(\mathbb P_S\) and \(\mathbb P_N\) are the
theory-dependent projections appropriate to the gravitini and
spin-\(\tfrac12\) fermions.

These quantities are called fermion shifts because they appear
without derivatives of the supersymmetry parameter:
\begin{align}
\delta\psi_{\mu A}
&=
\mathcal D_\mu\epsilon_A
+
g_{\rm gau}S_{AB}(\phi)\gamma_\mu\epsilon^B
+\cdots,
\nonumber\\
\delta\chi^I
&=
\cdots
+
g_{\rm gau}N^I{}_A(\phi)\epsilon^A.
\label{eq:q12-fermion-shifts}
\end{align}
The group-theoretic role of the \(T\)-tensor is thus to turn the
constant gauging data into scalar-dependent quantities entering the
fermionic transformations.

\narrativeheading{The scalar potential and the supersymmetry Ward identity}

Once the gauging has generated fermion shifts, local supersymmetry
does not allow the scalar potential to be chosen independently. The
variation of the shift and fermion-mass terms must be cancelled by
bosonic terms in the action, and this fixes the potential in terms of
the same gauging data.

The resulting supersymmetry Ward identity has the schematic form
\begin{equation}
V_{\rm gau}(\phi)\,\delta_A{}^B
=
g_{\rm gau}^2
\left[
\sum_I
c_I\,
N^I{}_A(\phi)\,
\overline N_I{}^B(\phi)
-
c_\psi\,
S_{AC}(\phi)\,
\overline S^{BC}(\phi)
+\cdots
\right].
\label{eq:q12-Ward-identity}
\end{equation}
The coefficients \(c_I\), \(c_\psi\), and the possible additional
shift matrices depend on the supergravity theory and on conventions.
The important structural feature is the relative sign between the
spin-\(\tfrac12\) and gravitino contributions.

Since the shift matrices are linear in the dressed gauging data, the
potential is quadratic in the embedding tensor. If the gauge coupling
is absorbed into \(\Theta\), the explicit factor
\(g_{\rm gau}^2\) in Eq.~\eqref{eq:q12-Ward-identity} disappears.

In four-dimensional \(\mathcal N=1\) supergravity the familiar
expression
\begin{equation}
V
=
F^i g_{i\bar j}\overline F^{\bar j}
+
\frac12 D^aD_a
-
3|m_{3/2}|^2
\label{eq:q12-N1-Ward}
\end{equation}
provides a simple realization of the same structure. The \(F\)- and
\(D\)-terms provide the positive contributions, while the gravitino
shift gives the negative term. This correlation between the
gravitino mass, the remaining fermion shifts and the scalar potential
will be central to the no-scale discussion in the next question.

\narrativeheading{Geometric Scherk--Schwarz reduction and frame algebras}

The circle is the simplest geometric example because its invariant
one-form \(d\theta\) is closed. A non-Abelian group manifold gives
the next natural step: its invariant one-forms rotate into one
another. For instance, on \(SU(2)\simeq S^3\), the left-invariant
one-forms satisfy Maurer--Cartan relations with constant coefficients.

More generally, suppose that the internal space admits a globally
defined coframe \(\eta^a\), \(a=1,\ldots,n_{\rm int}\), obeying
\begin{equation}
d\eta^a
=
-\frac12
f_{bc}{}^a\,
\eta^b\wedge\eta^c,
\qquad
f_{bc}{}^a=\text{constant}.
\label{eq:q12-frame-algebra}
\end{equation}
The constants \(f_{bc}{}^a\) measure the failure of the coframe to be
closed and encode how the internal frame rotates over the compact
space. The circle corresponds to the trivial case \(f_{bc}{}^a=0\).

Their constancy is essential: it allows the reduced theory to inherit
constant algebraic data rather than arbitrary functions of the
internal coordinates. Applying \(d\) once more and using \(d^2=0\)
gives
\begin{equation}
f_{[ab}{}^d f_{c]d}{}^e=0,
\label{eq:q12-Jacobi}
\end{equation}
which is the Jacobi identity. If \(E_a\) denotes the dual frame,
\(\eta^a(E_b)=\delta^a{}_b\), then
\[
[E_a,E_b]
=
f_{ab}{}^cE_c.
\]
The internal frame therefore defines a Lie algebra, and the constants
\(f_{ab}{}^c\) descend as part of the lower-dimensional gauging data.

This is only the local algebraic part of the compactification. Let
\(G_{\rm int}\) be the Lie group integrating the frame algebra. A
compact space may be obtained by a quotient
\[
K=\Gamma\backslash G_{\rm int},
\]
with \(\Gamma\) a suitable discrete subgroup, but such a cocompact
lattice need not exist. In particular, its existence requires
unimodularity,
\begin{equation}
f_{ab}{}^b=0,
\label{eq:q12-unimodularity}
\end{equation}
which is necessary but not sufficient in general. The Jacobi identity
therefore guarantees consistency of the local frame algebra, not the
existence of a compact global realization.

\narrativeheading{Generalized Scherk--Schwarz reduction}

The ordinary frame construction only sees the internal tangent
bundle. This becomes too restrictive once internal diffeomorphisms
and gauge transformations of higher-dimensional form fields mix under
dualities. Generalized geometry combines these transformations into a
single enlarged frame. In the NS--NS sector, for example, metric
twisting and \(H\)-flux can then appear as different components of
the same generalized frame algebra.

In double and exceptional field theory the ordinary frame \(E_a\) is
replaced by a generalized frame \(E_A(Y)\). The variables \(Y\) are
generalized internal coordinates: in double field theory the ordinary
coordinates are supplemented by dual coordinates to make T-duality
manifest, while exceptional field theory uses the corresponding
coordinate representation of the relevant U-duality group.

A generalized Scherk--Schwarz reduction is based on a generalized
parallelization satisfying
\begin{equation}
\mathbb L_{E_A}E_B
=
-
X_{AB}{}^C E_C,
\qquad
X_{AB}{}^C=\text{constant},
\label{eq:q12-generalized-frame-algebra}
\end{equation}
together with the appropriate section and consistency conditions.
The generalized frame contains both tangent and form-gauge
directions, so \(X_{AB}{}^C\) can package geometric twisting and
form-field flux in one algebraic object. In toroidal NS--NS examples,
the same language also accommodates the duality-related quantities
usually denoted \(Q\)- and \(R\)-flux.

The algebraic structure is different from the ordinary frame case.
The generalized Lie derivative is not antisymmetric in its two
arguments. Defining
\[
E_A\circ E_B
\equiv
\mathbb L_{E_A}E_B,
\]
the product need not obey
\(E_A\circ E_B=-E_B\circ E_A\). Instead, subject to the section and
consistency conditions, it satisfies
\begin{equation}
E_A\circ(E_B\circ E_C)
=
(E_A\circ E_B)\circ E_C
+
E_B\circ(E_A\circ E_C).
\label{eq:q12-Leibniz-identity}
\end{equation}
This is the Leibniz identity. The constants \(X_{AB}{}^C\) therefore
define, in general, a Leibniz algebra rather than an ordinary Lie
algebra.

This is the generalized-geometric counterpart of
Eq.~\eqref{eq:q12-symmetric-trivial}. The symmetric directions of
\(X_{AB}{}^C\) do not define additional independent Lie generators,
but they cannot simply be discarded from the gauge-covariant
formulation. Their realization involves higher-rank potentials and
generalized St\"uckelberg transformations. Vectors can mix with
two-forms, two-forms with three-forms, and so on; the resulting
sequence of fields and gauge transformations is the tensor hierarchy.
When the generalized frame, its density factors and its coordinate
dependence obey the required conditions, the constants
\(X_{AB}{}^C\) reproduce the embedding-tensor data of the
lower-dimensional gauged supergravity.

\narrativeheading{The gauged theory remains locally supersymmetric}

The algebraic machinery above does not by itself break local
supersymmetry. A consistent gauging modifies the covariant
derivatives, fermionic transformations, mass terms and scalar
potential in a correlated way such that the resulting supergravity
remains locally supersymmetric. Supersymmetry breaking is instead a
property of a bosonic solution.

On a given bosonic background, all fermionic variations must vanish
for the same non-zero supersymmetry parameter. It is convenient to
collect them into the Killing-spinor operator
\[
\mathscr K:
\mathcal S
\longrightarrow
\mathcal E_{\rm ferm},
\qquad
\mathscr K(\epsilon)
=
\left(
\delta_\epsilon\psi_{\mu A},
\delta_\epsilon\chi^I,
\ldots
\right),
\]
where \(\mathcal S\) is the space of supersymmetry parameters and
\(\mathcal E_{\rm ferm}\) denotes the collection of fermionic
variations evaluated on the chosen background.

Since these variations are linear in \(\epsilon\), the preserved
supersymmetries form the kernel
\begin{equation}
\mathcal S_{\rm preserved}
=
\ker\mathscr K.
\label{eq:q12-Killing-kernel}
\end{equation}
A non-zero gravitino shift \(S_{AB}\neq0\) is not by itself
a criterion for supersymmetry breaking. Supersymmetric anti-de Sitter
vacua provide the standard counterexample: a non-vanishing gravitino
shift can be required by the Killing-spinor equation.

For a consistent supersymmetric truncation, a lower-dimensional
Killing spinor uplifts through the same reduction ansatz. The converse
holds only for higher-dimensional Killing spinors retained by the
truncation, so an uplift can preserve additional supersymmetries
invisible in the reduced theory.

\narrativeheading{What the lower-dimensional gauging remembers}

The embedding tensor captures the local gauging data of the
lower-dimensional supergravity. Together with the scalar geometry,
it determines the gauge generators, covariant derivatives, fermion
shifts, mass matrices and scalar potential. It does not reconstruct
the complete higher-dimensional compactification.

The geometric reductions above already provide an example. The
constants \(f_{ab}{}^c\) determine the local frame algebra, but not
the discrete subgroup \(\Gamma\) used to form a compact quotient
\(K=\Gamma\backslash G_{\rm int}\). Different global compactifications
can therefore share the same local gauging data. Topology, lattices
and discrete identifications are not contained in \(\Theta\).

The converse problem also occurs. An embedding tensor may satisfy the
lower-dimensional linear and quadratic constraints without admitting
an uplift within a specified higher-dimensional construction. In a
generalized Scherk--Schwarz reduction, for example, the generalized
frame must also satisfy the section and consistency conditions of
double or exceptional field theory. These are higher-dimensional
requirements that are not encoded solely in \(\Theta\).

The embedding tensor is the local algebraic image of the
compactification data in the reduced theory. For an ordinary circle
reduction this information comes from the single twist generator
\(\mathcal M\); for geometric reductions it includes
\(f_{ab}{}^c\); for generalized reductions it is encoded in the
constant generalized-frame data \(X_{AB}{}^C\).

\narrativeheading{Bibliographical remarks}

Embedding-tensor formulations of gauged supergravity, including the
linear and quadratic constraints and the associated tensor hierarchy,
can be found in
Refs.~\cite{Nicolai:2000sc,deWit:2002vt,deWit:2007mt,
deWit:2008ta,Samtleben:2008pe}.
Geometric reductions in which constant internal frame data descend to
lower-dimensional gaugings are discussed in
Refs.~\cite{Kaloper:1999yr,Hull:2005hk}.
Generalized Scherk--Schwarz reductions in double and exceptional field
theory, in which generalized frame data reproduce the embedding
tensor of the reduced gauged supergravity, can be found in
Refs.~\cite{Aldazabal:2011nj,Geissbuhler:2011mx,
Lee:2014mla,Hohm:2014qga}.

\begin{takehome}
\begin{takehomeitems}

\item
On a circle, local translations of the compact coordinate are gauged
by the Kaluza--Klein vector, while the Scherk--Schwarz ansatz makes
this symmetry act through the twist generator \(\mathcal M\).

\item
The embedding tensor assigns a global-symmetry generator \(X_M\) to
each vector field \(A_\mu^M\). The elementary circle reduction is the
rank-one case.

\item
The quadratic constraint expresses gauge covariance and closure of
the selected generators. The linear constraint independently restricts
the embedding tensor to representations compatible with
supersymmetry.

\item
Dressing the embedding tensor with the scalar fields gives the
\(T\)-tensor. Its components determine the fermion shifts and, through
the supersymmetry Ward identity, the scalar potential.

\item
Geometric Scherk--Schwarz reductions are governed by constant Lie
algebras of internal frames. Generalized reductions replace them by
generalized frame algebras that are naturally Leibniz and are
accompanied by the tensor hierarchy.

\item
A consistent gauging remains locally supersymmetric. Supersymmetry
breaking is determined only after choosing a bosonic background and
solving the complete fermionic Killing-spinor system.

\item
The embedding tensor remembers the local lower-dimensional gauging,
not the complete global compactification that produced it.

\end{takehomeitems}
\end{takehome}

\nextquestion
{Why does Scherk--Schwarz breaking often lead to a no-scale structure?}

\end{reviewquestion}
% Question 14 v3, renumbered from Question13v2.tex.
% Original section/equation labels retained for cross-reference stability.
%%%%%%%%%%%%%%%%%%%%%%%%%%%%%%%%%%%%%%%%%%%%%%%%%%%%%%%%%%%%%%%

\begin{reviewquestion}
{Why does Scherk--Schwarz breaking often lead to a no-scale structure?}
\label{sec:ss-no-scale}

\begin{essentialidea}

In many classic Scherk--Schwarz compactifications, the
supersymmetry-breaking scale is set by a modulus while the classical
potential remains flat along the same direction. The gravitino mass
can therefore vary continuously along a family of Minkowski vacua.

The origin of this structure becomes particularly transparent in a
four-dimensional \(N=1\) description. Suppose the relevant
compactification moduli enter through
\[
K=-\log Y(r),
\qquad
r^a=T^a+\bar T^{\bar a},
\]
with \(Y\) homogeneous of degree \(p\). Homogeneity implies
\[
K^{a\bar b}K_{\bar b}=-r^a,
\qquad
K^{a\bar b}K_aK_{\bar b}=p.
\]
If the superpotential is independent of these moduli, the second
identity gives the no-scale cancellation for \(p=3\). For constant
non-zero \(W\), the first gives \(F^a\propto r^a\), so
supersymmetry is broken along the Euler direction that rescales the
compactification.

A Scherk--Schwarz spectrum naturally depends on this same scale
modulus. When
\[
m_i(r,\omega)
=
M_{\rm KK}(r)\,\mu_i(\omega),
\]
motion along the no-scale direction changes the gravitino and
Kaluza--Klein masses without costing classical potential energy.
This is the origin of the characteristic sliding spectrum of
Scherk--Schwarz no-scale models.

For the homogeneous geometry above, the Goldstino direction also has
a universal holomorphic sectional curvature,
\[
\kappa_f=-\frac{2}{p}
\]
in the curvature convention adopted below, and hence
\(\kappa_f=-2/3\) at the no-scale value \(p=3\).

Additional superpotential dependence, further gaugings, localized
interactions or quantum corrections can lift this structure. The
discussion below identifies the assumptions under which the standard
Scherk--Schwarz no-scale behaviour emerges.

\end{essentialidea}

\narrativeheading{Why the no-scale question arises}

A Scherk--Schwarz compactification can give the gravitino a non-zero
mass while leaving the modulus that sets the compactification scale
classically undetermined. This is initially surprising. As discussed
in Question~\ref{sec:ss-gauging}, the fermion shifts generated by a
gauging are tied by local supersymmetry to the scalar potential. One
might therefore expect the modulus controlling the
supersymmetry-breaking scale to acquire a potential as well.

The no-scale situation is the case in which the two effects separate.
The gravitino mass varies with the modulus, but the classical scalar
potential remains flat along the same direction. The
Scherk--Schwarz spectrum explains why the masses depend on that
modulus; the scalar geometry explains why the corresponding potential
can nevertheless cancel.

\narrativeheading{A sliding scale along a vacuum family}

Before specializing to supergravity, it is useful to state precisely
what is meant by a classically flat family. The elementary potential
\(V(x,y)=x^2\) has minima along
\(\mathcal N=\{x=0\}\). Along this family \(V=0\), and
\(dV=2x\,dx\) vanishes at every point. Thus \(y\) is a genuine flat
direction of a continuous family of stationary vacua.

More generally, let \(\mathcal M_{\rm scal}\) be the scalar manifold
of the lower-dimensional theory and let
\(\mathcal N\subset\mathcal M_{\rm scal}\) be a connected
submanifold satisfying
\begin{equation}
V\big|_{\mathcal N}=0,
\qquad
dV\big|_{\mathcal N}=0.
\label{eq:q13-stationary-family}
\end{equation}
We shall call \(\mathcal N\) a \emph{stationary Minkowski family}.
No assumption is made here about stability in directions transverse
to \(\mathcal N\).

The existence of such a family already implies massless scalar
directions. If \(q\in\mathcal N\) and
\(X\in T_q\mathcal N\), then \(dV\) vanishes all along
\(\mathcal N\), so for any
\(Y\in T_q\mathcal M_{\rm scal}\),
\[
(\nabla^2V)_q(X,Y)
=
(\nabla_XdV)_q(Y)
=
0.
\]
Hence
\begin{equation}
T_q\mathcal N
\subseteq
\ker(\nabla^2V)_q.
\label{eq:q13-hessian-kernel}
\end{equation}

The characteristic no-scale situation is obtained when
supersymmetry is broken on this family and the breaking scale itself
changes along it. At a generic point there is then a tangent
direction \(X\in T_q\mathcal N\) for which
\begin{equation}
X(m_{3/2})\neq0.
\label{eq:q13-sliding-condition}
\end{equation}
The potential remains unchanged while the gravitino mass slides.

\narrativeheading{The Scherk--Schwarz origin of the sliding spectrum}

The simplest example is again the circle. For a field of
Scherk--Schwarz charge \(q\),
\begin{equation}
m_{n,q}(R,\omega)
=
\frac{n+q\omega}{R},
\qquad
n\in\mathbb Z,
\label{eq:q13-circle-spectrum}
\end{equation}
up to the charge and normalization conventions used in the preceding
questions. The twist fixes the dimensionless displacement of the
tower, while \(R\) fixes its physical scale.

At the half-twist discussed in
Question~\ref{sec:ss-half-twist}, the light Dirac gravitino has
\(m_{3/2}=1/(2R)\). If \(R\) parametrizes a flat vacuum family,
its mass varies continuously along that family. The same is true for
the rest of the shifted Kaluza--Klein spectrum.

The mechanism generalizes whenever the internal operator factorizes
as
\begin{equation}
\mathscr D_{\rm int}(r,\omega)
=
M_{\rm KK}(r)\,
\widehat{\mathscr D}_{\omega},
\label{eq:q13-factorized-operator}
\end{equation}
with \(\widehat{\mathscr D}_{\omega}\) dimensionless. If
\(\widehat{\mathscr D}_{\omega}u_i=\mu_i(\omega)u_i\),
then
\begin{equation}
m_i(r,\omega)
=
M_{\rm KK}(r)\,\mu_i(\omega).
\label{eq:q13-factorized-spectrum}
\end{equation}

For the circle,
\(M_{\rm KK}=1/R\) and
\(\mu_{n,q}=n+q\omega\). Once the scale modulus is flat, the
whole shifted spectrum moves with it. What remains to be explained is
why the scalar potential can leave that scale modulus unfixed.

\narrativeheading{The homogeneous no-scale geometry}

The answer is particularly transparent in an important class of
four-dimensional \(N=1\) supergravities. Consider moduli \(T^a\),
define \(r^a=T^a+\bar T^{\bar a}\), and take
\begin{equation}
K
=
-\log Y(r),
\label{eq:q13-homogeneous-K}
\end{equation}
where \(Y>0\) is homogeneous of degree \(p\),
\begin{equation}
Y(\lambda r)
=
\lambda^pY(r).
\label{eq:q13-homogeneity}
\end{equation}

The familiar one-modulus example,
\[
Y(r)=r^3,
\qquad
r=T+\bar T,
\]
gives
\[
K=-3\log(T+\bar T).
\]
It is simply the case \(p=3\) of the more general geometry.

Let
\(Y_a=\partial Y/\partial r^a\) and
\(Y_{ab}=\partial_a\partial_bY\). Euler's theorem and its derivative
give
\begin{equation}
r^aY_a=pY,
\qquad
r^aY_{ab}=(p-1)Y_b.
\label{eq:q13-Euler}
\end{equation}
Since \(K\) depends on the complex moduli only through \(r^a\),
\[
K_a=-\frac{Y_a}{Y},
\qquad
K_{a\bar b}
=
-\frac{Y_{ab}}{Y}
+
\frac{Y_aY_b}{Y^2}.
\]
Contracting the metric with \(r^b\) gives
\(K_{a\bar b}r^b=-K_a\), and hence
\begin{equation}
K^{a\bar b}K_{\bar b}
=
-r^a.
\label{eq:q13-radial-identity}
\end{equation}
A further contraction gives
\begin{equation}
K^{a\bar b}K_aK_{\bar b}
=
p.
\label{eq:q13-homogeneous-noscale}
\end{equation}

The two identities follow from the same homogeneity property, but
they encode different physics.

Suppose that, in the K\"ahler frame under consideration, the
superpotential is independent of these moduli,
\begin{equation}
\partial_aW=0.
\label{eq:q13-W-independent}
\end{equation}
Then \(D_aW=K_aW\). If this modulus sector contains the
non-vanishing \(F\)-terms relevant to the vacuum family, and no
additional \(D\)-term or other tree-level contribution is present,
the \(F\)-term potential is
\begin{align}
V_F
&=
e^K
\left(
K^{a\bar b}D_aW D_{\bar b}\overline W
-
3|W|^2
\right)
\nonumber\\
&=
e^K(p-3)|W|^2.
\label{eq:q13-p-potential}
\end{align}

For \(p=3\), the positive moduli \(F\)-term contribution therefore
cancels the universal negative gravitino contribution. The potential
vanishes along these moduli even though supersymmetry can remain
broken.

This is the familiar no-scale cancellation. The crucial point is not
that the gravitino mass vanishes, but precisely that it need not:
\(m_{3/2}=e^{K/2}|W|\) can vary as the moduli move while \(V_F\)
remains zero.

\narrativeheading{The Goldstino points along the scale direction}

The first identity,
Eq.~\eqref{eq:q13-radial-identity}, determines the direction of
supersymmetry breaking.

For a constant non-zero superpotential \(W=W_0\),
\[
F^a
=
-
e^{K/2}
K^{a\bar b}
D_{\bar b}\overline W.
\]
Since
\(D_{\bar b}\overline W=K_{\bar b}\overline W_0\),
one obtains
\begin{equation}
F^a
=
e^{K/2}\overline W_0\,r^a.
\label{eq:q13-F-radial}
\end{equation}

The Goldstino direction is therefore parallel to the Euler vector
\(E=r^a\partial/\partial r^a\), which generates a common rescaling of
the homogeneous coordinates. In the compactification interpretation,
this is precisely the scale direction.

This gives a direct geometric interpretation of the radion
\(F\)-term discussed in
Question~\ref{sec:ss-wilson-radion}: in the homogeneous benchmark,
the auxiliary direction responsible for supersymmetry breaking is
aligned with the modulus that rescales the compactification.

The alignment itself does not require \(p=3\). It follows from
homogeneity. The special value \(p=3\) is what makes the scalar
potential cancel.

\newpage
\narrativeheading{The curvature seen by the Goldstino}

The same homogeneous structure also fixes the intrinsic K\"ahler
geometry along the supersymmetry-breaking direction.

Let \(g_{a\bar b}=K_{a\bar b}\). Since \(K=-\log Y\) and \(Y\)
has degree \(p\), the metric has degree \(-2\),
\begin{equation}
r^c\partial_cg_{a\bar b}
=
-2g_{a\bar b}.
\label{eq:q13-metric-homogeneity}
\end{equation}
Because \(K\) depends on \(T^a\) and \(\bar T^{\bar a}\) only through
\(r^a\), this implies
\[
K_{a\bar b c}r^c=-2K_{a\bar b},
\qquad
K_{ac\bar b}r^ar^c=2K_{\bar b},
\]
together with the complex-conjugate relation. Repeating the
differentiation and contraction gives
\begin{equation}
K_{a\bar b c\bar d}
r^ar^br^cr^d
=
6p,
\label{eq:q13-fourth-contraction}
\end{equation}
while
\(K_{a\bar b}r^ar^b=p\).

We adopt the curvature convention
\begin{equation}
R_{a\bar b c\bar d}
=
-\partial_a\partial_{\bar b}g_{c\bar d}
+
g^{e\bar f}
(\partial_a g_{c\bar f})
(\partial_{\bar b}g_{e\bar d}).
\label{eq:q13-curvature-convention}
\end{equation}
Contracting with four Euler vectors gives
\begin{align}
R_{a\bar b c\bar d}
r^ar^br^cr^d
&=
-6p
+
4K_eK^{e\bar f}K_{\bar f}
\nonumber\\
&=
-2p.
\label{eq:q13-radial-curvature-contraction}
\end{align}

Since the Euler direction has norm \(p\), define the corresponding
unit vector by \(f^a=r^a/\sqrt p\). Its holomorphic sectional
curvature is
\begin{equation}
\kappa_f
\equiv
R_{a\bar b c\bar d}
f^a\bar f^{\bar b}
f^c\bar f^{\bar d}
=
-\frac{2}{p}.
\label{eq:q13-goldstino-curvature}
\end{equation}
For the no-scale value \(p=3\), this gives
\(\kappa_f=-2/3\).

The sign of this quantity depends on the convention used for the
Riemann tensor. References employing the opposite convention must
therefore be translated before sectional-curvature bounds are
compared.

The relevance of \(\kappa_f\) is that the K\"ahler curvature along the
Goldstino direction enters the scalar mass matrix and the sgoldstino
sector. General analyses of non-supersymmetric Minkowski vacua make
this dependence explicit~\cite{Gomez-Reino:2006gtu}.
Here, however, the curvature is not being used to establish the
flatness. For the homogeneous \(p=3\) benchmark that flatness already
follows directly from Eq.~\eqref{eq:q13-p-potential}. The curvature
calculation gives an additional intrinsic property of the same
supersymmetry-breaking direction.

\narrativeheading{A comparison with O'Raifeartaigh pseudomoduli}

There is a useful analogy with global supersymmetry. In
renormalizable Wess--Zumino models with canonical K\"ahler metric, a
stable tree-level supersymmetry-breaking configuration possesses,
under the usual assumptions, a pseudomodulus aligned with the
supersymmetry-breaking direction~\cite{ORaifeartaigh:1975nky,Komargodski:2009jf}.

The resemblance is that the Goldstino direction is again associated
with a classically flat scalar direction. The origin of the flatness
is different. In the O'Raifeartaigh case it follows from the
structure of the renormalizable global-supersymmetry model. In
supergravity there is no general theorem requiring the scalar partner
of the Goldstino to remain massless: its behaviour depends on the
K\"ahler geometry and on the other couplings.

In the homogeneous no-scale model, the flat direction instead follows
from the cancellation derived above. The \(R\)-symmetry used for the
Scherk--Schwarz twist does not by itself protect the radial modulus.
The relation between \(R\)-symmetry and generic \(F\)-term breaking
in global supersymmetry is a separate issue~\cite{Nelson:1993nf}.

\narrativeheading{How the Scherk--Schwarz and no-scale structures meet}

We can now see why the two structures occur together so naturally in
the standard examples.

The Scherk--Schwarz monodromy provides the dimensionless deformation
\(\mu_i(\omega)\) of the Kaluza--Klein spectrum. The compactification
modulus supplies its overall scale \(M_{\rm KK}(r)\). The homogeneous
scalar geometry aligns the Goldstino with the Euler direction, so the
same modulus also controls the supersymmetry-breaking auxiliary
field. At degree \(p=3\), the supersymmetry Ward identity becomes the
no-scale cancellation and leaves this direction classically flat.

The result is a family of Minkowski vacua in which the
supersymmetry-breaking scale can slide:
\[
V=0,
\qquad
m_{3/2}=m_{3/2}(r),
\qquad
m_i=M_{\rm KK}(r)\mu_i(\omega).
\]
The same scale direction controls the Kaluza--Klein spectrum, the
gravitino mass and the Goldstino direction, while costing no
classical potential energy.

\narrativeheading{When the no-scale structure is lifted}

The preceding result relies on a definite set of assumptions. The
internal spectrum need not depend on a single overall scale; several
radii, warping or additional dimensionful parameters may enter. The
scalar geometry need not have the homogeneous form used above, or its
degree need not be \(p=3\).

The superpotential may also depend on the scale modulus. Additional
\(F\)-terms, \(D\)-terms, fluxes, internal curvature, localized
interactions or further gaugings can then contribute to the scalar
potential. Localized terms of the type discussed in
Question~\ref{sec:ss-localized-boundary-terms} provide one possible
source of such lifting. Quantum corrections can likewise generate a
potential for a direction that is flat at tree level and can modify
the effective K\"ahler geometry.

Thus the no-scale behaviour discussed here is characteristic of an
important class of Scherk--Schwarz compactifications rather than a
kinematical consequence of the twist alone. The mechanisms that lift
the flat direction and determine the supersymmetry-breaking scale are
the subject of the next question.

\newpage
\narrativeheading{Bibliographical remarks}

The no-scale structure of supergravity and its relation to
supersymmetry breaking are discussed in the early literature on
no-scale models. Homogeneous formulations and sliding-gravitino
supergravities provide a particularly useful language for the
Scherk--Schwarz setting; the structures used above are discussed in
Refs.~\cite{Ferrara:1994kg,Cremmer:1983bf,Ellis:1984bm}.

The relation between the Goldstino direction and K\"ahler curvature
in non-supersymmetric Minkowski vacua is discussed in
Ref.~\cite{Gomez-Reino:2006gtu}. The comparison with tree-level
pseudomoduli in global supersymmetry is related to
Refs.~\cite{ORaifeartaigh:1975nky,Komargodski:2009jf,Nelson:1993nf}.

\begin{takehome}
\begin{takehomeitems}

\item
In Scherk--Schwarz no-scale models, the same modulus can control the
compactification scale and remain classically flat. The shifted
spectrum then slides as
\[
m_i(r,\omega)
=
M_{\rm KK}(r)\mu_i(\omega).
\]

\item
For a homogeneous K\"ahler potential
\[
K=-\log Y(r),
\qquad
Y(\lambda r)=\lambda^pY(r),
\]
one has
\[
K^{a\bar b}K_{\bar b}=-r^a,
\qquad
K^{a\bar b}K_aK_{\bar b}=p.
\]

\item
If the superpotential is independent of these moduli and no
additional tree-level contribution is present, \(p=3\) gives the
no-scale cancellation \(V_F=0\).

\item
For constant non-zero \(W\), the supersymmetry-breaking auxiliary
direction satisfies \(F^a\propto r^a\). The Goldstino is therefore
aligned with the Euler scale direction.

\item
The homogeneous geometry fixes the holomorphic sectional curvature
along this direction to
\[
\kappa_f=-\frac2p,
\]
and hence \(\kappa_f=-2/3\) for \(p=3\).

\item
Additional moduli dependence, further gaugings, localized
interactions or quantum effects can lift the no-scale direction and
determine the supersymmetry-breaking scale.

\end{takehomeitems}
\end{takehome}

\nextquestion
{What can lift the no-scale direction and determine the
supersymmetry-breaking scale?}

\end{reviewquestion}
% Question 15 v3, renumbered from Question14v2.tex.
% Original section/equation labels retained for cross-reference stability.
%%%%%%%%%%%%%%%%%%%%%%%%%%%%%%%%%%%%%%%%%%%%%%%%%%%%%%%%%%%%%%%

\begin{reviewquestion}
{What can lift the no-scale direction and determine the
supersymmetry-breaking scale?}
\label{sec:ss-lifting}

\begin{essentialidea}

A no-scale theory contains a classically flat vacuum family. The
question is not simply whether a correction is present, but whether it
produces a non-constant potential along the former flat directions.

A useful picture is a flat valley. A perturbation can tilt the floor
of the valley directly, or it can first displace the minimum in the
transverse directions. These two effects generally occur at different
orders.

Let the unperturbed scalar potential be \(V_0\), with no-scale vacuum
manifold \(\mathcal N\), and perturb it as
\[
V_\varepsilon
=
V_0+\varepsilon V_1+\varepsilon^2V_2+\cdots .
\]
If the Hessian of \(V_0\) is non-degenerate transverse to
\(\mathcal N\), the transverse fields can be solved for
perturbatively. Their first displacement is
\[
y^A_\star
=
-\varepsilon
(H_\perp^{-1})^{AB}
\partial_BV_1\big|_{\mathcal N}
+
O(\varepsilon^2),
\]
while the induced potential along the former flat family is
\[
V_{\rm eff}^{(\mathcal N)}
=
\varepsilon V_1\big|_{\mathcal N}
+
\varepsilon^2
\left[
V_2\big|_{\mathcal N}
-
\frac12
(\partial_AV_1)
(H_\perp^{-1})^{AB}
(\partial_BV_1)
\right]
+
O(\varepsilon^3).
\]

The first-order lifting is therefore controlled by the restriction
\(V_1|_{\mathcal N}\), not by the mere presence of \(V_1\) in the full
scalar theory. If this restriction is constant, the degeneracy
survives at first order, but transverse backreaction can still
generate a non-constant potential at the next order.

Lifting does not yet mean stabilization. A lifted direction may run
away, possess only unstable extrema, or admit a stable minimum.
Moreover, fixing the compactification scale need not fix an effective
Scherk--Schwarz phase, and neither statement by itself determines the
amount of preserved supersymmetry or the vacuum energy.

The aim of this question is therefore not to review particular models
of modulus stabilization, but to isolate a general criterion that can
be applied to any classical or quantum deformation of a no-scale
Scherk--Schwarz compactification.

\end{essentialidea}

\narrativeheading{A flat valley and what a perturbation can do}

Consider two real scalar fields \(t\) and \(y\), with
\begin{equation}
V_0(t,y)
=
\frac12 M^2y^2,
\qquad
M^2>0.
\label{eq:q14-toy-V0}
\end{equation}
The minima form the continuous family \(\mathcal N=\{y=0\}\).
The coordinate \(t\) parametrizes the flat direction, while \(y\) is
a stable transverse field of mass \(M\).

Now perturb the potential by
\begin{equation}
V_\varepsilon(t,y)
=
\frac12M^2y^2
+
\varepsilon
\left[
U(t)+J(t)y
\right]
+
O(\varepsilon^2),
\label{eq:q14-toy-perturbation}
\end{equation}
where \(U(t)\) and \(J(t)\) are arbitrary smooth functions.

Two effects are immediately visible. Restricting the perturbation to
the original flat family gives \(V_1|_{\mathcal N}=U(t)\). If
\(U(t)\) is non-constant, the flat direction is lifted directly at
order \(\varepsilon\).

The transverse minimum also moves. The \(y\)-equation gives
\[
M^2y+\varepsilon J(t)+O(\varepsilon^2)=0,
\]
and hence
\begin{equation}
y_\star(t)
=
-\varepsilon
\frac{J(t)}{M^2}
+
O(\varepsilon^2).
\label{eq:q14-toy-shift}
\end{equation}
Substituting back into the potential gives
\begin{align}
V_{\rm eff}(t)
&=
\varepsilon U(t)
+
\frac12M^2y_\star^2
+
\varepsilon J(t)y_\star
+
O(\varepsilon^3)
\nonumber\\
&=
\varepsilon U(t)
-
\frac{\varepsilon^2}{2M^2}
J(t)^2
+
O(\varepsilon^3).
\label{eq:q14-toy-effective}
\end{align}

This elementary example already contains the two mechanisms that will
appear in the general result. If \(U(t)\) is non-constant, the valley
is tilted directly at first order. If \(U(t)\) is constant but
\(J(t)\) is not, the first-order degeneracy survives, while the
transverse readjustment generates a \(t\)-dependent potential at
second order.

The negative sign of the second contribution has a simple meaning:
allowing a stable transverse field to relax lowers the energy relative
to holding it fixed at its unperturbed value.

Nothing here guarantees stabilization. A non-constant \(U(t)\), or
the induced term \(-J(t)^2/(2M^2)\), may instead produce a runaway or
an unstable extremum. The same distinction will remain important in
the general case.

\narrativeheading{Lifting a no-scale vacuum manifold}

Let \(V_0\) possess a no-scale stationary Minkowski family in the
sense of Question~\ref{sec:ss-no-scale},
\begin{equation}
V_0\big|_{\mathcal N}=0,
\qquad
dV_0\big|_{\mathcal N}=0,
\label{eq:q14-unperturbed-family}
\end{equation}
with \(\mathcal N\subset\mathcal M_{\rm scal}\).

As shown there,
\[
T_q\mathcal N
\subseteq
\ker(\nabla^2V_0)_q
\]
for every \(q\in\mathcal N\). We now assume that these tangential zero
modes are the only zero modes of the Hessian.

Choose local coordinates \((t^a,y^A)\) adapted to the vacuum family,
so that \(\mathcal N=\{y^A=0\}\). The \(t^a\) parametrize the
vacuum manifold, while the \(y^A\) describe transverse directions.
Define the transverse Hessian by
\begin{equation}
(H_\perp)_{AB}(t)
=
\partial_A\partial_BV_0(t,0),
\label{eq:q14-transverse-hessian}
\end{equation}
and assume that it is invertible.

This is the Morse--Bott non-degeneracy condition appropriate to the
problem: the critical set is the smooth manifold \(\mathcal N\), and
the Hessian has no zero modes transverse to it
\cite{Nicolaescu:MorseTheory}. Since \(V_0\) and its
first derivatives vanish on \(\mathcal N\), the transverse gradient
starts linearly,
\[
\partial_AV_0(t,y)
=
(H_\perp)_{AB}(t)y^B
+
O(y^2).
\]

Now perturb the theory,
\begin{equation}
V_\varepsilon
=
V_0
+
\varepsilon V_1
+
\varepsilon^2V_2
+
O(\varepsilon^3).
\label{eq:q14-perturbation}
\end{equation}
The transverse fields readjust according to
\(\partial_AV_\varepsilon=0\). At first order this gives
\[
(H_\perp)_{AB}y^B
+
\varepsilon
\partial_AV_1\big|_{\mathcal N}
+
O(\varepsilon^2)
=
0,
\]
and therefore
\begin{equation}
y^A_\star(t)
=
-\varepsilon
(H_\perp^{-1})^{AB}
\partial_BV_1\big|_{\mathcal N}
+
O(\varepsilon^2).
\label{eq:q14-transverse-shift}
\end{equation}

The transverse fields can thus move already at first order even when
the former flat directions are not yet lifted at that order.

Substituting \(y=y_\star(t)\) into the full potential gives
\begin{align}
V_{\rm eff}^{(\mathcal N)}(t)
&=
\varepsilon
V_1\big|_{\mathcal N}
\nonumber\\
&\quad+
\varepsilon^2
\left[
V_2\big|_{\mathcal N}
-
\frac12
(\partial_AV_1)
(H_\perp^{-1})^{AB}
(\partial_BV_1)
\right]
+
O(\varepsilon^3),
\label{eq:q14-induced-potential}
\end{align}
with all quantities on the right-hand side evaluated on
\(\mathcal N\).

The factor \(1/2\) follows from the competition between the quadratic
energy cost associated with \(V_0\) and the linear gain from \(V_1\),
exactly as in the two-field example.

The result can be written intrinsically. The scalar metric defines the
normal space
\[
N_q\mathcal N=(T_q\mathcal N)^\perp ,
\]
and the transverse Hessian is the restriction of
\(\nabla^2V_0\) to this space. If \(d_\perp V_1\) denotes the
restriction of \(dV_1\) to the normal directions, the backreaction
term is
\begin{equation}
-\frac12
\left\langle
d_\perp V_1,
H_\perp^{-1}d_\perp V_1
\right\rangle .
\label{eq:q14-intrinsic-backreaction}
\end{equation}

If \(H_\perp\) is positive definite, this term is non-positive.
Integrating out a stable transverse sector therefore lowers, or leaves
unchanged, the induced potential at this order. No such sign statement
follows when the transverse Hessian is indefinite.

\narrativeheading{The first non-trivial lifting order}

Equation~\eqref{eq:q14-induced-potential} gives a direct criterion for
lifting. At first order, the induced potential is simply
\[
\mathcal V_1
=
V_1\big|_{\mathcal N}.
\]
The no-scale family is therefore lifted at order \(\varepsilon\) only
when this restriction is non-constant along \(\mathcal N\), or
equivalently when
\begin{equation}
d_{\mathcal N}
\left(
V_1\big|_{\mathcal N}
\right)
\not\equiv0,
\label{eq:q14-first-order-lifting}
\end{equation}
where \(d_{\mathcal N}\) denotes the intrinsic differential along the
vacuum manifold.

A deformation may thus be non-zero throughout the full scalar
manifold and still leave the no-scale degeneracy intact at first
order. A constant \(V_1|_{\mathcal N}\) can shift the vacuum energy
without selecting any point on the former flat family.

More generally, write
\begin{equation}
V_{\rm eff}^{(\mathcal N)}
=
\sum_{k\geq1}
\varepsilon^k\mathcal V_k.
\label{eq:q14-induced-expansion}
\end{equation}
The first two coefficients are
\[
\mathcal V_1
=
V_1\big|_{\mathcal N},
\]
and
\[
\mathcal V_2
=
V_2\big|_{\mathcal N}
-
\frac12
\left\langle
d_\perp V_1,
H_\perp^{-1}d_\perp V_1
\right\rangle.
\]

If the degeneracy is first lifted at some finite order, define
\begin{equation}
k_\star
=
\min
\left\{
k\geq1
\;\middle|\;
d_{\mathcal N}\mathcal V_k
\not\equiv0
\right\}.
\label{eq:q14-first-lifting-order}
\end{equation}
The leading lifting potential is then
\begin{equation}
V_{\rm lift}
=
\varepsilon^{k_\star}
\mathcal V_{k_\star}
+
O(\varepsilon^{k_\star+1}).
\label{eq:q14-leading-lift}
\end{equation}

The relevant perturbative order is not the first order at
which the microscopic action changes. It is the first order at which
the induced potential on the former vacuum manifold becomes
non-constant.

\narrativeheading{Lifting is not yet stabilization}

Once the former flat direction acquires a non-constant potential, the
problem is not finished. The resulting potential may be monotonic, in
which case the modulus runs toward the boundary of field space.

A finite scale can be selected only if the induced potential has a
critical point \(t_\star\),
\begin{equation}
d_{\mathcal N}
V_{\rm eff}^{(\mathcal N)}
\big|_{t_\star}
=
0.
\label{eq:q14-critical-point}
\end{equation}
Local stabilization additionally requires the Hessian along the
relevant former flat directions to be positive.

The transverse spectrum must also remain acceptable. If
\(H_\perp>0\) initially, sufficiently small perturbations preserve
transverse positivity by continuity. If the unperturbed family
already has a transverse instability, the perturbative construction
still computes the induced tangential potential, but it does not cure
that instability.

Thus a non-constant potential can produce a runaway, a critical point
can be a saddle, and a locally stable minimum can have positive, zero
or negative vacuum energy. Lifting, scale selection, stability and
vacuum characterization are separate questions.

\narrativeheading{What can generate the lifting potential?}

The general criterion allows very different microscopic effects to be
treated in the same way.

At tree level, additional gauging data can deform the Scherk--Schwarz
embedding tensor, schematically
\(\Theta=\Theta_{\rm SS}+\Delta\Theta\). The dressed \(T\)-tensor,
fermion shifts and scalar potential introduced in
Question~\ref{sec:ss-gauging} then change accordingly. The resulting
gauged supergravity can remain locally supersymmetric while its
no-scale family is lifted.

Internal curvature and background fluxes can generate
modulus-dependent terms after dimensional reduction and Weyl
rescaling. Localized interactions of the type discussed in
Question~\ref{sec:ss-localized-boundary-terms} can contribute directly
to the scalar potential or modify fields whose expectation values
enter it.

There is also a possibility intrinsic to the Scherk--Schwarz
construction. If the twist generator \(\mathcal M\) acts
non-trivially on a bosonic background field,
\(\mathcal M\Phi_0\neq0\), then the corresponding covariant internal
derivative does not vanish in the periodic-field description.
Internal kinetic energy can then contribute to the reduced scalar
potential.

The microscopic origins of these effects differ, but the test for
lifting is always the same: compute the induced potential on
\(\mathcal N\) and determine whether it varies along the former flat
directions.

\narrativeheading{The \(N=1\) no-scale sector under deformation}

The homogeneous \(N=1\) construction of
Question~\ref{sec:ss-no-scale} provides a useful laboratory for
applying the general criterion.

Let
\begin{equation}
K
=
K_0+\varepsilon\Delta K,
\qquad
W
=
W_0+\varepsilon\Delta W,
\label{eq:q14-N1-deformation}
\end{equation}
with
\[
K_0=-\log Y,
\qquad
Y(\lambda r)=\lambda^3Y(r),
\qquad
\partial_aW_0=0.
\]
The unperturbed modulus sector obeys the no-scale identity, while the
deformed \(F\)-term potential remains
\begin{equation}
V_F
=
e^K
\left(
K^{a\bar b}
D_aW
D_{\bar b}\overline W
-
3|W|^2
\right).
\label{eq:q14-N1-potential}
\end{equation}

Expanding it as
\[
V_F
=
V_0+\varepsilon V_1+\varepsilon^2V_2+\cdots
\]
reduces the problem to the general analysis above. At first order, the
no-scale family is lifted only if
\begin{equation}
d_{\mathcal N}
\left(
V_1\big|_{\mathcal N}
\right)
\not\equiv0.
\label{eq:q14-N1-lifting-criterion}
\end{equation}

A non-zero \(\Delta K\) or \(\Delta W\) therefore need not generate a
force along the original no-scale family at the same perturbative
order. What matters is the corresponding coefficient of the scalar
potential after restriction to \(\mathcal N\). If the first-order
restriction is constant, one proceeds to the next order, including
the transverse backreaction derived above.

This \(N=1\) sector is used only as a convenient supergravity
realization of the criterion. In string compactifications, modulus
stabilization is generally a coupled problem involving several
fields, flux and quantization conditions, localized sources and
global consistency constraints.

\narrativeheading{Rigid monodromy versus a dynamical spectral phase}

There is another way in which a supersymmetry-breaking scale can
become dynamical, logically distinct from stabilizing the
compactification radius.

Question~\ref{sec:ss-localized-boundary-terms} showed that localized
data can modify the phase entering a Kaluza--Klein spectrum. For the
commuting boundary problem considered there, one may write
\begin{equation}
\omega_{\rm eff}
=
\omega_{\rm rigid}
+
\frac{
\Omega_0(\varphi)
+
\Omega_\pi(\varphi)
}{\pi}
\qquad
(\mathrm{mod}\ 1).
\label{eq:q14-effective-phase}
\end{equation}
Here \(\omega_{\rm rigid}\) is the microscopic
Scherk--Schwarz monodromy parameter, whereas the boundary angles
depend on localized dynamical fields \(\varphi^I\).

The microscopic parameter is fixed, \(d\omega_{\rm rigid}=0\), while
\(d\omega_{\rm eff}\) can be non-zero when these localized fields
vary. The microscopic monodromy can therefore remain unchanged even
though the phase seen by a particular Kaluza--Klein spectrum becomes
dynamical.

Determining the compactification scale \(M_{\rm KK}(r)\) and
determining the fields entering \(\omega_{\rm eff}(\varphi)\) are distinct dynamical problems. They need not be solved at the
same order or by the same mechanism.

A measured spectral shift constrains \(\omega_{\rm eff}\), not
necessarily the microscopic \(\omega_{\rm rigid}\). As emphasized in
Question~\ref{sec:ss-localized-boundary-terms}, compactifications with
the same effective spectral phase can still differ in their localized
dynamics and in their complete supersymmetry-breaking structure.

\narrativeheading{Quantum scale selection}

The no-scale family may survive all classical effects retained in a
given effective theory. The first non-trivial potential can then arise
quantum mechanically.

Question~\ref{sec:ss-no-scale} showed that the classical spectrum can
take the form
\[
m_i(r,\omega)
=
M_{\rm KK}(r)\,\mu_i(\omega).
\]
These same masses enter the quantum functional determinant.
Schematically,
\begin{equation}
V_{\rm 1-loop}(r,\omega)
\sim
\frac12
\operatorname{Str}
\sum_n
\int
\frac{d^dp}{(2\pi)^d}
\log
\left[
p^2+m_n^2(r,\omega)
\right].
\label{eq:q14-oneloop-schematic}
\end{equation}
The spectrum whose scale slides classically can therefore generate a
quantum potential for the modulus controlling that scale.

The local and non-local pieces of this quantum potential, their
ultraviolet sensitivity, and the relevant supertraces are separate
questions. They will be treated in the quantum part of the review.

\narrativeheading{What has actually been fixed?}

Even after a stable critical point has been found, one must still
identify which quantities have actually been selected dynamically.
The compactification scale may be fixed while an effective spectral
phase remains dynamical, or conversely; other flat directions may
also survive.

The amount of preserved supersymmetry and the vacuum energy are
independent questions. A fixed non-zero gravitino mass does not
determine the former: as in Question~\ref{sec:ss-gauging}, preserved
supersymmetry is found from the kernel of the complete fermionic
Killing system. The value of the potential and the full scalar Hessian
are likewise required to characterize the resulting vacuum as
de Sitter, Minkowski or anti-de Sitter and to determine its local
stability.

\narrativeheading{Bibliographical remarks}

The no-scale supergravity framework and the relation between flat
moduli, vanishing classical vacuum energy and a sliding gravitino mass
are discussed in the early literature on no-scale models and
Scherk--Schwarz supergravity; see, for example,
Refs.~\cite{Ferrara:1994kg,Cremmer:1983bf,Ellis:1984bm}.

The perturbative treatment used here is formulated as a deformation
problem for a Morse--Bott \cite{Nicolaescu:MorseTheory} vacuum manifold. 
The leading lifting
potential is obtained by restricting the perturbation to the
unperturbed vacuum family, while integrating out non-degenerate
transverse directions generates the second-order backreaction term in
Eq.~\eqref{eq:q14-induced-potential}.

Corrections capable of lifting no-scale directions arise in many
forms in effective supergravity and string compactifications,
including additional gaugings, flux and curvature effects, localized
interactions and quantum corrections. Their detailed realization is
strongly model dependent and will not be reviewed here.

\newpage
\begin{takehome}
\begin{takehomeitems}

\item
A perturbation of a no-scale theory should first be evaluated on the
unperturbed vacuum manifold. At leading order,
\[
V_{\rm eff}^{(\mathcal N)}
=
\varepsilon V_1\big|_{\mathcal N}
+
O(\varepsilon^2),
\]
so the no-scale direction is lifted at first order only if this
restriction is non-constant.

\item
The transverse fields generally readjust already at order
\(\varepsilon\),
\[
y^A_\star
=
-\varepsilon
(H_\perp^{-1})^{AB}
\partial_BV_1\big|_{\mathcal N}
+
O(\varepsilon^2),
\]
and their backreaction contributes to the tangential potential one
order later.

\item
The first lifting order is the first order at which the induced
potential on \(\mathcal N\) becomes non-constant. It need not coincide
with the first order at which the microscopic theory is deformed.

\item
Lifting does not guarantee a finite critical point or a stable vacuum.
The tangential and transverse spectra must be analysed separately.

\item
Additional gaugings, curvature, fluxes, localized interactions,
bosonic Scherk--Schwarz charges and quantum corrections can all
contribute to the induced potential. In an \(N=1\) description, the
same criterion applies to corrections of \(K\) and \(W\).

\item
The compactification scale and the effective spectral phase are
distinct dynamical quantities. If the no-scale structure survives
classically, the Scherk--Schwarz spectrum itself can generate the
quantum potential that selects the remaining modulus.

\end{takehomeitems}
\end{takehome}

\nextquestion
{How are inequivalent Scherk--Schwarz twists classified, and how much
supersymmetry can they preserve?}

\end{reviewquestion}
% Question 16 v3, renumbered from Question15v2.tex.
% Original section/equation labels retained for cross-reference stability.
%%%%%%%%%%%%%%%%%%%%%%%%%%%%%%%%%%%%%%%%%%%%%%%%%%%%%%%%%%%%%%%

\begin{reviewquestion}
{How are inequivalent Scherk--Schwarz twists classified, and how much
supersymmetry can they preserve?}
\label{sec:ss-inequivalent-twists}

\begin{essentialidea}

A Scherk--Schwarz twist is not fundamentally a number. It is a
representation of the global identification group of the
compactification,
\[
\rho:\Gamma\longrightarrow G,
\]
where \(\Gamma\) contains the identifications of the internal space and
\(G\) is the group of admissible automorphisms of the parent theory.

Two representations related by conjugation describe the same gluing
data written in different field bases. The first classical
classification problem is therefore
\[
\mathcal X(\Gamma,G)
=
\operatorname{Hom}(\Gamma,G)/G,
\]
with the usual mathematical refinements understood when necessary.

Two different questions can then be asked at each point of this
representation space. The residual internal symmetry is
\[
G_{\rm residual}(\rho)
=
Z_G\!\left(\rho(\Gamma)\right),
\]
the centralizer of the complete image of \(\rho\), while the preserved
supersymmetries form
\[
\mathcal S_{\rm preserved}(\rho)
=
\bigcap_{\gamma\in\Gamma}
\ker\!\left[
\rho_{\mathcal S}(\gamma)-\mathbf1
\right],
\]
where \(\rho_{\mathcal S}\) is the action of the same gluing data on
the supersymmetry parameters.

These two quantities probe the same global data in different
representations. Residual symmetry is controlled by the adjoint action
on the internal symmetry algebra, while preserved supersymmetry is
controlled by the representation carried by the supercharges. Their
enhancement loci therefore need not coincide.

\end{essentialidea}

\narrativeheading{The representation space of twists}

Let \(\Gamma\) be the group generated by the global identifications of
the compactification. Familiar examples are
\(\Gamma=\mathbb Z\) for a circle,
\(\Gamma=\mathbb Z^r\) for an \(r\)-torus, and
\(\Gamma=D_\infty\) for the orbifold \(S^1/\mathbb Z_2\) discussed in
Question~\ref{sec:ss-orbifolds}.

Let \(G\) denote the group of transformations of the parent theory
that may consistently be used as gluing data, in the sense explained
in Question~\ref{sec:twist-symmetry}. A rigid Scherk--Schwarz twist is
then a homomorphism
\begin{equation}
\rho:\Gamma\longrightarrow G.
\label{eq:q15-rho}
\end{equation}
For each \(\gamma\in\Gamma\), the element \(\rho(\gamma)\) tells us how
the fields are reidentified under that global identification, while
the homomorphism condition
\(\rho(\gamma_1\gamma_2)=\rho(\gamma_1)\rho(\gamma_2)\)
ensures compatibility with composition.

Two representations related by a global change of field basis,
\begin{equation}
\rho'(\gamma)
=
g\,\rho(\gamma)\,g^{-1},
\qquad
g\in G,
\label{eq:q15-conjugation}
\end{equation}
describe the same rigid gluing data. The corresponding classical
representation space is therefore
\begin{equation}
\mathcal X(\Gamma,G)
=
\operatorname{Hom}(\Gamma,G)/G.
\label{eq:q15-character-variety}
\end{equation}

For compact \(G\), this orbit-space description is sufficient for the
semisimple holonomies considered below. We shall use
\(\mathcal X(\Gamma,G)\) without dwelling on refinements needed in more
general representation varieties.

Two qualifications are useful. First,
Eq.~\eqref{eq:q15-character-variety} classifies rigid global gluing
data. A dynamically generated effective phase
\(\omega_{\rm eff}\), of the type discussed in
Question~\ref{sec:ss-lifting}, is not a new point of this space unless
the rigid representation \(\rho\) itself changes.

Second, conjugation is only the minimal equivalence relation.
Mapping-class transformations and exact dualities of the complete
compactification may identify further points of
\(\mathcal X(\Gamma,G)\).

\narrativeheading{The circle: conjugacy classes and alcoves}

For a circle, \(\Gamma=\mathbb Z\), so a homomorphism is completely
determined by the image \(U\in G\) of one generator. Hence
\[
\mathcal X(\mathbb Z,G)
=
G/\!\operatorname{Ad}G,
\]
the space of conjugacy classes of \(G\).

If \(G\) is compact and connected, every element is conjugate to an
element of a maximal torus \(T\), and two elements of \(T\) are
conjugate precisely when they are related by the Weyl group
\(\mathcal W\). Thus
\begin{equation}
\mathcal X(\mathbb Z,G)
\simeq
T/\mathcal W.
\label{eq:q15-circle-classification}
\end{equation}

Choose generators \(\mathbf H\) of a Cartan subalgebra and write
\begin{equation}
U
=
\exp\!\left(
2\pi i\,\boldsymbol\omega\cdot\mathbf H
\right).
\label{eq:q15-Cartan-monodromy}
\end{equation}
The coordinate \(\boldsymbol\omega\) is not unique. The exponential
map identifies
\(\boldsymbol\omega\sim\boldsymbol\omega+\lambda\) with
\(\lambda\in\Lambda_{\rm cochar}\), where
\(\Lambda_{\rm cochar}\) is the cocharacter lattice of \(G\), while
the Weyl group further identifies
\(\boldsymbol\omega\sim w(\boldsymbol\omega)\), with
\(w\in\mathcal W\). These two identifications generate the
extended affine Weyl action,
\begin{equation}
\mathcal W_{\rm ext}
=
\mathcal W\ltimes\Lambda_{\rm cochar}.
\label{eq:q15-extended-affine-Weyl}
\end{equation}
For a simply connected semisimple group, the familiar affine Weyl
group is obtained using the coroot lattice. The cocharacter lattice is
the appropriate object when the global form of \(G\) is kept
explicit.

A fundamental domain of this action provides an alcove description of
the inequivalent circle twists. The lattice identification is the
group-theoretic origin of the Kaluza--Klein relabelling freedom
encountered in Questions~\ref{sec:ss-goldstino}
and~\ref{sec:ss-superhiggs}.

For \(G=SU(2)\), with
\[
U(\omega)=e^{2\pi i\omega\sigma_3},
\]
one has \(\omega\sim\omega+1\) and
\(\omega\sim-\omega\). A fundamental alcove is therefore
\begin{equation}
0\leq\omega\leq\frac12.
\label{eq:q15-su2-alcove}
\end{equation}
Its endpoints are the untwisted point and the half-twist.

\narrativeheading{Several cycles and commuting tuples}

For an \(r\)-torus, \(\Gamma=\mathbb Z^r\), so a homomorphism is
specified by \(r\) commuting elements of \(G\):
\begin{equation}
\operatorname{Hom}(\mathbb Z^r,G)
=
\left\{
(U_1,\ldots,U_r)
\;\middle|\;
U_iU_j=U_jU_i
\right\}.
\label{eq:q15-commuting-tuples}
\end{equation}
The quotient by \(G\) acts by simultaneous conjugation,
\begin{equation}
(U_1,\ldots,U_r)
\sim
(gU_1g^{-1},\ldots,gU_rg^{-1}).
\label{eq:q15-simultaneous-conjugation}
\end{equation}

In the component where all monodromies can be conjugated into a common
maximal torus, one may write
\[
U_i
=
\exp\!\left(
2\pi i\,\boldsymbol\omega_i\cdot\mathbf H
\right).
\]
Each \(\boldsymbol\omega_i\) has its own cocharacter-lattice
identifications, while one common Weyl transformation
\(w\in\mathcal W\) acts simultaneously on the whole tuple.

The toral component \(T^r/\mathcal W\) need not exhaust
\(\operatorname{Hom}(\mathbb Z^r,G)/G\). For compact groups with
non-trivial global structure, non-toral commuting tuples can occur,
and the representation space may possess additional components. This
richer structure will matter only when the global form of the symmetry
group and several independent cycles are simultaneously relevant.

The compactification also carries a marking of its cycles. A change
of primitive basis
\(\Lambda\in GL(r,\mathbb Z)\) acts on
\(\pi_1(T^r)=\mathbb Z^r\), and hence on the monodromy tuple.
Such a transformation is a physical redundancy only when the complete
torus data, including its lattice, metric and background fields, are
transformed as well. Acting only on the monodromies while keeping the
rest of the compactification fixed need not define an equivalence.

\narrativeheading{Residual symmetry as a centralizer}

Once a rigid twist \(\rho\) has been chosen, an internal
transformation survives the compactification only if it is compatible
with every gluing operation. The residual symmetry group is therefore
\begin{equation}
G_{\rm residual}(\rho)
=
Z_G\!\left(\rho(\Gamma)\right),
\label{eq:q15-centralizer}
\end{equation}
with
\[
Z_G\!\left(\rho(\Gamma)\right)
=
\left\{
g\in G
\;\middle|\;
g\rho(\gamma)=\rho(\gamma)g
\quad
\forall\gamma\in\Gamma
\right\}.
\]

The important point is that this is the centralizer of the complete
image \(\rho(\Gamma)\), not necessarily the centralizer of one derived
element.

For a circle there is only one independent generator, so the
distinction is invisible. A regular element in the interior of a Weyl
alcove has centralizer equal to a maximal torus, while at special
points the centralizer can enlarge.

On orbifolds the distinction becomes essential. A translation
monodromy can have a large centralizer even when the complete set of
orbifold generators has only a smaller common centralizer.
Residual-symmetry enhancement is therefore naturally described as a
jump in the conjugacy type of \(Z_G(\rho(\Gamma))\).

\narrativeheading{Preserved supersymmetry as a fixed subspace}

Supersymmetry is tested on the same global gluing data, but in a
different representation. Let \(\mathcal S\) be the vector space of
supersymmetry parameters and let
\[
\rho_{\mathcal S}:
\Gamma
\longrightarrow
GL(\mathcal S)
\]
be the representation induced by \(\rho\) on that space.

For a flat background in which the global gluing is the only
non-trivial obstruction to a Killing spinor, a supersymmetry parameter
survives precisely when it is invariant under every element of
\(\Gamma\):
\begin{equation}
\mathcal S_{\rm preserved}(\rho)
=
\mathcal S^{\rho_{\mathcal S}(\Gamma)}
=
\bigcap_{\gamma\in\Gamma}
\ker\!\left[
\rho_{\mathcal S}(\gamma)-\mathbf1
\right].
\label{eq:q15-fixed-subspace}
\end{equation}
This is the criterion of
Question~\ref{sec:twist-supersymmetry}, now written for an arbitrary
identification group. Partial breaking corresponds to
\(0<\dim\mathcal S_{\rm preserved}<\dim\mathcal S\).

For commuting monodromies lying in a common maximal torus, let
\(\lambda_A\) be the weight of a supercharge \(Q_A\). Then
\[
U_iQ_A
=
e^{2\pi i\langle\lambda_A,\boldsymbol\omega_i\rangle}
Q_A.
\]
The supercharge survives precisely when
\begin{equation}
e^{2\pi i
\langle\lambda_A,\boldsymbol\omega_i\rangle}
=
1
\qquad
\text{for every independent cycle }i.
\label{eq:q15-weight-condition}
\end{equation}

Several cycles can therefore reduce the preserved supersymmetry
because the final invariant subspace is the intersection of the fixed
subspaces of all monodromies.

\narrativeheading{Enhancement loci}

At each point \([\rho]\in\mathcal X(\Gamma,G)\), two quantities are
naturally associated with the same gluing data:
\begin{equation}
[\rho]
\longmapsto
\left(
Z_G\!\left(\rho(\Gamma)\right),
\;
\mathcal S^{\rho_{\mathcal S}(\Gamma)}
\right).
\label{eq:q15-two-functions}
\end{equation}
The first probes the adjoint action on the internal symmetry algebra;
the second probes the representation carried by the supercharges.
The two enhancement loci can differ.

Consider first
\[
G=SU(2)_R,
\qquad
U(\omega)=e^{2\pi i\omega\sigma_3},
\]
acting on a doublet of supercharges.

For \(0<\omega<1/2\), the element is regular and
\(Z_{SU(2)}(U)=U(1)\). Its eigenvalues on the doublet are
\(e^{\pm2\pi i\omega}\), so no supercharge survives for generic
\(\omega\) in this interval.

At \(\omega=0\), \(U=\mathbf1\): the centralizer becomes all of
\(SU(2)\), and the complete supercharge doublet is invariant.

At the other endpoint, \(\omega=1/2\), one has
\(U=-\mathbf1\). Since this element lies in the centre,
\[
Z_{SU(2)}(U)=SU(2),
\]
but \(U-\mathbf1=-2\mathbf1\) is invertible on the supercharge
doublet. Thus no supercharge survives. Maximal enhancement of the
residual internal symmetry can coexist with complete
supersymmetry breaking.

This circle result should not be confused with the half-twist orbifold
of Question~\ref{sec:ss-half-twist}. On \(S^1/\mathbb Z_2\), the
rigid datum is the full representation of \(D_\infty\), equivalently
the pair \((P_0,P_\pi)\). The translation monodromy
\(U=P_\pi P_0\) is only a derived element. At the half-twist,
\(U=-\mathbf1\) still has centralizer \(SU(2)_R\), whereas the symmetry
preserving the complete orbifold gluing data is
\begin{equation}
Z_{SU(2)_R}(P_0)
\cap
Z_{SU(2)_R}(P_\pi)
=
U(1)_R.
\label{eq:q15-half-orbifold-centralizer}
\end{equation}
This is the common centralizer found in
Question~\ref{sec:ss-half-twist}, and it is this \(U(1)_R\) that
protects the Dirac gravitino.

The converse phenomenon can also occur. Suppose the relevant
\(R\)-symmetry contains \(SU(4)_R\), with the supercharges in the
fundamental representation, and consider
\begin{equation}
U
=
\operatorname{diag}
\left(
1,\,
e^{i\alpha},\,
e^{i\beta},\,
e^{-i(\alpha+\beta)}
\right).
\label{eq:q15-su4-example}
\end{equation}
For generic \(\alpha\) and \(\beta\) such that all four eigenvalues are
distinct, \(U\) is regular and
\[
Z_{SU(4)}(U)=U(1)^3.
\]
Nevertheless, the first component of the fundamental has eigenvalue
one, so one supercharge survives. Partial supersymmetry can therefore
occur without any enhancement of the residual internal symmetry.

The four possibilities can be summarized as
\begin{equation}
\begin{array}{c|c|l}
Z_G
&
\dim\mathcal S_{\rm preserved}
&
\text{example}
\\[1mm]
\hline
\text{generic}
&
0
&
\text{generic }SU(2)_R\text{ twist}
\\
\text{enhanced}
&
\text{maximal}
&
\omega=0
\\
\text{enhanced}
&
0
&
\omega=\tfrac12
\\
\text{generic}
&
\text{partial}
&
\text{regular }SU(4)_R\text{ example}
\end{array}
\label{eq:q15-four-strata}
\end{equation}

Twist space is stratified independently by centralizer type
and by the dimension of the preserved-supercharge subspace.

\narrativeheading{Beyond the flat monodromy problem}

The fixed-subspace formula
Eq.~\eqref{eq:q15-fixed-subspace} is exact only when the global gluing
is the sole non-trivial obstruction to a Killing spinor.

In a curved or fluxed compactification, the supersymmetry parameters
must also satisfy the complete local Killing system,
schematically
\begin{equation}
\mathcal D_m\epsilon=0,
\qquad
\mathcal A(\phi,F,\ldots)\epsilon=0,
\label{eq:q15-full-Killing-system}
\end{equation}
where \(\mathcal D_m\) is the relevant supercovariant derivative and
the second equation denotes collectively the algebraic fermionic
supersymmetry conditions generated by background scalars, fluxes and
other bosonic fields.

The preserved supersymmetries are then the global solutions of this
complete system. Equation~\eqref{eq:q15-fixed-subspace} is the flat
special case in which the only non-trivial global information is the
representation \(\rho_{\mathcal S}\).

Rigid orbifold gluing fits naturally into the same language. For
\(S^1/\mathbb Z_2\), \(\Gamma=D_\infty\), with
\(\rho(r_0)=P_0\) and \(\rho(r_\pi)=P_\pi\). The relations
\[
U=P_\pi P_0,
\qquad
P_0UP_0=U^{-1}
\]
follow directly from the defining relations of \(D_\infty\).

Localized boundary dynamics of the type discussed in
Question~\ref{sec:ss-localized-boundary-terms} adds further matching
conditions and need not be encoded by the rigid representation
\(\rho\) alone.

\narrativeheading{Physical equivalences beyond conjugation}

The quotient by conjugation in
Eq.~\eqref{eq:q15-character-variety} is the minimal identification
associated with a global change of field basis. It need not be the
complete physical quotient.

Schematically, the physical parameter space takes the form
\begin{equation}
\mathcal M_{\rm physical}
=
\frac{
\{\text{consistent rigid compactification and gluing data}\}
}{
\{\text{physical equivalences of the complete theory}\}
}.
\label{eq:q15-physical-moduli}
\end{equation}
The denominator may contain mapping-class transformations, further
global identifications and exact quantum dualities.

Care is required not to quotient twice by the same redundancy. A gauge
transformation at a base point already acts on holonomies by
simultaneous conjugation and is therefore contained in
\(\operatorname{Hom}(\Gamma,G)/G\). Likewise, in a Cartan
description, the cocharacter-lattice identifications are already part
of the extended affine Weyl action.

A further transformation should be imposed only when it represents an
additional physical equivalence of the complete compactification. This
is theory dependent: a symmetry of the classical lower-dimensional
supergravity may fail to survive in the ultraviolet completion,
whereas an exact string duality may identify backgrounds that appear
different in a particular effective description.

\narrativeheading{What a spectrum can and cannot determine}

A Kaluza--Klein spectrum probes the twist only in the representation
carried by the states whose masses are being measured.

For a sector transforming in a representation \(\mathcal R\), the
spectral data contain information such as
\[
\operatorname{Spec}_{\mathcal R}
\!\left(\rho(\gamma)\right).
\]
This need not determine the full class
\([\rho]\in\mathcal X(\Gamma,G)\).

Two twists can, for example, have identical eigenvalues on one
multiplet while acting differently on other fields. That multiplet
then has the same Kaluza--Klein shifts in the two compactifications,
even though the complete representations
\(\rho:\Gamma\to G\) are inequivalent.

A spectrum measured in one representation therefore cannot in general
reconstruct the complete twist class, still less the global
compactification, its localized sectors or its ultraviolet
equivalences.

There is an additional subtlety from
Question~\ref{sec:ss-lifting}: the phase extracted from a spectrum may
be a dynamical \(\omega_{\rm eff}\), rather than the rigid parameter
\(\omega_{\rm rigid}\) entering \(\rho\).

The spectral problem and the classification problem are therefore
closely related, but they contain different amounts of information.

\narrativeheading{Bibliographical remarks}

The classification of flat holonomies on circles and higher tori is
closely related to standard results on compact Lie groups and flat
bundles. For a circle, conjugacy classes in a compact connected group
are described by a maximal torus modulo the Weyl group, with the
global form of the group encoded in the relevant cocharacter lattice;
see, for example,
Refs.~\cite{Borel:1991,Bourbaki:Lie4-6}.

Representation spaces
\(\operatorname{Hom}(\Gamma,G)/G\) and moduli spaces of flat
connections are discussed, for example, in
Refs.~\cite{Atiyah:1982fa,Goldman:1988}. For several commuting
holonomies, the representation space can contain components not
captured by the simple toral description; related structures are
discussed in Refs.~\cite{Witten:1997bs,Borel:1999bx}.

In Scherk--Schwarz compactification, the same mathematics classifies
rigid global gluing data. Preserved supersymmetry is obtained by
applying the gluing representation to the supersymmetry parameters and
taking the common invariant subspace, while the residual internal
symmetry is the centralizer of the full image of the identification
group. Partial supersymmetry breaking by non-trivial string
monodromies is discussed, for example, in
Ref.~\cite{Kiritsis:1997ca}. The distinction between the translation
monodromy and the complete orbifold data was illustrated explicitly in
Questions~\ref{sec:ss-orbifolds} and~\ref{sec:ss-half-twist}.

For compactifications with several cycles or non-trivial global forms
of the symmetry group, the topology of commuting-tuple spaces can be
richer than the toral description \(T^r/\mathcal W\). Exact mapping
classes and ultraviolet dualities may further identify points of the
classical representation space, but such identifications depend on
the complete theory.

\newpage
\begin{takehome}
\begin{takehomeitems}

\item
A rigid Scherk--Schwarz twist is a homomorphism
\[
\rho:\Gamma\to G,
\qquad
\mathcal X(\Gamma,G)
=
\operatorname{Hom}(\Gamma,G)/G,
\]
where the quotient implements global changes of field basis.

\item
For a circle and compact connected \(G\),
\[
\mathcal X(\mathbb Z,G)\simeq T/\mathcal W.
\]
In Cartan coordinates, the global form of \(G\) enters through the
cocharacter lattice and the corresponding affine or extended affine
Weyl identifications.

\item
For several cycles, the rigid data are commuting tuples modulo
simultaneous conjugation. Their toral component need not exhaust the
full representation space.

\item
Residual symmetry and preserved supersymmetry are determined by
\[
G_{\rm residual}
=
Z_G\!\left(\rho(\Gamma)\right),
\qquad
\mathcal S_{\rm preserved}
=
\bigcap_{\gamma\in\Gamma}
\ker\!\left[
\rho_{\mathcal S}(\gamma)-\mathbf1
\right].
\]
They probe different representations of the same gluing data and
their enhancement loci need not coincide.

\item
The \(SU(2)_R\) half-twist on a circle has maximal internal
centralizer but preserves no supercharge, while a regular \(SU(4)_R\)
twist can preserve one supercharge without enlarging its generic
centralizer.

\item
The full image \(\rho(\Gamma)\) matters. On \(S^1/\mathbb Z_2\), the
translation monodromy alone does not determine the residual symmetry;
the complete endpoint data must be included.

\item
Conjugation is only the minimal equivalence relation, and a spectrum
in one representation does not in general reconstruct the complete
twist class. Mapping classes, exact dualities and dynamical effective
phases must be treated separately when present.

\end{takehomeitems}
\end{takehome}

\nextquestion
{What do supertrace sum rules reveal about the ultraviolet softness of
Scherk--Schwarz breaking?}

\end{reviewquestion}
% Question 17 v3, renumbered from Question16v2.tex.
% Original section/equation labels retained for cross-reference stability.
%%%%%%%%%%%%%%%%%%%%%%%%%%%%%%%%%%%%%%%%%%%%%%%%%%%%%%%%%%%%%%%

\begin{reviewquestion}
{What do supertrace sum rules reveal about the ultraviolet softness of
Scherk--Schwarz breaking?}
\label{sec:ss-supertraces}

\begin{essentialidea}

Supertraces organize the ultraviolet sensitivity of a
supersymmetry-breaking spectrum, but spontaneous supersymmetry
breaking alone does not imply
\[
\operatorname{Str}M^2=0.
\]
A finite lower-dimensional supergravity can therefore display ordinary
quadratic sensitivity.

A genuine Scherk--Schwarz compactification contains additional
information: the complete Kaluza--Klein tower is correlated by one
global monodromy.  This correlation is not captured by an arbitrary
finite mass matrix.

The distinction becomes manifest after Poisson resummation.  The
Kaluza--Klein sum is rewritten as a sum over winding sectors.  The
zero-winding sector contains the local short-distance contribution and
is independent of the rigid Scherk--Schwarz phase.  All dependence on
the twist lies in non-zero winding sectors, which require propagation
around the compact direction and are exponentially suppressed at short
proper time.

The twist-dependent one-loop contribution is therefore finite and
non-local, of Casimir type.  For \(D\) non-compact dimensions,
\[
V_D(\omega)
=
\frac{1}{R^D}f_D(\omega),
\qquad
f_D(\omega)
\propto
\sum_{w=1}^{\infty}
\frac{1-\cos(2\pi w\omega)}{w^{D+1}}.
\]
For a fixed order-one twist this scales parametrically as
\(m_{3/2}^D\) when \(m_{3/2}\sim\omega/R\), although the full
\(\omega\)-dependence must be retained when the twist is small.

The ultraviolet softness of Scherk--Schwarz breaking is thus a
property of the complete higher-dimensional spectral problem and of
the locality of the breaking datum.  It does not imply the absence of
ordinary local counterterms in the parent theory or, on orbifolds, at
fixed loci.

\end{essentialidea}

\narrativeheading{Why begin with supertraces}

For a four-dimensional theory with field-dependent masses
\(m_s(\phi)\), a cutoff representation of the one-loop effective
potential has schematically the form
\begin{equation}
V_{\rm 1-loop}
=
c_0\Lambda^4\operatorname{Str}M^0
+
c_2\Lambda^2\operatorname{Str}M^2
+
c_4\operatorname{Str}
\left[
M^4\log\frac{M^2}{\Lambda^2}
\right]
+\cdots .
\label{eq:q16-cutoff-expansion}
\end{equation}
The numerical coefficients depend on the regulator and normalization
conventions, but the supertraces organize which combinations of the
spectrum can accompany the various ultraviolet sensitivities.

For states labelled by \(s\), with spin \(J_s\) and mass \(m_s\), we
define
\begin{equation}
\operatorname{Str}M^{2k}
=
\sum_s
(-1)^{2J_s}
(2J_s+1)\,
m_s^{2k}.
\label{eq:q16-supertrace-definition}
\end{equation}
The sign distinguishes bosons from fermions, while \(2J_s+1\) counts
the physical polarizations in the usual four-dimensional massive-state
notation.

An exactly supersymmetric massive spectrum has equal bosonic and
fermionic contributions within complete supermultiplets.  In
particular, \(\operatorname{Str}M^0=0\).  Once supersymmetry is
broken, however, the fate of higher supertraces depends on the theory
and on which spectrum is included in the trace.

\narrativeheading{A generic broken supergravity need not have
\(\operatorname{Str}M^2=0\)}

Spontaneous supersymmetry breaking by itself does not force the
quadratic supertrace to vanish.  In a four-dimensional \(N=1\)
supergravity with broken supersymmetry and vanishing classical vacuum
energy, one may write schematically
\begin{equation}
\operatorname{Str}M^2
=
2Q\,m_{3/2}^2,
\label{eq:q16-generic-StrM2}
\end{equation}
where \(Q\) is a dimensionless quantity determined by the scalar and
gauge kinetic geometry and by the supersymmetry-breaking direction.
There is no general reason for \(Q\) to vanish.

Particular no-scale or string-derived supergravities can satisfy
additional relations for which \(Q=0\), and hence
\(\operatorname{Str}M^2=0\).  Examples of this type occur in the
string-effective supergravities discussed in
Ref.~\cite{Ferrara:1994kg}.

This condition should not be confused with the no-scale cancellation
of Question~\ref{sec:ss-no-scale}.  The latter is a statement about
the classical scalar potential; the former is a statement about a
mass spectrum and its ultraviolet supertrace.

The Scherk--Schwarz problem adds another ingredient: the spectrum is
an infinite correlated Kaluza--Klein tower rather than an arbitrary
finite collection of split multiplets.

\narrativeheading{A Scherk--Schwarz tower contains more information
than a finite mass matrix}

Consider the elementary circle spectrum with \(N\) real bosonic
degrees of freedom and the same number of real fermionic degrees of
freedom,
\begin{equation}
m_{B,n}^2
=
\frac{n^2}{R^2},
\qquad
m_{F,n}^2
=
\frac{(n+\omega)^2}{R^2},
\qquad
n\in\mathbb Z .
\label{eq:q16-simple-spectrum}
\end{equation}
The masses at different Kaluza--Klein levels are not independent:
the complete fermionic tower is obtained from the bosonic one by the
single spectral displacement \(n\mapsto n+\omega\).

One may formally write
\begin{equation}
\operatorname{Str}M^{2k}
=
N
\sum_{n\in\mathbb Z}
\left[
\left(\frac nR\right)^{2k}
-
\left(\frac{n+\omega}{R}\right)^{2k}
\right].
\label{eq:q16-formal-KK-supertrace}
\end{equation}
This is not an ordinary convergent sum, so its ultraviolet meaning
depends on how the infinite tower is regulated.  A cutoff imposed
directly on the integer label \(n\), for example, need not respect the
spectral-flow relation between the shifted and unshifted towers.

The appropriate ultraviolet statement is instead obtained from the
complete one-loop spectral determinant.

\narrativeheading{The one-loop determinant and Poisson resummation}

Begin in four non-compact dimensions and subtract the supersymmetric
value at \(\omega=0\):
\begin{equation}
V_1(\omega)
=
\frac N2
\sum_{n\in\mathbb Z}
\int\frac{d^4p}{(2\pi)^4}
\left[
\log\!\left(
p^2+\frac{n^2}{R^2}
\right)
-
\log\!\left(
p^2+\frac{(n+\omega)^2}{R^2}
\right)
\right].
\label{eq:q16-one-loop-determinant}
\end{equation}
Here \(p\) is Euclidean momentum and the factor \(1/2\) is the usual
one-loop factor for a real degree of freedom.

Using
\[
\log A
=
-\int_0^\infty\frac{dt}{t}\,e^{-tA},
\qquad
\int\frac{d^4p}{(2\pi)^4}e^{-tp^2}
=
\frac{1}{16\pi^2t^2},
\]
gives
\begin{equation}
V_1(\omega)
=
-\frac{N}{32\pi^2}
\int_0^\infty\frac{dt}{t^3}
\sum_{n\in\mathbb Z}
\left[
e^{-tn^2/R^2}
-
e^{-t(n+\omega)^2/R^2}
\right].
\label{eq:q16-Schwinger}
\end{equation}
The Schwinger parameter \(t\) has the usual interpretation: the
short-proper-time region \(t\to0\) probes ultraviolet physics.

Now use Poisson resummation,
\begin{equation}
\sum_{n\in\mathbb Z}
e^{-t(n+\omega)^2/R^2}
=
R\sqrt{\frac{\pi}{t}}
\sum_{w\in\mathbb Z}
e^{-\pi^2R^2w^2/t}
e^{2\pi iw\omega}.
\label{eq:q16-Poisson}
\end{equation}
The integer \(w\) is the winding number in the resummed description.
Substitution gives
\begin{equation}
V_1(\omega)
=
-\frac{NR\sqrt\pi}{32\pi^2}
\sum_{w\in\mathbb Z}
\left(
1-e^{2\pi iw\omega}
\right)
\int_0^\infty
dt\,
t^{-7/2}
e^{-\pi^2R^2w^2/t}.
\label{eq:q16-Poisson-potential}
\end{equation}

\narrativeheading{Why the twist-dependent ultraviolet part vanishes}

The zero-winding sector is special.  For \(w=0\), the factor
multiplying the proper-time integral is exactly
\begin{equation}
1-e^{2\pi iw\omega}=0.
\label{eq:q16-zero-winding-cancellation}
\end{equation}
The complete zero-winding contribution to the
\(\omega\)-dependent difference therefore vanishes before the
proper-time integral is performed.  There is no limiting
\(0\times\infty\) ambiguity.

This has a direct geometric interpretation.  A zero-winding path does
not traverse the non-contractible circle.  It probes only a local
neighbourhood of the higher-dimensional spacetime and cannot detect a
global Scherk--Schwarz holonomy.

For \(w\neq0\), the integrand instead contains
\[
\exp\!\left(
-\frac{\pi^2R^2w^2}{t}
\right),
\]
which is exponentially suppressed as \(t\to0\).  Every non-zero
winding sector is ultraviolet finite.

The conclusion is precise: the divergent local part of the one-loop
effective action is insensitive to the rigid Scherk--Schwarz phase,
while the phase-dependent contribution comes entirely from finite
non-zero winding sectors.  This does not say that the parent
higher-dimensional theory has no ultraviolet divergences.  It says
that its local divergent coefficients cannot depend on a global flat
monodromy.

\narrativeheading{The general-\(D\) Casimir law}

The same argument works in any number \(D\) of non-compact dimensions.
The Euclidean Gaussian integral is
\begin{equation}
\int
\frac{d^Dp}{(2\pi)^D}
e^{-tp^2}
=
\frac{1}{(4\pi t)^{D/2}}.
\label{eq:q16-D-momentum-integral}
\end{equation}
After Poisson resummation, a non-zero winding sector contains
\begin{equation}
\int_0^\infty
dt\,
t^{-(D+3)/2}
e^{-\pi^2R^2w^2/t}
=
\frac{
\Gamma\!\left(\frac{D+1}{2}\right)
}{
\pi^{D+1}R^{D+1}|w|^{D+1}
}.
\label{eq:q16-general-D-integral}
\end{equation}

Combining this result with the factor produced by Poisson resummation
and pairing \(w\) with \(-w\) gives
\begin{equation}
V_D(\omega)
=
-
\frac{
N\,\Gamma\!\left(\frac{D+1}{2}\right)
}{
(4\pi)^{D/2}\pi^{D+1/2}
}
\,
\frac{1}{R^D}
\sum_{w=1}^{\infty}
\frac{
1-\cos(2\pi w\omega)
}{
w^{D+1}
}.
\label{eq:q16-general-D-exact}
\end{equation}
The sign corresponds to the convention used above, with the bosonic
tower unshifted and the fermionic tower shifted.

The structurally important result is
\begin{equation}
V_D(\omega)
=
\frac{1}{R^D}\,f_D(\omega),
\label{eq:q16-general-D-scaling}
\end{equation}
with \(f_D\) a finite periodic dimensionless function.  For a fixed
non-zero twist of order unity, \(m_{3/2}\sim\omega/R\), and therefore
\(V_D\sim m_{3/2}^D\) parametrically.

This last statement is not an identity in \(\omega\).  For small
\(\omega\) and \(D>2\),
\[
1-\cos(2\pi w\omega)
=
2\pi^2w^2\omega^2+O(\omega^4),
\]
so that
\begin{equation}
V_D(\omega)
\sim
\frac{\omega^2}{R^D}
\sum_{w=1}^{\infty}\frac1{w^{D-1}}
\sim
\frac{m_{3/2}^2}{R^{D-2}}.
\label{eq:q16-small-omega}
\end{equation}
Thus \(R^{-D}f_D(\omega)\), rather than a pure power of the gravitino
mass, is the invariant statement.

\narrativeheading{The explicit four-dimensional result}

For \(D=4\), Eq.~\eqref{eq:q16-general-D-exact} reduces to
\begin{equation}
V_4(\omega)
=
-\frac{3N}{64\pi^6R^4}
\sum_{w=1}^{\infty}
\frac{
1-\cos(2\pi w\omega)
}{
w^5}.
\label{eq:q16-finite-potential}
\end{equation}
The result is finite, periodic under \(\omega\to\omega+1\), vanishes
at the supersymmetric point, and has the expected \(R^{-4}\) Casimir
scaling.

If desired, the winding sum can be expressed in terms of
polylogarithms:
\begin{equation}
V_4(\omega)
=
-\frac{3N}{64\pi^6R^4}
\left[
\zeta(5)
-
\frac12
\operatorname{Li}_5(e^{2\pi i\omega})
-
\frac12
\operatorname{Li}_5(e^{-2\pi i\omega})
\right].
\label{eq:q16-polylog}
\end{equation}
This form is useful for analytic manipulations, but the winding sum
makes the locality properties more transparent.

\narrativeheading{The Casimir interpretation}

Poisson resummation has reorganized the determinant according to the
distance explored in the compact direction.  The zero-winding sector
is local: it can be represented by the usual short-time heat-kernel
coefficients of the higher-dimensional differential operator.
Non-zero winding sectors probe the global compact geometry.

This is the same physical separation encountered from the boundary
scattering picture in
Question~\ref{sec:ss-localized-boundary-terms}.  There the non-local
part arose from propagation over a finite compact distance and back;
here it appears as non-zero winding around the circle.  The two
descriptions are different organizations of the same Casimir
physics.

The periodic-field description of
Question~\ref{sec:ss-wilson-radion} provides a complementary view.  A
rigid Scherk--Schwarz monodromy can be represented by a flat
connection: its local curvature vanishes, while its gauge-invariant
information is carried by the holonomy around the circle.  Local
ultraviolet counterterms can depend on local field strengths,
couplings and background invariants, but they cannot intrinsically
depend on a flat holonomy that can only be measured by transporting a
state around a non-contractible cycle.

Thus the winding, Casimir and Wilson-line descriptions give three
ways of recognizing the same locality property.

\narrativeheading{What a finite Kaluza--Klein truncation misses}

The need for the complete tower can be seen directly.  Retain only
the levels \(|n|\leq N_{\rm max}\).  Then
\begin{align}
\operatorname{Str}_{|n|\leq N_{\rm max}}M^2
&=
\frac{N}{R^2}
\sum_{n=-N_{\rm max}}^{N_{\rm max}}
\left[
n^2-(n+\omega)^2
\right]
\nonumber\\
&=
-\frac{
N(2N_{\rm max}+1)\omega^2
}{R^2}.
\label{eq:q16-truncated-StrM2}
\end{align}
This diverges linearly as \(N_{\rm max}\to\infty\).

The divergence does not reproduce the ultraviolet behaviour of the
complete higher-dimensional determinant.  A symmetric cutoff on the
integer label truncates the shifted and unshifted towers at the same
values of \(n\), thereby breaking the spectral-flow relation precisely
at the edges of the sum.

A finite Kaluza--Klein truncation can therefore have supertraces that
look ultraviolet sensitive even though the twist-dependent part of
the complete spectral determinant is finite.

This also explains why the coefficient \(Q\) of a finite
lower-dimensional supergravity and the finite Scherk--Schwarz Casimir
potential answer different questions.  The former is a local
supertrace in the truncated effective theory; the latter is obtained
from the complete correlated Kaluza--Klein spectrum.

\narrativeheading{Rigid monodromy and effective spectral phases}

Question~\ref{sec:ss-lifting} distinguished a rigid microscopic
monodromy \(\omega_{\rm rigid}\) from an effective phase
\(\omega_{\rm eff}(\varphi)\) that can also depend on localized
dynamical fields.

The ultraviolet argument above applies directly to the rigid global
part: a contractible short-distance process cannot detect it.  If the
effective phase receives contributions from localized dynamical
sectors, however, those sectors are local microscopic degrees of
freedom and can renormalize local operators in the usual way.

Two theories can consequently display the same
\(\omega_{\rm eff}\) in a particular spectrum while having different
ultraviolet structures.  Equality of the spectral shift does not
imply equality of the underlying local dynamics.

\narrativeheading{From the smooth circle to boundaries}

A smooth circle has no distinguished local submanifold on which new
counterterms can be supported.  Orbifolds and intervals do.

A loop confined to an arbitrarily small neighbourhood of a fixed locus
can detect the local projection, the local field content and the
symmetries preserved there.  It cannot, however, detect a distant
boundary or the complete global Scherk--Schwarz holonomy.

The one-loop effective action can therefore be organized schematically
as
\begin{equation}
\Gamma_{\rm 1-loop}
=
\Gamma_{\rm bulk}^{\rm local}
+
\sum_f
\Gamma_f^{\rm local}
+
\Gamma_{\rm finite}^{\rm nonlocal},
\label{eq:q16-local-nonlocal-decomposition}
\end{equation}
where \(f\) labels fixed loci.  The precise local operators depend on
the field content, interactions and boundary conditions; the final
term contains the genuinely non-local dependence on the complete
compactification.

The ultraviolet softness of the global Scherk--Schwarz effect
therefore does not eliminate local bulk or fixed-locus counterterms.
It tells us instead that the divergent coefficients of such local
operators cannot depend on global information that is invisible in an
arbitrarily small neighbourhood.  The allowed local divergences are
the subject of the next question.

\narrativeheading{Bibliographical remarks}

Supertrace sum rules have long been used to characterize ultraviolet
sensitivity in spontaneously broken supersymmetric theories and
supergravities. Additional relations in no-scale and string-derived
effective supergravities can lead to cancellations such as
\(\operatorname{Str}M^2=0\); an example relevant to the present
discussion can be found in Ref.~\cite{Ferrara:1994kg}.

For Scherk--Schwarz compactifications, proper-time methods and Poisson
resummation make the ultraviolet softness of the twist-dependent
contribution manifest: the local zero-winding sector is separated from
the finite non-zero winding sectors that probe the global
compactification. On orbifolds and intervals, additional local
counterterms may be supported on fixed loci; these will be discussed in
the next question.

\begin{takehome}
\begin{takehomeitems}

\item
Supertraces diagnose the ultraviolet sensitivity of a specified
spectrum, but spontaneous supersymmetry breaking alone does not imply
\(\operatorname{Str}M^2=0\).  A finite lower-dimensional supergravity
and the complete Scherk--Schwarz Kaluza--Klein tower need not have the
same supertrace behaviour.

\item
After Poisson resummation, the zero-winding sector contains the local
short-distance contribution.  Its twist-dependent part vanishes
identically before the proper-time integral is performed.

\item
All dependence on a rigid Scherk--Schwarz phase resides in non-zero
winding sectors.  They are exponentially suppressed at short proper
time and give a finite non-local Casimir contribution.

\item
For \(D\) non-compact dimensions,
\[
V_D(\omega)
=
\frac{1}{R^D}f_D(\omega).
\]
For a fixed order-one twist this scales parametrically as
\(m_{3/2}^D\), while the small-\(\omega\) behaviour must be extracted
from the full periodic function \(f_D\).

\item
A cutoff on a finite set of Kaluza--Klein levels can destroy the
spectral-flow organization of the tower and generate divergent
supertraces that do not reproduce the ultraviolet behaviour of the
complete higher-dimensional determinant.

\item
The finiteness concerns the global twist-dependent contribution, not
all ultraviolet divergences.  Local bulk counterterms remain allowed,
and orbifolds or intervals can additionally generate local
fixed-locus counterterms.

\end{takehomeitems}
\end{takehome}

\nextquestion
{Which ultraviolet divergences and local counterterms remain allowed?}

\end{reviewquestion}
% Question 18 v4, renumbered from Question17v2.tex.
% Original section/equation labels retained for cross-reference stability.
%%%%%%%%%%%%%%%%%%%%%%%%%%%%%%%%%%%%%%%%%%%%%%%%%%%%%%%%%%%%%%%

\begin{reviewquestion}
{Which ultraviolet divergences and local counterterms remain allowed?}
\label{sec:ss-local-divergences}

\begin{essentialidea}

The ultraviolet softness established in
Question~\ref{sec:ss-supertraces} concerns the genuinely non-local,
Scherk--Schwarz-dependent part of the effective action. It does not
imply that the compactified theory has no local ultraviolet
divergences.

The correct organization is by support. At one loop, the
short-proper-time expansion of the heat kernel separates local
contributions supported in the regular bulk from those supported on
fixed loci. Bulk coefficients renormalize bulk operators, while
fixed-locus coefficients renormalize operators localized on the
corresponding strata.

The global Scherk--Schwarz information belongs to a different part of
the effective action. In an image or worldline representation, a
group element can contribute a local ultraviolet term only where it
has a fixed set. If it moves every point by a finite distance, its
short-proper-time contribution is exponentially suppressed as
\(\exp[-\ell^2/(4t)]\) and is ultraviolet finite.

For \(S^1/\mathbb Z_2\), the identity therefore produces bulk local
counterterms, the endpoint reflections produce fixed-point local
counterterms, and non-zero translations produce finite winding or
Casimir terms.

A divergent functional localized at one endpoint may depend on the
local boundary operator there, but not jointly on independently
variable data at a distant endpoint. Mixed dependence on disjoint
fixed loci is consequently non-local and ultraviolet finite.

Schematically,
\[
\Gamma^{(1)}
=
\Gamma_{\rm bulk}^{\rm local}
+
\sum_f\Gamma_f^{\rm local}
+
\Gamma_{\rm Casimir}^{\rm nonlocal}.
\]
The first terms may contain divergences and scheme-dependent finite
pieces. The last contains the dependence on relative boundary data,
transported holonomy and compactification size that cannot be altered
by local counterterms.

\end{essentialidea}

\narrativeheading{The one-loop problem as a heat-kernel problem}

Work in Euclidean signature on
\(\mathscr X=M_D\times K\), where \(M_D\) is the
\(D\)-dimensional non-compact spacetime and \(K\) the compact internal
space.

After gauge fixing, the quadratic fluctuation operators can be
organized into Laplace-type blocks
\begin{equation}
\Delta
=
-
\left(
g^{MN}\nabla_M\nabla_N+\mathcal E
\right),
\label{eq:q17-Laplace-type}
\end{equation}
where \(\nabla_M\) contains the relevant spin and gauge connections
and \(\mathcal E\) is a local endomorphism of the field bundle.
First-order fermionic operators may instead be treated through the
corresponding squared Laplace-type operator, with ghosts, statistics
and multiplicities included in the supertrace.

The one-loop effective action is formally
\begin{equation}
\Gamma^{(1)}
=
-\frac12
\int_0^\infty
\frac{dt}{t}\,
\operatorname{Str}e^{-t\Delta}.
\label{eq:q17-one-loop-heat}
\end{equation}
The ultraviolet region is the short-proper-time limit \(t\to0\).

Let the internal space be stratified into smooth pieces
\(\Sigma\), including the regular bulk and any fixed or singular
loci. The heat trace then has the asymptotic expansion
\begin{equation}
\operatorname{Str}e^{-t\Delta}
\sim
\sum_{\Sigma}
\frac{1}{
(4\pi t)^{(D+\dim\Sigma)/2}
}
\sum_{n\geq0}
t^{n/2}
A_{n,\Sigma},
\label{eq:q17-stratified-expansion}
\end{equation}
where each \(A_{n,\Sigma}\) is an integral over
\(\widehat\Sigma=M_D\times\Sigma\) of a local density constructed from
the fluctuation operator, the background geometry and fields, and the
local boundary or matching data.

These are the standard heat-kernel coefficients for manifolds with
boundaries or singular strata
\cite{Vassilevich:2003xt,DrydenEtAl:2008}. Hence the divergent part
is a sum of local functionals supported on the individual strata,
\begin{equation}
\Gamma_{\rm div}^{(1)}
=
\sum_{\Sigma}
\Gamma_{\Sigma,\rm div}^{\rm local}.
\label{eq:q17-stratified-divergences}
\end{equation}

Ultraviolet divergences are therefore \emph{stratum local}.

\narrativeheading{The fixed-set principle}

The geometric origin of this localization is especially transparent
when \(K=\widetilde K/\Gamma\) is a quotient of a covering space by a
discrete identification group. Let
\(\rho:\Gamma\to G\) denote the corresponding action on the field
fibres, as in Question~\ref{sec:ss-inequivalent-twists}.

In an image representation, the internal heat trace has the form
\begin{equation}
\operatorname{Tr}_{K,\rho}e^{-t\Delta}
=
\sum_{\gamma\in\Gamma}
\int_{\mathcal F}
d^dy\,
\operatorname{tr}
\left[
\rho(\gamma)\,
K_{\widetilde K}(t;y,\gamma y)
\right],
\label{eq:q17-image-trace}
\end{equation}
where \(\mathcal F\) is a fundamental domain.

At short proper time, the off-diagonal heat kernel behaves as
\begin{equation}
K_{\widetilde K}(t;y,\gamma y)
\sim
\frac{
e^{-d(y,\gamma y)^2/(4t)}
}{
(4\pi t)^{d/2}
}
\sum_{k\geq0}
t^k a_k(y,\gamma y).
\label{eq:q17-offdiagonal-heat}
\end{equation}

If \(\gamma\) has no fixed point and the action is properly
discontinuous, its displacement has a strictly positive minimum
\(\ell_\gamma\) on the fundamental domain. Its contribution is then
suppressed by \(e^{-\ell_\gamma^2/(4t)}\) as \(t\to0\), and is
ultraviolet finite.

If \(\gamma\) has a fixed set, the displacement vanishes there. The
short-time integral is no longer exponentially suppressed, and
expansion in the normal directions produces local heat-kernel
coefficients supported on \(\operatorname{Fix}(\gamma)\).

Thus fixed-point-free images contribute only to the finite non-local
part, whereas images with fixed sets can generate local ultraviolet
terms on those sets. This is the fixed-set form of the
local/non-local Casimir decomposition.

\narrativeheading{The image decomposition on \(S^1/\mathbb Z_2\)}

For the infinite dihedral group introduced in
Question~\ref{sec:ss-orbifolds}, take on the covering line
\(t:y\mapsto y+2\pi R\) and \(r_0:y\mapsto-y\), with
\(\rho(t)=U\) and \(\rho(r_0)=P_0\). The second endpoint involution is
\(P_\pi=UP_0\).

Every element of \(D_\infty\) is either \(t^w\) or \(t^wr_0\),
with \(w\in\mathbb Z\). For a flat scalar Laplacian, the image kernel
on \(0\leq y\leq\pi R\) can be written as
\begin{align}
K_\rho(t;y,y')
=
\sum_{w\in\mathbb Z}
\Big[
&
U^w
K_{\mathbb R}
\left(
t;y,y'+2\pi Rw
\right)
\nonumber\\
&
+
U^wP_0
K_{\mathbb R}
\left(
t;y,-y'+2\pi Rw
\right)
\Big].
\label{eq:q17-Dinfinity-kernel}
\end{align}

On the diagonal, the two image displacements are
\begin{align}
d\!\left(y,t^wy\right)
&=
2\pi R|w|,
\\
d\!\left(y,t^wr_0y\right)
&=
2\left|y-\pi Rw\right|.
\label{eq:q17-Dinfinity-displacements}
\end{align}

The identity translation fixes every point and produces the bulk
coefficients. Every non-zero translation has no fixed point and is
suppressed by
\(\exp[-\pi^2R^2w^2/t]\); these sectors contain the finite winding
dependence on \(U^w\).

The reflection \(r_0\) fixes \(y=0\) and carries \(P_0\), while
\(r_\pi=tr_0\) fixes \(y=\pi R\) and carries \(P_\pi\). The bulk,
fixed-point and winding contributions therefore arise directly from
the three geometrically distinct classes of images.

For the flat scalar block, the leading short-time trace is
\begin{equation}
\operatorname{Tr}_{I,\rho}
e^{t\partial_y^2}
=
\frac{\pi R}{\sqrt{4\pi t}}\,
\operatorname{tr}\mathbf1
+
\frac14
\operatorname{tr}
\left(
P_0+P_\pi
\right)
+
O\!\left(
e^{-\pi^2R^2/t}
\right).
\label{eq:q17-flat-interval-trace}
\end{equation}

For one scalar component, Neumann--Neumann conditions give the
constant term \(+1/2\), Dirichlet--Dirichlet gives \(-1/2\), and mixed
Dirichlet--Neumann gives zero. The first term is a bulk
invariant, the constant term contains local endpoint data, and
dependence requiring propagation over a finite distance lies beyond
the local short-time expansion.

\narrativeheading{Local boundary data}

The involutions \(P_b\) are only the simplest local boundary data. A
standard mixed boundary problem for a Laplace-type operator is
\begin{align}
\Pi_-^{(b)}\Phi\big|_b
&=
0,
\label{eq:q17-Dirichlet-projector}
\\
\left(
\nabla_n+\mathcal S_b
\right)
\Pi_+^{(b)}\Phi\big|_b
&=
0,
\label{eq:q17-Robin-projector}
\end{align}
where \(\nabla_n\) is the outward normal derivative and
\(\mathcal S_b\) a local endomorphism. The projectors are
complementary; for a pure orbifold parity,
\[
\Pi_\pm^{(b)}
=
\frac12
\left(
\mathbf1\pm P_b
\right).
\]
The elementary even Neumann condition corresponds to
\(\mathcal S_b=0\).

Localized masses and interactions can modify \(\mathcal S_b\) or
produce more general matching conditions, as discussed in
Question~\ref{sec:ss-localized-boundary-terms}. We denote collectively
by \(\mathcal B_b=(\Pi_-^{(b)},\Pi_+^{(b)},\mathcal S_b,\ldots)\) the
local data defining the endpoint problem.

Boundary heat-kernel coefficients are local polynomials in the germ
of \((\Delta,\mathcal B_b)\). They may involve the induced metric,
intrinsic and extrinsic curvature, \(\mathcal E\), the curvature
\(\Omega_{MN}=[\nabla_M,\nabla_N]\) of the bundle connection, the
projectors, \(\mathcal S_b\), and their local covariant derivatives.

They cannot involve independent data located at a disjoint endpoint.

\narrativeheading{The support principle}

Let \(\Sigma\) be the support of a local counterterm. Its divergent
coefficient is determined by a finite jet of the fluctuation operator
and of the local matching data on that support. Schematically,
\begin{equation}
\Gamma_{\Sigma,\rm div}^{\rm local}
=
\int_{\widehat\Sigma}
\mathcal L_{\Sigma,\rm div}
\left[
j^N
\left(
\Delta,\mathcal B_\Sigma
\right)
\right].
\label{eq:q17-support-functional}
\end{equation}

For a bulk counterterm, only bulk germs enter. For a counterterm at
\(y=0\), the coefficient may depend on \(\mathcal B_0\), but not on
an independently variable boundary operator \(\mathcal B_\pi\) at the
other endpoint. A radion field may of course enter if it is itself part of the local
field content. What cannot enter a divergent coefficient is
information whose definition requires comparing separated regions,
such as a transported product of independent endpoint data.

\narrativeheading{A disjoint-support finiteness criterion}

Let \(\mathcal B_0\) and \(\mathcal B_\pi\) denote independently
variable data at the two endpoints. Locality implies
\begin{equation}
\Gamma_{\rm div}
=
\Gamma_{\rm bulk}^{\rm local}
+
\Gamma_0^{\rm local}[\mathcal B_0]
+
\Gamma_\pi^{\rm local}[\mathcal B_\pi].
\label{eq:q17-local-sum}
\end{equation}
There can therefore be no connected divergent term depending jointly
on the two disjoint supports:
\begin{equation}
\frac{
\delta^2\Gamma_{\rm div}
}{
\delta\mathcal B_0\,
\delta\mathcal B_\pi
}
=
0.
\label{eq:q17-mixed-divergence-zero}
\end{equation}

The second functional derivative of a local functional is supported
on the diagonal, whereas the two variations above have disjoint
supports. Consequently,
\begin{equation}
\frac{
\delta^2\Gamma^{(1)}
}{
\delta\mathcal B_0\,
\delta\mathcal B_\pi
}
=
\frac{
\delta^2\Gamma_{\rm nonlocal}^{(1)}
}{
\delta\mathcal B_0\,
\delta\mathcal B_\pi
}
\label{eq:q17-mixed-response}
\end{equation}
is ultraviolet finite, apart from possible infrared singularities.

This is the analogue of the interaction part of a Casimir energy:
individual surface self-energies may be divergent and
scheme-dependent, while a response that necessarily correlates
separated supports belongs to the finite non-local sector.

A related effect appears explicitly in heterotic orbifold amplitudes.
Twisted states localized at fixed points renormalize amplitudes
involving Kaluza--Klein exchange differently from purely untwisted
contributions because localized interactions need not conserve
internal KK momentum
\cite{Antoniadis:1993jp}. This provides an early concrete example of
localized radiative effects; the general support argument above does
not rely on the model-specific resummation used there.

\narrativeheading{Local symmetries on each stratum}

The support principle also determines which symmetries constrain a
local counterterm. Let \(\mathcal B_b\) be the complete local boundary
operator at an endpoint \(b\). Its local stabilizer in the parent
internal symmetry group is
\begin{equation}
G_b^{\rm loc}
=
\operatorname{Stab}_G(\mathcal B_b).
\label{eq:q17-local-stabilizer}
\end{equation}

For a pure orbifold boundary condition this reduces to
\(G_b^{\rm loc}=Z_G(P_b)\), the centralizer encountered in
Questions~\ref{sec:ss-half-twist}
and~\ref{sec:ss-inequivalent-twists}. Additional localized masses,
kinetic terms or interactions can reduce this stabilizer further.

Let \(\mathcal S_b^{\rm loc}\) denote the supersymmetry parameters
satisfying the local boundary projector and all local algebraic
boundary Killing conditions at \(b\). Assuming these symmetries are
non-anomalous and the regulator preserves them, a divergent density
supported at \(b\) must respect the corresponding local internal and
supersymmetry constraints.

The local supersymmetry spaces at the two endpoints need not agree.
Global supersymmetry may vanish because they fail to match after
transport,
\begin{equation}
\mathcal S_{\rm global}
=
\mathcal S_0^{\rm loc}
\cap
\mathcal U(\pi R,0)^{-1}
\mathcal S_\pi^{\rm loc}
=
\{0\}.
\label{eq:q17-global-misalignment}
\end{equation}

A divergence at \(y=0\) is nevertheless constrained by the
supersymmetry locally realized there. It cannot detect its
misalignment with the supersymmetry preserved at \(y=\pi R\), because
that comparison requires finite propagation between the two
endpoints.

This is the ultraviolet counterpart of the distinction between local
endpoint supersymmetry and global Killing-spinor compatibility made in
Questions~\ref{sec:ss-orbifolds}
and~\ref{sec:ss-localized-boundary-terms}.

\narrativeheading{When can a boundary divergence depend on the twist?}

The answer must be phrased in terms of local boundary data rather than
in terms of a coordinate called \(\omega\).

In one description, a continuous family of compactifications may be
represented by a rotating endpoint involution
\(P_\pi=P_\pi(\omega)\). A counterterm localized at \(y=\pi R\) may
then depend on \(P_\pi(\omega)\), because the local boundary problem
itself has changed. This is ordinary sensitivity to a local boundary
condition, not ultraviolet sensitivity to a distant holonomy.

In another trivialization, part of the same compactification data may
be represented by a flat bulk connection together with transformed
endpoint operators. The local heat-kernel coefficients depend on the
local equivalence class of \((\Delta,\mathcal B_b)\), not on how the
same physical data are distributed between a bulk connection and an
endpoint matrix.

Thus, if \(\omega\) varies while the local germ at a support remains
fixed and only the global flat transport changes, the divergent
coefficient there cannot depend on \(\omega\). If varying \(\omega\)
changes the local boundary operator itself, the local counterterms can
change accordingly.

\narrativeheading{Boundary rotations and the non-local determinant}

For the commuting quadratic boundary problem of
Question~\ref{sec:ss-localized-boundary-terms}, the spectrum can take
the form
\begin{equation}
m_n
=
\frac{n+\omega_{\rm eff}}{R},
\qquad
\omega_{\rm eff}
=
\omega_{\rm rigid}
+
\frac{\Theta_0+\Theta_\pi}{\pi}.
\label{eq:q17-effective-spectrum}
\end{equation}

If the boundary data affect the quadratic problem only through this
uniform shift, without adding extra propagating fields,
momentum-dependent form factors or additional eigenvalue branches, the
non-local determinant is obtained from the smooth-circle result of
Question~\ref{sec:ss-supertraces} by replacing
\(\omega\) with \(\omega_{\rm eff}\).

This statement concerns only the non-local spectral determinant. The
local counterterms still depend separately on the microscopic
operators \(\mathcal B_0\) and \(\mathcal B_\pi\). Two theories with
the same \(\omega_{\rm eff}\) can share the same quadratic
Casimir dependence while having different local renormalizations,
localized degrees of freedom and Goldstino sectors.

\narrativeheading{Counterterms versus radiatively transmitted soft
terms}

The same lower-dimensional operator can receive both a locally
renormalized contribution and a finite non-local
Scherk--Schwarz contribution. For an operator \(\mathcal O_b\)
supported on a boundary, write schematically
\begin{equation}
c_{\mathcal O_b}^{\rm phys}
=
c_{\mathcal O_b}^{\rm local}(\mu)
+
c_{\mathcal O_b}^{\rm SS}
\left(
R,U,\mathcal B_0,\mathcal B_\pi
\right).
\label{eq:q17-local-plus-nonlocal}
\end{equation}

The first term contains ultraviolet matching information and depends
on the renormalization prescription. The second is generated by
propagation that detects the global Scherk--Schwarz configuration and
is ultraviolet finite.
A calculation of the finite term
\(c_{\mathcal O_b}^{\rm SS}\) does not determine the full physical
coefficient whenever the corresponding local counterterm is allowed
by the exact local symmetries.

A Scherk--Schwarz-induced soft term is therefore predicted without
additional ultraviolet input only when the corresponding local
counterterm is forbidden, related to other couplings by an exact local
symmetry, or fixed by an independent renormalization condition.

If the local term is forbidden while the global compactification
breaks the protecting symmetry non-locally, the leading contribution
can be a finite transmitted soft term. This is the mechanism studied
in the next question. If instead the local counterterm is allowed, the
finite Scherk--Schwarz contribution remains a calculable threshold,
but the total coefficient is not a UV-independent prediction of the
compactification alone.

\narrativeheading{Bibliographical remarks}

Heat-kernel methods on manifolds with boundaries and singular strata,
including mixed Dirichlet--Robin conditions and local boundary
invariants, are reviewed in Refs.~\cite{Vassilevich:2003xt,DrydenEtAl:2008}.

The generation of counterterms localized at orbifold fixed planes by
bulk quantum loops was exhibited explicitly in
Ref.~\cite{Georgi:2000ks}. Related one-loop bulk and brane
renormalization effects in gauge theories on \(S^1/\mathbb Z_2\) were
analysed in Ref.~\cite{vonGersdorff:2002as}.

An early heterotic-orbifold example in which fixed-point localized
twisted states affect the renormalization of Kaluza--Klein exchange
amplitudes, reflecting the loss of internal KK-momentum conservation
at localized interactions, can be found in
Ref.~\cite{Antoniadis:1993jp}.

\begin{takehome}
\begin{takehomeitems}

\item
The ultraviolet expansion on an orbifold is stratified: the bulk and
the fixed loci possess separate local heat-kernel coefficients and
therefore separate local counterterms.

\item
An image contributes to the local ultraviolet expansion only where it
has a fixed set. Fixed-point-free images are exponentially suppressed
at short proper time and belong to the finite non-local sector.

\item
For \(S^1/\mathbb Z_2\), the identity generates bulk terms, the two
endpoint reflections generate fixed-point terms, and non-zero
translations generate finite winding/Casimir contributions.

\item
A divergent coefficient depends only on the local fluctuation and
boundary data on its own support. In particular,
\[
\frac{\delta^2\Gamma_{\rm div}}
{\delta\mathcal B_0\,\delta\mathcal B_\pi}=0,
\]
so a response that necessarily correlates two disjoint endpoints is
ultraviolet finite.

\item
Boundary counterterms are constrained by the exact local symmetries
at their support. Global supersymmetry can nevertheless be broken by
a mismatch between locally supersymmetric boundary conditions.

\item
A finite Scherk--Schwarz contribution and a local counterterm are
different pieces of the effective theory. Their sum is a
UV-independent prediction only when the local term is forbidden,
symmetry related, or fixed independently.

\end{takehomeitems}
\end{takehome}

\nextquestion
{How is Scherk--Schwarz supersymmetry breaking radiatively transmitted
to boundary fields?}

\end{reviewquestion}
% Question 19 v3, renumbered from Question18v3.tex.
% Original section/equation labels retained for cross-reference stability.
%%%%%%%%%%%%%%%%%%%%%%%%%%%%%%%%%%%%%%%%%%%%%%%%%%%%%%%%%%%%%%%

\begin{reviewquestion}
{How is Scherk--Schwarz supersymmetry breaking radiatively transmitted
to boundary fields?}
\label{sec:ss-boundary-mediation}

\begin{essentialidea}

A field localized on a boundary has no Kaluza--Klein tower of its own
and does not wind around the compact dimension. It can nevertheless
feel Scherk--Schwarz supersymmetry breaking through bulk propagators
whose spectrum and boundary wave functions depend on the global
compactification data.

The cleanest benchmark is the pure anti-periodic point on
\(S^1/\mathbb Z_2\), with \(\omega=\tfrac12\) and
\(M_0=M_\pi=0\). The two light spin-\(\tfrac32\) states then form the
Dirac gravitino of Question~\ref{sec:ss-half-twist}, with
\(m_{3/2}=1/(2R)\), and the boundary-value problem has the residual
\(U(1)_R\) discussed there.

Radiative transmission is controlled by three separate pieces of
information: whether a boundary operator is allowed by the local
symmetries, whether boundary matter couples to the required bulk
channel, and whether the corresponding boundary-restricted propagator
contains a local short-distance contribution or probes the global
compactification.

These distinctions already separate scalar and fermion masses. The
ordinary gravitino couples universally to the boundary supercurrent,
and a boundary scalar receives a finite Scherk--Schwarz gravitational
threshold. A Majorana mass for an \(R\)-charged boundary fermion is
forbidden at the exact half-twist. A Dirac mass can be \(R\)-neutral,
but requires an additional matter current coupled to the companion
spin-\(\tfrac32\) channel; its existence and normalization are not
fixed by the Scherk--Schwarz geometry alone.

Near the half-twist, the protecting \(U(1)_R\) is weakly broken.
Scalar thresholds are even under reflection about
\(\omega=\tfrac12\), whereas Majorana responses are odd. The former
therefore change quadratically in the displacement from the
half-twist, while the latter can appear linearly.

\end{essentialidea}

\narrativeheading{Boundary mediation as a restricted-resolvent problem}

Let the compact interval be \(I=[0,\pi R]\), and denote by
\(\Delta_\omega\) the Euclidean quadratic bulk operator with the
Scherk--Schwarz boundary conditions labelled by \(\omega\). For fixed non-compact Euclidean momentum \(p\), define
\begin{equation}
\mathscr G_\omega(p)
=
\left(
p^2+\Delta_\omega
\right)^{-1}.
\label{eq:q18-bulk-resolvent}
\end{equation}

The field value itself is not always the appropriate datum at an
orbifold boundary. We therefore let \(\mathcal T_b\) denote the
boundary restriction map at the endpoint \(b=0,\pi\), including the
relevant parity projectors and normal derivatives.

The boundary-to-boundary kernels are then
\begin{equation}
\mathbb G_{bb'}(p;\omega)
=
\mathcal T_b\,
\mathscr G_\omega(p)\,
\mathcal T_{b'}^\dagger .
\label{eq:q18-boundary-resolvent}
\end{equation}
If boundary currents \(J_b\) couple linearly to the corresponding
restricted bulk fields, integrating out the bulk fluctuations gives,
at quadratic order,
\begin{equation}
\Gamma_{\rm ind}^{(2)}
=
-\frac12
\sum_{b,b'}
\int
\frac{d^4p}{(2\pi)^4}\,
J_b(-p)\,
\mathbb G_{bb'}(p;\omega)\,
J_{b'}(p),
\label{eq:q18-induced-boundary-action}
\end{equation}
up to statistics and index contractions.

The same restricted resolvents enter loop corrections to boundary
masses. The compactification fixes the kernels
\(\mathbb G_{bb'}\), while the microscopic boundary theory fixes
which currents exist and how strongly they couple. An operator allowed
by symmetry is therefore not necessarily generated.

\narrativeheading{The pure half-twist benchmark}

We now set \(\omega=\tfrac12\) and \(M_0=M_\pi=0\). The breaking is
then entirely in the Scherk--Schwarz gluing data, with no independent
localized gravitino masses.

On \(0\leq y\leq\pi R\), the two half-integer towers relevant to the
spin-\(\tfrac32\) system may be represented by
\begin{align}
f_n^{(+)}(y)
&=
\sqrt{\frac{2}{\pi R}}\,
\cos\!\left[
\frac{n+\tfrac12}{R}\,y
\right],
\nonumber\\
f_n^{(-)}(y)
&=
\sqrt{\frac{2}{\pi R}}\,
\sin\!\left[
\frac{n+\tfrac12}{R}\,y
\right],
\qquad n\geq0.
\label{eq:q18-half-modes}
\end{align}
At \(y=0\), the \(+\) profiles have non-zero boundary values, whereas
the \(-\) profiles vanish but have non-zero normal derivatives. Thus
the even channel couples directly to ordinary boundary currents,
while the odd channel can enter through derivative or bulk-mediated
couplings. This kinematic fact does not determine whether a suitable
matter current exists for the odd channel.

For \(p=|p_E|\), the corresponding Green functions are
\begin{align}
G_+(p;y,y')
&=
\frac{
\cosh(py_<)\,
\sinh\!\big[p(\pi R-y_>)\big]
}{
p\,\cosh(\pi pR)
},
\nonumber\\
G_-(p;y,y')
&=
\frac{
\sinh(py_<)\,
\cosh\!\big[p(\pi R-y_>)\big]
}{
p\,\cosh(\pi pR)
},
\label{eq:q18-half-Gpm}
\end{align}
where \(y_<\) and \(y_>\) denote respectively the smaller and larger
of \(y,y'\).

The even same-boundary propagator at \(y=0\) is
\(\tanh(\pi pR)/p\) at the half-twist, while in the unbroken
\(\omega=0\) problem it is \(\coth(\pi pR)/p\). Their
Scherk--Schwarz-dependent difference is therefore
\begin{equation}
\Delta G_{00}^{\rm SS}(p)
\equiv
G_+^{(1/2)}(p;0,0)
-
G_+^{(0)}(p;0,0)
=
-\frac{2}{
p\,\sinh(2\pi pR)
}.
\label{eq:q18-half-same-difference}
\end{equation}
At large momentum this behaves as
\(\Delta G_{00}^{\rm SS}(p)\sim-4e^{-2\pi pR}/p\).

This provides an instructive qualification to the locality discussion
of Question~\ref{sec:ss-local-divergences}. A same-boundary amplitude
need not be ultraviolet divergent: here the local short-distance
singularity is identical in the twisted and untwisted theories and
cancels in their difference. By contrast, a contribution that
necessarily propagates between opposite endpoints contains factors
such as \(1/\cosh(\pi pR)\), and is exponentially suppressed because
the two supports are separated by the finite distance \(\pi R\).

\narrativeheading{The universal scalar threshold}

Consider a canonically normalized chiral multiplet
\(Q=(\phi,\psi,F)\) localized at one endpoint. The ordinary gravitino
couples universally to the local supersymmetry current,
\begin{equation}
\mathcal L_{\psi J}
=
-\frac{1}{2M_4}\,
\psi_\mu J^\mu_{\rm SUSY}
+\text{h.c.},
\qquad
J^\mu_{\rm SUSY}
\supset
\sqrt2\,
(\partial_\nu\bar\phi)\,
\sigma^\nu\bar\sigma^\mu\psi+\cdots .
\label{eq:q18-supercurrent}
\end{equation}

The scalar threshold is not obtained from one isolated
gravitino-exchange diagram. Local supersymmetry relates the cubic
supercurrent vertices to quartic contact, or seagull, interactions,
and the complete locally supersymmetric set must be combined before
subtracting the unbroken reference background.

For the minimal boundary-coupled five-dimensional supergravity
benchmark, the resulting matching coefficient is
\begin{equation}
\mathcal C_A
=
\frac{2\pi R}{3M_4^2}
=
\frac{2}{3M_5^3},
\qquad
M_4^2=\pi R\,M_5^3.
\label{eq:q18-scalar-matching}
\end{equation}
The Scherk--Schwarz contribution to the scalar mass is then
\begin{equation}
\delta m_\phi^2
=
\mathcal C_A
\int
\frac{d^4p_E}{(2\pi)^4}\,
p_E^2\,
\Delta G_{00}^{\rm SS}(p).
\label{eq:q18-scalar-threshold-integral}
\end{equation}
Using Eq.~\eqref{eq:q18-half-same-difference} and performing the
angular integral gives
\begin{equation}
\delta m_\phi^2
=
-\frac{\mathcal C_A}{4\pi^2}
\int_0^\infty
dp\,
\frac{p^4}{
\sinh(2\pi pR)
}.
\label{eq:q18-scalar-radial}
\end{equation}
With \(x=2\pi Rp\) and
\begin{equation}
\int_0^\infty
dx\,
\frac{x^4}{\sinh x}
=
\frac{93}{2}\,\zeta(5),
\label{eq:q18-zeta5-integral}
\end{equation}
one obtains
\begin{equation}
\delta m_\phi^2
=
-
\frac{31\zeta(5)}{128\pi^6}
\frac{1}{R^4M_4^2}
=
-
\frac{31\zeta(5)}{128\pi^7}
\frac{1}{R^5M_5^3}.
\label{eq:q18-universal-scalar}
\end{equation}

The sign is negative in this minimal anti-periodic benchmark. This is
a statement about one finite Scherk--Schwarz threshold, not about the
sign of the complete physical scalar mass. Local renormalized terms,
radion effects and additional interactions can contribute to the full
mass matrix, as emphasized in Question~\ref{sec:ss-local-divergences}.

The gravitational scalar threshold in this class of
Scherk--Schwarz compactifications has been analysed in
Refs.~\cite{Gherghetta:2001sa,Rattazzi:2003rj,Antoniadis:1997ic,Antoniadis:2015chx}.

\narrativeheading{Fermion masses and the residual \(U(1)_R\)}

At the exact half-twist, the residual \(U(1)_R\) constrains the
fermionic operators that can be generated on the boundary.

For a localized Weyl fermion \(\psi\) of \(R\)-charge \(q_\psi\), a
Majorana bilinear \(\psi\psi\) carries charge \(2q_\psi\).  An exact
\(U(1)_R\) therefore forbids such a mass whenever \(q_\psi\neq0\).

A Dirac bilinear \(\psi\xi\), by contrast, is compatible with the
symmetry when \(q_\psi+q_\xi=0\).  This condition is necessary but not
sufficient.  The existence of the Dirac gravitino of
Question~\ref{sec:ss-half-twist} does not by itself imply that both of
its spin-\(\tfrac32\) components couple to the boundary matter sector
in the way required to generate a Dirac mass.

Denoting the two gravitini by \(\psi_\mu\) and \(\chi_\mu\), the
relevant couplings have schematically the form
\begin{equation}
\mathcal L_{\rm matter}
\supset
-\frac{1}{2M_4}\,
\psi_\mu J^\mu
-
\chi_\mu j_\chi^\mu
+\text{h.c.}
\label{eq:q18-companion-current}
\end{equation}
The ordinary supercurrent \(J^\mu\) is fixed by local supersymmetry.
The existence and normalization of a companion current
\(j_\chi^\mu\) depend instead on how the boundary matter sector is
embedded into the higher-dimensional or extended-supergravity theory.

A radiative Dirac mass therefore requires more than the
spin-\(\tfrac32\) spectrum: an oppositely charged fermion must be
present, a companion matter current must couple to \(\chi_\mu\), and
the loop must connect the two matter channels.  These ingredients are
not fixed by the Scherk--Schwarz geometry alone.

\narrativeheading{The companion-channel kernels}

Once the companion current is specified, its propagation is fixed by
the half-twist boundary problem. The mixed even--odd propagator with
one bulk Dirac-mass insertion can be written
\begin{equation}
G_{+-}^{(1)}(p;y,y')
=
m_{3/2}
\int_0^{\pi R}
dz\,
G_+(p;y,z)\,
G_-(p;z,y').
\label{eq:q18-mixed-propagator}
\end{equation}

For two currents on the same boundary,
\begin{equation}
\left.
\partial_{y'}
G_{+-}^{(1)}(p;0,y')
\right|_{y'=0}
=
\frac{m_{3/2}}{2p^2}\,
\tanh^2(\pi pR).
\label{eq:q18-same-companion-kernel}
\end{equation}
The kernel approaches a power-law local term at large momentum.
Accordingly, its complete coefficient requires a boundary
counterterm. After that local piece is subtracted, the finite
Scherk--Schwarz remainder contains
\begin{equation}
\int_0^\infty
dp\,p
\left[
\tanh^2(\pi pR)-1
\right]
=
-\frac{\ln2}{\pi^2R^2}.
\label{eq:q18-same-finite-integral}
\end{equation}

For currents on opposite boundaries one finds instead
\begin{equation}
G_{+-}^{(1)}(p;0,\pi R)
=
\frac{m_{3/2}}
{2p^3\cosh(\pi pR)}
\left[
\pi pR-\tanh(\pi pR)
\right].
\label{eq:q18-opposite-companion-kernel}
\end{equation}
This kernel is exponentially suppressed at large momentum and is
intrinsically ultraviolet finite. The corresponding dimensionless
integral is
\begin{equation}
\int_0^\infty
dx\,
\frac{x-\tanh x}{\cosh x}
=
2G_{\rm Cat}-1,
\label{eq:q18-Catalan-integral}
\end{equation}
where \(G_{\rm Cat}\) is Catalan's constant.

To separate these geometrical results from the microscopic matter
couplings, introduce coefficients
\(\mathcal A_{\rm same}\) and \(\mathcal A_{\rm opp}\), of mass
dimensions \(-2\) and \(-1\), respectively. They contain the
companion-current normalization, tensor contractions and the factors
required to close the matter loop. The finite same-boundary part and
the opposite-boundary contribution then take the form
\begin{align}
m_{D,\rm same}^{\rm fin}
&=
-
\frac{
\mathcal A_{\rm same}\ln2
}{
4\pi^4
}
m_{3/2}^3,
\nonumber\\
m_{D,\rm opp}
&=
\frac{
\mathcal A_{\rm opp}(2G_{\rm Cat}-1)
}{
8\pi^3
}
m_{3/2}^2.
\label{eq:q18-Dirac-thresholds}
\end{align}
The first must be supplemented by an independently renormalized local
term,
\(m_{D,\rm same}^{\rm phys}(\mu)
=m_{D,\rm local}(\mu)+m_{D,\rm same}^{\rm fin}\).

For a conservative supergravity normalization one may write
\(\mathcal A_{\rm same}=a_{\rm same}/M_4^2\) and
\(\mathcal A_{\rm opp}=a_{\rm opp}m_{3/2}/M_4^2\), with dimensionless
microscopic coefficients. Both topologies then give parametrically
\begin{equation}
m_D^{\rm matter}
\sim
\frac{1}{16\pi^2}
\frac{m_{3/2}^3}{M_4^2},
\label{eq:q18-Dirac-scaling}
\end{equation}
up to the model-dependent coefficients and the numerical factors
displayed above. A parametrically larger result requires a stronger
normalization of the companion current and is an additional effective
theory assumption, not a consequence of the Scherk--Schwarz
boundary conditions.

\narrativeheading{Small departures from the half-twist}

Consider now an effective spectral phase
\(\omega_{\rm eff}=\tfrac12-\epsilon_\omega\), with
\(|\epsilon_\omega|\ll1\). The two light spin-\(\tfrac32\) masses are
\((1/2\mp\epsilon_\omega)/R\), so their splitting is
\(2|\epsilon_\omega|/R\). The residual \(U(1)_R\) of the exact
half-twist is then weakly broken and Majorana operators are no longer
forbidden.

The expansion around the half-twist is nevertheless constrained by
the reflection \(\omega\leftrightarrow1-\omega\). Scalar thresholds
are even under this reflection, whereas the Majorana response is odd:
\begin{equation}
m_{\rm scalar}^2
\left(
\frac12-\epsilon_\omega
\right)
=
m_{\rm scalar}^2
\left(
\frac12
\right)
+
O(\epsilon_\omega^2),
\qquad
m_{\rm Majorana}
\left(
\frac12-\epsilon_\omega
\right)
=
O(\epsilon_\omega).
\label{eq:q18-half-expansion}
\end{equation}

For the gravitationally transmitted Majorana benchmark, the relevant
spectral function may be written
\begin{equation}
f_{1/2}(\omega)
=
\frac{3i}{8\pi^3}
\left[
\operatorname{Li}_4
\left(
e^{-2\pi i\omega}
\right)
-
\operatorname{Li}_4
\left(
e^{2\pi i\omega}
\right)
\right],
\label{eq:q18-majorana-function}
\end{equation}
which gives
\begin{equation}
f_{1/2}
\left(
\frac12-\epsilon_\omega
\right)
=
\frac{9\zeta(3)}{8\pi^2}\,
\epsilon_\omega
+
O(\epsilon_\omega^3).
\label{eq:q18-majorana-half-expansion}
\end{equation}
With gravitational-strength mediation this corresponds parametrically
to
\begin{equation}
M_{1/2}^{\rm grav}
\sim
\epsilon_\omega\,
\frac{1}{16\pi^2}
\frac{m_{3/2}^3}{M_4^2}.
\label{eq:q18-majorana-scaling}
\end{equation}

The exact half-twist is a useful expansion point: the scalar
threshold is already non-zero there but has no linear displacement,
whereas the Majorana response starts at first order. The
Dirac/pseudo-Dirac organization and the corresponding matter
selection rules are discussed further in
Ref.~\cite{Benakli:2026zwx}.

\narrativeheading{What the spin-\(\tfrac32\) spectrum does not fix}

The distinction between rigid monodromy and an effective spectral
phase made in Questions~\ref{sec:ss-localized-boundary-terms},
\ref{sec:ss-lifting} and~\ref{sec:ss-local-divergences} remains
important here.

Two compactifications can have the same spin-\(\tfrac32\) eigenvalues
and the same \(\omega_{\rm eff}\) while differing in their localized
Goldstino sectors, boundary counterterms, companion currents, or
direct local \(R\)-breaking interactions. The spin-\(\tfrac32\)
spectrum fixes the propagation problem only after the complete
quadratic boundary conditions have been specified; it does not fix
the microscopic boundary vertices.

The pure half-twist geometry therefore determines the even and odd
bulk wave functions, the corresponding restricted kernels and the
residual \(U(1)_R\). Local supersymmetry also fixes the ordinary
supercurrent coupling and hence the universal scalar threshold.
Companion-current normalizations and allowed local counterterms remain
additional information.

\narrativeheading{Bibliographical remarks}

Finite radiative transmission of supersymmetry breaking through an
extra dimension, including applications to localized matter and
soft terms, was studied in
Refs.~\cite{Antoniadis:1998sd,Kaplan:1999ac,Chacko:1999mi,Barbieri:2000vh,Arkani-Hamed:2001mi}.
Related Scherk--Schwarz orbifold calculations can be found in
Ref.~\cite{vonGersdorff:2002tj}.

The Dirac and pseudo-Dirac organization near the half-twist, together
with the companion-current analysis used above, is discussed in
Ref.~\cite{Benakli:2026zwx}.

\begin{takehome}
\begin{takehomeitems}

\item
Boundary fields can feel Scherk--Schwarz breaking through
boundary-restricted bulk propagators even though they have no
Kaluza--Klein momentum of their own.

\item
Same-boundary and opposite-boundary effects can both be ultraviolet
finite for different reasons. In the former, the local singularity
can cancel after subtraction of the supersymmetric reference theory;
in the latter, finite separation gives exponential suppression.

\item
For the minimal pure half-twist benchmark, the locally supersymmetric
gravitational coupling gives
\[
\delta m_\phi^2
=
-
\frac{31\zeta(5)}{128\pi^6}
\frac{1}{R^4M_4^2}.
\]
This is one finite contribution to the scalar mass matrix, not a
prediction of its final sign.

\item
At the exact half-twist, the residual \(U(1)_R\) forbids a Majorana
mass for an \(R\)-charged boundary fermion.  An \(R\)-invariant Dirac
mass is not generated merely because the bulk gravitino is Dirac:
the second spin-\(\tfrac32\) component must couple to an appropriate
companion matter current so that a non-vanishing loop can actually be
closed.  The existence and normalization of this current are
additional microscopic data, not consequences of the
Scherk--Schwarz spectrum.

\item
Once such a companion current exists, the half-twist fixes the
propagation kernels connecting it to the ordinary matter current.
Same-boundary diagrams contain a local piece that must be
renormalized, leaving a finite remainder proportional to \(\ln2\);
opposite-boundary diagrams necessarily propagate across the interval
and are intrinsically ultraviolet finite, with the corresponding
kernel giving the constant \(2G_{\rm Cat}-1\).

\item
Near the half-twist, scalar thresholds are even in the displacement
from \(\omega=\tfrac12\), while Majorana responses are odd. The same
spin-\(\tfrac32\) spectrum can nevertheless correspond to different
boundary soft spectra when the localized currents and counterterms
differ.

\end{takehomeitems}
\end{takehome}

\nextquestion
{What quantum consistency conditions must a Scherk--Schwarz twist
satisfy?}

\end{reviewquestion}
% Question 20 v3, renumbered from Question19v2.tex.
% Original section/equation labels retained for cross-reference stability.
%%%%%%%%%%%%%%%%%%%%%%%%%%%%%%%%%%%%%%%%%%%%%%%%%%%%%%%%%%%%%%%

\begin{reviewquestion}
{What quantum consistency conditions must a Scherk--Schwarz twist
satisfy?}
\label{sec:ss-quantum-consistency}

\begin{essentialidea}

Classically, a Scherk--Schwarz compactification is specified by
telling the fields how they are reidentified under the global
identifications of the compact space.  Question~\ref{sec:ss-inequivalent-twists}
organized these data into a homomorphism
\(\rho:\Gamma\to G\), where \(\Gamma\) is the identification group
and \(G\) an admissible symmetry group of the parent theory.

Quantum mechanically, two further questions arise.

The first concerns fermions.  Two transformations that act
identically on all bosonic operators can still differ by fermion
parity \((-1)^F\).  The same bosonic gluing can therefore admit
different fermionic lifts, or fail to admit one at all.

The second concerns anomalies.  A transformation may preserve the
classical action while changing the phase of the quantum functional
integral.  Such a transformation can still be used as a prescribed
background, but it cannot automatically define an autonomous quotient
or orbifold.

The classical twist must therefore first be lifted to the full
fermionic symmetry,
\(\widetilde\rho:\Gamma\to G_{\rm f}\), and the anomaly of the parent
theory must then be restricted to this lifted action.  If the parent
anomaly is denoted by \(\mathfrak A\), the relevant anomaly is
\[
\mathfrak A_{\widetilde\rho}
=
(B\widetilde\rho)^*\mathfrak A .
\]
An autonomous quantum compactification requires a coherent
trivialization of this anomaly.  At this level of description, the
quantum refinement of the classical twist is specified by
the fermionic lift together with such a trivialization.

A second layer appears when the twist is varied continuously.  The
fermionic spectrum changes, and particular Kaluza--Klein branches can
cross zero.  These resonance walls depend on the representation being
examined.  The adjoint representation detects residual-symmetry
enhancement, the supercharge representation detects supersymmetry
enhancement, and physical fermion representations detect their own
zero-mode walls.  These loci need not coincide.

Spectral flow and the eta invariant describe the passage through these
walls.  Globally, the fermion determinants form an anomaly line over
twist space.  Its curvature contains the perturbative anomaly, while
its holonomy detects global anomalies.

\end{essentialidea}

\narrativeheading{Why classical gluing is not yet quantum gluing}

The classical twist of Question~\ref{sec:ss-inequivalent-twists}
tells us how fields are identified when one moves around the compact
space.  At the quantum level, however, two extra consistency checks
appear before one can regard the compactification as a stand-alone
theory.

The first check is already visible on a circle.  Suppose all bosonic
fields are periodic.  A fermion can nevertheless satisfy either
\(\psi(y+2\pi R)=+\psi(y)\) or
\(\psi(y+2\pi R)=-\psi(y)\).  The bosonic gluing is identical, but
the fermionic theories are different: these are the two spin
structures on the circle.

The reason is fermion parity.  The transformation \((-1)^F\) acts
trivially on every bosonic operator and by a minus sign on every
fermionic operator.  Bosonic data alone therefore cannot distinguish
a transformation \(g\) from \((-1)^Fg\).

The first quantum question is consequently whether the classical
gluing admits a globally consistent action on the fermionic theory.

A second question remains even after such an action has been found.
A transformation can leave the classical action invariant but change
the phase of the functional integral.  The symmetry is then
anomalous.  Using it as a fixed external background remains
meaningful, but quotienting or gauging it requires additional
anomaly-cancellation data.

The two issues are logically distinct.  A twist may fail before
anomalies are considered because it cannot be lifted consistently to
the fermions, or the fermionic lift may exist but be anomalous.

\narrativeheading{Lifting the twist to the fermionic symmetry}

Let \(G_{\rm b}\) denote the symmetry group seen by bosonic operators.
The full symmetry acting on fermions can be a non-trivial central
extension
\begin{equation}
1
\longrightarrow
\mathbb Z_2^F
\longrightarrow
G_{\rm f}
\overset{\pi}{\longrightarrow}
G_{\rm b}
\longrightarrow
1,
\label{eq:q19-fermionic-extension}
\end{equation}
where \(\mathbb Z_2^F=\{\mathbf1,(-1)^F\}\).

Suppose the classical bosonic gluing is
\(\rho_{\rm b}:\Gamma\to G_{\rm b}\).  A fermionic lift is a
homomorphism \(\widetilde\rho:\Gamma\to G_{\rm f}\) satisfying
\begin{equation}
\pi\circ\widetilde\rho=\rho_{\rm b}.
\label{eq:q19-lift-condition}
\end{equation}

The central extension
Eq.~\eqref{eq:q19-fermionic-extension} is characterized by a class
\(e\in H^2(BG_{\rm b},\mathbb Z_2)\), where \(BG_{\rm b}\) is the
classifying space of \(G_{\rm b}\).  The bosonic twist induces a map
\(B\rho_{\rm b}:B\Gamma\to BG_{\rm b}\), and hence the pulled-back
class
\begin{equation}
(B\rho_{\rm b})^*e
\in
H^2(B\Gamma,\mathbb Z_2).
\label{eq:q19-pulled-extension}
\end{equation}
A strict fermionic lift exists precisely when the pulled-back
extension splits,
\begin{equation}
(B\rho_{\rm b})^*e=0.
\label{eq:q19-fermionic-lift-obstruction}
\end{equation}

This is the abstract statement that all fermionic signs assigned to
the generators of \(\Gamma\) must be compatible with the relations
among those generators.

Even when a lift exists, it need not be unique.  For a central
extension on which \(\Gamma\) acts trivially on \(\mathbb Z_2^F\),
the possible lifts form a torsor under
\begin{equation}
H^1(B\Gamma,\mathbb Z_2)
\simeq
\operatorname{Hom}(\Gamma,\mathbb Z_2).
\label{eq:q19-lift-torsor}
\end{equation}
A torsor has the same relative structure as a group but no preferred
origin: two lifts have a well-defined difference in
\(H^1(B\Gamma,\mathbb Z_2)\), without one lift being singled out as
canonical.

\narrativeheading{The circle as the elementary lifting problem}

For a circle, \(\Gamma=\mathbb Z\) and \(B\mathbb Z\simeq S^1\).
Consequently,
\[
H^2(B\mathbb Z,\mathbb Z_2)
=
H^2(S^1,\mathbb Z_2)
=
0.
\]
There is therefore no obstruction of the type
Eq.~\eqref{eq:q19-fermionic-lift-obstruction}: once the bosonic image
of the generator has been chosen, it can be lifted to the fermionic
symmetry.

The ambiguity is instead measured by
\[
H^1(B\mathbb Z,\mathbb Z_2)
=
H^1(S^1,\mathbb Z_2)
\simeq
\mathbb Z_2.
\]
Let \(1\in\mathbb Z\) denote the generator and let
\(g\in G_{\rm f}\) be one lift of its bosonic image.  The other is
\((-1)^Fg\).  On a fermion,
\begin{align}
\widetilde\rho_+(1)=g
&\quad\Longrightarrow\quad
\psi(y+2\pi R)=g\,\psi(y),
\nonumber\\
\widetilde\rho_-(1)=(-1)^Fg
&\quad\Longrightarrow\quad
\psi(y+2\pi R)=-g\,\psi(y).
\label{eq:q19-circle-two-lifts}
\end{align}

For trivial internal bosonic gluing, \(g=\mathbf1\), these are simply
the periodic and anti-periodic spin structures.  The elementary sign
choice encountered at the beginning of the review is the
simplest example of the general statement that an unobstructed
bosonic twist can possess inequivalent fermionic lifts.

The fermionic lift must also be compatible with the geometric spin
structure.  For an ordinary orientation-preserving unitary symmetry,
the combined local structure is schematically
\begin{equation}
\widehat G_d
=
\operatorname{Spin}(d)
\times_{\mathbb Z_2^F}
G_{\rm f},
\label{eq:q19-combined-spin-structure}
\end{equation}
where the central element
\(-1\in\operatorname{Spin}(d)\) is identified with fermion parity.
If the compactification includes orientation reversal or an
antiunitary operation, the Spin structure is replaced by the
appropriate Pin or more general twisted tangential structure.

The first quantum consistency condition is therefore already
non-trivial: the classical gluing must lift coherently to the complete
fermionic and geometric symmetry data.

\narrativeheading{A symmetry of the action can still be anomalous}

Assume now that a fermionic lift
\(\widetilde\rho:\Gamma\to G_{\rm f}\) has been chosen.  Classical
symmetry means that the action is invariant.  Quantum symmetry
requires more: the functional measure must transform consistently.

Schematically, one can have \(S[g\Phi]=S[\Phi]\) while
\[
D(g\Phi)
=
e^{i\alpha[g,A]}\,D\Phi,
\]
where the phase can depend on the transformation and on background
fields \(A\).  When this phase cannot be removed consistently by
local counterterms, the symmetry is anomalous.

This distinction matters directly for Scherk--Schwarz compactification.
A transformation can be an admissible classical monodromy and
nevertheless fail to define an autonomous quantum quotient.

There are several equivalent languages for anomalies.  The one most
useful for treating perturbative and global anomalies together is to
regard the anomaly of a \(d\)-dimensional quantum field theory as an
invertible \((d+1)\)-dimensional field theory
\(\mathfrak A_{G_{\rm f}}\).  The detailed categorical construction
will not be needed here.  Its immediate physical consequence is that,
on a closed \(d\)-dimensional spacetime \(X\) with symmetry
background \(A\), the anomalous partition function is not naturally
a complex number but an element of a one-dimensional complex vector
space,
\begin{equation}
Z_{\mathcal Q}(X,A)
\in
\mathcal L_{X,A}.
\label{eq:q19-line-valued-partition}
\end{equation}
The space \(\mathcal L_{X,A}\) is the anomaly line.

Locally in the space of backgrounds one may choose a phase convention
and represent the fermion determinant by an ordinary complex
function.  On an overlap, two such choices need agree only up to a
non-zero phase.  These local functions therefore assemble naturally
into a section of a complex line bundle rather than into one globally
defined function.

A trivialization identifies these lines coherently with \(\mathbb C\)
and turns the partition function into an ordinary number.
Determinant lines and exponentiated eta invariants provide concrete
realizations of this structure for fermionic theories
\cite{Dai:1994kq,Freed:2014iua,
Garcia-Etxebarria:2018ajm}.

\narrativeheading{Only the anomaly seen by the twist matters}

The parent symmetry group can be anomalous even though the particular
subgroup used in the Scherk--Schwarz compactification is not.

The lifted twist induces a map
\(B\widetilde\rho:B\Gamma\to BG_{\rm f}\).  The anomaly relevant to
the compactification is the anomaly of the parent theory restricted
to this action,
\begin{equation}
\mathfrak A_{\widetilde\rho}
=
(B\widetilde\rho)^*
\mathfrak A_{G_{\rm f}}.
\label{eq:q19-pulled-anomaly}
\end{equation}

It is therefore neither necessary that the whole group \(G_{\rm f}\)
be anomaly free nor sufficient that one infinitesimal generator look
harmless.  An anomalous continuous symmetry can contain an
anomaly-free discrete subgroup suitable for Scherk--Schwarz gluing.
Conversely, a continuous generator can have vanishing perturbative
anomaly while a corresponding discrete subgroup retains a torsion or
global anomaly.

The relevant object is always the complete pulled-back anomaly
\(\mathfrak A_{\widetilde\rho}\).

\narrativeheading{Fixed backgrounds, compactifications, and orbifolds}

An anomaly does not forbid every use of the corresponding symmetry.
What is required depends on what one wants to do with the symmetry
background.

If the Scherk--Schwarz configuration is treated as a prescribed
external background, the anomalous theory can still be evaluated.
Its partition function remains line valued,
\(Z(X,\widetilde\rho)\in\mathcal L_{X,\widetilde\rho}\).

A stand-alone compactification requires more.  Its partition function
must be an ordinary number, so the pulled-back anomaly must be
trivialized.  An orbifold or gauging is still more restrictive:
different symmetry backgrounds must be summed over or identified, and
the trivialization must be coherent over the corresponding family of
\(\Gamma\)-bundles and compatible with cutting and gluing.

A trivialization is a natural isomorphism
\begin{equation}
\tau:
\mathfrak A_{\widetilde\rho}
\overset{\sim}{\longrightarrow}
\mathbf1,
\label{eq:q19-anomaly-trivialization}
\end{equation}
where \(\mathbf1\) denotes the trivial invertible theory.

At this level of description, the quantum Scherk--Schwarz data may
therefore be summarized by
\begin{equation}
(\widetilde\rho,\tau).
\label{eq:q19-quantum-twist-datum}
\end{equation}
The lift specifies how the gluing acts on fermions, while the
trivialization supplies the anomaly-cancellation data needed to make
that action an autonomous quantum compactification.

\narrativeheading{Obstruction and choice are different}

The anomaly can affect the classical twist space in two distinct
ways.  For some lifted representations,
\(\operatorname{Triv}(\mathfrak A_{\widetilde\rho})\) is empty.  The
twist is then obstructed as an autonomous quantum compactification.

For other representations a trivialization exists but is not unique.
Different choices can define distinct quantum theories even when the
classical monodromy is identical.

The quantum refinement of the classical twist space can therefore be
written schematically as
\begin{equation}
\mathfrak X_{\rm q}
(\Gamma,G_{\rm f};\mathfrak A)
=
\frac{
\left\{
(\widetilde\rho,\tau)
\;\middle|\;
\widetilde\rho:\Gamma\to G_{\rm f},
\quad
\tau\in
\operatorname{Triv}
\bigl(
(B\widetilde\rho)^*\mathfrak A
\bigr)
\right\}
}{
\text{quantum equivalences}
}.
\label{eq:q19-quantum-twist-space}
\end{equation}

A simple cohomological model makes the distinction transparent.
Suppose that a finite-group anomaly is represented by
\(\alpha\in Z^{d+1}(B\Gamma,U(1))\).  A trivialization is a
\(d\)-cochain \(\beta\) satisfying
\begin{equation}
\delta\beta=\alpha.
\label{eq:q19-cochain-trivialization}
\end{equation}

There is a direct physical interpretation of this equation.  The
anomalous transformation of the original quantum theory produces the
phase represented by \(\alpha\).  Multiplying the path integral by the
local topological phase represented by \(\beta\) changes its
transformation law by \(\delta\beta\).  Equation
\eqref{eq:q19-cochain-trivialization} says precisely that the added
phase cancels the anomaly.

If \([\alpha]\neq0\) in
\(H^{d+1}(B\Gamma,U(1))\), no such \(\beta\) exists and the anomaly is
an obstruction.

If \([\alpha]=0\), however, the trivialization need not be unique.
Two solutions \(\beta\) and \(\beta'\) obey
\(\delta(\beta'-\beta)=0\), so their difference is a \(d\)-cocycle.
Modulo equivalence, the possible choices form a torsor under
\begin{equation}
H^d(B\Gamma,U(1)).
\label{eq:q19-trivialization-torsor}
\end{equation}

Thus the same mathematics distinguishes two physically different
questions: whether an anomaly can be cancelled, and which
anomaly-cancelling phase has been chosen.

The distinction is familiar in two-dimensional finite-group
orbifolds: \(H^3(B\Gamma,U(1))\) contains the obstruction to gauging,
while \(H^2(B\Gamma,U(1))\) parametrizes discrete-torsion choices.

For general fermionic systems, ordinary group cohomology is replaced
by the appropriate generalized cohomology or bordism theory.  The
logical separation between obstruction and choice remains the same.

\narrativeheading{Varying the twist: where do fermion zero modes
appear?}

Once a quantum-consistent twist has been identified, one can ask how
the quantum theory changes when the twist is varied continuously.

Let a circle monodromy lie in a maximal torus,
\[
U(\boldsymbol\omega)
=
\exp\!\left(
2\pi i\,\boldsymbol\omega\cdot\mathbf H
\right),
\]
and let \(\mathcal R\) be a representation with weight set
\({\rm Wt}(\mathcal R)\).  Let
\(\nu\in\{0,\tfrac12\}\) encode the periodic or anti-periodic spin
structure.

For a state of weight \(\mu\), the signed internal Dirac eigenvalues
are
\begin{equation}
\lambda_{n,\mu}
=
\frac1R
\left(
n+\nu+
\langle\mu,\boldsymbol\omega\rangle
\right),
\qquad
n\in\mathbb Z.
\label{eq:q19-weight-spectrum}
\end{equation}

A zero mode appears when
\begin{equation}
\nu+
\langle\mu,\boldsymbol\omega\rangle
\in
\mathbb Z.
\label{eq:q19-resonance-condition}
\end{equation}

The union of these loci defines the resonance arrangement of the
representation,
\begin{equation}
\mathscr H_{\mathcal R}
=
\bigcup_{\mu\in{\rm Wt}(\mathcal R)}
\left\{
\boldsymbol\omega
\;\middle|\;
\nu+
\langle\mu,\boldsymbol\omega\rangle
\in\mathbb Z
\right\}.
\label{eq:q19-resonance-arrangement}
\end{equation}

Away from \(\mathscr H_{\mathcal R}\), the internal Dirac operator is
invertible.  On a resonance wall its kernel jumps and one or more
Kaluza--Klein branches cross zero.

\narrativeheading{The \(SU(2)_R\) example revisited}

Take
\begin{equation}
U(\omega)
=
e^{2\pi i\omega\sigma_3},
\qquad
0\leq\omega\leq\frac12.
\label{eq:q19-SU2-family}
\end{equation}

In the adjoint representation of \(SU(2)_R\), the non-zero weights are
\(\alpha=\pm2\).  For \(\nu=0\), the resonance condition is
\(2\omega\in\mathbb Z\), so inside the fundamental alcove
\begin{equation}
\mathscr H_{\rm ad}
=
\left\{
0,\frac12
\right\}.
\label{eq:q19-SU2-adjoint-walls}
\end{equation}
These are exactly the two points where the centralizer of the circle
monodromy is enhanced.

The supercharge doublet instead has weights \(\lambda=\pm1\).
For periodic supersymmetry parameters, the condition is
\(\omega\in\mathbb Z\), and hence
\begin{equation}
\mathscr H_{\mathcal S}
=
\{0\}.
\label{eq:q19-SU2-SUSY-wall}
\end{equation}

The half-twist \(\omega=1/2\) is a
symmetry-enhancement wall in the adjoint representation but not a
supersymmetry-enhancement wall.  A physical fermion can behave
differently again, depending on its representation.  An adjoint
fermion has a resonance at \(\omega=1/2\), while a doublet fermion
with the same spin structure does not.

Thus
\[
\mathscr H_{\rm ad},
\qquad
\mathscr H_{\mathcal S},
\qquad
\mathscr H_{\mathcal R_f}
\]
are different representation-dependent kernel strata of the same
family of internal Dirac operators.

This spectral statement reproduces the group-theoretic distinction
found in Questions~\ref{sec:ss-half-twist}
and~\ref{sec:ss-inequivalent-twists} from a different viewpoint.

\narrativeheading{Spectral flow and the eta invariant}

A resonance wall is a point at which a fermionic eigenvalue crosses
zero.  The eta invariant keeps track of the resulting spectral
asymmetry.

For one weight, write
\(a=\nu+\langle\mu,\boldsymbol\omega\rangle\).  Away from an integer
value of \(a\), let \(\{a\}\in(0,1)\) denote its fractional part.
The eta function of the signed circle spectrum is
\begin{equation}
\eta_a(s)
=
\zeta\!\left(s,\{a\}\right)
-
\zeta\!\left(s,1-\{a\}\right),
\label{eq:q19-eta-Hurwitz}
\end{equation}
where \(\zeta(s,q)\) is the Hurwitz zeta function.

Using \(\zeta(0,q)=\tfrac12-q\), one finds
\begin{equation}
\eta_a(0)
=
1-2\{a\}.
\label{eq:q19-sawtooth-eta}
\end{equation}

The eta invariant is piecewise linear in the twist and jumps by
\(\pm2\) when one eigenvalue crosses zero.  The corresponding spectral
flow is \(\pm1\).

For anomaly questions the relevant quantity is the reduced eta
invariant
\begin{equation}
\bar\eta
=
\frac{\eta+h}{2},
\qquad
h=\dim\ker\mathscr D.
\label{eq:q19-reduced-eta}
\end{equation}
At the crossing the kernel dimension changes, and the change of
\(\bar\eta\) is integral.  The exponentiated quantity
\begin{equation}
\exp\!\left(
2\pi i\bar\eta
\right)
\label{eq:q19-exponentiated-eta}
\end{equation}
therefore remains continuous.

The resonance wall is not a singularity of the anomaly data.  What
changes is the local description in which the crossing mode has been
integrated out.

\narrativeheading{From spectral flow to the anomaly line}

Fix a spacetime \(X\) and consider a smooth family of lifted twists
over the classical representation space
\(\mathcal X(\Gamma,G_{\rm f})\).

Locally, away from a resonance wall, the fermion determinant can be
represented as an ordinary complex function of the twist parameters.
Across different patches these functions need not agree by equality;
they can differ by phases.  The determinants therefore assemble into
a section of a line bundle
\begin{equation}
\mathcal L_{\rm SS}
\longrightarrow
\mathcal X(\Gamma,G_{\rm f}),
\label{eq:q19-anomaly-line-twist-space}
\end{equation}
and the partition function is a section
\begin{equation}
Z_X
\in
\Gamma\!\left(
\mathcal X(\Gamma,G_{\rm f}),
\mathcal L_{\rm SS}
\right).
\label{eq:q19-section-over-twist-space}
\end{equation}

The perturbative anomaly determines the curvature of the natural
connection on this line.  Global anomalies determine its holonomy
around non-contractible loops.

Let
\(\ell:S^1\to\mathcal X(\Gamma,G_{\rm f})\) be such a loop.
Transport the background continuously around \(\ell\).  To compare
the initial and final theories, introduce an auxiliary interval
parametrizing the path and identify its ends after the corresponding
symmetry transformation.  The resulting closed
\((d+1)\)-dimensional manifold is the mapping torus \(Y_\ell\).

The holonomy of the anomaly line is
\begin{equation}
\operatorname{Hol}_{\ell}
\left(
\mathcal L_{\rm SS}
\right)
=
\mathfrak A_{G_{\rm f}}(Y_\ell).
\label{eq:q19-holonomy-mapping-torus}
\end{equation}

For fermionic theories, this phase can be represented by the
exponentiated reduced eta invariant together with local inflow terms,
schematically
\begin{equation}
\mathfrak A(Y)
\sim
\exp\!\left(
2\pi i\,\bar\eta(D_Y)
\right)
\exp\!\left(
2\pi i\,S_{\rm inflow}(Y)
\right).
\label{eq:q19-eta-phase}
\end{equation}
The overall sign convention in the exponent depends on orientation
and chirality conventions.  The distinction between trivial and
non-trivial holonomy does not.

The resonance walls are therefore zeros of particular determinant
sections, not singularities of the anomaly line itself.  The anomaly
line tells us how the local descriptions on the different sides of
those walls fit together globally.

\narrativeheading{Local and global anomalies are different tests}

For chiral fermions in even dimension \(d\), the perturbative anomaly
is encoded by a degree-\((d+2)\) characteristic form.  For a complex
Weyl fermion in representation \(\mathcal R\),
\begin{equation}
I_{d+2}
=
\left[
\widehat A(TX)
\operatorname{ch}_{\mathcal R}(F)
\right]_{d+2},
\label{eq:q19-anomaly-polynomial}
\end{equation}
with the appropriate chirality sign and normalization.  This local
index density determines the infinitesimal anomalous variation and the
curvature of the anomaly line.

If the local anomaly polynomial cancels, the anomaly line is flat.
Flat does not mean trivial: a flat line bundle may still carry
non-trivial holonomy around a non-contractible loop.

\narrativeheading{A familiar example: the \(SU(2)\) global anomaly}

A standard four-dimensional example makes the distinction concrete.
Consider an odd number of left-handed Weyl fermions in the doublet of
a gauged \(SU(2)\).

There is no ordinary cubic \(SU(2)\) gauge anomaly.  Group
theoretically, the symmetric invariant \(d^{abc}\) vanishes for
\(SU(2)\), so the familiar perturbative triangle test gives zero.

Nevertheless,
\[
\pi_4(SU(2))
\simeq
\mathbb Z_2.
\]
Under a gauge transformation in the non-trivial homotopy class, the
fermion path integral changes sign when the number of Weyl doublets
is odd.  The theory is quantum mechanically inconsistent
despite the complete absence of the perturbative cubic anomaly.  This
is the \(SU(2)\) global anomaly of Witten
\cite{Witten:1982fp}.

The example shows sharply what is meant by a global anomaly.  Even
after every ordinary local anomaly coefficient has vanished, the
fermion determinant can retain a discrete phase under a
topologically non-trivial transformation.  No perturbative expansion
around the identity can detect that phase.

The bordism formulation is the modern generalization of this
phenomenon.

\narrativeheading{From global anomalies to bordism}

After perturbative anomalies have been cancelled, the remaining
global anomaly is characterized by a homomorphism
\begin{equation}
\alpha_{\widetilde\rho}:
\Omega_{d+1}^{\mathfrak s}
(B\Gamma)_{\rm tors}
\longrightarrow
U(1),
\label{eq:q19-global-anomaly-character}
\end{equation}
where \(\mathfrak s\) denotes the relevant Spin, Pin or other
tangential structure.

The notation has a direct geometric meaning.  A closed
\((d+1)\)-dimensional manifold equipped with the required tangential
structure and a \(\Gamma\)-background defines a possible global
anomaly probe.  Two such backgrounds are bordant when their disjoint
union is the boundary of a \((d+2)\)-dimensional manifold carrying
compatible structures.  Bordism identifies probes that can be
continuously filled in this generalized sense.

The torsion subgroup is the part not detected by integration of
ordinary local characteristic forms.  It is therefore precisely where
purely global phases can survive after the perturbative anomaly
polynomial has vanished.

Thus cancellation of \(I_{d+2}\) does not imply
\(\alpha_{\widetilde\rho}=1\).  The first statement says that the
anomaly line is flat; the second says that its remaining global
holonomies are trivial.

\narrativeheading{Checking one generator at a time is not enough}

For a non-cyclic identification group, anomaly freedom of each
generator separately does not guarantee anomaly freedom of the full
twist.

Suppose
\[
\Gamma
=
\left\langle
\gamma_1,\ldots,\gamma_r
\;\middle|\;
R_a(\gamma_1,\ldots,\gamma_r)=1
\right\rangle .
\]
Classically, the images
\(\widetilde\rho(\gamma_i)\) must satisfy the same relations.

Quantum mechanically, mixed anomaly phases can involve several
generators at once.  It is therefore possible for every cyclic
subgroup \(\langle\gamma_i\rangle\) to be individually anomaly free
while the full \(\Gamma\)-action remains obstructed.

The simplest finite-dimensional analogue is a projective
representation,
\begin{equation}
\widehat\rho(\gamma_1)
\widehat\rho(\gamma_2)
=
c(\gamma_1,\gamma_2)\,
\widehat\rho(\gamma_1\gamma_2),
\label{eq:q19-projective-action}
\end{equation}
where \(c(\gamma_1,\gamma_2)\in U(1)\).  If this phase cannot be
removed by rephasing the operators, the group law is realized only
projectively.

In field theory, the anomaly theory is the higher-dimensional
generalization of this phenomenon.  The trivialization \(\tau\)
supplies the coherence required for all relations of \(\Gamma\) to
hold quantum mechanically.

For \(\Gamma=\mathbb Z^r\), this includes mixed anomalies among
commuting monodromies.  For \(\Gamma=D_\infty\), it includes the
compatibility of the two orbifold reflections, not merely the separate
consistency of each endpoint involution.

\narrativeheading{Fixed loci and anomaly inflow}

Orbifolds add another layer because anomalies can be localized on
fixed loci.

Let \(\Sigma_f\) denote the fixed submanifolds.  The anomalous
variation of the effective action can have the schematic form
\begin{equation}
\delta_\lambda\Gamma_{\rm eff}
=
2\pi i
\left[
\int_{\rm bulk}
I_{\rm bulk}^{(1)}(\lambda)
+
\sum_f
\int_{\Sigma_f}
I_f^{(1)}(\lambda)
\right].
\label{eq:q19-stratified-anomaly}
\end{equation}
Here \(I_{\rm bulk}^{(1)}\) and \(I_f^{(1)}\) are descent forms and
\(S_{\rm inflow}\) denotes a bulk term whose variation can compensate
the localized anomaly.

Quantum consistency requires the total variation to vanish,
\begin{equation}
\delta_\lambda
\left(
\Gamma_{\rm eff}
+
S_{\rm inflow}
\right)
=
0.
\label{eq:q19-inflow-condition}
\end{equation}

The anomaly therefore need not vanish separately in every microscopic
sector.  A fixed-locus anomaly can be cancelled by the variation of a
bulk Chern--Simons, Green--Schwarz or other invertible inflow term.

Conversely, cancellation only after integrating over all fixed loci
is not enough if gauge invariance fails locally before the inflow
contribution is included.

The distribution of anomaly between bulk and fixed loci is partly
scheme dependent.  Local Bardeen counterterms or bulk Chern--Simons
terms can move where the anomaly appears.  The invariant information
is the total relative anomaly class and its trivialization, not one
particular assignment of anomaly density to individual fixed points.

This mirrors the support analysis of
Question~\ref{sec:ss-local-divergences}: local contributions can live
on different strata, while consistency is a statement about the
complete system.

\narrativeheading{The Kaluza--Klein tower participates in anomaly
matching}

Scherk--Schwarz masses do not remove an anomaly.

After compactification, one may separate the light fields from the
massive Kaluza--Klein tower.  The anomaly of the parent theory
restricted to the Scherk--Schwarz background must be reproduced by
the complete lower-dimensional effective description.  Schematically,
\begin{equation}
\mathfrak A_{\rm parent,\widetilde\rho}
\simeq
\mathfrak A_{\rm light}
\otimes
\mathfrak A_{\rm massive}.
\label{eq:q19-KK-anomaly-matching}
\end{equation}

The massive contribution need not appear as a light anomalous
fermion.  Integrating out Kaluza--Klein states can generate
Chern--Simons, Wess--Zumino or eta-invariant terms whose variation
supplies the missing anomaly.

A lower-dimensional truncation can look anomalous even when
the complete compactification is consistent, because the omitted
tower also contains the topological term required for anomaly
matching.

The converse is equally important.  Giving anomalous fermions
Scherk--Schwarz masses does not make the anomaly disappear.  Without
the corresponding matching topological contribution, the quantum
theory remains inconsistent.

Massive states can therefore remain essential to anomaly consistency
even when they are absent from the low-energy particle spectrum.

\narrativeheading{Mapping tori are useful tests, but not the whole
classification}

A loop in twist space naturally produces a mapping torus, and
Eq.~\eqref{eq:q19-holonomy-mapping-torus} evaluates the anomaly on
that manifold.

This makes mapping tori particularly useful for testing the
transformations that actually appear in a Scherk--Schwarz
compactification.  They do not necessarily generate the entire
relevant bordism group.

Once the perturbative anomaly has been cancelled, complete absence of
global anomalies requires
\begin{equation}
\alpha_{\widetilde\rho}(Y)
=
1
\qquad
\text{for every }
[Y]
\in
\Omega_{d+1}^{\mathfrak s}(B\Gamma)_{\rm tors}.
\label{eq:q19-full-bordism-test}
\end{equation}

Passing the mapping-torus tests associated with a selected set of
loops proves only that the anomaly vanishes on those particular
bordism classes.  The full bordism problem depends on the spacetime
dimension, the fermionic symmetry extension, the global form of the
gauge group and the relevant tangential structure.

\narrativeheading{The quantum consistency hierarchy}

The quantum consistency problem can now be read as a sequence of
increasingly refined questions.

One starts from the classical bosonic gluing
\(\rho_{\rm b}:\Gamma\to G_{\rm b}\).  The first question is whether
this gluing admits a coherent fermionic lift
\(\widetilde\rho:\Gamma\to G_{\rm f}\).

The anomaly of the parent quantum theory is then restricted to the
lifted twist,
\(\mathfrak A_{\widetilde\rho}
=(B\widetilde\rho)^*\mathfrak A\), and one asks whether it admits a
coherent trivialization
\(\tau:\mathfrak A_{\widetilde\rho}\simeq\mathbf1\).

In practice, this can require checking the perturbative anomaly
polynomial, mixed anomalies among several identification generators,
fixed-locus anomalies and inflow on orbifolds, and the remaining
global anomaly on the relevant bordism group.

Only after these steps does the classical gluing define an autonomous
quantum Scherk--Schwarz compactification.  Physically equivalent
choices of \((\widetilde\rho,\tau)\) must finally be identified.

Quantum consistency therefore does not simply delete anomalous points
from the classical twist space.  It refines each classical point by
asking whether a fermionic lift exists, whether its anomaly can be
trivialized, and, when it can, which trivialization has been chosen.

\narrativeheading{Bibliographical remarks}

The relation between spectral asymmetry, eta invariants and index
theory originates in the Atiyah--Patodi--Singer framework
\cite{Atiyah:1975jf}.  The relation between eta invariants,
determinant lines and fermionic anomalies was developed further in
the Dai--Freed construction
\cite{Dai:1994kq}.  The interpretation of anomalies in terms of
invertible field theories is discussed in
Ref.~\cite{Freed:2014iua}, while a particle-physics oriented
introduction to Dai--Freed anomalies and their applications can be
found in Ref.~\cite{Garcia-Etxebarria:2018ajm}.

The four-dimensional \(SU(2)\) example used above is the global gauge
anomaly of Ref.~\cite{Witten:1982fp}.  It provides the standard
illustration that cancellation of perturbative gauge anomalies need
not eliminate global anomalies.

The resonance-wall construction used here combines this anomaly
framework with the Scherk--Schwarz twist space of
Question~\ref{sec:ss-inequivalent-twists}.  The adjoint, supercharge
and physical-fermion resonance loci are different
representation-dependent kernel strata of the same family of internal
Dirac operators.  Spectral flow and the eta invariant then organize
the fermion determinant as these strata are crossed.

As a concrete string-theory realization of the pullback-anomaly
criterion,  cyclic asymmetric
heterotic orbifolds were analysed in \cite{Cheng:2026abk}. It was shown 
that the pulled-back worldsheet
Dai--Freed anomaly reproduces the familiar level-matching conditions
and, for appropriate even-order actions, additional mixed mod-two
constraints.  This example illustrates the general
framework but does not determine the anomaly theory appropriate to
other dimensions, symmetry groups or tangential structures.

\begin{takehome}
\begin{takehomeitems}

\item
A classical bosonic twist must first lift to the full fermionic
symmetry, including \((-1)^F\) and the relevant geometric Spin or Pin
structure.  On a circle the lift is unobstructed, but the two choices
differing by \((-1)^F\) give the two spin structures.

\item
A classical symmetry can remain anomalous quantum mechanically.  The
relevant anomaly is the anomaly of the parent theory pulled back to
the actual lifted Scherk--Schwarz action,
\[
\mathfrak A_{\widetilde\rho}
=
(B\widetilde\rho)^*\mathfrak A .
\]
An autonomous quotient requires a coherent trivialization of this
anomaly.

\item
Obstruction and choice are different.  A twist may admit no quantum
completion, or several inequivalent ones.  In the elementary
cohomological description, \(\delta\beta=\alpha\) expresses
cancellation of the anomalous phase, while closed shifts of \(\beta\)
parametrize different trivializations.

\item
Different representations encounter zero modes on different loci in
twist space.  A state of weight \(\mu\) becomes massless when
\[
\nu+\langle\mu,\boldsymbol\omega\rangle\in\mathbb Z.
\]
Thus residual-symmetry enhancement, supersymmetry enhancement and
physical-fermion resonance walls need not coincide.

\item
Crossing a resonance wall produces spectral flow.  The eta invariant
measures the spectral asymmetry, while its exponentiated reduced form
fits continuously through a zero crossing.  Globally, the different
local determinants are organized as a section of the anomaly line
over twist space.

\item
Vanishing of the perturbative anomaly polynomial is not sufficient
for quantum consistency.  Global anomalies can survive as discrete
holonomies, as illustrated by the four-dimensional \(SU(2)\) anomaly.
Mixed anomalies, orbifold fixed-locus anomalies and inflow, the
massive Kaluza--Klein tower, and the relevant bordism classes must all
be included in the complete test.

\end{takehomeitems}
\end{takehome}

\nextquestion
{What additional restrictions arise when Scherk--Schwarz breaking is
embedded in string theory?}

\end{reviewquestion}
% Question 21 v3, renumbered from Question20v2.tex.
% Original section/equation labels retained for cross-reference stability.
%%%%%%%%%%%%%%%%%%%%%%%%%%%%%%%%%%%%%%%%%%%%%%%%%%%%%%%%%%%%%%%

\begin{reviewquestion}
{What additional restrictions arise when Scherk--Schwarz breaking is
embedded in string theory?}
\label{sec:ss-string-restrictions}

\begin{essentialidea}

A Scherk--Schwarz twist that is perfectly consistent in a
higher-dimensional effective field theory need not define a consistent
string background.

The reason is physical.  In field theory one specifies how a set of
fields is reidentified around a non-contractible cycle.  Perturbative
string theory must instead define a complete two-dimensional quantum
theory.  The twist must act consistently not only on the massless
fields but also on the massive string states, momentum and winding
charges, worldsheet fermions, spin structures, GSO projections,
operator products and all twisted sectors required by the quotient.

If \(\mathcal C\) denotes the parent worldsheet CFT and
\(\mathcal T_{\rm EFT}\) its low-energy effective theory, a candidate
action of a finite group \(\mathcal G\) on the light fields must first
lift to an exact action on \(\mathcal C\).  Such a microscopic lift
can fail to exist, and inequivalent microscopic actions can induce the
same low-energy Scherk--Schwarz twist.

If \(\mathcal G\) is actually quotiented, the twisted sectors are not
optional.  On a torus they are labelled by commuting pairs of
holonomies, and modular transformations mix these sectors.  Modular
closure leads to level-matching conditions, while higher-genus
consistency requires factorization and sewing.

String theory also turns many continuous symmetries of the
low-energy equations into arithmetic symmetries of an integral charge
lattice.  A transformation that acts consistently on the massless
supergravity fields can therefore fail to be an exact symmetry of the
string spectrum.

Finally, the usual spontaneous Scherk--Schwarz interpretation requires
more than a non-supersymmetric quotient.  The breaking should belong
to a dynamical family in which moduli control the supersymmetry-breaking
scale and an appropriate limit restores the complete supersymmetric
parent background.

Thus the string-theory problem is not merely to choose an admissible
monodromy.  One must construct its microscopic realization, its
twisted sectors and amplitudes, and verify the global quantum
consistency of the complete string background.

\end{essentialidea}

\narrativeheading{Why a string is more demanding than a point
particle}

The additional string conditions can already be understood from the
difference between a particle loop and a string loop.

For a point particle moving on a target-space circle of radius \(R\),
a Scherk--Schwarz boundary condition may be written
\begin{equation}
\Phi(y+2\pi R)
=
U\,\Phi(y),
\label{eq:q20-EFT-monodromy}
\end{equation}
where \(U\) is an internal symmetry acting on the field \(\Phi\).
Once \(U\) is specified, the Kaluza--Klein spectral problem is
defined.

A closed string propagating at one loop sweeps out a worldsheet
torus.  The torus itself has two independent non-contractible cycles.
A string configuration can carry one orbifold
transformation around one worldsheet cycle and another around the
second.

This is why a condition such as
\(\psi(y+2\pi R)=-\psi(y)\) does not by itself define a string
vacuum.  It specifies how one spacetime fermion behaves around a
target-space cycle, but says nothing about the required worldsheet
twisted sectors, their modular transformations, their operator
products or their factorization properties.

Thus specifying the anti-periodicity of a spacetime fermion fixes only
one piece of the construction.  A string background must also supply
the worldsheet sectors and amplitudes required to realize that
anti-periodicity consistently.

\narrativeheading{From an EFT symmetry to an exact CFT symmetry}

Let \(\mathcal C\) be the complete worldsheet conformal field theory
of the supersymmetric parent background, and let
\(\mathcal T_{\rm EFT}\) denote the low-energy theory obtained from
its light string states.

By an automorphism of \(\mathcal C\) we mean a transformation that
preserves the structures required to define the CFT: the stress
tensor, operator products, state space, charge assignments,
spin-structure data and projections.

Restriction to the light sector defines schematically
\begin{equation}
\operatorname{Res}_{\rm IR}:
\operatorname{Aut}(\mathcal C)
\longrightarrow
\operatorname{Aut}(\mathcal T_{\rm EFT}).
\label{eq:q20-IR-restriction}
\end{equation}

Suppose that the low-energy theory admits a candidate action
\(\rho_{\rm EFT}:\mathcal G\to
\operatorname{Aut}(\mathcal T_{\rm EFT})\).
A microscopic realization is an action
\(\rho_{\rm ws}:\mathcal G\to\operatorname{Aut}(\mathcal C)\)
satisfying
\begin{equation}
\operatorname{Res}_{\rm IR}\circ\rho_{\rm ws}
=
\rho_{\rm EFT}.
\label{eq:q20-string-lift-condition}
\end{equation}

It is useful to denote the inequivalent microscopic realizations by
\begin{equation}
\operatorname{Lift}_{\mathcal C}(\rho_{\rm EFT})
=
\left\{
\rho_{\rm ws}
\;\middle|\;
\operatorname{Res}_{\rm IR}\circ\rho_{\rm ws}
=
\rho_{\rm EFT}
\right\}
/\text{equivalence}.
\label{eq:q20-lift-space}
\end{equation}

This set can be empty, contain one element, or contain several.

If it is empty, the candidate EFT twist does not exist in the chosen
string background.  If several lifts exist, the same action on the
light fields has more than one microscopic completion.

This possibility is physically natural.  Two transformations can act
identically on every massless state while acting differently on
massive oscillators, momentum--winding states or internal
gauge-lattice states.  The corresponding string compactifications can
therefore share the same leading Scherk--Schwarz EFT while differing
in their massive spectra, twisted sectors and threshold corrections.

The low-energy twist retains only part of the microscopic information.

\narrativeheading{Twisted sectors are worldsheet gauge backgrounds}

Assume now that the finite symmetry \(\mathcal G\) is actually used
to form an orbifold.

A twisted sector on a connected closed worldsheet \(\Sigma\) can be
described as a flat principal \(\mathcal G\)-bundle.  After choosing a
base point, its holonomies define a homomorphism
\begin{equation}
\varphi:
\pi_1(\Sigma)
\longrightarrow
\mathcal G.
\label{eq:q20-worldsheet-hom}
\end{equation}
Changing the trivialization at the base point conjugates all
holonomies simultaneously.  The topological sectors are therefore
classified by
\begin{equation}
\mathfrak B_{\mathcal G}(\Sigma)
=
\operatorname{Hom}
\left(
\pi_1(\Sigma),\mathcal G
\right)
/\mathcal G .
\label{eq:q20-sector-space}
\end{equation}

This representation problem should not be confused with the
target-space Scherk--Schwarz representation discussed earlier.
There, \(\pi_1(K)\) describes transport through the compact target
space.  Here, \(\pi_1(\Sigma)\) labels orbifold holonomies carried by
the worldsheet.

If the action of \(\mathcal G\) on the parent CFT is not faithful,
different elements of \(\mathcal G\) can act identically on the local
CFT fields.  It is nevertheless not always legitimate to quotient
\(\mathcal G\) by this kernel.  The apparently redundant elements can
still label distinct twisted sectors or modify the global orbifold
data.  We therefore keep track of the microscopic orbifold group
together with its specified action on the CFT.

For a closed oriented surface of genus \(g\),
\begin{equation}
\pi_1(\Sigma_g)
=
\left\langle
a_1,b_1,\ldots,a_g,b_g
\;\middle|\;
\prod_{i=1}^{g}[a_i,b_i]=1
\right\rangle .
\label{eq:q20-genus-group}
\end{equation}
A sector is specified by
\((g_1,h_1,\ldots,g_g,h_g)\in\mathcal G^{2g}\) satisfying
\begin{equation}
\prod_{i=1}^{g}[g_i,h_i]=1,
\label{eq:q20-genus-relation}
\end{equation}
modulo simultaneous conjugation.

\narrativeheading{The torus explains why twisted sectors are not
optional}

At one loop the worldsheet is a torus, with
\(\pi_1(T^2)=\mathbb Z^2\).  Let \(g\) and \(h\) denote the
holonomies around two oriented primitive cycles.  They must commute.

We denote the corresponding torus amplitude by
\(Z_{g,h}(\tau,\bar\tau)\), where \(\tau\) is the complex structure
modulus.  The element \(g\) specifies the spatially twisted Hilbert
space, while \(h\) is inserted in the trace.

For a finite orbifold group, the torus partition function has the
schematic form
\begin{equation}
Z_{\rm orb}
=
\frac1{|\mathcal G|}
\sum_{\substack{g,h\in\mathcal G\\gh=hg}}
\epsilon(g,h)\,
Z_{g,h},
\label{eq:q20-orbifold-partition}
\end{equation}
where \(\epsilon(g,h)\) denotes any allowed relative phase.

The two torus cycles are not physically distinguished.  Large
orientation-preserving diffeomorphisms change their primitive basis
and generate \(SL(2,\mathbb Z)\).  With the convention adopted here,
\begin{align}
S:\quad
(g,h)
&\longmapsto
(h,g^{-1}),
\label{eq:q20-S-action}
\\
T:\quad
(g,h)
&\longmapsto
(g,gh).
\label{eq:q20-T-action}
\end{align}
Other orientation conventions redistribute the inverses without
changing the underlying physics.

A twisted sector cannot therefore be considered in isolation.
Modular transformations generate an orbit of sectors, and a
consistent orbifold must contain the complete orbit with mutually
compatible amplitudes.

This is the first place where the string problem is visibly more
restrictive than a field-theory monodromy.  The latter specifies how
the retained fields transform.  The worldsheet construction must
supply every additional sector required for modular closure~\cite{Dixon:1985jw,Dixon:1986jc,Vafa:1986wx}.

\narrativeheading{Why modular invariance requires level matching}

Consider an element \(g\in\mathcal G\) of finite order \(N\), and let
\(\mathcal H_g\) denote its twisted Hilbert space.

A state in this sector has left- and right-moving conformal
Hamiltonians \(L_0\) and \(\bar L_0\).  Their difference measures the
worldsheet momentum around the closed string.  Under
\(T:\tau\mapsto\tau+1\), the state acquires the relative phase
\(e^{2\pi i(L_0-\bar L_0)}\).

After \(N\) applications of \(T\), the topological sector returns to
itself.  In the ordinary cyclic case this gives
\begin{equation}
N(L_0-\bar L_0)
\in
\mathbb Z.
\label{eq:q20-level-matching}
\end{equation}

For a twisted ground state of conformal weights
\(h_g,\bar h_g\), one similarly requires
\begin{equation}
N(h_g-\bar h_g)
\in
\mathbb Z.
\label{eq:q20-ground-state-level-matching}
\end{equation}

Fractional oscillator frequencies, zero-point energies, shifted
momenta, winding contributions and internal gauge charges all enter
\(L_0-\bar L_0\).  Level matching is a condition on the
complete twisted string spectrum rather than on its massless sector.

Its physical meaning is simple: it allows the modular \(T\)-orbit of
the twisted sector to close consistently.

\narrativeheading{Modular invariance is only the first sewing test}

A modularly invariant torus amplitude is necessary, but it is not by
itself a definition of a consistent orbifold CFT.

One must first construct the twisted Hilbert spaces
\(\mathcal H_g\), specify how the centralizer of \(g\) acts in each
sector, choose the allowed phases and define operator products
involving twisted fields.

String amplitudes must also exist on worldsheets of arbitrary genus.
When a surface degenerates or is cut along a cycle, the amplitude must
factorize into amplitudes of the resulting pieces, with the correct
sum over intermediate states.

Schematically, a consistent orbifold assigns an amplitude
\begin{equation}
(\Sigma,P)
\longmapsto
Z_{\mathcal C}[\Sigma,P]
\label{eq:q20-amplitude-assignment}
\end{equation}
to every allowed closed worldsheet equipped with a flat
\(\mathcal G\)-bundle \(P\).  These amplitudes must transform
consistently under mapping-class groups and satisfy the required
sewing relations.

The torus partition function is an initial quantum test of the
construction, not its complete definition.

For open or unoriented strings, further amplitudes enter.  D-branes
and orientifold planes require Klein-bottle, annulus and M\"obius-strip
amplitudes, together with consistent boundary and crosscap states.
Their modular channels and factorization properties must be
compatible with the Scherk--Schwarz action, and tadpole cancellation
can impose additional constraints.  We shall not develop those
conditions here.

\narrativeheading{The charge lattice makes the symmetry problem
arithmetic}

String theory also restricts which transformations can be exact
symmetries in the first place.

In a toroidal compactification, left- and right-moving charges belong
to an integral lattice
\begin{equation}
\Lambda
\subset
\mathbb R^{d_L,d_R}.
\label{eq:q20-charge-lattice}
\end{equation}
For a Narain compactification, this lattice is even and self-dual with
respect to a bilinear form \(\eta\) of signature \((d_L,d_R)\).

The continuum equations admit \(O(d_L,d_R;\mathbb R)\), but an exact
string symmetry must preserve the spectrum of allowed charges.
Hence it must satisfy \(g\Lambda=\Lambda\) and
\(g^T\eta g=\eta\).  The exact symmetry group is therefore reduced to
the lattice automorphism group
\begin{equation}
O(\Lambda)
=
\left\{
g\in O(d_L,d_R;\mathbb R)
\;\middle|\;
g\Lambda=\Lambda
\right\}.
\label{eq:q20-lattice-automorphism-group}
\end{equation}

In an integral basis this becomes the appropriate arithmetic group
usually denoted \(O(d_L,d_R;\mathbb Z)\).

The physical origin of the restriction is immediate.  A continuous
transformation may act perfectly on every massless supergravity field
while sending a massive momentum--winding state to a charge vector
that does not belong to the string spectrum.  Such a transformation
is not an exact symmetry of the microscopic theory.

Lattice preservation is still not sufficient.  The symmetry must
also lift consistently to vertex operators and their cocycle phases~\cite{Narain:1985jj,Narain:1986am}.

The role of the complete momentum--winding charge lattice, including
its relation to Scherk--Schwarz shifts, is the subject of the next
question. Discrete restrictions on rigid twists will be discussed
thereafter.

\narrativeheading{Fermions, GSO projections, and asymmetric actions}

The microscopic symmetry must also act consistently on the
worldsheet fermionic sectors.

Worldsheet fermions possess different spin structures around the
cycles of the worldsheet, and perturbative superstring theories sum
over them with phases determined by the GSO projection.  Spacetime
supercharges arise from particular Ramond-sector states.  A
transformation acting on spacetime supersymmetry must have
a consistent action on the worldsheet fermions, the Ramond and
Neveu--Schwarz sectors, the GSO projection and any accompanying gauge
or lattice degrees of freedom.

This is the microscopic reason why imposing an anti-periodic
boundary condition on a spacetime fermion is not yet a string
construction.

Exact string symmetries need not, however, be geometric.  A toroidal
CFT can admit an asymmetric action
\begin{equation}
(X_L,X_R)
\longmapsto
(g_LX_L,g_RX_R),
\qquad
g_L\neq g_R,
\label{eq:q20-asymmetric-action}
\end{equation}
where \(X_L\) and \(X_R\) are the left- and right-moving components of
an internal coordinate.

Such a transformation may have no interpretation as an isometry of a
classical target-space manifold and yet be an exact automorphism of
the full CFT.  String symmetry is therefore broader than geometric
symmetry.  At the same time, it remains more restrictive than symmetry
of the low-energy equations.

\narrativeheading{The anomaly condition of
Question~\ref{sec:ss-quantum-consistency} reappears microscopically}

Question~\ref{sec:ss-quantum-consistency} asked whether the symmetry
used for Scherk--Schwarz gluing survives as a consistent quantum
operation.

In a worldsheet orbifold this becomes concrete: quotienting by
\(\mathcal G\) gauges the corresponding finite symmetry of the CFT.
An anomalous action can obstruct the quotient even if each
group element preserves the classical operator algebra.

Failures of modular covariance, level matching or higher-genus sewing
can be manifestations of such a microscopic obstruction.  In a class
of cyclic asymmetric heterotic orbifolds, for example, a Dai--Freed
analysis reproduces the familiar level-matching condition and, for
even-order actions, additional mixed mod-two constraints
\cite{Cheng:2026abk}.

This gives a concrete realization of the relation between the general
anomaly language of Question~\ref{sec:ss-quantum-consistency} and
worldsheet consistency.  It does not imply that every string
consistency condition is encoded in one anomaly invariant.

\narrativeheading{When does the string construction describe
spontaneous Scherk--Schwarz breaking?}

A consistent non-supersymmetric quotient is not automatically a
spontaneously broken Scherk--Schwarz background.

The first question is whether the quotient acts non-trivially on the
spacetime supercharges at all.  If some supercharges remain invariant,
the corresponding supersymmetry survives.

If all are removed, supersymmetry is broken, but the spontaneous
interpretation requires additional dynamical information, as discussed
in Question~\ref{sec:ss-spontaneous}.  In particular, restoration of
supersymmetry in some limiting background is not by itself sufficient;
the parameter controlling the breaking should belong to the dynamical
moduli of the same underlying theory.

The standard perturbative string realization combines an internal
action \(g_R\) with a translation \(\delta\) along a compact
direction,
\begin{equation}
g_{\rm SS}
=
g_R\,\delta .
\label{eq:q20-freely-acting-generator}
\end{equation}
The combined quotient is then freely acting.

If \(g_R\) acts non-trivially on all spacetime supercharges, the
background breaks supersymmetry.  When the translation radius \(R\)
is a modulus, the breaking scale typically behaves as
\(m_{\rm SS}\sim1/R\).  In the decompactification limit
\(R\to\infty\), finite-energy spectra and amplitudes approach those of
the supersymmetric parent background.

Free action and supersymmetry restoration therefore play different
roles.  The free translation ensures that the breaking is not tied to
a fixed locus, while the dynamical radius provides the modulus that
controls the breaking scale.  When supersymmetry is restored as this
modulus is taken to the appropriate limit, the construction has the
standard string-theoretic interpretation of spontaneous
Scherk--Schwarz breaking~\cite{Rohm:1983aq,Kounnas:1989dk,Antoniadis:1988jn}.

\narrativeheading{What can the low-energy EFT test, and what can it
miss?}

A Wilsonian effective theory derived from a consistent string
background retains information about heavy string states through
higher-dimension operators, thresholds and topological terms.
Schematically,
\begin{equation}
S_{\rm W}
=
S_{2\partial}
+
\sum_i
\frac{c_i}{M_s^{\Delta_i-D}}
\int d^Dx\,\mathcal O_i
+
S_{\rm top}
+
\cdots .
\label{eq:q20-Wilsonian-action}
\end{equation}

An exact Scherk--Schwarz symmetry must preserve the complete
Wilsonian data.  A transformation that is only a symmetry of the
two-derivative action can still provide a useful approximate
low-energy description, but does not thereby define an exact string
quotient.

Some microscopic consistency conditions are visible in the EFT.
Anomaly matching, quantized topological terms, charges of retained
states and symmetry relations among Wilson coefficients can all rule
out proposed twists.
Other constraints require information that a finite truncation does
not contain: the complete momentum--winding lattice, massive
oscillator states, twisted Hilbert spaces, modular completion or
higher-genus sewing data.

The logical asymmetry is therefore important.  Failure of a
low-energy consistency test can rule out a proposed microscopic
realization.  Passing every test visible in a finite EFT does not
prove that a microscopic string lift exists.
Likewise, when several string lifts induce the same leading
Scherk--Schwarz EFT, they can still be distinguished by their heavy
spectra, threshold corrections, topological terms and
higher-dimension operators.  The map from microscopic string
backgrounds to finite low-energy truncations is many-to-one and does
not admit a general inverse.

\narrativeheading{Bibliographical remarks}

The construction of string orbifolds, including twisted sectors,
modular covariance, level matching and discrete-torsion phases, was
developed in the classic orbifold literature; see
Refs.~\cite{Dixon:1985jw,Dixon:1986jc,Vafa:1986wx}.
The description in terms of commuting worldsheet holonomies makes
explicit why a single target-space monodromy does not define the
complete string quotient. Toroidal compactifications provide the standard setting in which
continuous symmetries of the low-energy equations are reduced to
arithmetic automorphisms of an integral momentum--winding lattice
\cite{Narain:1985jj,Narain:1986am}. Freely acting orbifolds provide 
the standard perturbative string realizations of Scherk--Schwarz 
supersymmetry breaking and its
restoration along a modulus-dependent family
\cite{Rohm:1983aq,Kounnas:1989dk,Antoniadis:1988jn}.
The anomaly framework of
Question~\ref{sec:ss-quantum-consistency} gives a complementary
description of some microscopic orbifold constraints.  In suitable
asymmetric constructions, the pulled-back worldsheet anomaly
reproduces level matching and additional discrete consistency
conditions \cite{Cheng:2026abk}.

\begin{takehome}
\begin{takehomeitems}

\item
A Scherk--Schwarz symmetry of the low-energy theory must lift to an
exact automorphism of the complete parent worldsheet CFT.  Such a lift
can fail to exist, and different microscopic lifts can induce the same
action on the light fields.

\item
Orbifolding requires complete twisted sectors.  On a closed
worldsheet they are flat \(\mathcal G\)-bundles, classified by
\[
\operatorname{Hom}
\left(
\pi_1(\Sigma),\mathcal G
\right)
/\mathcal G .
\]
On the torus this reduces to commuting pairs \((g,h)\), which are
mixed by modular transformations and therefore occur in complete
modular orbits.

\item
For a cyclic twist of order \(N\), modular closure requires level
matching, \(N(L_0-\bar L_0)\in\mathbb Z\).  Genus-one modular
covariance is necessary but not sufficient: factorization and sewing
must also hold at higher genus.

\item
Exact toroidal string symmetries must preserve the integral charge
lattice.  Continuous symmetries of the low-energy equations are reduced to arithmetic automorphism groups, and lattice
preservation must be supplemented by consistent vertex-operator,
fermionic and GSO data.

\item
String symmetries can be asymmetric and need not admit a geometric
target-space interpretation.  Exact CFT symmetry is therefore more
general than geometric symmetry but more restrictive than symmetry of
the low-energy equations.

\item
Orbifolding gauges the microscopic finite symmetry, so the anomaly
conditions of Question~\ref{sec:ss-quantum-consistency} remain part of
the string consistency problem.

\item
The usual spontaneous Scherk--Schwarz interpretation arises when the
supersymmetry-breaking scale is controlled by dynamical moduli and an
appropriate limit restores the supersymmetric parent background.
Freely acting orbifolds provide the standard perturbative realization
of this mechanism.

\item
A finite low-energy EFT can rule out a proposed string lift when one
of its necessary consistency conditions fails, but passing every test
visible below the cutoff cannot establish that the required charge
lattice, twisted sectors and worldsheet amplitudes exist.

\end{takehomeitems}
\end{takehome}

\nextquestion
{How is Scherk--Schwarz breaking encoded in the string charge lattice?}

\end{reviewquestion}
% Question 22 v3, renumbered from Question21v2.tex.
% Original section/equation labels retained for cross-reference stability.
%%%%%%%%%%%%%%%%%%%%%%%%%%%%%%%%%%%%%%%%%%%%%%%%%%%%%%%%%%%%%%%

\begin{reviewquestion}
{How is Scherk--Schwarz breaking encoded in the string charge lattice?}
\label{sec:ss-string-charge-lattice}

\begin{essentialidea}

In field theory, a Scherk--Schwarz deformation is often summarized by
the shift
\[
m
\longrightarrow
m+\xi q,
\]
where \(m\) is a Kaluza--Klein momentum number and \(q\) an internal
charge.

A string state carries more information.  Momentum, winding, gauge
charges and the charges of worldsheet currents belong to one
Lorentzian charge system.  A consistent deformation cannot shift one
entry while leaving the others untouched, because the quadratic form
controlling level matching and the string mass formula must be
preserved.

An ordinary gauge Wilson line gives the simplest example.  If a string
of winding number \(n\) experiences the gauge shift \(P\to P+nA\),
preservation of the charge-lattice norm forces
\[
m
\longrightarrow
m-A\cdot P-\frac12A^2n.
\]
The linear and quadratic terms are therefore parts of one orthogonal
transformation, rather than independent corrections to the mass
formula.

The same charge-space geometry applies to a coordinate-dependent
Scherk--Schwarz deformation once the charge acting on spacetime
supersymmetry is included among the internal charges.  The familiar
field-theory momentum shift is then the zero-winding limit of a
Lorentzian transformation of the complete string charge vector.

Gauge Wilson lines and Scherk--Schwarz deformations are thus described
by the same type of parabolic isometry of an enlarged charge space,
although they act on different internal directions and have different
physical meanings.

Not every low-energy \(R\)-symmetry charge can be used in this way.
The charge must arise from an admissible dimension-one worldsheet
current, act non-trivially on the sector generating spacetime
supersymmetry, and transform consistently under the orbifold.  These
worldsheet conditions are logically distinct from the integrality
condition determining whether the real charge-space isometry preserves
the exact string lattice.

This question develops the charge-space geometry.  The additional
conditions that make the deformation rigid or discrete are treated in
the next question.

\end{essentialidea}

\narrativeheading{Why the field-theory shift is incomplete}

Consider a field-theory tower whose masses depend on a Kaluza--Klein
number \(m\) and an internal charge \(q\) through
\(m_{\rm KK}\sim(m+\xi q)/R\).  At this level, it is natural to think
that the Scherk--Schwarz deformation simply translates one momentum
quantum number.

A closed string on the same circle also carries winding.  In a
heterotic compactification it may carry an internal gauge-lattice
charge as well.  The labels \(m\), \(n\) and \(P\) are not
independent decorations of a field-theory state.  They are components
of a single string charge vector.

The question consequently changes.  One must ask not only how the
Kaluza--Klein momentum is shifted, but whether the complete
transformation preserves the quadratic form of the string charge
lattice.

\narrativeheading{The charge vector behind a toroidal string state}

For definiteness, consider the heterotic string compactified on a
\(d\)-dimensional torus.  A state carries integer Kaluza--Klein
numbers \(m_i\), integer winding numbers \(n^i\), and a gauge charge
\(P\) in the sixteen-dimensional gauge lattice.

The corresponding left- and right-moving momenta combine with \(P\)
into a Narain charge vector
\(\mathcal P=(P;p_L,p_R)\).  The allowed vectors form an even
self-dual lattice
\(\Lambda_{\rm N}\subset\mathbb R^{d+16,d}\).

Evenness means that the norm of every lattice vector is an even
integer in the conventional normalization.  Self-duality means that
every vector pairing integrally with the lattice already belongs to
the lattice.

The Lorentzian quadratic form is
\begin{equation}
\langle\mathcal P,\mathcal P\rangle
=
P^2+p_L^2-p_R^2.
\label{eq:q21-Narain-form}
\end{equation}
It enters both the level-matching condition and the string mass
formula.

For one compact circle, a basis adapted to momentum and winding
reduces the relevant part of the norm to
\begin{equation}
P^2+2mn,
\label{eq:q21-one-circle-form}
\end{equation}
up to the overall normalization convention.

This simple expression already shows why a shift of \(P\) or \(m\)
cannot be considered in isolation.

\narrativeheading{A Wilson line shows why the quadratic term is
necessary}

Let \(A\) be a gauge Wilson line along the compact circle.  A string
with winding number \(n\) goes around the circle \(n\) times and
experiences the gauge holonomy on each traversal.  Its gauge charge is
therefore shifted to
\begin{equation}
P_A
=
P+nA.
\label{eq:q21-WL-gauge-shift}
\end{equation}

If this were the only change, the gauge contribution to the lattice
norm would become
\[
(P+nA)^2
=
P^2+2nA\cdot P+n^2A^2,
\]
and the original norm would not be preserved.

The compact momentum must change at the same time:
\begin{equation}
m_A
=
m-A\cdot P-\frac12A^2n.
\label{eq:q21-WL-momentum-shift}
\end{equation}
Indeed,
\begin{align}
P_A^2+2nm_A
&=
(P+nA)^2
+
2n
\left(
m-A\cdot P-\frac12A^2n
\right)
\nonumber\\
&=
P^2+2mn.
\label{eq:q21-WL-norm-check}
\end{align}

The coefficient of the quadratic term is fixed by this equality.  It
is not an optional string correction added to the linear
field-theory shift.

In terms of \(p_L\) and \(p_R\), the same combination
\(m-A\cdot P-\tfrac12A^2n\) appears in both momenta, with the winding
contribution entering them with opposite signs.

\narrativeheading{The parabolic isometry behind the Wilson-line
formula}

The preceding transformation has a useful coordinate-independent
form.

Let \(W\) be an internal charge space with bilinear form
\((\,\cdot\,,\,\cdot\,)_W\).  Add two null vectors \(e\) and \(f\)
satisfying
\begin{equation}
e^2=f^2=0,
\qquad
e\cdot f=1,
\qquad
e\cdot W=f\cdot W=0.
\label{eq:q21-hyperbolic-plane}
\end{equation}
The pair \((e,f)\) represents one momentum--winding plane, and the
full real charge space is
\(V=W\oplus\operatorname{span}_{\mathbb R}\{e,f\}\).

For \(a\in W\), define
\begin{align}
T_a(e)
&=
e,
\nonumber\\
T_a(x)
&=
x-(a,x)_W\,e,
\qquad x\in W,
\nonumber\\
T_a(f)
&=
f+a-\frac12(a,a)_W\,e.
\label{eq:q21-transvection}
\end{align}

The coefficient of \(e\) in the last line is fixed by requiring the
image of \(f\) to remain null:
\begin{align}
T_a(f)^2
&=
\left(
f+a-\frac12a^2e
\right)^2
\nonumber\\
&=
a^2
-
2\left(\frac12a^2\right)
(f\cdot e)
\nonumber\\
&=
0.
\label{eq:q21-null-check}
\end{align}
Here \(f^2=e^2=f\cdot a=e\cdot a=0\) and \(f\cdot e=1\).
The remaining scalar products are preserved directly, so
\(T_a\in O(V)\).

The transformation is a unipotent parabolic isometry fixing the null
direction \(e\), and may be viewed as an
Eichler-transvection-type transformation associated with that null
direction.

Writing a charge vector as \(v=P+me+nf\) and taking \(a=A\),
Eq.~\eqref{eq:q21-transvection} reproduces
\[
P\to P+nA,
\qquad
m\to m-A\cdot P-\frac12A^2n.
\]

The component Wilson-line formula is therefore simply the coordinate
expression of one orthogonal transformation.

\narrativeheading{Adding the charge that acts on spacetime
supersymmetry}

A Scherk--Schwarz deformation requires a charge that distinguishes
states belonging to the same spacetime supermultiplet.

Let \(J_{\rm SS}(z)\) be a dimension-one worldsheet current and
\(Q_{\rm SS}\) its zero-mode charge.  The microscopic requirement is
that this charge act non-trivially on the worldsheet sector from which
the spacetime supercharges are constructed.

In the heterotic constructions considered below, the infinitesimal
current charge therefore acts non-trivially on the relevant
two-dimensional supercurrent \(G_{\rm ws}\).  At the level of the
local generator this can be expressed as
\begin{equation}
[Q_{\rm SS},G_{\rm ws}]
\neq
0.
\label{eq:q21-current-supercurrent}
\end{equation}

This does not mean that the finite quotient violates worldsheet
supersymmetry.  The finite transformation used in the compactification
must preserve the super-Virasoro structure.  When the local generator
does not commute with \(G_{\rm ws}\), only selected finite rotations
can have this property, as discussed in
Question~\ref{sec:ss-string-discreteness}.

The charge space must now be enlarged to include the eigenvalues of
\(Q_{\rm SS}\).  Denote the resulting internal charge vector by
\(Q\), and let \(\xi\) specify the Scherk--Schwarz direction in this
space.

For one compact circle, the complete transformation is
\begin{align}
Q'
&=
Q-n\xi,
\nonumber\\
m'
&=
m+\xi\cdot Q-\frac12\xi^2n,
\nonumber\\
n'
&=
n.
\label{eq:q21-SS-charge-transformation}
\end{align}

With the convention of Eq.~\eqref{eq:q21-transvection}, this is
\(T_{-\xi}\).  Setting \(a=-\xi\) gives
\(Q'=Q+na=Q-n\xi\) and
\(m'=m-a\cdot Q-\tfrac12a^2n\).

The transformation preserves the extended quadratic form:
\begin{equation}
(Q-n\xi)^2
+
2n
\left(
m+\xi\cdot Q-\frac12\xi^2n
\right)
=
Q^2+2mn.
\label{eq:q21-SS-norm-check}
\end{equation}

In the zero-winding sector, \(n=0\), the charge is unchanged and the
momentum becomes \(m'=m+\xi\cdot Q\).  For a single \(U(1)\) charge
\(q\), this is precisely the field-theory rule
\(m\to m+\xi q\).

The winding-dependent terms are the additional string information
required by preservation of the full charge-space norm.

\narrativeheading{Wilson lines and Scherk--Schwarz deformations share
one charge-space geometry}

With the conventions used above, a gauge Wilson line is represented by
\(T_A\), while the Scherk--Schwarz deformation is represented by
\(T_{-\xi}\).  Once gauge and Scherk--Schwarz charges are included in
the same internal charge space \(W\), both are parabolic
transvections fixing the same momentum--winding null direction.

For any \(a,b\in W\),
\begin{equation}
T_aT_b
=
T_{a+b}
=
T_bT_a.
\label{eq:q21-transvection-group-law}
\end{equation}
No orthogonality assumption on \(a\) and \(b\) is required.  The mixed
term \(a\cdot b\) is already contained in
\(-\tfrac12(a+b)^2e\).  Thus \(T_A\) and \(T_{-\xi}\) commute even
when their charge-space directions are not orthogonal.

Their commutativity in this setting is a property of the
underlying charge-space transformations, not an accidental
cancellation in the component mass formula.

This conclusion relies on both deformations belonging to the Abelian
unipotent subgroup associated with one common null direction.  More
general backgrounds can involve several null directions or
non-Abelian current algebras, for which the corresponding
transformations need not commute.

\narrativeheading{Which worldsheet currents are admissible?}

The existence of an \(R\)-symmetry generator in the low-energy
supergravity is not sufficient to produce a string
Scherk--Schwarz deformation.

The candidate charge must be represented by a dimension-one
worldsheet current.  It must act non-trivially on the sector from
which spacetime supersymmetry is constructed, while the finite
transformation used in the quotient must remain an exact operation of
the worldsheet theory.

An orbifold imposes a further compatibility condition.  Suppose an
orbifold generator \(\vartheta\) acts on a compact coordinate
direction \(i\) through a representation \(R_i(\vartheta)\).  A
Scherk--Schwarz current attached to that direction must transform in
the corresponding representation:
\begin{equation}
\vartheta J_{\rm SS}\vartheta^{-1}
=
R_i(\vartheta)J_{\rm SS}.
\label{eq:q21-current-covariance}
\end{equation}

This allows the coordinate-dependent deformation and the orbifold
projection to be imposed simultaneously.

Current covariance and lattice integrality answer different
questions.  Equation~\eqref{eq:q21-current-covariance} tests whether
the candidate current fits the orbifold action.  Requiring the
associated transvection to preserve the integral string lattice tests
whether it defines an exact microscopic charge transformation.  The
example below verifies the first condition, not the second.

\narrativeheading{A concrete \(\mathbb Z_4\) eigen-current}

Consider a heterotic \(\mathbb Z_4\) orbifold with twist vector
\begin{equation}
v
=
\frac14(1,1,-2).
\label{eq:q21-Z4-twist-vector}
\end{equation}
Since the three components sum to zero, the rotation lies in
\(SU(3)\), as required for the standard supersymmetric parent
orbifold.

Let \(f_1,f_2,f_3\) be three complex internal worldsheet fermions, and
let \(\psi\) be a worldsheet fermion associated with a non-compact
spacetime direction.  The orbifold acts as
\begin{equation}
f_1\mapsto if_1,
\qquad
f_2\mapsto if_2,
\qquad
f_3\mapsto-f_3,
\qquad
\psi\mapsto\psi.
\label{eq:q21-Z4-fermion-action}
\end{equation}

A current associated with the first compact complex direction is
\begin{equation}
J^1
=
\frac1{\sqrt2}
\left[
:f_1^\dagger\,\operatorname{Re}f_3:
+
:f_1\psi:
\right],
\label{eq:q21-Z4-current}
\end{equation}
where the colons denote normal ordering.

Each fermion has conformal weight \(1/2\), so both bilinears have
conformal weight one.

Under the orbifold,
\(f_1^\dagger\mapsto-if_1^\dagger\) and
\(\operatorname{Re}f_3\mapsto-\operatorname{Re}f_3\).  Hence
\begin{align}
:f_1^\dagger\operatorname{Re}f_3:
&\longmapsto
(-i)(-1)
:f_1^\dagger\operatorname{Re}f_3:
\nonumber\\
&=
i:f_1^\dagger\operatorname{Re}f_3:.
\end{align}
The second bilinear transforms in the same way,
\(:f_1\psi:\mapsto i:f_1\psi:\).  Therefore
\begin{equation}
\vartheta J^1\vartheta^{-1}
=
iJ^1.
\label{eq:q21-Z4-current-phase}
\end{equation}

This is the phase carried by the first compact complex direction,
\(e^{2\pi i v_1}=i\).  The current therefore has precisely the
covariance required to associate the Scherk--Schwarz charge with that
direction.

The current has been written in a complex eigenbasis.  Its Hermitian
conjugate has eigenvalue \(-i\); in a real basis the corresponding two
current components are rotated into one another by the
\(\mathbb Z_4\) generator.

The occurrence of the spacetime fermion \(\psi\) is also important,
because it allows the current charge to act on the sector involved in
the construction of spacetime supersymmetry
\cite{Benakli:1995ut}.

\narrativeheading{What the massive charge lattice can remember}

Projection to the massless effective theory loses much of the
charge-lattice information.

A Wilson line can remove a gauge-lattice state from the massless
spectrum while leaving it at a mass of order \(1/R\) when the
corresponding compact direction is large.  Its charge remains the
original lattice vector \(P\), and need not coincide with the gauge
quantum numbers of any massless field.

For zero winding, its momentum is shifted by \(A\cdot P\).  After
choosing the nearest integer Kaluza--Klein number, the lightest state
in that charge sector has a mass of order
\[
\frac{\operatorname{dist}(A\cdot P,\mathbb Z)}{R}.
\]
A fractional value of \(A\cdot P\) can therefore produce states below
the first integer Kaluza--Klein excitation of the massless spectrum.

Knowledge of the massless gauge algebra is not sufficient to
reconstruct these states.  One must know how the Wilson line is
embedded in the complete charge lattice.

The same loss of information occurs for supersymmetry.  A
four-dimensional orbifold may leave only \(N=1\) supersymmetry in the
massless sector, while an untwisted subsector descends from a parent
sector with \(N=2\) or \(N=4\).  Its Kaluza--Klein excitations can
still assemble into extended supermultiplets before the
Scherk--Schwarz splitting is applied.

The complete tower can retain cancellations and
representation-theoretic information absent from the massless
four-dimensional spectrum, providing a microscopic counterpart of
the tower-level cancellations discussed in
Question~\ref{sec:ss-supertraces}.

\narrativeheading{Where the lattice formula applies}

The affine charge transformation acts directly in sectors whose
states are described by the momentum, winding and current charges
included in the extended lattice.

In the symmetric Abelian heterotic orbifolds for which the
construction above was developed, ordinary fixed-point twisted states
do not carry the same internal momentum quantum numbers as the
untwisted sector.  They therefore do not receive the same tree-level
Scherk--Schwarz shift
\cite{Benakli:1995ut}.

This is a property of that class of models, not a theorem about every
string orbifold.  Freely acting orbifolds, asymmetric constructions
and more general CFT realizations can assign different momentum,
winding or fractional charges to their twisted sectors.

The action of the Scherk--Schwarz deformation must consequently be
determined sector by sector from the complete worldsheet construction.
The charge-lattice formula is exact in the sectors to which its charge
vector applies; it should not be extended to other sectors by analogy
alone.

\narrativeheading{From the real charge space to the exact string
lattice}

The analysis above identifies the real Lorentzian transformation
behind the string Scherk--Schwarz shift.  The linear momentum shift
and its quadratic winding correction are not independent ingredients:
they are fixed components of one parabolic isometry.

This is not yet sufficient to define an exact string operation.  The
isometry must preserve the integral lattice of physical charges, and
the associated current must remain compatible with the orbifold, GSO
projection and operator algebra.

These arithmetic and worldsheet conditions determine which
continuous transformations of the real charge space survive as exact
string operations.  They are the subject of the next question.

\narrativeheading{Bibliographical remarks}

The organization of momentum, winding and gauge charges into an even
self-dual Lorentzian lattice is the basis of the Narain description of
toroidal string compactification
\cite{Narain:1985jj,Narain:1986am}.

Coordinate-dependent compactifications can be described in terms of
transformations acting on enlarged charge and momentum data; related
formulations can be found in
Refs.~\cite{Kounnas:1988ye,Ferrara:1988jx,Kounnas:1989dk,
Antoniadis:1990ew}.

A treatment in which gauge Wilson lines and Scherk--Schwarz
supersymmetry breaking act simultaneously on an extended heterotic
charge vector, together with explicit orbifold restrictions on the
supersymmetry-breaking current and the dependence of light
Kaluza--Klein quantum numbers on the Wilson-line embedding, can be
found in Ref.~\cite{Benakli:1995ut}.

The same coordinate-dependent compactification language has also been
used in later heterotic constructions, where the deformation acts on
spacetime, internal \(R\)-symmetry and gauge charges and determines
the shifted Kaluza--Klein spectrum
\cite{Abel:2017vos}.

\begin{takehome}
\begin{takehomeitems}

\item
Momentum, winding and internal charges form a single Lorentzian string
charge vector.  A Scherk--Schwarz deformation must therefore preserve
the complete charge-space quadratic form, not only the Kaluza--Klein
spectrum of the zero-winding sector.

\item
For a gauge Wilson line,
\[
P\to P+nA,
\qquad
m\to m-A\cdot P-\frac12A^2n.
\]
The linear and quadratic shifts are fixed components of one parabolic
isometry of the momentum--winding charge space.

\item
After the charge acting on spacetime supersymmetry is included, the
same geometry gives
\[
Q\to Q-n\xi,
\qquad
m\to m+\xi\cdot Q-\frac12\xi^2n.
\]
At zero winding this reduces to the field-theory shift
\(m\to m+\xi q\).

\item
Gauge Wilson lines and Scherk--Schwarz deformations belong to the same
Abelian family of transvections when they act through a common
momentum--winding null direction, although they act on different
internal charges and have different physical interpretations.

\item
The supersymmetry-breaking charge must arise from an admissible
dimension-one worldsheet current, act non-trivially on the
supersymmetry sector and transform consistently under the orbifold.
The explicit \(\mathbb Z_4\) current illustrates these worldsheet
conditions.

\item
The massive string tower can retain gauge charges and extended
supersymmetry structures absent from the massless EFT, and different
twisted sectors need not experience the same Scherk--Schwarz shift.

\item
A Lorentzian isometry of the real charge space is not yet an exact
string symmetry.  It must also preserve the integral charge lattice
and the complete worldsheet data; these additional restrictions are
the subject of the next question.

\end{takehomeitems}
\end{takehome}

\nextquestion
{When is a rigid Scherk--Schwarz twist discrete, and how can the
effective phase nevertheless vanish continuously?}

\end{reviewquestion}
% Question 23 v3, renumbered from Question22v2.tex.
% Original section/equation labels retained for cross-reference stability.
%%%%%%%%%%%%%%%%%%%%%%%%%%%%%%%%%%%%%%%%%%%%%%%%%%%%%%%%%%%%%%%

\begin{reviewquestion}
{When is a rigid Scherk--Schwarz twist discrete, and how can the
effective phase nevertheless vanish continuously?}
\label{sec:ss-string-discreteness}

\begin{essentialidea}

The mass scale remains continuous even for a quantized twist.

The half-twist gives the simplest example.  If the microscopic
identification acts as \(-1\) on the gravitino, the Kaluza--Klein
tower is
\begin{equation}
m_n
=
\frac{\left|n+\tfrac12\right|}{R},
\qquad
m_{3/2}
=
\frac1{2R}.
\label{eq:q22-half-twist-spectrum}
\end{equation}
The twist remains exactly fixed as the radius varies, while
\(m_{3/2}\to0\) for \(R\to\infty\).

The dimensionless twist and the dimensionful Kaluza--Klein scale are
therefore different data.  More generally, the physical
Scherk--Schwarz gap seen by a given branch is
\begin{equation}
m_{\rm SS}
=
M_{\rm KK}\,
\operatorname{dist}
\left(
\omega_{\rm eff},
\mathbb Z
\right),
\label{eq:q22-physical-gap}
\end{equation}
where \(M_{\rm KK}\) is the relevant Kaluza--Klein spacing and
\(\omega_{\rm eff}\in\mathbb R/\mathbb Z\) is the phase actually seen
by that branch.

The microscopic twist \(\omega_{\rm rigid}\) can be quantized by the
finite-order group action, the integral string charge lattice, the
worldsheet operator algebra and modular consistency.  The effective
phase need not coincide with this rigid datum: continuous Wilson-line
moduli, dynamical backgrounds or localized interactions can also
contribute to the spectral boundary condition.

A small Scherk--Schwarz gap can arise in two different
ways.  The Kaluza--Klein scale may tend to zero while the effective
phase remains of order one, as in decompactification.  Alternatively,
continuous dynamics can drive \(\omega_{\rm eff}\) close to an
integer while the rigid microscopic twist remains fixed.

These two mechanisms can have very different relations to the tower
of states that limits the lower-dimensional description.

\end{essentialidea}

\narrativeheading{A fixed twist and a variable mass scale}

The half-twist already resolves the most immediate apparent paradox.
The microscopic identification remains
\(U_{\rm SS}|\psi_{3/2}\rangle=-|\psi_{3/2}\rangle\); nothing is
continuously dialled between \(U_{\rm SS}=-\mathbf1\) and the
identity.  The mass nevertheless depends continuously on the radius
through Eq.~\eqref{eq:q22-half-twist-spectrum}.

If \(R\) is regarded as an external parameter, the limit
\(R\to\infty\) compares a family of compactification backgrounds.  If
the radius is a dynamical modulus, its expectation value controls the
breaking scale within the lower-dimensional theory, as discussed in
Questions~\ref{sec:ss-wilson-radion}
and~\ref{sec:ss-string-restrictions}.

The restoration limit also changes the range of validity of the
lower-dimensional description.  The Kaluza--Klein tower behaves as
\begin{equation}
m_n^{\rm KK}
=
\frac{|n|}{R}
\longrightarrow
0
\qquad
(R\to\infty).
\label{eq:q22-KK-tower-decompactification}
\end{equation}
A finite four-dimensional truncation therefore ceases to be an
appropriate description near the limit.

Supersymmetry is restored because the physical scale multiplying the
fixed identification tends to zero, not because the microscopic twist
approaches the identity.

\narrativeheading{Why finite-order string twists are discrete}

Let \(U_\xi\) be the group element implementing the microscopic
Scherk--Schwarz identification.  In a quotient of order \(N\), it must
satisfy
\begin{equation}
U_\xi^N
=
\mathbf1.
\label{eq:q22-finite-order}
\end{equation}

If a string state \(|\Psi\rangle\) has eigenphase
\(U_\xi|\Psi\rangle=e^{2\pi i\Omega_\Psi}|\Psi\rangle\), then
Eq.~\eqref{eq:q22-finite-order} requires
\begin{equation}
N\Omega_\Psi
\in
\mathbb Z
\label{eq:q22-eigenphase-condition}
\end{equation}
for every state in the complete string Hilbert space.

For a one-parameter action with
\(\Omega_\Psi=\omega_{\rm rigid}q_\Psi\), this becomes
\begin{equation}
N\omega_{\rm rigid}q_\Psi
\in
\mathbb Z
\qquad
\text{for every charge }q_\Psi
\text{ in the string spectrum}.
\label{eq:q22-charge-quantization}
\end{equation}
The massive charge spectrum can therefore impose restrictions that are
invisible in the massless effective theory.

The charge-by-charge condition above can be expressed intrinsically
as a statement about the group element itself.  The charges
\(q_\Psi\) are the weights with which a maximal torus acts on the
string Hilbert space, so requiring every eigenphase to have order
dividing \(N\) is equivalent to requiring the corresponding point of
the torus to be \(N\)-torsion.

Let
\(T=\mathfrak t/\Lambda_{\rm cochar}\) be a compact maximal torus,
where \(\mathfrak t\) is its Lie algebra and
\(\Lambda_{\rm cochar}\) its cocharacter lattice.  If
\[
U_{\rm SS}
=
\exp
\left(
2\pi i\,\boldsymbol\omega_{\rm rigid}\cdot H
\right),
\]
then \(U_{\rm SS}\) has order dividing \(N\) precisely when
\begin{equation}
N\boldsymbol\omega_{\rm rigid}
\in
\Lambda_{\rm cochar}.
\label{eq:q22-torsion-condition}
\end{equation}

For fixed \(N\), the allowed points are the \(N\)-torsion points of
the torus,
\(T[N]=N^{-1}\Lambda_{\rm cochar}/\Lambda_{\rm cochar}\).

This is only the first microscopic test.  The transformation must also
preserve the integral string charge lattice, lift consistently to the
vertex operators and their cocycles, act consistently in every
twisted sector, and satisfy the modular, sewing and anomaly conditions
discussed in Questions~\ref{sec:ss-quantum-consistency}
and~\ref{sec:ss-string-restrictions}.

Thus the set of twists is discrete for a fixed finite-order
construction.  Allowing the order \(N\), the orbifold group or the
parent CFT to change does not create a continuous modulus within one
fixed theory.

\narrativeheading{A worldsheet current need not generate a continuous
string symmetry}

The same quantization can be understood directly on the worldsheet.

Suppose a dimension-one current has zero-mode charge \(Q_{\rm SS}\).
One can formally write \(e^{i\alpha Q_{\rm SS}}\) for any real
\(\alpha\).  This does not imply that every \(\alpha\) defines an
automorphism of the complete string theory.

For a Scherk--Schwarz charge, \(Q_{\rm SS}\) must act non-trivially on
the sector from which spacetime supersymmetry is constructed.  In the
coordinate-dependent heterotic constructions discussed in
Question~\ref{sec:ss-string-charge-lattice}, this can be expressed at
the level of the local generator as
\begin{equation}
[Q_{\rm SS},G_{\rm ws}]
\neq
0,
\end{equation}
where \(G_{\rm ws}\) is the relevant worldsheet supercurrent.

A generic value of \(\alpha\) therefore need not define an
automorphism of the full worldsheet theory.  Isolated finite values can
nevertheless do so, because their action may close on the complete
operator algebra and preserve the GSO and super-Virasoro structures.

In the simplest example, the local fields entering the worldsheet
supercurrent have integer charge under \(Q_{\rm SS}\), whereas the
spin fields creating spacetime fermions have half-integer charge.
For
\[
U=e^{2\pi iQ_{\rm SS}},
\]
a field of integer charge \(q\) acquires the phase
\(e^{2\pi iq}=1\), while a field of half-integer charge acquires
\(e^{2\pi iq}=-1\).  The same finite transformation can therefore act
trivially on the local worldsheet algebra and as \(-1\) on spacetime
fermions.

This produces the anti-periodic Scherk--Schwarz spectrum without
requiring a continuous \(U(1)\) symmetry of the complete string
theory.

The \(\mathbb Z_4\) calculation in
Question~\ref{sec:ss-string-charge-lattice} tested a related but
different condition: the candidate current had the required orbifold
covariance.  Current covariance, finite-order exponentiation and
preservation of the integral charge lattice are distinct microscopic
requirements.

\narrativeheading{Continuous Wilson-line moduli and discrete orbifold
embeddings}

The term ``Wilson line'' can describe two objects with different
microscopic status.

Before orbifolding, a surviving Wilson-line zero mode can be a
continuous Narain modulus.  Its holonomy can be written
\[
W(\boldsymbol\theta)
=
\exp
\left(
2\pi i\,\boldsymbol\theta\cdot H
\right),
\qquad
\boldsymbol\theta
\in
\mathfrak t/\Lambda_{\rm cochar},
\]
with a further Weyl identification when conjugate holonomies are
identified.  The parameter \(\boldsymbol\theta\) is then the
expectation value of a dynamical scalar field, and moving it changes
the background continuously within the same moduli space.

A discrete Wilson line used in an orbifold embedding has a different
status.  If it is associated with an element of order \(N\), it obeys
a lattice condition of the form
\begin{equation}
NA
\in
\Lambda_{\rm gauge},
\label{eq:q22-discrete-Wilson-line}
\end{equation}
together with model-dependent orbifold and modular constraints.  It
is a rigid microscopic choice rather than a continuously variable zero
mode.

A continuous Wilson-line modulus can nevertheless contribute to the
phase seen by a particular state.  Its contribution belongs to the
continuous background data, not to the discrete label of the
microscopic quotient.

\narrativeheading{The spectrum can see more than the rigid twist}

The statement that a quantized \(\omega_{\rm rigid}\) admits no
continuous small-\(\omega_{\rm rigid}\) limit is correct when the
spectrum depends only on that rigid datum.  It need not remain true
for the effective phase entering the complete spectral problem.

Localized interactions give a concrete lower-dimensional example.
In the conventions of
Question~\ref{sec:ss-localized-boundary-terms}, let \(M_0\) and
\(M_\pi\) be the dimensionless coefficients of localized gravitino
bilinears at the two ends of an interval.  The tower is
\begin{equation}
m_n
=
\frac1R
\left|
n
+
\omega_{\rm rigid}
+
\frac1\pi
\left[
\arctan(M_0)
+
\arctan(M_\pi)
\right]
\right|.
\label{eq:q22-boundary-spectrum}
\end{equation}
The phase seen by this spectrum is therefore
\begin{equation}
\omega_{\rm eff}
=
\omega_{\rm rigid}
+
\frac1\pi
\left[
\arctan(M_0)
+
\arctan(M_\pi)
\right]
\pmod 1.
\label{eq:q22-effective-phase}
\end{equation}

The rigid monodromy can remain fixed while the localized contributions
vary through dynamical expectation values.  There need not be a
continuous small-\(\omega_{\rm rigid}\) limit.  What can become
continuously small is the spectral mismatch produced by the sum of
rigid and dynamical contributions.

Equation~\eqref{eq:q22-effective-phase} concerns the quadratic
spectrum.  It does not imply that a rigid twist and a localized
supersymmetry-breaking sector define the same microscopic theory.
Their Goldstino content, local interactions and ultraviolet completion
can remain different even when they give the same
\(\omega_{\rm eff}\).

\narrativeheading{Approaching a cancellation locus
non-perturbatively}

Let \(\Phi\) denote the moduli and other dynamical fields controlling
the additional contribution to the phase, and write
\[
\omega_{\rm eff}(\Phi)
=
\omega_{\rm rigid}
+
\omega_{\rm dyn}(\Phi)
\pmod 1.
\]

A supersymmetric spectral locus is reached when
\(\omega_{\rm rigid}+\omega_{\rm dyn}(\Phi)\in\mathbb Z\).
Near such a locus, let \(k\in\mathbb Z\) denote the integer approached
by \(\omega_{\rm eff}\), and write
\begin{equation}
\omega_{\rm dyn}(\Phi)
=
k-\omega_{\rm rigid}
+
\delta\omega_{\rm np}(\Phi).
\label{eq:q22-near-cancellation}
\end{equation}

The exponentially small term is not, in general, what cancels an
order-one rigid twist.  The perturbative or order-one dynamical
contributions must first place the system close to the cancellation
locus.  The residual displacement can then be non-perturbatively
small,
\begin{equation}
\left|
\delta\omega_{\rm np}(\Phi)
\right|
\sim
\exp
\left[
-\operatorname{Re}S_{\rm np}(\Phi)
\right].
\label{eq:q22-nonperturbative-residual}
\end{equation}

A familiar source of exponentially small terms is non-perturbative
gauge dynamics,
\begin{equation}
W_{\rm np}
\sim
A(\Phi)\,
e^{-a f(\Phi)}.
\label{eq:q22-nonperturbative-superpotential}
\end{equation}
The gauge kinetic function \(f(\Phi)\) can depend on the dilaton,
geometric moduli or several fields.  How this contribution feeds into
a localized mass or an effective Scherk--Schwarz phase is
model dependent, but whenever it does, the residual displacement from
the cancellation locus can inherit the exponential suppression
\(e^{-\operatorname{Re}S_{\rm np}}\). This suppression is therefore not universally controlled by the string coupling alone.

When \(\omega_{\rm rigid}=0\), a small non-perturbative contribution
can instead generate the entire effective phase.  Near a cancellation
locus one then has parametrically
\begin{equation}
m_{\rm SS}
\sim
M_{\rm KK}\,
e^{-\operatorname{Re}S_{\rm np}(\Phi)}.
\label{eq:q22-nonperturbative-SS-scale}
\end{equation}

\narrativeheading{Comparing the supersymmetry-breaking scale with the tower}

To distinguish the possible restoration mechanisms without committing
to a particular modulus, it is useful to compare the
supersymmetry-breaking gap directly with the first infinite tower that
limits the lower-dimensional description.

Let \(m_{\rm tower}\) be the characteristic scale of that tower along
the trajectory under consideration.  Define
\begin{equation}
\Xi
=
\frac{m_{\rm tower}}{m_{\rm SS}}
=
\frac{m_{\rm tower}}{M_{\rm KK}}\,
\frac1{
\operatorname{dist}
\left(
\omega_{\rm eff},
\mathbb Z
\right)
}.
\label{eq:q22-tower-ratio}
\end{equation}

For ordinary decompactification, the light tower is the Kaluza--Klein
tower itself, with
\(m_{\rm tower}\sim M_{\rm KK}\sim1/R\).  If the dimensionless twist
remains of order one, then
\begin{equation}
\Xi
=
O(1).
\label{eq:q22-tower-ratio-decompactification}
\end{equation}
The tower and the Scherk--Schwarz gap collapse at the same parametric
rate.

At a finite-distance cancellation locus, the tower scale can remain
finite while \(m_{\rm SS}\to0\), and \(\Xi\) diverges.

Along an asymptotic non-perturbative trajectory,
Eq.~\eqref{eq:q22-nonperturbative-residual} instead gives
\begin{equation}
\Xi
\sim
\frac{m_{\rm tower}}{M_{\rm KK}}\,
e^{\operatorname{Re}S_{\rm np}(\Phi)}.
\label{eq:q22-tower-ratio-nonperturbative}
\end{equation}
If \(m_{\rm tower}/M_{\rm KK}\) varies more slowly than the
exponential, supersymmetry is restored parametrically faster than the
tower descends.

The ratio above can become parametrically large in a concrete
asymptotic regime.  To illustrate this possibility, consider a
weak-coupling trajectory in which the non-perturbative action is
controlled by the dilaton.  Suppose
\(\operatorname{Re}S_{\rm np}\sim a/g_s^p\), with \(a,p>0\), while
the dimensionless internal volume \(v=V_6M_s^6\) is held fixed.  Then
\begin{equation}
M_{\rm Pl}^2
\sim
\frac{v}{g_s^2}\,
M_s^2,
\qquad
\frac{M_s}{M_{\rm Pl}}
\sim
\frac{g_s}{\sqrt v}.
\label{eq:q22-Planck-string-scaling}
\end{equation}

Along this trajectory, the perturbative string tower becomes light
only as a power of \(g_s\) in Planck units, while the residual
Scherk--Schwarz phase can be exponentially small.  A parametrically
large hierarchy can therefore open,
\begin{equation}
m_{\rm SS}
\ll
m_{\rm tower}
\ll
M_{\rm Pl}.
\label{eq:q22-hierarchical-window}
\end{equation}

This example is deliberately specific.  If the non-perturbative
action is controlled mainly by another modulus, the accompanying
tower and its scaling in Planck units must be determined along that
trajectory.  The invariant statement is the comparison in
Eq.~\eqref{eq:q22-tower-ratio}, not a universal identification of
\(S_{\rm np}\) with the dilaton.

\narrativeheading{What is and is not quantized}

The finite-order group element defining a rigid string quotient can be
quantized, while the moduli controlling the Kaluza--Klein scale,
surviving Wilson lines or localized dynamical contributions remain
continuous.  The phase entering a particular spectrum can 
contain both discrete and continuous data.

A small Scherk--Schwarz gap can consequently arise from
decompactification, from cancellation in the effective phase, or from
both.  These mechanisms are distinguished by the accompanying tower
of states and by the microscopic origin of the dynamical contribution.
If additional independent supersymmetry-breaking sectors are present,
closing the Scherk--Schwarz gap alone does not restore the complete
supersymmetry.

\narrativeheading{Bibliographical remarks}

The distinction between continuous Narain deformations and discrete
coordinate-dependent compactifications is discussed in early string
implementations of Scherk--Schwarz breaking; related formulations can
be found in
Refs.~\cite{Kounnas:1988ye,Ferrara:1988jx,Kounnas:1989dk,
Antoniadis:1990ew}.

A treatment combining gauge Wilson lines and coordinate-dependent
Scherk--Schwarz deformations in an enlarged heterotic charge lattice,
including discrete restrictions imposed by Abelian orbifolds, can be
found in Ref.~\cite{Benakli:1995ut}.

Coordinate-dependent compactifications and discrete charge vectors
also appear in later heterotic constructions.  For example,
Ref.~\cite{Abel:2017vos} uses a CDC vector acting on spacetime,
internal \(R\)-symmetry and gauge charges to determine the shifted
Kaluza--Klein spectrum.

The distinction emphasized here between a quantized rigid twist and a
modulus-dependent effective spectral phase is more general than these
specific constructions.  Localized masses provide one explicit
lower-dimensional realization; their microscopic origin and possible
non-perturbative dependence are model dependent.

\newpage
\begin{takehome}
\begin{takehomeitems}

\item
A finite-order microscopic twist is discrete because its action must
close on the complete string spectrum.  For an order-\(N\) quotient,
\(N\omega_{\rm rigid}q_\Psi\in\mathbb Z\) for every string charge
\(q_\Psi\), together with the lattice, worldsheet and modular
consistency conditions.

\item
A discrete twist does not imply a discrete supersymmetry-breaking
scale.  At the half-twist, \(m_{3/2}=1/(2R)\), so the mass vanishes
in the decompactification limit while the microscopic twist remains
fixed.

\item
A dimension-one worldsheet current need not generate a continuous
exact string symmetry.  The full operator algebra can preserve only
selected finite rotations.

\item
Continuous Wilson-line zero modes are moduli, whereas finite-order
orbifold Wilson lines are rigid lattice embeddings.  Both can
contribute to the spectral phase without having the same microscopic
status.

\item
The spectrum can depend on an effective phase containing both rigid
and dynamical contributions.  Localized gravitino masses provide the
explicit example
\[
\omega_{\rm eff}
=
\omega_{\rm rigid}
+
\frac1\pi
\left[
\arctan(M_0)+\arctan(M_\pi)
\right]
\pmod1 .
\]

\item
Near a dynamical cancellation locus, the residual mismatch can be
non-perturbatively small.  For an order-one rigid twist, the
exponential term controls the residual only after order-one
contributions have already brought the system close to that locus.

\item
The ratio \(m_{\rm tower}/m_{\rm SS}\) distinguishes restoration by
decompactification from restoration by cancellation.  It is typically
\(O(1)\) for a fixed order-one twist in decompactification, but can
become parametrically large when the effective phase approaches an
integer much faster than the relevant tower descends.

\end{takehomeitems}
\end{takehome}

\nextquestion
{How does a freely acting orbifold realize stringy Scherk--Schwarz
breaking?}

\end{reviewquestion}
% Question 24 v3, renumbered from Question23v2.tex.
% Original section/equation labels retained for cross-reference stability.
%%%%%%%%%%%%%%%%%%%%%%%%%%%%%%%%%%%%%%%%%%%%%%%%%%%%%%%%%%%%%%%

\begin{reviewquestion}
{How does a freely acting orbifold realize stringy
Scherk--Schwarz breaking?}
\label{sec:ss-freely-acting}

\begin{essentialidea}

A freely acting orbifold gives an exact string realization of the
Scherk--Schwarz mechanism by combining an internal transformation that
acts on spacetime supercharges with a translation along a compact
direction.

Take a covering circle of radius \(NR\), with
\(y\sim y+2\pi NR\), and let
\(\delta_N:y\mapsto y+2\pi R\) be its order-\(N\) translation.
If \(g_R\) is an internal symmetry satisfying \(g_R^N=\mathbf1\), the
combined generator is
\begin{equation}
g_{\rm SS}
=
g_R\,\delta_N.
\label{eq:q23-generator}
\end{equation}

Every non-trivial power of the translation moves every point of the
covering circle.  In the elementary product construction, the
Scherk--Schwarz subgroup therefore has no fixed points, even when
\(g_R\) has fixed points on another factor of the compactification.
The breaking is global rather than localized on a defect.

The quotient of the covering circle by \(\delta_N\) is a smooth circle
of radius \(R\).  The internal action survives as a flat holonomy.
If a field component obeys
\(g_R\Phi=e^{2\pi iq/N}\Phi\), then one circuit around the quotient
circle gives
\begin{equation}
\Phi(y+2\pi R)
=
e^{-2\pi iq/N}\Phi(y),
\label{eq:q23-equivariance}
\end{equation}
and hence
\begin{equation}
p_{y,m}
=
\frac{m-q/N}{R},
\qquad
m\in\mathbb Z.
\label{eq:q23-shifted-spectrum}
\end{equation}

The Scherk--Schwarz shift is therefore not added after the orbifold
has been constructed: it is the exact spectrum implied by the
equivariance condition on the smooth quotient.

Freeness does not remove twisted sectors.  It changes their geometric
character: their non-trivial information is carried by winding along
the translated direction rather than by localization at a fixed
point.

At large radius the holonomy remains fixed while the
Scherk--Schwarz mass splittings vanish.  The supersymmetric parent is
then recovered at finite energy even though the microscopic orbifold
generator never approaches the identity.

\end{essentialidea}

\narrativeheading{Why adding a translation changes the orbifold}

An ordinary orbifold transformation can leave points of the internal
space fixed.  Its twisted sectors may then contain states localized
near those fixed loci, and local operators can be supported there.

A translation behaves differently.  For \(1\leq k<N\),
\(\delta_N^k\) shifts the covering coordinate by \(2\pi kR\).  This
displacement is non-zero modulo the covering circumference
\(2\pi NR\), so
\begin{equation}
\operatorname{Fix}
\left(
g_{\rm SS}^k
\right)
=
\varnothing,
\qquad
1\leq k<N.
\label{eq:q23-no-fixed-points}
\end{equation}

The statement concerns the complete action
\(g_R^k\delta_N^k\).  The internal transformation \(g_R^k\) can have
fixed points on another factor of the compactification, but no point
of the product is fixed while its circle coordinate is translated.

The quotient therefore introduces no fixed-point singularity
associated with the Scherk--Schwarz subgroup.  It replaces the
covering circle of radius \(NR\) by a circle of radius \(R\), while
retaining the internal action through a global boundary condition on
the fields.

\narrativeheading{How the twist survives on the smooth quotient}

Consider a component of a field in the \(g_R\)-eigenspace with
eigenvalue \(e^{2\pi iq/N}\).  Invariance under the combined
identification requires the translation to compensate this internal
phase, giving Eq.~\eqref{eq:q23-equivariance}.

Transport the field once around the quotient circle.  It returns to
the same spacetime point but to a value differing by
\(e^{-2\pi iq/N}\).

Mathematically, the field is a section of a flat complex line bundle,
or equivalently of the corresponding one-dimensional local system.
Its holonomy is
\begin{equation}
\rho_q(1)
=
e^{-2\pi iq/N},
\label{eq:q23-holonomy}
\end{equation}
where \(1\) denotes one positively oriented circuit of the quotient
circle.

The underlying complex line bundle is topologically trivial, since
\(H^2(S^1,\mathbb Z)=0\).  Its flat connection can nevertheless have
non-trivial holonomy, which cannot be removed by a single-valued field
redefinition unless \(q=0\) modulo \(N\).

The compatible modes are
\begin{equation}
\Phi_m(y)
=
\exp
\left[
i\frac{m-q/N}{R}\,y
\right],
\qquad
m\in\mathbb Z.
\label{eq:q23-equivariant-modes}
\end{equation}
Their shifted Kaluza--Klein spectrum is the direct spectral
consequence of the flat holonomy.

\narrativeheading{The covering-space projection gives the same
spectrum}

The same result can be derived directly on the covering circle.

A covering-space momentum eigenmode is
\[
\exp
\left(
i\frac{n}{NR}y
\right),
\qquad
n\in\mathbb Z.
\]
Under \(\delta_N\), it acquires the phase \(e^{2\pi in/N}\).
Multiplication by the internal eigenphase gives the total phase
\(e^{2\pi i(n+q)/N}\).

Orbifold invariance requires
\begin{equation}
n+q
=
0
\pmod N.
\label{eq:q23-projection-congruence}
\end{equation}
Writing \(n=Nm-q\) gives
\[
\frac{n}{NR}
=
\frac{m-q/N}{R},
\]
which is exactly Eq.~\eqref{eq:q23-shifted-spectrum}.

The covering-space projection and the quotient-space holonomy are
therefore two exact descriptions of the same orbifold spectrum.

\narrativeheading{The half-shift benchmark}

The simplest supersymmetry-breaking example has \(N=2\).  The
covering circle has radius \(2R\), and the generator combines a
half-period translation with spacetime fermion parity,
\begin{equation}
g_{\rm SS}
=
(-1)^{F_s}\delta_2,
\qquad
\delta_2:
y\longmapsto y+2\pi R.
\label{eq:q23-half-shift}
\end{equation}

Here \(F_s\) denotes spacetime fermion number.  Its difference from a continuous \(R\)-symmetry twist is
explained in Question~\ref{sec:ss-spin-fermion-parity-R}.
Only the resulting spectral shift is needed below.

Bosons have \(q=0\), so their momenta are \(p_{B,m}=m/R\).  Fermions
have \(q=1\), giving \(p_{F,m}=(m-\tfrac12)/R\).  After an integer
relabelling of \(m\), this becomes
\begin{equation}
p_{F,m}
=
\frac{m+\tfrac12}{R}.
\label{eq:q23-half-fermions}
\end{equation}

The lightest fermionic mass is therefore \(1/(2R)\), exactly as in
the half-twist analysis of
Question~\ref{sec:ss-half-twist}.  The same spectrum is now understood
as the difference between trivial holonomy for spacetime bosons and
holonomy \(-1\) for spacetime fermions on the same smooth quotient
circle.

\narrativeheading{Twisted sectors carry fractional winding}

A freely acting orbifold still has twisted sectors.  What changes is
their geometric interpretation.

In the sector twisted by \(g_{\rm SS}^k\), with \(1\leq k<N\), the
full worldsheet fields obey the corresponding \(g_R^k\) boundary
condition.  Along the translated circle,
\begin{equation}
y(\sigma+2\pi)
=
y(\sigma)
+
2\pi NR
\left(
w+\frac{k}{N}
\right),
\qquad
w\in\mathbb Z.
\label{eq:q23-fractional-winding-bc}
\end{equation}

The winding is fractional when measured relative to the covering
circle.  On the quotient circle the same string winds \(Nw+k\) times,
so the sector label \(k\) selects a congruence class of ordinary
quotient windings.

The state is not localized on a fixed locus of the full
Scherk--Schwarz element.  It can still be localized in internal
directions fixed by \(g_R^k\), but it necessarily extends or winds
along the translated circle.

The winding contribution to its mass scales as
\begin{equation}
M_{\rm wind}
\sim
\frac{NR}{\alpha'}
\left|
w+\frac{k}{N}
\right|
=
\frac{R}{\alpha'}
\left|
Nw+k
\right|,
\label{eq:q23-winding-mass}
\end{equation}
up to oscillator, zero-point and additional lattice contributions.
For non-trivial \(k\), the minimum of \(|Nw+k|\) is non-zero, so these
sectors are generally driven to infinite mass as \(R\to\infty\).

Freeness therefore changes the nature of the twisted sectors, not
their existence.  Their non-trivial information is carried by winding
along the translated direction rather than by localization at a fixed
locus.

\narrativeheading{Why the twist-dependent ultraviolet part is
non-local}

Question~\ref{sec:ss-local-divergences} showed that local ultraviolet
terms are associated with group elements that leave some point fixed.
The freely acting construction provides a direct geometric
realization of that principle.

For \(1\leq k<N\), the shortest displacement between a point and its
image under \(\delta_N^k\) is
\begin{equation}
\ell_k
=
2\pi R\,
\min(k,N-k)
>
0.
\label{eq:q23-minimal-displacement}
\end{equation}

In a target-space heat-kernel representation, the corresponding image
contributes at short proper time through
\begin{equation}
\exp
\left(
-\frac{\ell_k^2}{4t}
\right),
\label{eq:q23-heat-kernel-suppression}
\end{equation}
and is exponentially suppressed as \(t\to0\).

The identity sector still contains the ordinary local ultraviolet
structure of the parent theory.  What is absent is a local divergent
coefficient depending on a non-trivial freely acting
Scherk--Schwarz image.

After Poisson resummation, the same statement appears as the fact that
the twist dependence is carried by non-zero winding sectors.  The
ultraviolet softness of the Scherk--Schwarz-dependent contribution is a direct consequence of the finite displacement.

\narrativeheading{Large-radius restoration at fixed holonomy}

For a fixed sector \(q\), the holonomy
\(\rho_q(1)=e^{-2\pi iq/N}\) does not vary with the radius.  The
physical gap does:
\begin{equation}
\Delta_{\rm SS}^{(q)}
=
\frac{1}{R}
\operatorname{dist}
\left(
\frac{q}{N},
\mathbb Z
\right).
\label{eq:q23-physical-gap}
\end{equation}
At fixed \(q\) and \(N\), it vanishes as \(R\to\infty\) even though
the microscopic holonomy remains non-trivial.

Different parts of the string spectrum behave differently in this
limit.  The ordinary Kaluza--Klein tower becomes light,
\(m_n^{\rm KK}\sim |n|/R\), while the non-trivial fractionally wound
sectors of Eq.~\eqref{eq:q23-winding-mass} are generally driven to
infinite mass.

There is no requirement that every finite-radius state approach a
finite-mass state of the supersymmetric parent.  What matters is that,
at any fixed physical energy scale, the spectrum and amplitudes of the
states that remain accessible approach those of the supersymmetric
higher-dimensional background.

If \(R\) is a dynamical modulus, this large-radius family provides the
standard spontaneous Scherk--Schwarz interpretation adopted in this
review.  If \(R\) is only an external parameter, the same equations
describe a family of backgrounds without by themselves establishing
spontaneous breaking.

\narrativeheading{Scope and limitations}

The elementary construction used a geometric circle and a cyclic
group, but neither restriction is fundamental.

In an asymmetric orbifold, left- and right-moving worldsheet fields
can transform differently.  A classical quotient geometry may then be
unavailable, while the shift, its twisted sectors and its modular
completion remain well defined in the CFT.  Whenever such a model has
a large-radius field-theory limit, the same equivariance mechanism can
still produce a Scherk--Schwarz spectrum.

For an orbifold with several generators, freeness must be checked for
the complete group rather than only for a generating set.  Two
generators can each contain a non-trivial translation while their
product cancels the translations and leaves a pure internal action
with fixed loci.

Freeness also says nothing by itself about perturbative stability.  It
removes fixed-point localization associated with the freely acting
subgroup, but the complete string mass formula still contains
winding, oscillator, zero-point and internal-lattice contributions.
Tachyons can therefore appear in other regions of moduli space,
particularly when the translated circle becomes small.

These qualifications identify precisely what follows from free action
and what must still be determined from the complete worldsheet
spectrum.

\narrativeheading{Bibliographical remarks}

Early string realizations of Scherk--Schwarz breaking through
anti-periodic, coordinate-dependent or freely acting constructions can
be found in
Refs.~\cite{Rohm:1983aq,Kounnas:1988ye,Ferrara:1988jx,
Kounnas:1989dk,Antoniadis:1990ew}.

A charge-lattice treatment of coordinate-dependent breaking in
heterotic orbifolds, including momentum and winding sectors and the
interplay with Wilson-line backgrounds, can be found in
Ref.~\cite{Benakli:1995ut}.

Freely acting orbifolds with open-string descendants were studied in
Ref.~\cite{Antoniadis:1999ux}.  In these constructions the orbifold
twists are accompanied by shifts acting on momentum or winding states,
and the resulting brane sectors can preserve different fractions of
the supersymmetry of the parent theory.  They provide explicit
open-string realizations of the distinction between the global
freely acting quotient and the supersymmetry seen by individual
localized sectors.

The flat-holonomy description used above is the quotient-space
reformulation of the same construction: an internal phase on the
covering space becomes the holonomy of a flat local system on the
smooth quotient.  This description makes particularly transparent the
distinction between the topological triviality of the underlying line
bundle and the non-trivial global information carried by its flat
connection.
\newpage
\begin{takehome}
\begin{takehomeitems}

\item
A freely acting Scherk--Schwarz orbifold combines an internal action
with a translation.  On a covering circle of radius \(NR\),
\(g_{\rm SS}=g_R\delta_N\), and every non-trivial element of the
elementary cyclic Scherk--Schwarz subgroup moves the circle
coordinate.

\item
The quotient circle has radius \(R\) and is smooth, but the internal
action survives as a flat holonomy.  A component with
\(g_R\)-eigenvalue \(e^{2\pi iq/N}\) has
\[
p_{y,m}
=
\frac{m-q/N}{R}.
\]
The same spectrum follows exactly from the covering-space orbifold
projection.

\item
For \(g_{\rm SS}=(-1)^{F_s}\delta_2\), bosons have integer momenta and
fermions half-integer momenta, reproducing the half-twisted
Scherk--Schwarz spectrum.

\item
Free action does not remove twisted sectors.  In the
\(g_{\rm SS}^k\)-twisted sector the string carries fractional winding
on the covering circle; these sectors are not localized at a fixed
point and are generally driven to large mass as \(R\to\infty\).

\item
A non-trivial freely acting image has a finite displacement, so its
short-proper-time heat kernel is exponentially suppressed.  The
Scherk--Schwarz-dependent ultraviolet contribution is 
non-local.

\item
At large radius the holonomy remains fixed while the physical
Scherk--Schwarz gap vanishes as \(1/R\).  The Kaluza--Klein tower
descends, whereas non-trivial fractionally wound sectors can leave the
finite-energy spectrum.

\item
For asymmetric or multi-generator orbifolds, freeness must be checked
for the complete group action.  Free action alone does not guarantee
supersymmetry breaking or perturbative stability.

\end{takehomeitems}
\end{takehome}

\nextquestion
{How does Scherk--Schwarz breaking act on D-branes and open strings?}

\end{reviewquestion}
% Question 25 v3, renumbered from Question24v2.tex.
% Original section/equation labels retained for cross-reference stability.
%%%%%%%%%%%%%%%%%%%%%%%%%%%%%%%%%%%%%%%%%%%%%%%%%%%%%%%%%%%%%%%

\begin{reviewquestion}
{How does Scherk--Schwarz breaking act on D-branes and open strings?}
\label{sec:ss-dbranes-open-strings}

\begin{essentialidea}

The main low-energy mechanisms discussed in this question are already
visible in higher-dimensional supergravity with localized sectors.
In the present review, the string-theory discussion is therefore used
to show how these ingredients can arise from D-branes and orientifolds
and which additional microscopic consistency conditions accompany
them, rather than to survey the many possible string constructions of
supersymmetry breaking.

The supergravity description of
Question~\ref{sec:ss-localized-boundary-terms} already provides a useful
starting point.  For the simple interval problem, localized gravitino
masses rotate the allowed supersymmetry subspaces at the boundaries.
Writing
\[
\Theta_b=\arctan M_b ,
\]
the gravitino spectrum depends on the effective phase
\begin{equation}
\omega_{\rm eff}
=
\omega
+
\frac{\Theta_0+\Theta_\pi}{\pi}
\pmod1 .
\label{eq:q24-effective-phase}
\end{equation}
The bulk Scherk--Schwarz twist and the localized sectors can 
contribute to the same spectral mismatch without having the same
microscopic origin.

The limit \(M_b\to\infty\) gives
\[
\Theta_b\to\frac{\pi}{2},
\qquad
\frac{\Theta_b}{\pi}\to\frac12 ,
\]
so a maximally rotated local projector produces a half-unit
contribution to the spectral shift.  This is a useful
effective-supergravity description of projector misalignment, not an
identification of a D-brane with a localized gravitino mass.

More generally, several localized sectors can preserve different
local supersymmetries.  The supersymmetry problem is then one of
compatibility between bulk transport and the local projectors.  String
theory realizes such sectors microscopically by D-branes and
orientifold planes and adds data that are absent from the supergravity
description: brane images and orbits, Chan--Paton actions, RR and
NS--NS charges, and the consistency of annulus, M\"obius and
Klein-bottle amplitudes.

A D-brane can consequently be locally supersymmetric at tree level
while the complete compactification admits no exact global
supercharge.  Conversely, the brane itself can break part of the local
parent supersymmetry and realize it non-linearly through a brane
Goldstino.  These are distinct mechanisms.

Open strings also need not feel the Scherk--Schwarz deformation
directly at tree level.  The relevant question is whether their
boundary conditions allow them to carry the compact quantum number on
which the Scherk--Schwarz action operates.  When no direct tree-level
splitting is present, finite bulk propagation can nevertheless detect
the global monodromy and transmit supersymmetry breaking radiatively
to the brane.

\end{essentialidea}

\narrativeheading{From localized masses to local projector rotations}

Question~\ref{sec:ss-localized-boundary-terms} showed that a localized
fermionic bilinear can rotate the subspace of fields allowed at a
boundary.  The supersymmetry parameter obeys the corresponding
matching condition, so the same rotation acts on the locally allowed
supersymmetry subspace.

For the flat \(SU(2)_R\) interval example, the signed gravitino
eigenvalues can be written
\begin{equation}
m_n^{\rm eig}
=
\frac1R
\left[
n+\omega+\frac{\Theta_0+\Theta_\pi}{\pi}
\right],
\qquad
\Theta_b=\arctan M_b .
\label{eq:q24-gravitino-spectrum}
\end{equation}
Only the total relative rotation enters this spectrum.

This does not mean that all microscopic realizations of the same
\(\omega_{\rm eff}\) are equivalent.  A genuine localized
supersymmetry-breaking sector can contain fermions that transform
inhomogeneously and provide additional Goldstino directions.  The
gravitino spectrum alone does not reconstruct these local degrees of
freedom, their interactions or their ultraviolet counterterms.

The saturated limit
\begin{equation}
M_b\longrightarrow\infty,
\qquad
\Theta_b\longrightarrow\frac{\pi}{2},
\qquad
\frac{\Theta_b}{\pi}\longrightarrow\frac12
\label{eq:q24-saturated-rotation}
\end{equation}
is nevertheless instructive.  It shows explicitly how a maximal
local projector rotation produces the same half-unit spectral
displacement that appeared in the pure Scherk--Schwarz benchmark of
Question~\ref{sec:ss-half-twist}.

\narrativeheading{Localized sectors need not sit only at the endpoints}

The interval language can be generalized to localized sectors placed
at arbitrary positions in the bulk.  Let \(y_n\) denote a sequence of
localized branes or defects, and let \(\theta_n\) characterize the
local supersymmetry projector selected at the \(n\)-th position.

Between two neighboring localized sectors, the bulk transports the
supersymmetry parameter, while crossing a localized mass produces an
additional rotation.  In the simple commuting \(SU(2)_R\) example,
the condition for a common Killing spinor can be expressed as
\begin{equation}
\theta_{n+1}-\theta_n
+
\arctan M_n
+
\arctan M_{n+1}
=
k\pi ,
\qquad
k\in\mathbb Z ,
\label{eq:q24-suspended-branes}
\end{equation}
up to the sign conventions used for the localized rotations.

The significance of this equation is more general than the particular
parametrization.  Several localized sectors do not impose independent
supersymmetry conditions.  Their projectors must be compared after
transport through the bulk.

Thus the basic supergravity problem already has the structure that
will reappear for D-branes:
\[
\text{bulk transport}
\quad+\quad
\text{local projector data}
\quad\longrightarrow\quad
\text{global supersymmetry}.
\]

\narrativeheading{What the microscopic D-brane description adds}

In string theory, the localized sectors above can be realized by
D-branes and orientifold planes.  The effective quantities \(M_b\) and
the local projector angles summarize only part of the microscopic
information.

A complete open-string construction must also specify how the
Scherk--Schwarz quotient acts on the D-branes themselves.  A brane can
be invariant under the quotient, or it can be mapped to another brane,
in which case the complete orbit of images must be included.  The
orbifold action must also be represented on the Chan--Paton labels of
open-string endpoints.

When orientifold planes are present, further data include the
orientifold projection and the RR and NS--NS charges carried by the
localized objects.  The one-loop vacuum amplitude has the schematic
form
\begin{equation}
\mathcal Z_{\rm 1\text{-}loop}
=
\mathcal T+\mathcal K+\mathcal A+\mathcal M ,
\label{eq:q24-open-one-loop}
\end{equation}
where the four terms denote the torus, Klein-bottle, annulus and
M\"obius-strip contributions.  Their transverse-channel factorization
and the associated tadpole conditions constrain which brane images,
Chan--Paton representations and charge assignments are allowed.

These requirements are genuinely microscopic.  They restrict which
localized supergravity boundary data admit a consistent string
completion.  They do not, by themselves, define a new low-energy
mechanism of Scherk--Schwarz supersymmetry breaking.

\narrativeheading{What does the open-string spectrum feel at tree level?}

Consider a Scherk--Schwarz circle \(y\sim y+2\pi R\).  An open string
with Neumann boundary conditions along this direction carries
Kaluza--Klein momentum.  If the Scherk--Schwarz action couples to a
charge \(q\), its momentum can take the shifted form
\begin{equation}
p_y
=
\frac{n+\omega q}{R},
\qquad
n\in\mathbb Z .
\label{eq:q24-open-momentum}
\end{equation}
The resulting boson--fermion splitting is then present already at tree
level.

For Dirichlet boundary conditions, the same open string does not carry
this Kaluza--Klein momentum tower.  A brane localized in the
Scherk--Schwarz direction can therefore possess a massless open-string
spectrum that is supersymmetric at tree level even when the full
compactification breaks supersymmetry.

This should not be turned into a universal ``parallel versus
transverse'' rule.  A Dirichlet sector can still carry stretched-string
or image-separation data, and after T-duality momentum can become
winding or boundary-position information.

The invariant statement is that a direct tree-level
Scherk--Schwarz splitting requires the open-string boundary conditions
to admit a compact quantum number on which the Scherk--Schwarz action
acts.

\narrativeheading{The brane orbit is part of the open-string data}

Let \(B_a\) denote a D-brane boundary condition.  The
Scherk--Schwarz generator acts not only on closed-string states but
also on this boundary condition,
\[
g_{\rm SS}: B_a\longmapsto B_{a'} .
\]

If \(B_a\) is invariant, the Scherk--Schwarz group acts directly on
the corresponding open-string Hilbert space, including its
Chan--Paton indices.  If \(B_a\) is moved, the open-string theory must
contain the complete orbit
\[
B_a,\quad
g_{\rm SS}B_a,\quad
g_{\rm SS}^2B_a,\quad\ldots .
\]

Strings can then stretch between different members of this orbit, and
their masses depend on the relative position, Wilson lines and the
Scherk--Schwarz action.

The same closed-string background can admit inequivalent
open-string completions.  The closed-string monodromy does not, by
itself, determine the brane orbit or the Chan--Paton embedding.

\narrativeheading{Local supersymmetry is not a global supercharge}

The distinction between local and global supersymmetry becomes
particularly transparent in the presence of localized branes.

Let \(\mathcal S_y\) denote the space of supersymmetry parameters
preserved locally by the bulk equations near \(y\).  After choosing a
reference trivialization,
\begin{equation}
\mathcal S_y
=
U(y,0)\mathcal S_0 .
\label{eq:q24-local-susy-fibre}
\end{equation}

A globally defined supercharge must return to itself after transport
around the compact direction.  Thus
\[
\mathcal S_{\rm global}
=
\operatorname{Fix}(U_{\rm SS}) .
\]

A D-brane localized at \(y_b\) imposes an additional local projector.
Let \(\mathcal S_B\) denote its preserved subspace, and
\(\mathcal S_O\) the subspace selected by an orientifold projection
when one is present.  The supersymmetry respected by the local
tree-level brane action is
\begin{equation}
\mathcal S_B^{\rm tree,loc}(y_b)
=
\mathcal S_{y_b}
\cap
\mathcal S_B
\cap
\mathcal S_O .
\label{eq:q24-local-tree-susy}
\end{equation}

This is only a local statement.  To determine exact supersymmetry, all
subspaces must first be expressed in the same local trivialization and
the global monodromy condition must also be imposed.

One can therefore have
\begin{equation}
\mathcal S_B^{\rm tree,loc}(y_b)\neq\{0\},
\qquad
\mathcal S_B^{\rm exact}=\{0\}.
\label{eq:q24-local-not-global}
\end{equation}

The brane then has supersymmetric local couplings at tree level, but
there is no globally conserved supercharge protecting its complete
quantum effective action.

\narrativeheading{Several branes require transported intersections}

For two localized stacks at \(y_a\) and \(y_b\), the supersymmetry
conditions cannot be intersected before transporting them to the same
fibre.

If \(\mathcal S_a^{\rm tree,loc}\) and
\(\mathcal S_b^{\rm tree,loc}\) denote the respective local
supersymmetry subspaces, then, after transporting the second to
\(y_a\), the supersymmetry compatible with both endpoints is
\begin{equation}
\mathcal S_{ab}^{\rm tree}
=
\mathcal S_a^{\rm tree,loc}
\cap
U(y_a,y_b)\,
\mathcal S_b^{\rm tree,loc}.
\label{eq:q24-transported-intersection}
\end{equation}

When both projectors are already expressed in the same trivialization,
this reduces to an ordinary intersection.

It is therefore possible for the \(aa\) and \(bb\) sectors to look
supersymmetric separately while the strings stretching between the
two stacks do not.  Their relative projector orientation, separation,
Wilson lines and Scherk--Schwarz transport all enter the same
compatibility problem.

\narrativeheading{Non-linear supersymmetry is a different local statement}

A second distinction is needed.  A D-brane can itself break part of
the supersymmetry available in its local parent background.

The supersymmetries broken specifically by the brane projector are
then realized non-linearly on its worldvolume.  The corresponding
Goldstino has an inhomogeneous transformation of the form
\begin{equation}
\delta\lambda_G
=
f_G\,\epsilon_{\rm broken}
+\cdots .
\label{eq:q24-brane-goldstino}
\end{equation}

This is different from a brane whose local projector preserves the
available local supersymmetry while the Scherk--Schwarz monodromy
removes all global supercharges.

The two situations can be summarized as follows.  If the brane
projector itself eliminates a locally available supersymmetry, the
brane carries the corresponding non-linear Goldstino realization.  If
the projector is locally compatible with that supersymmetry but
\(\operatorname{Fix}(U_{\rm SS})=\{0\}\), the brane need not carry an
independent Goldstino associated with the global Scherk--Schwarz
obstruction.

String constructions in which these two mechanisms coexist provide a
useful illustration: a local brane sector can realize one
supersymmetry non-linearly while an independent
Scherk--Schwarz deformation removes the globally defined
supersymmetries of the bulk.

\narrativeheading{Goldstinos and pseudo-Goldstinos}

When several localized sectors break supersymmetry independently,
each sector can contain a local would-be Goldstino.  The complete
super-Higgs mechanism is nevertheless controlled by the broken local
supersymmetries of the full compactification, not by the separate
counting performed one sector at a time.

Only the combinations aligned with the true Goldstino directions are
absorbed by the massive gravitini.  Orthogonal combinations can remain
as physical spin-\(\tfrac12\) states, conventionally called
pseudo-Goldstinos.

This is another reason why the spectral parameter
\(\omega_{\rm eff}\) does not characterize the full localized
supersymmetry-breaking system.  Two configurations with the same
gravitino spectrum can have different Goldstino and pseudo-Goldstino
content.

\narrativeheading{The half-shift and T-duality}

The standard half-shift provides a simple check of the preceding
statements.  Consider
\[
g_{\rm SS}
=
(-1)^{F_s}\delta_{1/2},
\]
where \(F_s\) denotes spacetime fermion number.

For a brane wrapping the Scherk--Schwarz circle, open strings carry
the shifted momentum directly, so the fermionic tower differs from
the bosonic one by half a Kaluza--Klein unit.

After T-duality along the circle, Neumann and Dirichlet boundary
conditions are exchanged and momentum data are converted into winding
or boundary-position data.  The same half-shift can then appear as a
half-period displacement or exchange of brane images.

The physical criterion is therefore not whether a brane is called
parallel or transverse in one geometric frame.  It is whether the
boundary conditions and brane orbit allow the open sector to carry the
data on which the Scherk--Schwarz action operates.

\narrativeheading{Why a locally supersymmetric brane can be broken radiatively}

A brane localized at \(y_b\) can satisfy
\[
\mathcal S_B^{\rm tree,loc}(y_b)\neq\{0\}
\]
even when the compactification has no exact global supersymmetry.

An ultraviolet process confined to an arbitrarily small neighbourhood
of the brane cannot detect the global Scherk--Schwarz monodromy.  Its
local divergent counterterms are constrained by the local
symmetry structure described in
Question~\ref{sec:ss-local-divergences}.

Finite bulk propagation is different.  A bulk line can leave the
brane, propagate through regions in which the locally preserved
supersymmetry has rotated, and return.  Such a diagram can probe the
full transport around the compact direction and therefore the
monodromy \(U_{\rm SS}\).

When no global supercharge survives, finite non-local diagrams can
then generate supersymmetry-breaking operators on a brane that was
supersymmetric at tree level:
\begin{equation}
m_{\rm soft}^{\rm tree}=0,
\qquad
m_{\rm soft}^{\rm loop}\neq0 .
\label{eq:q24-radiative-brane-breaking}
\end{equation}

Whether a particular operator is actually generated still depends on
the local symmetries and on the existence of the required couplings.
The mechanism is the microscopic counterpart of the boundary
mediation discussed in
Question~\ref{sec:ss-boundary-mediation}.

\narrativeheading{Scherk--Schwarz breaking and intrinsic brane breaking}

Open and unoriented string constructions can contain supersymmetry
breaking that is independent of the Scherk--Schwarz monodromy.

A brane--orientifold configuration can fail to possess a common local
supersymmetry projector already before the Scherk--Schwarz
identification is imposed.  This is an intrinsic local source of
supersymmetry breaking.  It must be distinguished from a brane that is
locally supersymmetric at tree level and receives breaking only
through the global Scherk--Schwarz background.

The two mechanisms can coexist.  Schematically, a low-energy
observable may receive contributions of the form
\begin{equation}
m_{\rm soft}
=
m_{\rm SS}^{\rm tree}
+
m_{\rm SS}^{\rm loop}
+
m_{\rm brane}^{\rm local}
+\cdots .
\label{eq:q24-soft-decomposition}
\end{equation}

The separation is physical because the terms need not share the same
Goldstino structure, ultraviolet behaviour or restoration limit.

\narrativeheading{A finite-energy restoration test}

A splitting generated purely by the Scherk--Schwarz mechanism must
vanish in a Scherk--Schwarz restoration limit.  The converse is not
true: an independent local contribution may happen to vanish along
the same trajectory.  A splitting that remains finite therefore
signals an additional source of supersymmetry breaking.

This statement concerns the finite-energy theory.  Open-string sectors
that become infinitely heavy along the restoration trajectory need not
survive in the limiting spectrum.  Tadpole cancellation and
perturbative stability are separate consistency conditions of the
complete string background.

\narrativeheading{Bibliographical remarks}

Explicit open-string realizations of Scherk--Schwarz supersymmetry
breaking were studied in
Refs.~\cite{Antoniadis:1998ki,Antoniadis:1998ep}.
These constructions exhibit the different response of open-string
sectors to a bulk Scherk--Schwarz deformation: sectors carrying the
shifted compact momentum can be split already at tree level, whereas
localized sectors can remain supersymmetric at tree level and feel
the breaking only through the global compactification.  Related
boundary and orientifold realizations, including supersymmetric
open-string sectors embedded in non-supersymmetric Scherk--Schwarz
backgrounds, can be found in
Refs.~\cite{Fabinger:2000jd,Bossard:2024mls}.

A distinct mechanism, in which supersymmetry is broken intrinsically
by the brane--orientifold configuration rather than by the
Scherk--Schwarz monodromy, is discussed in
Ref.~\cite{Antoniadis:1999xk}.  We use this distinction here to
separate local brane supersymmetry breaking, with its Goldstino
realization, from global Scherk--Schwarz breaking that can be
transmitted to a locally supersymmetric brane.

The purpose of the present discussion is not to review the broader
class of non-supersymmetric D-brane and orientifold constructions.
Rather, these examples identify which parts of the
localized-supergravity description survive in string theory. A
consistent open-string completion also requires brane orbits,
Chan--Paton actions, orientifold projections and tadpole conditions.

\begin{takehome}
\begin{takehomeitems}

\item
Localized gravitino masses provide an effective-supergravity measure
of local projector rotation.  In the simple interval problem,
\[
\omega_{\rm eff}
=
\omega
+
\frac{\arctan M_0+\arctan M_\pi}{\pi}
\pmod1 ,
\]
while \(M_b\to\infty\) gives a half-unit contribution to the spectral
shift.

\item
Several localized sectors define a global compatibility problem:
their supersymmetry projectors must be compared after bulk transport.
The same structure applies to branes placed inside the interval, not
only at its endpoints.

\item
D-branes and orientifold planes give microscopic realizations of such
localized sectors, but add independent data such as brane orbits,
Chan--Paton embeddings, charges and tadpole constraints.  These data
restrict the supergravity description rather than defining a new
low-energy Scherk--Schwarz mechanism.

\item
A direct tree-level open-string Scherk--Schwarz splitting occurs only
when the boundary conditions allow the open sector to carry a compact
quantum number on which the Scherk--Schwarz transformation acts.
Wrapped versus transverse is therefore an example, not an invariant
criterion.

\item
A brane can preserve supersymmetry locally at tree level while no
global supercharge survives.  This must be distinguished from a brane
that itself breaks the locally available supersymmetry and realizes it
non-linearly through a brane Goldstino.

\item
With several localized breaking sectors, only the true Goldstino
combinations are absorbed by the gravitini; orthogonal combinations
can remain as pseudo-Goldstinos.

\item
A locally supersymmetric brane is not protected from global
Scherk--Schwarz breaking at the quantum level.  Finite bulk
propagation can probe the monodromy and generate radiative
supersymmetry-breaking operators even when the tree-level brane
spectrum is supersymmetric.

\item
Purely Scherk--Schwarz-induced effects vanish in the restoration
limit.  A finite-energy splitting that survives this limit proves that
an additional source of supersymmetry breaking is present, while
vanishing alone does not prove that the source was purely
Scherk--Schwarz.

\end{takehomeitems}
\end{takehome}

\nextquestion
{What is the relation between spin structures, \( (-1)^F \), and
\(R\)-symmetry twists?}

\end{reviewquestion}
% Question 26 v3, renumbered from Question25v2.tex.
% Original section/equation labels retained for cross-reference stability.
%%%%%%%%%%%%%%%%%%%%%%%%%%%%%%%%%%%%%%%%%%%%%%%%%%%%%%%%%%%%%%%

\begin{reviewquestion}
{What is the relation between spin structures, spacetime fermion
parity, and \texorpdfstring{\(R\)}{R}-symmetry twists?}
\label{sec:ss-spin-fermion-parity-R}

\begin{essentialidea}

A minus sign acquired by a fermion around a compact circle can have
different origins.  It may follow from the choice of spin structure,
from a holonomy of spacetime fermion parity \((-1)^{F_s}\), or from an
\(R\)-symmetry transformation that acts as \(-1\) on the
supercharges.

These possibilities can give the same gravitino boundary condition
without defining the same compactification.  The comparison must be
made on the complete set of physical states.

There is, however, one exact spacetime relation.  Spin--statistics
implies
\begin{equation}
(-1)^{F_s}=(-1)^{2J}
\label{eq:q25-spin-statistics}
\end{equation}
on every physical spacetime state.  Changing the spin structure by a
minus sign can therefore be compensated by inserting
\((-1)^{F_s}\).

A central \(R\)-symmetry element is different.  It can act in the
same way as fermion parity on the supercharges while acting
differently on matter fields.  A five-dimensional hypermultiplet will
give an explicit example.

The distinction becomes even sharper in string theory.  Two
operations that are indistinguishable on the light fields of an
effective theory can still act differently on massive string states,
momentum or winding sectors, twisted sectors or vertex operators.

\end{essentialidea}

\narrativeheading{Three ways of producing the same fermionic sign}

Consider a circle \(y\sim y+2\pi R\).  A spacetime spinor may satisfy
\begin{equation}
\psi(y+2\pi R)=-\psi(y).
\label{eq:q25-antiperiodic-spinor}
\end{equation}

The sign can originate directly from the spin structure.  Alternatively,
one may choose the periodic spin structure and accompany one circuit
of the circle by an internal transformation that acts as \(-1\) on
the spinor.  Spacetime fermion parity \((-1)^{F_s}\) provides a
universal example.  A central element of an \(R\)-symmetry group can
also act as \(-1\) on the gravitino or supercharges.

Agreement on this one field is not sufficient to identify the three
constructions.  To decide whether two descriptions define the same
background, their relative action must be tested on all physical
representations.

\narrativeheading{Spin structure and spacetime fermion parity}

A spin structure specifies the global gluing of local spinors.  Once
a reference spin structure has been chosen on a spin manifold \(X\),
the remaining choices form a torsor for
\(H^1(X,\mathbb Z_2)\).  On a circle,
\[
H^1(S^1,\mathbb Z_2)\simeq\mathbb Z_2,
\]
so there are periodic and anti-periodic choices.

Writing their holonomy as
\(\sigma_{\rm spin}=\pm1\), a spinor obeys
\begin{equation}
\psi(y+2\pi R)
=
\sigma_{\rm spin}\psi(y).
\label{eq:q25-spin-holonomy}
\end{equation}
Its Kaluza--Klein momenta are
\begin{equation}
p_y
=
\frac{n+\nu_{\rm spin}}{R},
\qquad
\nu_{\rm spin}
=
\begin{cases}
0,&\sigma_{\rm spin}=+1,\\[1mm]
\tfrac12,&\sigma_{\rm spin}=-1.
\end{cases}
\label{eq:q25-spin-spectrum}
\end{equation}

The spin-structure holonomy should not be confused with the central
element \(-\mathbf1_{\rm Spin}\) of the spin group.  On a state of
spacetime spin \(J\),
\[
\rho_J(-\mathbf1_{\rm Spin})=(-1)^{2J}.
\]
Spacetime fermion parity is independently defined as the grading of
the physical Hilbert space, \(+1\) on bosons and \(-1\) on fermions.
Equation~\eqref{eq:q25-spin-statistics} says that these two central
actions coincide on physical spacetime states.

It follows that an anti-periodic spin structure with no internal
twist and a periodic spin structure accompanied by a
\((-1)^{F_s}\) holonomy have the same action on the complete
spacetime field theory.  Their relative transformation
\[
\left(-\mathbf1_{\rm Spin},(-1)^{F_s}\right)
\]
acts trivially on every physical spacetime state.

\narrativeheading{Relation to finite temperature}

Antiperiodic spacetime fermions also occur in finite-temperature
theory.  The relation is not accidental: coordinate-dependent
compactifications were used in string theory to formulate and analyse
thermal backgrounds.  The same compactification technology can
therefore appear in both problems.

The physical interpretation is nevertheless different.  At
temperature \(T=\beta^{-1}\), Euclidean time is compact with period
\(\beta\), and the ordinary thermal partition function is
\begin{equation}
Z(\beta)
=
\operatorname{Tr}e^{-\beta H}.
\label{eq:q25-thermal-trace}
\end{equation}
Its Euclidean path integral has periodic bosons and anti-periodic
spacetime fermions.  Their half-integer frequencies are Matsubara
frequencies.

For a spatial Scherk--Schwarz compactification, the compact coordinate
is instead a spatial direction of the vacuum theory.  The shifted
momentum along it contributes to the physical mass spectrum of the
lower-dimensional Lorentzian theory.  Thus the same boundary
condition
\[
\psi(y+2\pi R)=-\psi(y)
\]
can occur in both settings without making the two physical problems
identical.

The operator formulation also fixes the role of fermion parity.
Inserting \((-1)^{F_s}\) in a thermal trace gives
\begin{equation}
\operatorname{Tr}
\left[
(-1)^{F_s}e^{-\beta H}
\right],
\label{eq:q25-graded-thermal-trace}
\end{equation}
whose path-integral representation has periodic spacetime fermions
around the Euclidean time circle.  A fermion-parity holonomy on a
spatial compactification and a fermion-parity insertion in a thermal
trace therefore use closely related boundary-condition technology but
compute different physical quantities.

\narrativeheading{Why a central \(R\)-twist can be different}

A central \(R\)-symmetry element need not agree with fermion parity on
every field.  A five-dimensional hypermultiplet gives a simple
counterexample.

The bulk \(R\)-symmetry is \(SU(2)_R\).  The supercharges \(Q^i\)
form a doublet.  The hypermultiplet contains a scalar doublet
\(\phi^i\) and an \(SU(2)_R\)-singlet fermion \(\psi\).

Let
\[
z_R=-\mathbf1_{SU(2)_R}.
\]
The central spin rotation acts as \(-1\) on \(Q^i\) and \(\psi\), and
as \(+1\) on \(\phi^i\).  The central \(SU(2)_R\) element acts as
\(-1\) on \(Q^i\) and \(\phi^i\), but as \(+1\) on \(\psi\).
Their product
\begin{equation}
\kappa_R
=
\left(
-\mathbf1_{\rm Spin},
-\mathbf1_{SU(2)_R}
\right)
\label{eq:q25-hyper-relative}
\end{equation}
therefore acts as
\begin{equation}
Q^i\mapsto Q^i,
\qquad
\phi^i\mapsto-\phi^i,
\qquad
\psi\mapsto-\psi.
\label{eq:q25-hyper-action}
\end{equation}

The supercharges cannot distinguish this relative transformation, but
the hypermultiplet can.

The same difference is visible in the spectrum.  With periodic spin
structure and
\[
U_R=e^{2\pi i\omega T_R},
\]
where the scalar doublet has weights \(\pm1\),
\begin{equation}
m_{\phi_\pm,n}
=
\frac{|n\pm\omega|}{R},
\qquad
m_{\psi,n}
=
\frac{|n|}{R}.
\label{eq:q25-hyper-R-spectrum}
\end{equation}
At \(\omega=\tfrac12\), the scalars are half-shifted while the
fermion is not.  With anti-periodic spin structure and no
\(R\)-twist, the fermion is half-shifted while the scalars remain
periodic.

The two compactifications therefore agree on the supercharges but
have different matter spectra.

\narrativeheading{Two different notions of triviality}

The hypermultiplet example shows that there are two distinct
questions one can ask about a relative central transformation.

First, does it act trivially on the parent supercharges?  Second, does
it act trivially on the complete physical theory?

To formulate this precisely, let \(\widehat G_R\) denote the
fermionic internal symmetry group relevant for the comparison.  It
contains the \(R\)-symmetry group \(G_R\) and, when necessary, the
central grading \((-1)^{F_s}\).  Fermion parity need not belong to
the connected \(R\)-symmetry group.

Consider
\begin{equation}
\widetilde G
=
\operatorname{Spin}_{\rm loc}
\times
\widehat G_R .
\label{eq:q25-covering-group}
\end{equation}
For central pairs
\(\kappa=(z_{\rm spin},z_R)\), define
\begin{equation}
\Gamma_Q
=
\left\{
\kappa\in Z(\widetilde G)
\;\middle|\;
\rho_Q(\kappa)=\mathbf1
\text{ on all parent supercharges}
\right\},
\label{eq:q25-Gamma-Q}
\end{equation}
and
\begin{equation}
\Gamma_{\rm triv}
=
\left\{
\kappa\in Z(\widetilde G)
\;\middle|\;
\kappa
\text{ acts trivially on the complete physical theory}
\right\}.
\label{eq:q25-Gamma-triv}
\end{equation}

Since the supercharges probe only part of the representation content,
\begin{equation}
\Gamma_{\rm triv}
\subseteq
\Gamma_Q.
\label{eq:q25-kernel-inclusion}
\end{equation}

For the hypermultiplet,
\[
\kappa_R\in\Gamma_Q,
\qquad
\kappa_R\notin\Gamma_{\rm triv}.
\]
Thus a relative transformation in
\(\Gamma_Q\setminus\Gamma_{\rm triv}\) preserves exactly the same
supercharges while changing the physical compactification.

\narrativeheading{The faithfully acting combined symmetry}

A central transformation belonging to \(\Gamma_{\rm triv}\) is
invisible on the complete theory.  Descriptions differing only by
such an element represent the same physical holonomy.

The faithfully acting combined group is therefore
\begin{equation}
G_{\rm comb}
=
\frac{
\operatorname{Spin}_{\rm loc}\times\widehat G_R
}{
\Gamma_{\rm triv}
}.
\label{eq:q25-combined-group}
\end{equation}

A physical compactification holonomy takes values in \(G_{\rm comb}\).
Writing it separately as a spin-structure contribution and an
internal contribution amounts to choosing a lift to
\(\widetilde G\).

This explains why the decomposition into ``spin structure'' and
``internal twist'' need not itself be physical.  Only the resulting
element of the faithfully acting group is intrinsic.

\narrativeheading{A practical test for a half-\(R\) twist}

Let \(U_{1/2}\in Z(G_R)\) be a central order-two element used as a
half-\(R\) twist.  To compare it with a spin-structure change, consider
its relative action on a physical representation \(\Phi\):
\begin{equation}
\mathscr D_{1/2}^{(\Phi)}
=
(-1)^{2J_\Phi}
\rho_R^\Phi(U_{1/2}).
\label{eq:q25-half-mismatch}
\end{equation}

If
\(\mathscr D_{1/2}^{(Q)}=\mathbf1\) for every supercharge, the two
descriptions preserve the same supersymmetry.  They describe the same
physical compactification only if
\(\mathscr D_{1/2}^{(\Phi)}=\mathbf1\) for every physical
representation.

For integer-normalized \(R\)-charges,
\[
\rho_R^\Phi(U_{1/2})=(-1)^{q_\Phi},
\]
so complete equivalence requires
\begin{equation}
q_\Phi
\equiv
F_s(\Phi)
\pmod2
\qquad
\text{for every physical representation }\Phi.
\label{eq:q25-half-charge-condition}
\end{equation}

The equality of an \(R\)-half-twist with fermion parity is a
property of the complete representation content of the parent theory,
not a consequence of their common action on the gravitino.

\narrativeheading{A global qualification: lifting the combined holonomy}

On a single circle the preceding comparison can be made directly.
With several non-contractible loops, one must also ask whether a
combined physical holonomy can be decomposed globally into separate
spin and internal pieces.

This is a lifting problem.  Equation~\eqref{eq:q25-combined-group}
defines the central extension
\begin{equation}
1
\longrightarrow
\Gamma_{\rm triv}
\longrightarrow
\widetilde G
\longrightarrow
G_{\rm comb}
\longrightarrow
1.
\label{eq:q25-central-extension}
\end{equation}
Let
\begin{equation}
\varrho:
\pi_1(X)
\longrightarrow
G_{\rm comb}
\label{eq:q25-combined-holonomy}
\end{equation}
be a flat combined holonomy.  Pulling back the extension gives an
obstruction
\begin{equation}
\varrho^*[\widetilde G]
\in
H^2\!\left(
\pi_1(X),\Gamma_{\rm triv}
\right).
\label{eq:q25-lift-obstruction}
\end{equation}
A global lift to
\(\operatorname{Spin}_{\rm loc}\times\widehat G_R\) exists when this
class vanishes.

For \(X=S^1\), \(\pi_1(S^1)=\mathbb Z\) has cohomological dimension
one, so this obstruction vanishes.

When a lift exists, it need not be unique.  Choose a reference spin
structure.  A split flat description can then be written as
\begin{equation}
\alpha\in H^1(X,\mathbb Z_2),
\qquad
h_R:
\pi_1(X)\longrightarrow\widehat G_R,
\label{eq:q25-split-data}
\end{equation}
where \(\alpha\) measures the spin structure relative to the reference
and \(h_R\) is the internal flat holonomy.

A second lift can differ by a homomorphism
\begin{equation}
c:
\pi_1(X)
\longrightarrow
\Gamma_{\rm triv}.
\label{eq:q25-kernel-map}
\end{equation}
Writing
\[
c(\gamma)
=
\bigl(
c_{\rm spin}(\gamma),
c_R(\gamma)
\bigr),
\]
the change of lift acts as
\begin{align}
\alpha
&\longmapsto
\alpha+[c_{\rm spin}],
\label{eq:q25-spin-lift-shift}
\\
h_R(\gamma)
&\longmapsto
c_R(\gamma)\,h_R(\gamma).
\label{eq:q25-internal-lift-shift}
\end{align}
Since \(c_R(\gamma)\) is central, the transformed \(h_R\) remains a
homomorphism.

Thus the ambiguity of a lift is measured by
\[
\operatorname{Hom}
\left(
\pi_1(X),\Gamma_{\rm triv}
\right).
\]

This still concerns ordinary holonomies around one-cycles.  The next
question asks whether the Scherk--Schwarz idea itself has a meaningful
analogue for higher-dimensional cycles.

\narrativeheading{Microscopic string equivalence is stronger}

The meaning of ``acts trivially'' depends on which theory is being
tested.

In a low-energy effective theory, one tests only the fields and
operators retained in the EFT.  Denote the corresponding kernel by
\(\Gamma_{\rm triv}^{\rm EFT}\).

In a perturbative string background, the same transformation must be
trivial on the complete physical string spectrum and operator
algebra.  Denote this stronger kernel by
\(\Gamma_{\rm triv}^{\rm CFT}\).  For the same candidate central
symmetry,
\begin{equation}
\Gamma_{\rm triv}^{\rm CFT}
\subseteq
\Gamma_{\rm triv}^{\rm EFT}.
\label{eq:q25-CFT-EFT-kernels}
\end{equation}

The inclusion can be strict.  A transformation invisible on all light
spacetime fields can still act on massive oscillator states, momentum
or winding sectors, twisted sectors, vertex operators or their
cocycles.

For example, the boundary condition
\[
\psi(y+2\pi R)=-\psi(y)
\]
specifies only the monodromy of one spacetime field.  A freely acting
string construction such as
\begin{equation}
g_{\rm SS}
=
(-1)^{F_s}\delta_{1/2}
\label{eq:q25-string-half-shift}
\end{equation}
also specifies the orbifold projection, momentum and winding sectors,
twisted sectors and their modular completion.

Operators such as \((-1)^{F_L}\) and \((-1)^{F_R}\) require the same
care.  They act separately on left- and right-moving worldsheet
sectors and should not be identified automatically with total
spacetime fermion parity.

\narrativeheading{Three levels of comparison}

The preceding discussion gives three increasingly strong tests.

Two descriptions may first be compared only on the parent
supercharges.  Equality at this level determines whether they preserve
the same supersymmetry.

One may next compare their action on every physical state retained in
the effective field theory.  This determines whether they define the
same EFT compactification.

Finally, in a string embedding, one must compare their action on the
complete microscopic spectrum and operator algebra.  Equality at this
level is stronger than either of the previous two.

Thus microscopic equivalence implies EFT equivalence, and EFT
equivalence implies the same preserved supersymmetry.  Neither
converse holds in general.

\narrativeheading{Does this tell us whether supersymmetry is broken
spontaneously?}

Not by itself.  Knowing the monodromy of the supercharges, or even
establishing that two descriptions are microscopically equivalent,
does not determine how the missing supersymmetries are realized
dynamically.

The relevant question is whether a given supercharge remains a
symmetry of the physical theory.  If it does but the background does
not preserve it, the symmetry can be spontaneously broken and must be
realized accordingly, through a Goldstino or, in supergravity, the
super-Higgs mechanism.  A supercharge removed by an orbifold or
orientifold projection is instead absent from the symmetry algebra of
the projected theory and is not a spontaneously broken generator of
that theory.

The D-brane discussion of
Question~\ref{sec:ss-dbranes-open-strings} gives a concrete example.
An isolated D-brane is a solitonic object of the parent theory: it
preserves part of the bulk supersymmetry and spontaneously breaks the
complementary part, which is realized non-linearly on its worldvolume.
After an orientifold projection, the locally available supersymmetry
is instead determined by the common brane and orientifold projectors.
A supersymmetry that would have been linearly realized on the isolated
brane can therefore be projected out, while another can remain
non-linearly realized with its Goldstino.  Such local projection and
spontaneous breaking can coexist~\cite{Antoniadis:1999xk}.

This local effect is distinct from the global Scherk--Schwarz
monodromy.  A locally allowed supersymmetry must still return to itself
after transport around the compact space.  The two mechanisms can remove different parent supercharges in the same
compactification.

This distinction is particularly relevant for the freely acting
fermion-parity construction of
Question~\ref{sec:ss-freely-acting}.  A generator
\[
(-1)^{F_s}\delta
\]
can produce mass splittings that vanish when the translation radius
becomes large.  The existence of this supersymmetric
decompactification limit does not by itself establish that every
supersymmetry absent at finite radius is spontaneously broken in the
microscopic theory.

This situation should be distinguished from the continuous
\(R\)-symmetry realizations discussed in
Question~\ref{sec:ss-wilson-radion}.  There the relevant
supersymmetry transformations remain part of the local gauge
structure, and the Goldstino and super-Higgs descriptions provide the
standard dynamical realization of spontaneous Scherk--Schwarz
breaking.

A restoration limit, the existence of a Goldstino for a particular
broken generator, and the statement that all missing parent
supercharges are spontaneously broken symmetries of one microscopic
theory are therefore distinct statements.

\narrativeheading{Bibliographical remarks}

Periodic and anti-periodic boundary conditions, together with internal
transformations acting on supersymmetry generators, occur already in
the original Scherk--Schwarz constructions
\cite{Scherk:1978ta,Scherk:1979zr}.  Their implementation in string
theory and the relation between coordinate-dependent compactifications
and fermionic boundary conditions were developed further in early
string constructions
\cite{Rohm:1983aq,Kounnas:1989dk,Antoniadis:1990ew}.

Coordinate-dependent compactification techniques were also used in
the study of finite-temperature string theory.  This historical
connection underlies the close formal relation between thermal
fermionic boundary conditions and Scherk--Schwarz shifts discussed
above, while the thermal and spatial compactifications retain
different physical interpretations.

Open-string and D-brane realizations make the distinction between
global Scherk--Schwarz breaking and supersymmetry breaking intrinsic
to localized brane/orientifold sectors particularly explicit
\cite{Antoniadis:1998ki,Antoniadis:1998ep,Antoniadis:1999xk}.
More recent constructions also illustrate that the fermionic action
of the twist can involve a non-trivial cover of the microscopic
duality group, and that supersymmetric open-string sectors can coexist
with a non-supersymmetric Scherk--Schwarz closed-string background
\cite{Bossard:2024mls}.  The detailed classification of these
constructions is not needed here; their role is to show how
projections, local supersymmetry realizations and global monodromies
can coexist in the same microscopic theory.

A complementary M-theory example is provided by opposite chirality
conditions at the two \(E_8\) boundaries, which combine into an
anti-periodic gravitino spin structure on the covering circle
\cite{Fabinger:2000jd}.

\begin{takehome}
\begin{takehomeitems}

\item
The same minus sign on a gravitino can arise from an anti-periodic spin
structure, spacetime fermion parity or a central \(R\)-symmetry
element.  Agreement on the gravitino alone does not establish physical
equivalence.

\item
Spin--statistics gives
\[
(-1)^{F_s}=(-1)^{2J}
\]
on physical spacetime states.  An anti-periodic spin structure and a
periodic spin structure accompanied by a \((-1)^{F_s}\) holonomy are
therefore equivalent at the level of the complete spacetime field
theory.

\item
A central \(R\)-element can agree with fermion parity on the
supercharges while acting differently on matter.  The five-dimensional
hypermultiplet provides an explicit example.

\item
\(\Gamma_Q\) contains central transformations invisible to all
supercharges, while \(\Gamma_{\rm triv}\) contains those invisible to
the complete physical theory.  Hence
\[
\Gamma_{\rm triv}\subseteq\Gamma_Q.
\]

\item
For a central half-\(R\) twist, complete equivalence with fermion
parity requires
\[
q_\Phi\equiv F_s(\Phi)\pmod2
\]
for every physical representation, not only for the gravitino.

\item
The physical combined holonomy takes values in
\[
G_{\rm comb}
=
\frac{
\operatorname{Spin}_{\rm loc}\times\widehat G_R
}{
\Gamma_{\rm triv}
}.
\]
Separate spin and internal holonomies are lifts of this combined datum,
and on a general compact space such a lift can have a genuine global
obstruction.

\item
Microscopic string equivalence is stronger than EFT equivalence:
\[
\Gamma_{\rm triv}^{\rm CFT}
\subseteq
\Gamma_{\rm triv}^{\rm EFT}.
\]
Massive string states, winding sectors and twisted sectors can
distinguish backgrounds that are identical in a low-energy
truncation.

\item
Finite-temperature compactification and spatial Scherk--Schwarz
compactification can use closely related fermionic boundary
conditions, but the former computes a thermal trace whereas the latter
defines the mass spectrum of a lower-dimensional Lorentzian theory.

\item
A supersymmetric restoration limit does not by itself prove
spontaneous breaking.  A supercharge projected out of the physical
theory is not a spontaneously broken generator, whereas a supercharge
that remains a symmetry and is broken dynamically has the associated
Goldstino or super-Higgs realization.

\end{takehomeitems}
\end{takehome}

\nextquestion
{Can Scherk--Schwarz data be associated with \(p\)-cycles for
\(p>1\)?}

\end{reviewquestion}
% Question 27 v3, renumbered from Question26v2.tex.
% Original section/equation labels retained for cross-reference stability.
%%%%%%%%%%%%%%%%%%%%%%%%%%%%%%%%%%%%%%%%%%%%%%%%%%%%%%%%%%%%%%%

\begin{reviewquestion}
{Can Scherk--Schwarz data be associated with \texorpdfstring{\(p\)}{p}-cycles for
\texorpdfstring{\(p>1\)}{p>1}?}
\label{sec:ss-higher-cycles}

\begin{essentialidea}

An ordinary Scherk--Schwarz twist is associated with a closed loop.
A flat one-form connection produces a holonomy around that loop, and
supersymmetry is broken only if the resulting transformation acts
non-trivially on the supercharges.

Higher-dimensional cycles admit analogous global data.  A two-form
connection can have a holonomy on a closed surface, and more generally
a \(p\)-form connection can have higher holonomy on a closed
\(p\)-cycle.  The physical probes are then extended objects: strings
for two-form connections, membranes for three-form connections, and
so on.

This higher holonomy is not, by itself, an ordinary
Scherk--Schwarz twist.  A \(p\)-form connection with \(p>1\) is
associated with a higher-form symmetry, whose charged operators are
extended rather than local.  For an ordinary pure higher-form
symmetry, the local supercurrent is neutral, so a non-trivial higher
holonomy need not rotate the supercharges.

Dimensional reduction can change the situation.  If the
compactification contains an internal \((p-1)\)-cycle
\(C_{p-1}\), a component of the \(p\)-form connection can reduce to an
ordinary one-form connection,
\[
A_1^{(C)}
=
\int_{C_{p-1}}B_p .
\]
Its Wilson line can then become Scherk--Schwarz data, but only if the
descendant holonomy is non-trivial and acts non-trivially on the
supercharges.

Thus higher holonomy, the existence of a one-form descendant, and
Scherk--Schwarz supersymmetry breaking are three different
statements.

\end{essentialidea}

\narrativeheading{Why the familiar construction lives on a loop}

The usual Scherk--Schwarz construction starts with an ordinary
\(0\)-form symmetry \(G\).  Its charged states are pointlike, and a
background gauge field for the symmetry is an ordinary one-form
connection.

On a circle,
\begin{equation}
\Phi(y+2\pi R)
=
\rho_\Phi(U)\Phi(y),
\qquad
U\in G,
\label{eq:q26-ordinary-SS}
\end{equation}
where \(\rho_\Phi\) denotes the representation carried by the field.

For a flat Abelian connection, \(U\) is its Wilson holonomy.  For a
non-Abelian connection,
\begin{equation}
U_\gamma
=
P\exp\!\left(
i\oint_\gamma A_1
\right).
\label{eq:q26-ordinary-holonomy}
\end{equation}

More generally, a flat \(G\)-bundle on a compact space \(X\) is
described by a homomorphism
\[
\rho:
\pi_1(X)\longrightarrow G
\]
modulo conjugation.

The appearance of \(\pi_1(X)\) reflects the degree of the gauge
connection: a one-form is integrated along a one-dimensional path.
There is no direct expression
\[
\int_{\Sigma_p}A_1
\]
for a closed \(p\)-cycle with \(p>1\).  To define a holonomy directly
on such a cycle, the gauge datum must itself have degree \(p\).

\narrativeheading{A two-cycle example}

The case \(p=2\) already contains the essential idea.

Let \(B_2\) be a two-form \(U(1)\) connection and
\(\Sigma_2\) a closed two-cycle.  A string whose Euclidean worldsheet
wraps \(\Sigma_2\) acquires the phase
\begin{equation}
\operatorname{Hol}_{B_2}(\Sigma_2)
=
\exp\!\left(
2\pi i q\int_{\Sigma_2}B_2
\right),
\label{eq:q26-Wilson-surface}
\end{equation}
when a globally defined representative \(B_2\) is available.  This
observable is usually called a Wilson surface.

Suppose that
\[
\Sigma_2
=
\gamma_1\times C_1,
\]
where \(C_1\) is an internal circle and \(\gamma_1\) a loop in the
remaining space.  Define
\begin{equation}
A_1^{(C)}
=
\int_{C_1}B_2 .
\label{eq:q26-B2-descendant}
\end{equation}
Then
\begin{equation}
\oint_{\gamma_1}A_1^{(C)}
=
\int_{\gamma_1\times C_1}B_2 .
\label{eq:q26-B2-factorization}
\end{equation}

A string wrapped on \(C_1\) becomes a pointlike particle after
dimensional reduction.  The Wilson surface seen by the wrapped string
therefore becomes an ordinary Wilson line seen by the lower-dimensional
particle.

This explains how a higher-cycle datum can descend to the familiar
one-cycle language.  It does not yet imply Scherk--Schwarz breaking:
the resulting ordinary holonomy must still act on the supercharges.

\narrativeheading{From extended probes to higher-form symmetries}

The preceding example changes both the gauge field and the object
that detects it.

A particle traces a worldline and couples to a one-form connection.
A string sweeps out a worldsheet and couples to a two-form connection.
More generally, a \(p\)-form connection \(B_p\) couples electrically
to an extended object of spatial dimension \(p-1\), whose worldvolume
has dimension \(p\).

The corresponding symmetry is called a
\emph{\((p-1)\)-form global symmetry}.  More generally, a
\(q\)-form global symmetry has \(q\)-dimensional charged operators on
a spatial slice and a background gauge connection of degree \(q+1\).

The first few cases are
\begin{equation}
\begin{array}{c|c|c|c}
\text{symmetry}
&
\text{background connection}
&
\text{holonomy}
&
\text{electric probe}
\\
\hline
0\text{-form}
&
A_1
&
\text{loop}
&
\text{particle worldline}
\\[1mm]
1\text{-form}
&
B_2
&
\text{surface}
&
\text{string worldsheet}
\\[1mm]
2\text{-form}
&
C_3
&
\text{three-cycle}
&
\text{membrane worldvolume}
\\[1mm]
\vdots
&
\vdots
&
\vdots
&
\vdots
\\[1mm]
(p-1)\text{-form}
&
B_p
&
p\text{-cycle}
&
(p-1)\text{-brane worldvolume}.
\end{array}
\label{eq:q26-degree-table}
\end{equation}

For \(p=2\) one often says \emph{Wilson surface}; for general degree
we shall use the broader term \emph{higher holonomy}.

We restrict here to Abelian higher connections.  This is the natural
setting for ordinary higher-form global symmetries in local quantum
field theory and already contains the structures needed below.

\narrativeheading{Why a differential form is not the complete global
datum}

A local expression such as
\[
\exp\!\left(
2\pi i\int_{\Sigma_p}B_p
\right)
\]
does not capture the full global structure, because a single
globally defined \(p\)-form potential need not exist.

For \(p=1\), local gauge potentials patch into a line bundle with
connection.  For \(p=2\), the corresponding global object is a
bundle gerbe with connection.  Higher values of \(p\) are described
uniformly by differential cohomology.

We use the convention
\begin{equation}
\check B
\in
\widehat H^{p+1}(X,\mathbb Z)
\label{eq:q26-differential-class}
\end{equation}
for a \(p\)-form \(U(1)\) gauge connection on \(X\).

Two pieces of information will be especially important for us: its
curvature
\[
H_{p+1}
=
\operatorname{curv}(\check B)
\]
and its holonomy on closed integral \(p\)-cycles,
\[
\operatorname{Hol}_{\check B}(\Sigma_p)
\in U(1).
\]

In the language of Cheeger--Simons differential characters, the
holonomy is a homomorphism on the group \(Z_p(X)\) of closed integral
\(p\)-cycles satisfying
\begin{equation}
\operatorname{Hol}_{\check B}
\left(
\partial C_{p+1}
\right)
=
\exp\!\left[
2\pi i
\int_{C_{p+1}}
H_{p+1}
\right].
\label{eq:q26-differential-character}
\end{equation}

The curvature is a closed form with integral periods.  The relevant
exact sequence is
\begin{equation}
0
\longrightarrow
H^p(X,U(1))
\longrightarrow
\widehat H^{p+1}(X,\mathbb Z)
\overset{\operatorname{curv}}{\longrightarrow}
\Omega_{\mathbb Z}^{p+1}(X)
\longrightarrow
0,
\label{eq:q26-differential-exact-sequence}
\end{equation}
where \(\Omega_{\mathbb Z}^{p+1}(X)\) denotes closed
\((p+1)\)-forms with integral periods.

Its kernel is the flat sector:
\begin{equation}
\widehat H_{\rm flat}^{p+1}(X,\mathbb Z)
\simeq
H^p(X,U(1)).
\label{eq:q26-flat-sector}
\end{equation}

For a flat class, the holonomy depends only on the homology class
\([\Sigma_p]\in H_p(X,\mathbb Z)\).  Since \(U(1)\) is divisible,
the \(\operatorname{Ext}\) term in the universal-coefficient theorem
vanishes, and
\begin{equation}
H^p(X,U(1))
\simeq
\operatorname{Hom}
\left(
H_p(X,\mathbb Z),
U(1)
\right).
\label{eq:q26-flat-character-group}
\end{equation}

A flat higher background is therefore characterized by its holonomy
character
\begin{equation}
\chi_{\check B}:
H_p(X,\mathbb Z)
\longrightarrow
U(1).
\label{eq:q26-higher-character}
\end{equation}

This is the higher-degree analogue of the character defined by an
ordinary flat Wilson line.

\narrativeheading{What physically detects the higher character}

The character \(\chi_{\check B}\) need not be visible to pointlike
fields.

An electrically charged \((p-1)\)-brane sweeps out a
\(p\)-dimensional worldvolume \(W_p\) and acquires
\begin{equation}
\exp\!\left(
2\pi i q
\int_{W_p}B_p
\right).
\label{eq:q26-extended-probe-phase}
\end{equation}
For a globally non-trivial connection, this expression means
\(\operatorname{Hol}_{\check B}(W_p)^q\).

A flat higher background can therefore produce a non-trivial phase for
wrapped strings or branes even when every local point-particle field
has ordinary periodic boundary conditions.  Its physical effect need
not appear as a shifted Kaluza--Klein momentum of a local field.

\narrativeheading{Why a pure higher-form symmetry does not twist the
supercharges}

An ordinary Scherk--Schwarz transformation must act on the
supersymmetry generators.

Let \(S^\mu_\alpha(x)\) denote the local supercurrent and
\begin{equation}
Q_\alpha
=
\int_{\Sigma_{D-1}}
d\Sigma_\mu\,
S^\mu_\alpha
\label{eq:q26-supercharge}
\end{equation}
the corresponding supercharge.

A conventional \(R\)-symmetry is a \(0\)-form symmetry.  It acts on
local operators and can rotate both the supercurrent and the
supercharges.  This is the microscopic condition emphasized in
Question~\ref{sec:ss-string-charge-lattice}: a Scherk--Schwarz
symmetry must act on the sector from which the spacetime supercharges
are constructed.

For an ordinary pure \(q\)-form global symmetry with \(q>0\), local
operators are neutral; the charged operators are extended.  In
particular, the local supercurrent is neutral.  In the absence of
boundary contributions, anomalies or higher-group mixing that changes
the charge algebra,
\begin{equation}
\rho_Q\!\left(G^{(q)}\right)
=
\mathbf1,
\qquad
q>0.
\label{eq:q26-higher-Q-neutrality}
\end{equation}

A flat higher-form holonomy can be physically non-trivial
without producing a monodromy of the supercharges.

The qualification \emph{pure} is essential.  Boundaries, anomalies,
higher-group structures, duality actions or other mixings can relate
ordinary and higher-form symmetries.  In such cases one must analyse
the full charge algebra rather than infer the action on the
supercharges from form degree alone.

Under a straightforward dimensional reduction, the lower-dimensional
supercharges descend from local supercharges of the parent theory.
A decoupled pure higher-form symmetry that was invisible to the parent
supercurrent will therefore normally remain invisible to them.
A non-trivial action after reduction requires additional mixing,
gauging or an equivalent mechanism.

As in Question~\ref{sec:ss-spin-fermion-parity-R}, the final question
is always how the induced holonomy acts on the supercharges.

\narrativeheading{Flat higher holonomy is not flux}

The direct higher analogue of a flat Wilson line has
\begin{equation}
\operatorname{curv}(\check B)
=
H_{p+1}
=
0,
\qquad
\operatorname{Hol}_{\check B}(\Sigma_p)
\neq1
\label{eq:q26-flat-higher-holonomy}
\end{equation}
for at least one closed \(p\)-cycle.

A flux background instead has \(H_{p+1}\neq0\).

The difference is local.  A non-zero field strength can enter Bianchi
identities, carry stress energy, source other fields and contribute
directly to fermionic supersymmetry variations.  A flat higher
holonomy has no such local field-strength contribution; it is detected
through global phases of extended probes.

Flux compactifications can break supersymmetry, but that mechanism is
not the flat-holonomy mechanism considered here.

\narrativeheading{How dimensional reduction lowers the form degree}

Consider a product compactification
\[
X=Y\times K
\]
and an oriented closed cycle \(C_{p-1}\subset K\).

A component of \(B_p\) with one leg along \(Y\) and \(p-1\) legs
along \(C_{p-1}\) gives an ordinary one-form connection on \(Y\),
\begin{equation}
A_1^{(C)}
=
\int_{C_{p-1}}B_p.
\label{eq:q26-reduced-one-form}
\end{equation}

For a closed loop \(\gamma_1\subset Y\),
\begin{equation}
\oint_{\gamma_1}A_1^{(C)}
=
\int_{\gamma_1\times C_{p-1}}B_p.
\label{eq:q26-factorized-integral}
\end{equation}
Hence
\begin{equation}
\operatorname{Hol}_{A^{(C)}}(\gamma_1)
=
\operatorname{Hol}_{\check B}
\left(
\gamma_1\times C_{p-1}
\right).
\label{eq:q26-holonomy-reduction}
\end{equation}

An electrically charged \((p-1)\)-brane wrapped on \(C_{p-1}\)
becomes a pointlike particle in \(Y\).  Its higher holonomy becomes an
ordinary Wilson-line phase.

Dimensional reduction has therefore converted a higher-form
background into a possible ordinary \(0\)-form background.  Whether
that ordinary symmetry acts on the supercharges remains a separate
question.

\narrativeheading{Fiber integration, transgression and the slant
product}

The preceding product-space calculation has a global formulation.

Suppose that \(X\) is the total space of a fibration
\[
\pi:X\longrightarrow Y
\]
with compact oriented fibres of dimension \(p-1\).
Differential cohomology admits integration over the fibre,
\begin{equation}
\pi_!:
\widehat H^{p+1}(X,\mathbb Z)
\longrightarrow
\widehat H^2(Y,\mathbb Z).
\label{eq:q26-fiber-integration}
\end{equation}
The degree drops by \(p-1\), and a class in
\(\widehat H^2(Y,\mathbb Z)\) is an ordinary \(U(1)\) line bundle
with connection.

Fiber integration is one standard degree-lowering operation in
differential cohomology.  Closely related constructions are often
described as transgression.  When one works with a fixed homology
class rather than a global fibration, the corresponding topological
operation is the slant product.

For a fixed \((p-1)\)-cycle \(C_{p-1}\), the flat descendant can be
written
\begin{equation}
a_C
=
[\check B]_{\rm flat}/[C_{p-1}]
\in
H^1(Y,U(1)).
\label{eq:q26-slant-descendant}
\end{equation}
We shall also denote this degree-lowering operation by
\begin{equation}
a_C
=
\tau_C
\left(
[\check B]_{\rm flat}
\right).
\label{eq:q26-flat-transgression}
\end{equation}

In the product case, the corresponding characters satisfy
\begin{equation}
\chi_C([\gamma_1])
=
\chi_{\check B}
\left(
[\gamma_1]\times[C_{p-1}]
\right).
\label{eq:q26-transgressed-character}
\end{equation}

These operations provide the global version of the local expression
\(A_1^{(C)}=\int_C B_p\).

\narrativeheading{When does the descendant become Scherk--Schwarz?}

Let
\[
a_C
=
\tau_C
\left(
[\check B]_{\rm flat}
\right)
\in
H^1(Y,U(1))
\]
be the one-form descendant associated with an internal cycle
\(C_{p-1}\).

If \(a_C=0\), this particular reduction does not produce an ordinary
Wilson line.  If \(a_C\neq0\), its holonomy around a loop
\(\gamma_1\subset Y\) is
\begin{equation}
U_C(\gamma_1)
=
\operatorname{Hol}_{a_C}(\gamma_1).
\label{eq:q26-descendant-holonomy}
\end{equation}

The descendant is Scherk--Schwarz data only if this ordinary holonomy
acts non-trivially on at least one parent supercharge,
\begin{equation}
\rho_Q
\left(
U_C(\gamma_1)
\right)
\neq
\mathbf1 .
\label{eq:q26-SS-test}
\end{equation}

Any spin, orbifold, boundary or other internal holonomies already
present must be included in the full monodromy before deciding which
supercharges survive.

The complete sequence is
\begin{equation}
[\check B]_{\rm flat}
\;\xrightarrow{\ \tau_C\ }\;
a_C
\;\xrightarrow{\ \operatorname{Hol}_{\gamma_1}\ }\;
U_C(\gamma_1)
\;\xrightarrow{\ \rho_Q\ }\;
\rho_Q\!\left(U_C(\gamma_1)\right).
\label{eq:q26-master-chain}
\end{equation}

A non-trivial higher character, a non-trivial one-form descendant and
a non-trivial action on the supercharges are distinct
conditions.

\narrativeheading{Torsion cycles and discrete higher holonomy}

Differential cohomology becomes especially important in the presence
of torsion cycles.

Consider the lens space \(L(N,1)\), for which
\[
H_1\!\left(
L(N,1),\mathbb Z
\right)
\simeq
\mathbb Z_N .
\]
Let \(\gamma\) generate this torsion group.  For \(p>1\), take
\[
X
=
L(N,1)\times S^{p-1},
\qquad
\Sigma_p
=
\gamma\times S^{p-1}.
\]
The class \([\Sigma_p]\) has order \(N\),
\[
N[\Sigma_p]=0.
\]

De~Rham cohomology does not detect this torsion class, but a flat
differential character does.  Since
\[
\chi_{\check B}:
H_p(X,\mathbb Z)
\longrightarrow
U(1)
\]
is a homomorphism,
\[
\chi_{\check B}\!\left(N[\Sigma_p]\right)=1,
\]
and therefore
\begin{equation}
\chi_{\check B}([\Sigma_p])^N
=
1.
\label{eq:q26-torsion-holonomy}
\end{equation}

The allowed higher holonomies on this torsion cycle are thus discrete
\(N\)-th roots of unity.

This is precisely the kind of information that is lost if one keeps
only the local notation
\(\exp(2\pi i\int_{\Sigma_p}B_p)\).

\narrativeheading{A genuinely higher datum can remain physically
non-trivial}

A flat higher class may be non-trivial even when every one-form
descendant relevant to a chosen reduction vanishes.

Its character \(\chi_{\check B}\) can still be detected by wrapped
strings, branes or torsion cycles and can affect topological phases or
non-perturbative amplitudes.  It simply does not produce an ordinary
Wilson line for that reduction.

Whether a higher datum has one-form descendants is therefore relative
to the compactification and to the internal cycles available for
reduction.

\narrativeheading{The higher-cycle descent criterion}

The preceding construction can be summarized for all relevant
internal cycles at once.

Let \(\mathcal C_{p-1}\) denote the set of oriented internal
\((p-1)\)-cycles or fibres relevant to the reduction.  For every
\(C\in\mathcal C_{p-1}\), compute
\[
a_C
=
\tau_C
\left(
[\check B]_{\rm flat}
\right).
\]

Define
\begin{equation}
\mathfrak T_1
\left(
\check B;\mathcal C_{p-1}
\right)
=
\left\{
(C,a_C)
\;\middle|\;
C\in\mathcal C_{p-1},
\quad
a_C\neq0
\right\},
\label{eq:q26-descendant-set}
\end{equation}
and
\begin{equation}
\begin{aligned}
\mathfrak T_{\rm SS}
\left(
\check B;\mathcal C_{p-1}
\right)
=
\bigl\{
(C,a_C)\in
\mathfrak T_1
\left(
\check B;\mathcal C_{p-1}
\right)
\ \big|\
&\text{there exists }
[\gamma_1]\in H_1(Y,\mathbb Z)
\\[-1mm]
&\text{with }
\rho_Q
\left(
\operatorname{Hol}_{a_C}(\gamma_1)
\right)
\neq
\mathbf1
\bigr\}.
\end{aligned}
\label{eq:q26-SS-descendant-set}
\end{equation}

The first set records the non-trivial one-form descendants available
in the chosen reduction.  The second retains only those whose
holonomy acts non-trivially on at least one parent supercharge.

Accordingly,
\begin{equation}
\mathfrak T_{\rm SS}
\left(
\check B;\mathcal C_{p-1}
\right)
\neq
\varnothing
\label{eq:q26-SS-descent-criterion}
\end{equation}
is the condition that the higher background possess an ordinary
Scherk--Schwarz descendant.

Both sets depend on the reduction data
\(\mathcal C_{p-1}\).  A differential-cohomology class can 
remain genuinely higher for one reduction and acquire an ordinary
Scherk--Schwarz descendant in another.

\narrativeheading{Relation to generalized Scherk--Schwarz reductions}

The phrase ``generalized Scherk--Schwarz reduction'' is also used for
a broader class of constructions.

Group-manifold reductions, geometric fluxes, doubled and exceptional
field theory, and generalized parallelizations can generate constant
lower-dimensional gaugings from coordinate-dependent frames.  These
constructions were discussed in
Question~\ref{sec:ss-gauging}.

They need not be described by flat higher connections: their local
field strengths, generalized torsion or frame algebra can be
non-zero.

The present question concerns the narrower problem of whether a flat
higher holonomy on a \(p\)-cycle can descend to an ordinary holonomy
that acts on the supercharges.  In this review, we reserve the phrase
\emph{higher-cycle Scherk--Schwarz datum} for a higher background
that possesses such a supersymmetry-twisting one-form descendant.

\narrativeheading{Bibliographical remarks}

Higher-form global symmetries and the relation between the degree of
the symmetry, its charged extended operators and its background gauge
field are developed systematically in
Ref.~\cite{Gaiotto:2014kfa}.

For \(p=2\), the global geometry of a two-form connection can be
described in terms of bundle gerbes.  Differential characters and
differential cohomology provide a uniform description in arbitrary
degree, including curvature, higher holonomy and torsion data
\cite{Murray1996,CheegerSimons1985,HopkinsSinger2005}.

Fiber integration, transgression and slant products provide the
standard mathematical language for the degree-lowering operations
used in dimensional reduction.  In the present context they make
precise the passage from a higher holonomy to an ordinary one-form
descendant.

The term ``generalized Scherk--Schwarz reduction'' is also widely
used for reductions involving non-zero frame algebra, flux or
generalized torsion.  Those constructions are broader than the flat
higher-holonomy problem considered here and should not be identified
with it.

\begin{takehome}
\begin{takehomeitems}

\item
Ordinary Scherk--Schwarz twisting is associated with one-cycles
because an ordinary \(0\)-form symmetry has a one-form background
connection.

\item
A \(p\)-form connection can have higher holonomy on a closed
\(p\)-cycle and is naturally associated with a
\((p-1)\)-form symmetry whose charged objects are extended.

\item
The global higher gauge datum is a differential-cohomology class
\[
\check B\in\widehat H^{p+1}(X,\mathbb Z).
\]
For a flat background,
\[
H^p(X,U(1))
\simeq
\operatorname{Hom}
\left(
H_p(X,\mathbb Z),
U(1)
\right),
\]
so its holonomy is naturally a character of \(p\)-cycle homology.

\item
An ordinary pure higher-form symmetry with positive form degree does
not act on local operators and therefore does not directly twist the
local supercurrent or supercharges.

\item
Flat higher holonomy and flux are different.  A flat higher
background has zero local curvature but can remain globally
non-trivial on extended probes or torsion cycles.

\item
Reduction on an internal \((p-1)\)-cycle can produce an ordinary
one-form connection.  Globally, this is described by fibre
integration, transgression or the corresponding slant product.

\item
A one-form descendant becomes Scherk--Schwarz data only if its
holonomy acts non-trivially on the parent supercharges.

\item
Torsion cycles can support discrete higher holonomies that are
invisible to de~Rham cohomology but are detected by differential
characters.

\item
The condition
\[
\mathfrak T_{\rm SS}
\left(
\check B;\mathcal C_{p-1}
\right)
\neq
\varnothing
\]
states that the higher background has at least one genuine ordinary
Scherk--Schwarz descendant in the chosen reduction.

\end{takehomeitems}
\end{takehome}

\nextquestion
{Can a non-trivial Scherk--Schwarz compactification preserve supersymmetry?}

\end{reviewquestion}
% Question 28 v3, renumbered from Question27v2.tex.
% Original section/equation labels retained for cross-reference stability.
%%%%%%%%%%%%%%%%%%%%%%%%%%%%%%%%%%%%%%%%%%%%%%%%%%%%%%%%%%%%%%%

\begin{reviewquestion}
{Can a non-trivial Scherk--Schwarz compactification preserve
supersymmetry?}
\label{sec:ss-without-susy-breaking}

\begin{essentialidea}

Yes.  A non-trivial Scherk--Schwarz gluing need not break
supersymmetry.  What matters is not whether the compactification
holonomy is non-trivial on the fields, but whether it acts
non-trivially on the supercharges.

A simple example is provided by an axion \(a\) with a shift symmetry
\(a\to a+c\).  On a circle of radius \(R\), one may impose
\begin{equation}
a(x,y+2\pi R)
=
a(x,y)+2\pi Rm,
\qquad
a(x,y)=a(x)+my .
\label{eq:q27-axion-ansatz}
\end{equation}
After dimensional reduction, the Kaluza--Klein vector gauges this
shift,
\begin{equation}
D_\mu a
=
\partial_\mu a-mA_\mu .
\label{eq:q27-axion-covariant-derivative}
\end{equation}
The reduction is therefore physically non-trivial and generates a
St\"uckelberg coupling even when the axionic shift acts trivially on
all supercharges.

This establishes that non-trivial gluing, gauging and mass generation
do not by themselves imply Scherk--Schwarz supersymmetry breaking.
Whether a particular background of the resulting gauged theory is
supersymmetric is a further question, determined by its fermionic
Killing-spinor conditions.

More generally, let
\[
\rho:
\pi_1(X)\longrightarrow G_{\rm sym}
\]
be the Scherk--Schwarz holonomy and let \(R_Q\) denote the action of
\(G_{\rm sym}\) on the parent supercharges.  Supersymmetry sees only
the induced holonomy
\[
\rho_Q=R_Q\circ\rho .
\]
The surviving supercharges are the vectors left invariant by every
element of \(\rho_Q(\pi_1(X))\).  A non-trivial compactification can preserve all, some or none of the parent supersymmetry.

\end{essentialidea}

\newpage
\narrativeheading{The axion already separates twisting from
supersymmetry breaking}

Assume that the higher-dimensional theory contains an axion \(a\)
whose classical action is invariant under the continuous shift
\begin{equation}
a\longrightarrow a+c .
\label{eq:q27-axion-shift}
\end{equation}

The ansatz in Eq.~\eqref{eq:q27-axion-ansatz} follows the orbit of this
symmetry as one moves around the compact direction.  The field is not
periodic as an ordinary real-valued function, but its values at
\(y\) and \(y+2\pi R\) are related by a symmetry of the
higher-dimensional theory.

A local reparametrization of the compact coordinate induces the
lower-dimensional transformations
\[
A_\mu\longrightarrow A_\mu-\partial_\mu\lambda,
\qquad
a\longrightarrow a-m\lambda ,
\]
so that
\[
D_\mu a=\partial_\mu a-mA_\mu
\]
is gauge covariant.  The axion kinetic term contains
\begin{equation}
-\frac12
\left(
\partial_\mu a-mA_\mu
\right)^2 .
\label{eq:q27-axion-kinetic-term}
\end{equation}
In unitary gauge the axion is absorbed and the Kaluza--Klein vector
becomes massive.

None of this requires the shift symmetry itself to rotate the
supercharges.  The example therefore shows that a
Scherk--Schwarz reduction can generate a non-trivial gauging and a
mass while its gluing remains invisible to supersymmetry.

It does not by itself prove that the field configuration selected by
the reduction is a supersymmetric vacuum.  That requires the
fermionic supersymmetry variations of the reduced theory to vanish.
We shall return to this distinction below.

\narrativeheading{What supersymmetry sees}

Let \(G_{\rm sym}\) be the exact symmetry group used in the
compactification gluing.  A flat Scherk--Schwarz background on a
compact space \(X\) is described by
\begin{equation}
\rho:
\pi_1(X)
\longrightarrow
G_{\rm sym},
\label{eq:q27-holonomy-representation}
\end{equation}
up to simultaneous conjugation and any further exact identifications
of the theory.

Let \(\mathcal S_Q\) be the vector space spanned by the parent
supercharges, and
\begin{equation}
R_Q:
G_{\rm sym}
\longrightarrow
\operatorname{GL}(\mathcal S_Q)
\label{eq:q27-Q-representation}
\end{equation}
their representation.  The induced holonomy on supersymmetry is
\begin{equation}
\rho_Q
=
R_Q\circ\rho .
\label{eq:q27-induced-Q-holonomy}
\end{equation}

A parent supercharge survives precisely when it returns to itself
around every non-contractible loop.  The preserved subspace is
therefore
\begin{equation}
\mathcal S_{\rm pres}(\rho)
=
\bigcap_{\gamma\in\pi_1(X)}
\ker
\left[
\rho_Q(\gamma)-\mathbf1
\right].
\label{eq:q27-preserved-subspace}
\end{equation}

Spin, orbifold, boundary and other internal transformations acting on
the supersymmetry parameters must be included in the complete
monodromy before this intersection is evaluated.

It is convenient to denote
\begin{equation}
\nu_Q([\rho])
=
\dim\mathcal S_{\rm pres}(\rho).
\label{eq:q27-nuQ}
\end{equation}
Then full, partial and vanishing supersymmetry correspond respectively
to
\[
\nu_Q=\dim\mathcal S_Q,
\qquad
0<\nu_Q<\dim\mathcal S_Q,
\qquad
\nu_Q=0.
\]

For example, if
\[
R_Q(U)
=
\begin{pmatrix}
1&0\\
0&e^{2\pi i\omega}
\end{pmatrix},
\qquad
\omega\notin\mathbb Z,
\]
one of the two supercharges survives.  With several independent
cycles, the surviving space is the common invariant subspace of all
their holonomies.

\narrativeheading{Twists invisible to all supercharges}

To characterize the twists that preserve every parent supercharge,
define
\begin{equation}
K_Q
=
\ker R_Q
=
\left\{
g\in G_{\rm sym}
\;\middle|\;
R_Q(g)=\mathbf1
\right\}.
\label{eq:q27-KQ}
\end{equation}

If
\begin{equation}
\rho\!\left(\pi_1(X)\right)
\subseteq
K_Q,
\label{eq:q27-holonomy-in-KQ}
\end{equation}
then the compactification holonomy is invisible to every parent
supercharge.

The holonomy can nevertheless act non-trivially on other fields or
states.  To separate these possibilities, let \(R_{\rm phys}\) be the
action of \(G_{\rm sym}\) on the complete physical theory and define
\begin{equation}
K_{\rm phys}
=
\ker R_{\rm phys}.
\label{eq:q27-Kphys}
\end{equation}
Since the supercharges belong to the physical theory,
\begin{equation}
K_{\rm phys}
\subseteq
K_Q.
\label{eq:q27-kernel-inclusion}
\end{equation}

An element of \(K_{\rm phys}\) is physically trivial.  By contrast,
an element of
\[
K_Q\setminus K_{\rm phys}
\]
acts non-trivially somewhere in the theory while remaining invisible
to every supercharge.  Holonomies valued in this set provide the
sharpest examples of physically non-trivial
Scherk--Schwarz twists that preserve all parent supersymmetry.

Within the flat-holonomy space
\begin{equation}
\mathcal X(X,G_{\rm sym})
=
\operatorname{Hom}
\left(
\pi_1(X),G_{\rm sym}
\right)
/G_{\rm sym},
\label{eq:q27-character-space}
\end{equation}
the locus preserving all parent supercharges is therefore
\begin{equation}
\mathcal X_{\rm full\,SUSY}(X)
=
\frac{
\operatorname{Hom}
\left(
\pi_1(X),K_Q
\right)
}{
G_{\rm sym}\text{ acting by conjugation}
}.
\label{eq:q27-KQ-character}
\end{equation}

The identity compactification is only one point of this locus whenever
\(K_Q\) contains physically non-trivial transformations.

\narrativeheading{The infinitesimal criterion for continuous
symmetries}

For a continuous symmetry group, the same distinction can be made at
the Lie-algebra level.

Let
\[
\mathfrak g_{\rm sym}
=
\operatorname{Lie}(G_{\rm sym}),
\]
and let
\[
dR_Q:
\mathfrak g_{\rm sym}
\longrightarrow
\operatorname{End}(\mathcal S_Q)
\]
be the infinitesimal action on the supercharges.  Define
\begin{equation}
\mathfrak k_Q
=
\ker dR_Q .
\label{eq:q27-lie-kernel}
\end{equation}

This is the Lie algebra of \(K_Q\).  A generator
\(M\in\mathfrak k_Q\) therefore generates a continuous subgroup whose
action is trivial on every supercharge.

Similarly,
\(\mathfrak k_{\rm phys}=\ker dR_{\rm phys}\) satisfies
\(\mathfrak k_{\rm phys}\subseteq\mathfrak k_Q\).
A generator in
\(\mathfrak k_Q\setminus\mathfrak k_{\rm phys}\) is infinitesimally
non-trivial on the physical theory while remaining invisible to
supersymmetry.

\narrativeheading{From the axion to a general scalar gauging}

The axionic example is the simplest member of a broader class.

Let scalar fields \(\phi^i\) parametrize a target space on which a
generator \(M\in\mathfrak g_{\rm sym}\) acts through a Killing vector
\(k_M^i(\phi)\).  A Scherk--Schwarz ansatz following the symmetry
orbit satisfies
\begin{equation}
\partial_y\phi^i
=
k_M^i(\phi).
\label{eq:q27-flow-equation}
\end{equation}

A local shift of the compact coordinate,
\(y\to y+\lambda(x)\), induces
\[
A_\mu\longrightarrow A_\mu-\partial_\mu\lambda,
\qquad
\delta\phi^i
=
-\lambda\,k_M^i(\phi),
\]
and the Kaluza--Klein vector gauges the target-space isometry:
\begin{equation}
D_\mu\phi^i
=
\partial_\mu\phi^i
-
A_\mu k_M^i(\phi).
\label{eq:q27-general-gauging}
\end{equation}

For the axionic translation, \(k_M^a=m\), reproducing
Eq.~\eqref{eq:q27-axion-covariant-derivative}.

If
\begin{equation}
M\in\mathfrak k_Q,
\qquad
k_M\neq0,
\label{eq:q27-susy-gauging-generator}
\end{equation}
the Scherk--Schwarz generator produces a non-trivial scalar gauging
while remaining invisible to every parent supercharge.

This condition is sufficient for a non-trivial scalar gauging of this
type.  More generally, an element of
\(\mathfrak k_Q\setminus\mathfrak k_{\rm phys}\) can act on tensors,
gauge fields or other sectors even when no scalar Killing vector is
involved.

\narrativeheading{Mass generation is not supersymmetry breaking}

In the axionic example, the vector mass follows from the St\"uckelberg
coupling
\[
D_\mu a=\partial_\mu a-mA_\mu.
\]
It is a consequence of the gauging, not of a non-trivial action of the
twist on the supercharges.

If the reduced theory admits a supersymmetric vacuum, the resulting
massive states organize into complete supermultiplets.  The presence
of a mass, a covariant derivative or a non-trivial gauging therefore
does not diagnose Scherk--Schwarz supersymmetry breaking.

\narrativeheading{Preserving the supercharges is not the same as
having a supersymmetric vacuum}

The condition
\[
\rho\!\left(\pi_1(X)\right)
\subseteq
K_Q
\]
states that the compactification gluing itself removes no parent
supercharge.

The reduced theory may nevertheless contain non-trivial gaugings,
fermion shifts and a scalar potential.  A particular lower-dimensional
background is supersymmetric only if its fermionic supersymmetry
variations vanish,
\begin{equation}
\delta\psi_\mu
=
0,
\qquad
\delta\lambda
=
0,
\qquad
\delta\chi
=
0,
\qquad
\ldots
\label{eq:q27-lower-SUSY-conditions}
\end{equation}
for at least one non-zero supersymmetry parameter.

Thus a Scherk--Schwarz holonomy can preserve all parent supercharges
while a particular vacuum of the resulting gauged theory is
non-supersymmetric.  Conversely, a non-trivial gauging need not
exclude supersymmetric vacua.

In this question, ``the twist preserves supersymmetry'' means that the
compactification holonomy admits invariant parent supercharges.  The
supersymmetry of an actual solution of the reduced theory must be
checked separately.

\narrativeheading{Classical symmetries need not be exact quantum
twists}

The axionic discussion so far assumes a symmetry of the classical
higher-dimensional theory.  Quantum effects can reduce the set of
allowed Scherk--Schwarz transformations.

For example, instantons and charge quantization can reduce a
continuous classical axionic shift to a discrete exact subgroup.  If
\[
a\sim a+2\pi f
\]
is the surviving identification, the monodromy in
Eq.~\eqref{eq:q27-axion-ansatz} must obey
\begin{equation}
2\pi Rm
\in
2\pi f\,\mathbb Z.
\label{eq:q27-axion-quantization}
\end{equation}

More generally, an admissible microscopic twist must belong to the
exact symmetry group of the quantum theory and preserve its charge
lattice and operator algebra.  This is the restriction emphasized in
Questions~\ref{sec:ss-string-charge-lattice}
and~\ref{sec:ss-string-discreteness}.

Quantum consistency can therefore reduce the allowed subset of
\(\mathcal X(X,G_{\rm sym})\), but once an exact holonomy is specified,
its action on supersymmetry is still determined by
\(R_Q\circ\rho\).

\narrativeheading{Bibliographical remarks}

The original Scherk--Schwarz construction already showed that twisted
dimensional reduction can turn symmetries of the higher-dimensional
theory into masses and non-trivial couplings in the reduced theory
\cite{Scherk:1978ta,Scherk:1979zr}.  Generalized reductions leading
to massive and gauged supergravities were developed further, for
example, in Ref.~\cite{Kaloper:1999yr}.

Duality-twisted compactifications provide a broader setting in which
the monodromy can act non-trivially on the physical spectrum while
leaving a non-zero subspace of supercharges invariant; see
Ref.~\cite{Dabholkar:2002sy}.  The resulting amount of preserved
supersymmetry is determined by the invariant subspace of the
monodromy acting on the supercharges, as in
Eq.~\eqref{eq:q27-preserved-subspace}.

More general twisted-torus and flux reductions lead to gauged
supergravities with non-trivial scalar potentials and fermion shifts.
Examples can be found in
Refs.~\cite{DAuria:2005npq,DallAgata:2005zff}.  These constructions
also illustrate the distinction emphasized above between
supersymmetry preserved by the compactification gluing and
supersymmetry of a particular vacuum of the resulting gauged theory.

At the quantum level, the symmetry available for a
Scherk--Schwarz twist must be an exact symmetry of the microscopic
theory.  Continuous symmetries present in a classical effective
description can be reduced to discrete identifications or restricted
by the quantum charge lattice, as discussed in
Questions~\ref{sec:ss-string-charge-lattice}
and~\ref{sec:ss-string-discreteness}.

\begin{takehome}
\begin{takehomeitems}

\item
A non-trivial Scherk--Schwarz holonomy need not break supersymmetry.
For
\[
\rho:\pi_1(X)\to G_{\rm sym},
\]
the surviving supercharges are determined by the induced action
\(\rho_Q=R_Q\circ\rho\).

\item
The preserved supersymmetry is the common invariant subspace
\[
\mathcal S_{\rm pres}(\rho)
=
\bigcap_{\gamma\in\pi_1(X)}
\ker
\left[
\rho_Q(\gamma)-\mathbf1
\right].
\]
It can contain all, some or none of the parent supercharges.

\item
The kernel \(K_Q=\ker R_Q\) contains transformations invisible to all
parent supercharges.  Elements of
\(K_Q\setminus K_{\rm phys}\) are physically non-trivial while
preserving the full parent supersymmetry.

\item
For continuous symmetries, the corresponding infinitesimal condition
is \(M\in\mathfrak k_Q=\ker dR_Q\).  Such a generator can still act
non-trivially on scalars, tensors or other fields.

\item
An axionic Scherk--Schwarz ansatz generates
\[
D_\mu a=\partial_\mu a-mA_\mu
\]
and a St\"uckelberg mass for the Kaluza--Klein vector even when the
shift symmetry acts trivially on the supercharges.

\item
Compatibility of the compactification holonomy with supersymmetry
does not guarantee that every vacuum of the reduced gauged theory is
supersymmetric.  The lower-dimensional fermionic variations must be
solved separately.

\item
The twist must be an exact quantum symmetry.  A continuous classical
shift can be reduced by quantum effects to a discrete subgroup, which
restricts the allowed Scherk--Schwarz monodromies without changing
the invariant-subspace criterion for the supercharges.

\end{takehomeitems}
\end{takehome}

\par\medskip\noindent\textit{The review concludes with:}\par
\noindent\textit{Scherk--Schwarz in a nutshell}\par

\end{reviewquestion}
% Question 29 v1, renumbered from Question28v2.tex.
% Original section/equation labels retained for cross-reference stability.
%%%%%%%%%%%%%%%%%%%%%%%%%%%%%%%%%%%%%%%%%%%%%%%%%%%%%%%%%%%%%%%

\begin{reviewquestion}
{Scherk--Schwarz in a nutshell}
\label{sec:ss-nutshell}

\begin{essentialidea}

Scherk--Schwarz reduction consists in compactifying a theory with a
non-trivial global identification by an admissible symmetry of the
parent theory.

For a flat geometric compactification, this datum can be represented
by a homomorphism
\[
\rho:
\pi_1(X)
\longrightarrow
G_{\rm adm},
\]
where \(G_{\rm adm}\) is the exact symmetry group available for the
compactification.  Orbifolds and string constructions replace this
simple geometric description by the appropriate identification group
and, when necessary, by exact automorphisms of the microscopic theory.

A physical sector in a representation \(R\) sees the induced holonomy
\(R\circ\rho\).  On a circle, if the corresponding eigenphase is
\(e^{2\pi i\alpha^{\rm eff}_{R,a}}\), the associated tower has
\begin{equation}
m_{R,a,n}
=
M_{\rm KK}
\left|
n+\alpha^{\rm eff}_{R,a}
\right|,
\qquad
n\in\mathbb Z.
\label{eq:q28-master-spectrum}
\end{equation}

Supersymmetry is determined by the same construction applied to the
supercharges.  After including the local Killing-spinor conditions,
spin structure, orbifold projections, boundary projectors and any
localized rotations, the preserved supercharges form
\begin{equation}
\mathcal S_{\rm preserved}
=
\bigcap_{\gamma}
\ker
\left[
\mathcal U_Q(\gamma)-\mathbf1
\right].
\label{eq:q28-master-susy}
\end{equation}

The construction therefore separates three pieces of data: the global
identification, its representation on each physical sector, and the
geometric scale that converts the resulting phases into masses.

\end{essentialidea}

\narrativeheading{The twist is global data}

A Scherk--Schwarz twist is not a local supersymmetry-breaking term
added to the parent Lagrangian.  It specifies how fields and states
are identified after transport around a non-contractible direction or
under the corresponding quotient action.

On a circle,
\[
\Phi(x,y+2\pi R)
=
R_\Phi(U)\Phi(x,y),
\]
while on an interval the same global information can be distributed
among bulk transport, endpoint projectors and localized matching
conditions.  In a freely acting orbifold, the translation can
disappear from the quotient geometry while its action survives in the
global boundary condition.

The identification is common to the theory, but the phase extracted
from it is representation dependent.  A single number called a
``Scherk--Schwarz angle'' is therefore sufficient only when the
relevant representations all reduce to the same one-parameter
description.

\narrativeheading{Supersymmetry is determined by global transport}

A gravitino mass is not the definition of Scherk--Schwarz breaking.
The relevant question is whether a local supersymmetry parameter can
be transported consistently through the complete compactification.

Equation~\eqref{eq:q28-master-susy} treats full preservation, partial
preservation and complete breaking in the same way: they correspond
to different dimensions of the common invariant subspace.

A non-trivial twist need not break supersymmetry.  It can act on
axions, tensors, gauge fields or other sectors while remaining
invisible to the supercharges.  Conversely, two compactifications can
have the same action on the supercharges while acting differently on
matter fields.

The same hierarchy persists in string theory.  Agreement on the light
effective fields is weaker than equivalence of the complete
microscopic background, since massive string states, winding sectors,
twisted sectors and operator data can still distinguish the two
constructions.

\narrativeheading{Rigid data, effective phases and physical scales}

The microscopic identification need not coincide with the phase seen
by a particular spectrum.

A rigid twist is part of the definition of the compactification.
The effective phase can receive additional contributions from
continuous Wilson lines, moduli or localized interactions.  For the
interval example discussed in
Question~\ref{sec:ss-localized-boundary-terms},
\begin{equation}
\omega_{\rm eff}
=
\omega_{\rm rigid}
+
\frac1\pi
\left[
\arctan(M_0)+\arctan(M_\pi)
\right]
\pmod1 .
\label{eq:q28-effective-phase-example}
\end{equation}

Different microscopic sources can therefore produce the same
quadratic spectrum without defining the same theory.

For a given supercharge branch, the dimensionful splitting is
\begin{equation}
\Delta_{\rm SS}
=
M_{\rm KK}\,
\operatorname{dist}
\left(
\alpha_Q^{\rm eff},
\mathbb Z
\right).
\label{eq:q28-physical-gap}
\end{equation}
It can vanish because the effective phase approaches an integer,
because the Kaluza--Klein scale tends to zero, or through both effects.
Spectral restoration therefore need not coincide with microscopic
untwisting.

\narrativeheading{Local ultraviolet physics and non-local
Scherk--Schwarz dependence}

The dependence on a pure Scherk--Schwarz twist is global.  A process
confined to an arbitrarily small region cannot determine how fields
are identified after transport around the compact space.

This accounts for the ultraviolet softness of the
twist-dependent part of many loop amplitudes.  In proper-time or
position-space descriptions, a non-trivial image requires propagation
over a finite compact distance and is suppressed in the local
short-distance limit.  After Poisson resummation, the same dependence
is carried by non-zero winding sectors.

This statement concerns the twist-dependent non-local contribution,
not the complete theory.  Local bulk counterterms remain allowed, and
boundaries or fixed loci can support additional local counterterms.
The finite Scherk--Schwarz-dependent remainder must be separated from
these local contributions.

\narrativeheading{Restoration does not determine the microscopic
realization}

At fixed non-trivial holonomy, decompactification can send
\(M_{\rm KK}\) to zero and restore the supersymmetric spectrum while a
Kaluza--Klein tower becomes light.  Alternatively, dynamical
contributions can drive the effective phase toward an integer while
the rigid microscopic twist remains fixed.

A restoration limit does not by itself establish that the breaking is
spontaneous.  In continuous \(R\)-symmetry realizations, a dynamical
modulus together with the Goldstino and super-Higgs mechanism provides
the standard spontaneous interpretation.  For discrete
fermion-parity constructions, the microscopic projection must be
analysed separately because some parent supercharges can be absent
from the projected theory rather than spontaneously broken.

\narrativeheading{What string theory adds}

The low-energy Scherk--Schwarz mechanism can be formulated entirely
in field theory or supergravity.  A string embedding imposes stronger
conditions on which twists actually exist.

The candidate transformation must extend to an exact automorphism of
the complete microscopic theory.  It must act consistently on the
charge lattice, massive states, vertex operators and cocycles; in an
orbifold construction the twisted sectors must satisfy level matching,
modular covariance and factorization.  Open and unoriented sectors add
brane images, Chan--Paton actions, crosscap conditions and tadpole
constraints.

These requirements can quantize or exclude twists that appear
admissible in a low-energy description and can distinguish
microscopic backgrounds with the same leading effective spectrum.

They do not, however, guarantee vacuum stability.  Tachyons, runaway
directions, moduli stabilization and backreaction remain separate
questions.

\narrativeheading{Scherk--Schwarz as a method}

Scherk--Schwarz reduction means compactifying with a non-trivial
symmetry identification.  Whether the identification breaks
supersymmetry is determined separately by its action on the
supercharges.

The same construction can therefore generate masses and gaugings
while preserving all or part of the parent supersymmetry.
Scherk--Schwarz reduction and Scherk--Schwarz supersymmetry breaking
are not synonymous.

\newpage
\narrativeheading{Bibliographical remarks}

The twisted-reduction mechanism originates in
Refs.~\cite{Scherk:1978ta,Scherk:1979zr}.  References for the
orbifold, supergravity, quantum-consistency and string-theory
developments summarized above have been given in the corresponding
questions of this review.

Related developments of Scherk--Schwarz compactification in field
theory, supergravity, orbifold constructions and string theory can be
found, for example, in
Refs.~\cite{Dudas:1997jn,Cotrone:1999xs,Angelantonj:2003hr,Angelantonj:2006ut,Abel:2018zyt,Kounnas:2016gmz}.

\begin{takehome}
\begin{takehomeitems}

\item
The irreducible Scherk--Schwarz datum is an admissible global
identification of the parent theory.

\item
Each physical representation sees the induced holonomy through its
own eigenphases, while the compactification geometry converts those
dimensionless phases into physical mass scales.

\item
Preserved supersymmetry is the common invariant subspace of the
complete transport acting on the supercharges, including spin,
orbifold, boundary and internal contributions.

\item
The rigid microscopic twist, the effective spectral phase and the
physical supersymmetry-breaking gap are different objects.  Equal
spectra need not imply equal microscopic compactifications.

\item
The ultraviolet softness concerns the twist-dependent non-local part
of amplitudes; it does not remove the local bulk or boundary
counterterms allowed by the compactification.

\item
String theory constrains and completes the Scherk--Schwarz data
through its charge lattice, operator algebra, twisted sectors and
worldsheet consistency conditions.  Microscopic consistency does not
by itself imply vacuum stability.

\item
Scherk--Schwarz reduction is a method of twisted compactification.  It
may preserve all, part or none of the parent supersymmetry;
supersymmetry breaking is one possible outcome, not the definition of
the construction.

\end{takehomeitems}
\end{takehome}

\section*{Acknowledgements}

I would like to thank I. Antoniadis, N. Benakli, A. Delgado, E. Dudas,
M. Goodsell, A. Goudeau, M. Quiros and P. Slavich for useful discussions that,
at various stages, helped shape some of the ideas developed in this
review.

\end{reviewquestion}

\newpage

\printbibliography

@article{Scherk:1978ta,
    author = "Scherk, Joel and Schwarz, John H.",
    title = "{Spontaneous Breaking of Supersymmetry Through Dimensional Reduction}",
    reportNumber = "LPTENS 78/28",
    doi = "10.1016/0370-2693(79)90425-8",
    journal = "Phys. Lett. B",
    volume = "82",
    pages = "60--64",
    year = "1979"
}

@article{Scherk:1979zr,
    author = "Scherk, Joel and Schwarz, John H.",
    title = "{How to Get Masses from Extra Dimensions}",
    reportNumber = "LPTENS-79-2",
    doi = "10.1016/0550-3213(79)90592-3",
    journal = "Nucl. Phys. B",
    volume = "153",
    pages = "61--88",
    year = "1979"
}

@article{Cremmer:1979uq,
    author = "Cremmer, E. and Scherk, Joel and Schwarz, John H.",
    title = "{Spontaneously Broken N=8 Supergravity}",
    reportNumber = "LPTENS 79/11",
    doi = "10.1016/0370-2693(79)90654-3",
    journal = "Phys. Lett. B",
    volume = "84",
    pages = "83--86",
    year = "1979"
}

@article{Kaluza:1921tu,
    author = "Kaluza, Th.",
    title = {{Zum Unit{\"a}tsproblem der Physik}},
    eprint = "1803.08616",
    archivePrefix = "arXiv",
    primaryClass = "physics.hist-ph",
    reportNumber = "HUPD-8401",
    doi = "10.1142/S0218271818700017",
    journal = "Sitzungsber. Preuss. Akad. Wiss. Berlin (Math. Phys. )",
    volume = "1921",
    pages = "966--972",
    year = "1921"
}

@article{Klein:1926tv,
    author = "Klein, Oskar",
    title = "{Quantum Theory and Five-Dimensional Theory of Relativity. (In German and English)}",
    doi = "10.1007/BF01397481",
    journal = "Z. Phys.",
    volume = "37",
    pages = "895--906",
    year = "1926"
}

@article{Klein:1926fj,
    author = "Klein, O.",
    title = "{The Atomicity of Electricity as a Quantum Theory Law}",
    doi = "10.1038/118516a0",
    journal = "Nature",
    volume = "118",
    pages = "516",
    year = "1926"
}

@article{Einstein:1938fk,
    author = "Einstein, Albert and Bergmann, P.",
    title = "{On a Generalization of Kaluza's Theory of Electricity}",
    doi = "10.2307/1968642",
    journal = "Annals Math.",
    volume = "39",
    pages = "683--701",
    year = "1938"
}

@inproceedings{Witten:2014fma,
    author = "Witten, Edward",
    title = "{A Note On Einstein, Bergmann, and the Fifth Dimension}",
    booktitle = "{Einstein Symposium 2005}",
    eprint = "1401.8048",
    archivePrefix = "arXiv",
    primaryClass = "physics.hist-ph",
    month = "1",
    year = "2014"
}

@article{Fayet:1985kt,
    author = "Fayet, Pierre",
    title = "{Six-dimensional Supersymmetric {QED}, R Invariance and $N=2$ Supersymmetry Breaking by Dimensional Reduction}",
    reportNumber = "LPTENS 85/12",
    doi = "10.1016/0550-3213(86)90280-4",
    journal = "Nucl. Phys. B",
    volume = "263",
    pages = "649--686",
    year = "1986"
}

@article{Fayet:1985ua,
    author = "Fayet, Pierre",
    title = "{Supersymmetric Grand Unification in a Six-dimensional Space-time}",
    reportNumber = "LPTENS-85/4",
    doi = "10.1016/0370-2693(85)90869-X",
    journal = "Phys. Lett. B",
    volume = "159",
    pages = "121--126",
    year = "1985"
}

@article{Rohm:1983aq,
    author = "Rohm, Ryan",
    title = "{Spontaneous Supersymmetry Breaking in Supersymmetric String Theories}",
    reportNumber = "Print-83-0932 (PRINCETON)",
    doi = "10.1016/0550-3213(84)90007-5",
    journal = "Nucl. Phys. B",
    volume = "237",
    pages = "553--572",
    year = "1984"
}

@article{Antoniadis:1988jn,
    author = "Antoniadis, Ignatios and Bachas, Constantin and Lewellen, David C. and Tomaras, T. N.",
    title = "{On Supersymmetry Breaking in Superstrings}",
    reportNumber = "SLAC-PUB-4572, CERN-TH-5007/88",
    doi = "10.1016/0370-2693(88)90679-X",
    journal = "Phys. Lett. B",
    volume = "207",
    pages = "441--446",
    year = "1988"
}

@article{Benakli:2007zza,
    author = "Benakli, Karim and Moura, Cesar",
    title = "{Brane-Worlds Pseudo-Goldstinos}",
    eprint = "0706.3127",
    archivePrefix = "arXiv",
    primaryClass = "hep-th",
    doi = "10.1016/j.nuclphysb.2007.09.010",
    journal = "Nucl. Phys. B",
    volume = "791",
    pages = "125--163",
    year = "2008"
}

@article{Banks:1988yz,
    author = "Banks, Tom and Dixon, Lance J.",
    title = "{Constraints on String Vacua with Space-Time Supersymmetry}",
    reportNumber = "PUPT-1086, SCIPP-8805",
    doi = "10.1016/0550-3213(88)90523-8",
    journal = "Nucl. Phys. B",
    volume = "307",
    pages = "93--108",
    year = "1988"
}

@article{Dine:1987vf,
    author = "Dine, Michael and Seiberg, Nathan",
    title = "{Microscopic Knowledge From Macroscopic Physics in String Theory}",
    reportNumber = "IASSNS-HEP-87-50",
    doi = "10.1016/0550-3213(88)90433-6",
    journal = "Nucl. Phys. B",
    volume = "301",
    pages = "357--380",
    year = "1988"
}

@article{Kounnas:1988ye,
    author = "Kounnas, Costas and Porrati, Massimo",
    title = "{Spontaneous Supersymmetry Breaking in String Theory}",
    reportNumber = "UCB-PTH-88/1, LBL-24789, LPTENS-88/3",
    doi = "10.1016/0550-3213(88)90153-8",
    journal = "Nucl. Phys. B",
    volume = "310",
    pages = "355--370",
    year = "1988"
}

@article{Ferrara:1988jx,
    author = "Ferrara, Sergio and Kounnas, Costas and Porrati, Massimo and Zwirner, Fabio",
    title = "{Superstrings with Spontaneously Broken Supersymmetry and their Effective Theories}",
    reportNumber = "UCB/PTH/88/19, LBL-25776, LPTENS-88/28, UCLA/88/TEP/27",
    doi = "10.1016/0550-3213(89)90048-5",
    journal = "Nucl. Phys. B",
    volume = "318",
    pages = "75--105",
    year = "1989"
}

@article{Kounnas:1989dk,
    author = "Kounnas, Costas and Rostand, Bruno",
    title = "{Coordinate Dependent Compactifications and Discrete Symmetries}",
    reportNumber = "LPTENS-89-22",
    doi = "10.1016/0550-3213(90)90543-M",
    journal = "Nucl. Phys. B",
    volume = "341",
    pages = "641--665",
    year = "1990"
}

@article{Dine:1986vd,
    author = "Dine, Michael and Seiberg, N.",
    title = "{Nonrenormalization Theorems in Superstring Theory}",
    reportNumber = "WIS/86/53-Ph",
    doi = "10.1103/PhysRevLett.57.2625",
    journal = "Phys. Rev. Lett.",
    volume = "57",
    pages = "2625",
    year = "1986"
}

@book{Husemoller:1994,
    author = "Husemoller, Dale",
    title = "{Fibre Bundles}",
    series = "Graduate Texts in Mathematics",
    volume = "20",
    edition = "3rd",
    publisher = "Springer-Verlag",
    address = "New York",
    year = "1994",
    isbn = "978-0-387-94087-8",
    doi = "10.1007/978-1-4612-6288-6"
}

@book{Lawson:1989,
    author = "Lawson, H. Blaine and Michelsohn, Marie-Louise",
    title = "{Spin Geometry}",
    series = "Princeton Mathematical Series",
    volume = "38",
    publisher = "Princeton University Press",
    address = "Princeton, NJ",
    year = "1989",
    isbn = "978-0-691-08542-5"
}

@article{Fayet:1974jb,
    author = "Fayet, Pierre and Iliopoulos, J.",
    title = "{Spontaneously Broken Supergauge Symmetries and Goldstone Spinors}",
    doi = "10.1016/0370-2693(74)90310-4",
    journal = "Phys. Lett. B",
    volume = "51",
    pages = "461--464",
    year = "1974"
}

@article{Volkov:1973jd,
    author = "Volkov, D. V. and Soroka, V. A.",
    title = "{Higgs Effect for Goldstone Particles with Spin 1/2}",
    journal = "JETP Lett.",
    volume = "18",
    pages = "312--314",
    year = "1973"
}

@article{Deser:1977uq,
    author = "Deser, Stanley and Zumino, B.",
    title = "{Broken Supersymmetry and Supergravity}",
    reportNumber = "Print-77-0426 (BRANDEIS)",
    doi = "10.1103/PhysRevLett.38.1433",
    journal = "Phys. Rev. Lett.",
    volume = "38",
    pages = "1433--1436",
    year = "1977"
}

@article{Cremmer:1978hn,
    author = "Cremmer, E. and Julia, B. and Scherk, Joel and Ferrara, S. and Girardello, L. and van Nieuwenhuizen, P.",
    title = "{Spontaneous Symmetry Breaking and Higgs Effect in Supergravity Without Cosmological Constant}",
    reportNumber = "CERN-TH-2554",
    doi = "10.1016/0550-3213(79)90417-6",
    journal = "Nucl. Phys. B",
    volume = "147",
    pages = "105",
    year = "1979"
}

@article{Fayet:1977vd,
    author = "Fayet, Pierre",
    title = "{Mixing Between Gravitational and Weak Interactions Through the Massive Gravitino}",
    reportNumber = "Print-77-0787 (ECOLE POLY)",
    doi = "10.1016/0370-2693(77)90414-2",
    journal = "Phys. Lett. B",
    volume = "70",
    pages = "461",
    year = "1977"
}

@article{Casalbuoni:1988qd,
    author = "Casalbuoni, R. and De Curtis, S. and Dominici, D. and Feruglio, F. and Gatto, Raoul",
    title = "{High-Energy Equivalence Theorem in Spontaneously Broken Supergravity}",
    reportNumber = "UGVA-DPT-1988-12-599",
    doi = "10.1103/PhysRevD.39.2281",
    journal = "Phys. Rev. D",
    volume = "39",
    pages = "2281",
    year = "1989"
}

@article{Casalbuoni:1988kv,
    author = "Casalbuoni, R. and De Curtis, S. and Dominici, D. and Feruglio, F. and Gatto, Raoul",
    title = "{A Gravitino--Goldstino High-Energy Equivalence Theorem}",
    reportNumber = "UGVA-DPT-1988/09-591",
    doi = "10.1016/0370-2693(88)91439-6",
    journal = "Phys. Lett. B",
    volume = "215",
    pages = "313--316",
    year = "1988"
}

@article{Antoniadis:1997ic,
    author = "Antoniadis, Ignatios and Quir{\'o}s, Mariano",
    title = "{Supersymmetry breaking in M theory and gaugino condensation}",
    eprint = "hep-th/9705037",
    archivePrefix = "arXiv",
    reportNumber = "CPTH-S521-0597, IEM-FT-156-97",
    doi = "10.1016/S0550-3213(97)00473-5",
    journal = "Nucl. Phys. B",
    volume = "505",
    pages = "109--122",
    year = "1997"
}

@article{Dudas:1997jn,
    author = "Dudas, Emilian and Grojean, Christophe",
    title = "{Four-dimensional M-theory and supersymmetry breaking}",
    eprint = "hep-th/9704177",
    archivePrefix = "arXiv",
    reportNumber = "CPTH-PC516-0497, SACLAY-SPHT-97-040",
    doi = "10.1016/S0550-3213(97)00523-6",
    journal = "Nucl. Phys. B",
    volume = "507",
    pages = "553--570",
    year = "1997"
}

@article{Marti:2001iw,
    author = "Marti, Daniel and Pomarol, Alex",
    title = "{Supersymmetric theories with compact extra dimensions in $N=1$ superfields}",
    eprint = "hep-th/0106256",
    archivePrefix = "arXiv",
    reportNumber = "UAB-FT-508",
    doi = "10.1103/PhysRevD.64.105025",
    journal = "Phys. Rev. D",
    volume = "64",
    pages = "105025",
    year = "2001"
}

@article{Kaplan:1999ac,
    author = "Kaplan, David E. and Kribs, Graham D. and Schmaltz, Martin",
    title = "{Supersymmetry breaking through transparent extra dimensions}",
    eprint = "hep-ph/9911293",
    archivePrefix = "arXiv",
    reportNumber = "SLAC-PUB-8294, BUHEP-99-27, FERMILAB-PUB-99-317-T",
    doi = "10.1103/PhysRevD.62.035010",
    journal = "Phys. Rev. D",
    volume = "62",
    pages = "035010",
    year = "2000"
}

@article{Chacko:1999mi,
    author = "Chacko, Zackaria and Luty, Markus A. and Nelson, Ann E. and Pont{\'o}n, Eduardo",
    title = "{Gaugino mediated supersymmetry breaking}",
    eprint = "hep-ph/9911323",
    archivePrefix = "arXiv",
    reportNumber = "UW-PT-99-22, UMD-PP-00-028",
    doi = "10.1088/1126-6708/2000/01/003",
    journal = "JHEP",
    volume = "01",
    pages = "003",
    year = "2000"
}

@article{Arkani-Hamed:2001mi,
    author = "Arkani-Hamed, Nima and Hall, Lawrence J. and Nomura, Yasunori and Smith, David Tucker and Weiner, Neal",
    title = "{Finite radiative electroweak symmetry breaking from the bulk}",
    eprint = "hep-ph/0102090",
    archivePrefix = "arXiv",
    reportNumber = "LBNL-47407, UCB-PTH-01-05",
    doi = "10.1016/S0550-3213(01)00213-1",
    journal = "Nucl. Phys. B",
    volume = "605",
    pages = "81--115",
    year = "2001"
}

@article{vonGersdorff:2001ak,
    author = "von Gersdorff, Gero and Quir{\'o}s, Mariano",
    title = "{Supersymmetry breaking on orbifolds from Wilson lines}",
    eprint = "hep-th/0110132",
    archivePrefix = "arXiv",
    reportNumber = "IEM-FT-220-01",
    doi = "10.1103/PhysRevD.65.064016",
    journal = "Phys. Rev. D",
    volume = "65",
    pages = "064016",
    year = "2002"
}

@article{Gherghetta:2001sa,
    author = "Gherghetta, Tony and Riotto, Antonio",
    title = "{Gravity mediated supersymmetry breaking in the brane world}",
    eprint = "hep-th/0110022",
    archivePrefix = "arXiv",
    reportNumber = "CERN-TH-2001-267, UNIL-IPT-01-14, NEIP-01-008",
    doi = "10.1016/S0550-3213(01)00609-8",
    journal = "Nucl. Phys. B",
    volume = "623",
    pages = "97--112",
    year = "2002"
}

@article{Rattazzi:2003rj,
    author = "Rattazzi, Riccardo and Scrucca, Claudio A. and Strumia, Alessandro",
    title = "{Brane to brane gravity mediation of supersymmetry breaking}",
    eprint = "hep-th/0305184",
    archivePrefix = "arXiv",
    reportNumber = "CERN-TH-2003-108, IFUP-TH-2003-24",
    doi = "10.1016/j.nuclphysb.2003.09.043",
    journal = "Nucl. Phys. B",
    volume = "674",
    pages = "171--216",
    year = "2003"
}

@article{Bagger:2001qi,
    author = "Bagger, Jonathan A. and Feruglio, Ferruccio and Zwirner, Fabio",
    title = "{Generalized symmetry breaking on orbifolds}",
    eprint = "hep-th/0107128",
    archivePrefix = "arXiv",
    reportNumber = "JHU-TIPAC-20013, DFPD-01-TH-29, CERN-TH-2001-182",
    doi = "10.1103/PhysRevLett.88.101601",
    journal = "Phys. Rev. Lett.",
    volume = "88",
    pages = "101601",
    year = "2002"
}

@article{Biggio:2002rb,
    author = "Biggio, Carla and Feruglio, Ferruccio and Wulzer, Andrea and Zwirner, Fabio",
    title = "{Equivalent effective Lagrangians for Scherk-Schwarz compactifications}",
    eprint = "hep-th/0209046",
    archivePrefix = "arXiv",
    reportNumber = "DFPD-02-TH-20",
    doi = "10.1088/1126-6708/2002/11/013",
    journal = "JHEP",
    volume = "11",
    pages = "013",
    year = "2002"
}

@article{vonGersdorff:2002tj,
    author = "von Gersdorff, Gero and Quir{\'o}s, Mariano and Riotto, Antonio",
    title = "{Radiative Scherk-Schwarz supersymmetry breaking}",
    eprint = "hep-th/0204041",
    archivePrefix = "arXiv",
    reportNumber = "CERN-TH-2002-075, IEM-FT-222-02, SNS-PH-02-05",
    doi = "10.1016/S0550-3213(02)00344-1",
    journal = "Nucl. Phys. B",
    volume = "634",
    pages = "90--104",
    year = "2002"
}

@article{Delgado:2002xf,
    author = "Delgado, Antonio and von Gersdorff, Gero and Quir{\'o}s, Mariano",
    title = "{Brane assisted Scherk-Schwarz supersymmetry breaking in orbifolds}",
    eprint = "hep-th/0210181",
    archivePrefix = "arXiv",
    reportNumber = "IEM-FT-228-02, JHU-TIPAC-20028",
    doi = "10.1088/1126-6708/2002/12/002",
    journal = "JHEP",
    volume = "12",
    pages = "002",
    year = "2002"
}

@article{Meissner:1999ja,
    author = "Meissner, Krzysztof A. and Nilles, Hans Peter and Olechowski, Marek",
    title = "{Supersymmetry breakdown at distant branes: The SuperHiggs mechanism}",
    eprint = "hep-th/9905139",
    archivePrefix = "arXiv",
    reportNumber = "BONN-TH-99-05, IFT-11-99",
    doi = "10.1016/S0550-3213(99)00407-4",
    journal = "Nucl. Phys. B",
    volume = "561",
    pages = "30--42",
    year = "1999"
}

@article{DeCurtis:2003hs,
    author = "De Curtis, Stefania and Dominici, Daniele and Pelaez, J. R.",
    title = "{Equivalence between Kaluza-Klein modes of gravitinos and Goldstinos in brane induced supersymmetry breaking}",
    eprint = "hep-th/0311226",
    archivePrefix = "arXiv",
    reportNumber = "DFF-412-11-03, FT-UCM-106-2003",
    doi = "10.1088/1126-6708/2004/01/052",
    journal = "JHEP",
    volume = "01",
    pages = "052",
    year = "2004"
}

@article{Bagger:2004rr,
    author = "Bagger, Jonathan A. and Belyaev, Dmitry V.",
    title = "{Brane-localized Goldstone fermions in bulk supergravity}",
    eprint = "hep-th/0406126",
    archivePrefix = "arXiv",
    reportNumber = "JHU-TIPAC-2004-03",
    doi = "10.1103/PhysRevD.72.065007",
    journal = "Phys. Rev. D",
    volume = "72",
    pages = "065007",
    year = "2005"
}

@article{Dixon:1985jw,
    author = "Dixon, Lance J. and Harvey, Jeffrey A. and Vafa, C. and Witten, Edward",
    title = "{Strings on Orbifolds}",
    reportNumber = "PRINT-85-0616 (PRINCETON)",
    doi = "10.1016/0550-3213(85)90593-0",
    journal = "Nucl. Phys. B",
    volume = "261",
    pages = "678--686",
    year = "1985"
}

@article{Dixon:1986jc,
    author = "Dixon, Lance J. and Harvey, Jeffrey A. and Vafa, C. and Witten, Edward",
    title = "{Strings on Orbifolds. 2.}",
    reportNumber = "PRINT-86-0246 (PRINCETON)",
    doi = "10.1016/0550-3213(86)90287-7",
    journal = "Nucl. Phys. B",
    volume = "274",
    pages = "285--314",
    year = "1986"
}

@article{Kaplan:2001cg,
    author = "Kaplan, David Elazzar and Weiner, Neal",
    title = "{Radion mediated supersymmetry breaking as a Scherk-Schwarz theory}",
    eprint = "hep-ph/0108001",
    archivePrefix = "arXiv",
    reportNumber = "ANL-HEP-PR-01-074, UW-PT-01-20",
    month = "8",
    year = "2001"
}

@article{Ferrara:1994kg,
    author = "Ferrara, Sergio and Kounnas, Costas and Zwirner, Fabio",
    title = "{Mass formulae and natural hierarchy in string effective supergravities}",
    eprint = "hep-th/9405188",
    archivePrefix = "arXiv",
    reportNumber = "CERN-TH-7192-94, LPTENS-94-12, UCLA-94-TEP-13",
    doi = "10.1016/0550-3213(94)00494-Y",
    journal = "Nucl. Phys. B",
    volume = "429",
    pages = "589--625",
    year = "1994",
    note = "[Erratum: Nucl.Phys.B 433, 255--255 (1995)]"
}

@article{Benakli:2026zwx,
    author = "Benakli, Karim and Goudeau, Arno",
    title = "{(Pseudo-)Dirac Gravitinos}",
    journal = "JHEP",
    volume = "09",
    pages = "030",
    year = "2026",
    doi = "10.1007/JHEP09(2026)030",
    eprint = "2605.04146",
    archivePrefix = "arXiv",
    primaryClass = "hep-th"
}

@article{Nicolai:2000sc,
    author = "Nicolai, Hermann and Samtleben, Henning",
    title = "{Maximal Gauged Supergravity in Three-Dimensions}",
    eprint = "hep-th/0010076",
    archivePrefix = "arXiv",
    reportNumber = "AEI-2000-064",
    doi = "10.1103/PhysRevLett.86.1686",
    journal = "Phys. Rev. Lett.",
    volume = "86",
    pages = "1686--1689",
    year = "2001"
}

@article{deWit:2002vt,
    author = "de Wit, Bernard and Samtleben, Henning and Trigiante, Mario",
    title = "{On Lagrangians and Gaugings of Maximal Supergravities}",
    eprint = "hep-th/0212239",
    archivePrefix = "arXiv",
    reportNumber = "ITP-UU-02-60, SPIN-02-36",
    doi = "10.1016/S0550-3213(03)00059-2",
    journal = "Nucl. Phys. B",
    volume = "655",
    pages = "93--126",
    year = "2003"
}

@article{deWit:2007mt,
    author = "de Wit, Bernard and Samtleben, Henning and Trigiante, Mario",
    title = "{The Maximal $\mathcal{N}=8$ Supergravity with Local $E_{7(7)}$ Invariance}",
    eprint = "0705.2101",
    archivePrefix = "arXiv",
    primaryClass = "hep-th",
    reportNumber = "ITP-UU-07-22, SPIN-07-16, LAPTH-1185-07",
    doi = "10.1088/1126-6708/2007/06/049",
    journal = "JHEP",
    volume = "06",
    pages = "049",
    year = "2007"
}

@article{deWit:2008ta,
    author = "de Wit, Bernard and Nicolai, Hermann and Samtleben, Henning",
    title = "{Gauged Supergravities, Tensor Hierarchies, and M-Theory}",
    eprint = "0801.1294",
    archivePrefix = "arXiv",
    primaryClass = "hep-th",
    reportNumber = "AEI-2008-003, ITP-UU-08-03, SPIN-08-03, LAPTH-1224-08",
    doi = "10.1088/1126-6708/2008/02/044",
    journal = "JHEP",
    volume = "02",
    pages = "044",
    year = "2008"
}

@article{Samtleben:2008pe,
    author = "Samtleben, Henning",
    title = "{Lectures on Gauged Supergravity and Flux Compactifications}",
    eprint = "0808.4076",
    archivePrefix = "arXiv",
    primaryClass = "hep-th",
    reportNumber = "ENS-LYON-LPENSL-08-36",
    doi = "10.1002/prop.200810579",
    journal = "Fortsch. Phys.",
    volume = "56",
    pages = "843--878",
    year = "2008"
}

@article{Kaloper:1999yr,
    author = "Kaloper, Nemanja and Myers, Robert C.",
    title = "{The Odd Story of Massive Supergravity}",
    eprint = "hep-th/9901045",
    archivePrefix = "arXiv",
    reportNumber = "MCGILL-98-38, SU-ITP-98-68",
    doi = "10.1088/1126-6708/1999/05/010",
    journal = "JHEP",
    volume = "05",
    pages = "010",
    year = "1999"
}

@article{Hull:2005hk,
    author = "Hull, Christopher M. and Reid-Edwards, Richard A.",
    title = "{Flux Compactifications of String Theory on Twisted Tori}",
    eprint = "hep-th/0503114",
    archivePrefix = "arXiv",
    reportNumber = "IMPERIAL-TP-05-CH-01",
    doi = "10.1002/prop.200900076",
    journal = "Fortsch. Phys.",
    volume = "57",
    pages = "862--944",
    year = "2009"
}

@article{Aldazabal:2011nj,
    author = "Aldazabal, Gerardo and Baron, Walter and Marqu{\'e}s, Diego and N{\'u}{\~n}ez, Carmen",
    title = "{The Effective Action of Double Field Theory}",
    eprint = "1109.0290",
    archivePrefix = "arXiv",
    primaryClass = "hep-th",
    reportNumber = "IAFE-11-04",
    doi = "10.1007/JHEP11(2011)052",
    journal = "JHEP",
    volume = "11",
    pages = "052",
    year = "2011",
    note = "[Erratum: JHEP 11, 109 (2011)]"
}

@article{Hohm:2014qga,
    author = "Hohm, Olaf and Samtleben, Henning",
    title = "{Consistent Kaluza-Klein Truncations via Exceptional Field Theory}",
    eprint = "1410.8145",
    archivePrefix = "arXiv",
    primaryClass = "hep-th",
    reportNumber = "MIT-CTP-4609",
    doi = "10.1007/JHEP01(2015)131",
    journal = "JHEP",
    volume = "01",
    pages = "131",
    year = "2015"
}

@article{Geissbuhler:2011mx,
    author = "Geissb{\"u}hler, David",
    title = "{Double Field Theory and N=4 Gauged Supergravity}",
    eprint = "1109.4280",
    archivePrefix = "arXiv",
    primaryClass = "hep-th",
    reportNumber = "LPTENS-11-37",
    doi = "10.1007/JHEP11(2011)116",
    journal = "JHEP",
    volume = "11",
    pages = "116",
    year = "2011"
}

@article{Lee:2014mla,
    author = "Lee, Kanghoon and Strickland-Constable, Charles and Waldram, Daniel",
    title = "{Spheres, Generalised Parallelisability and Consistent Truncations}",
    eprint = "1401.3360",
    archivePrefix = "arXiv",
    primaryClass = "hep-th",
    reportNumber = "IMPERIAL-TP-14-DW-01, ZMP-HH-14-3",
    doi = "10.1002/prop.201700048",
    journal = "Fortsch. Phys.",
    volume = "65",
    number = "10-11",
    pages = "1700048",
    year = "2017"
}

@article{Cremmer:1983bf,
    author = "Cremmer, Eugene and Ferrara, Sergio and Kounnas, Costas and Nanopoulos, Dimitri V.",
    title = "{Naturally Vanishing Cosmological Constant in $\mathcal{N}=1$ Supergravity}",
    reportNumber = "CERN-TH-3727",
    doi = "10.1016/0370-2693(83)90106-5",
    journal = "Phys. Lett. B",
    volume = "133",
    number = "1-2",
    pages = "61--66",
    year = "1983"
}

@article{Ellis:1984bm,
    author = "Ellis, John R. and Kounnas, Costas and Nanopoulos, Dimitri V.",
    title = "{No Scale Supersymmetric GUTs}",
    reportNumber = "CERN-TH-3824",
    doi = "10.1016/0550-3213(84)90555-8",
    journal = "Nucl. Phys. B",
    volume = "247",
    number = "2",
    pages = "373--395",
    year = "1984"
}

@article{Gomez-Reino:2006gtu,
    author = "Gomez-Reino, Marta and Scrucca, Claudio A.",
    title = "{Locally stable non-supersymmetric Minkowski vacua in supergravity}",
    eprint = "hep-th/0602246",
    archivePrefix = "arXiv",
    reportNumber = "LPENS-06-08",
    doi = "10.1088/1126-6708/2006/05/015",
    journal = "JHEP",
    volume = "05",
    pages = "015",
    year = "2006"
}

@article{ORaifeartaigh:1975nky,
    author = "O'Raifeartaigh, L.",
    title = "{Spontaneous Symmetry Breaking for Chiral Scalar Superfields}",
    reportNumber = "DIAS-TP-75-13",
    doi = "10.1016/0550-3213(75)90585-4",
    journal = "Nucl. Phys. B",
    volume = "96",
    number = "2",
    pages = "331--352",
    year = "1975"
}

@article{Komargodski:2009jf,
    author = "Komargodski, Zohar and Shih, David",
    title = "{Notes on SUSY and R-Symmetry Breaking in Wess-Zumino Models}",
    eprint = "0902.0030",
    archivePrefix = "arXiv",
    primaryClass = "hep-th",
    reportNumber = "IASSNS-HEP-09-17",
    doi = "10.1088/1126-6708/2009/04/093",
    journal = "JHEP",
    volume = "04",
    pages = "093",
    year = "2009"
}

@article{Nelson:1993nf,
    author = "Nelson, Ann E. and Seiberg, Nathan",
    title = "{R symmetry breaking versus supersymmetry breaking}",
    eprint = "hep-ph/9309299",
    archivePrefix = "arXiv",
    reportNumber = "RU-93-43, UW-PT-93-18",
    doi = "10.1016/0550-3213(94)90103-1",
    journal = "Nucl. Phys. B",
    volume = "416",
    number = "1",
    pages = "46--62",
    year = "1994"
}

@book{Nicolaescu:MorseTheory,
    author    = "Nicolaescu, Liviu I.",
    title     = "{An Invitation to Morse Theory}",
    publisher = "Springer",
    series    = "Universitext",
    edition   = "2",
    year      = "2011",
    doi       = "10.1007/978-1-4614-1105-5"
}

@book{Borel:1991,
    author = "Borel, Armand",
    title = "{Linear Algebraic Groups}",
    series = "Graduate Texts in Mathematics",
    volume = "126",
    edition = "2nd",
    publisher = "Springer-Verlag",
    address = "New York",
    year = "1991",
    isbn = "978-1-4612-6947-2",
    doi = "10.1007/978-1-4612-0941-6"
}

@book{Bourbaki:Lie4-6,
    author = "Bourbaki, Nicolas",
    title = "{Lie Groups and Lie Algebras: Chapters 4--6}",
    series = "Elements of Mathematics",
    publisher = "Springer-Verlag",
    address = "Berlin, Heidelberg",
    year = "2002",
    isbn = "978-3-540-42650-9",
    doi = "10.1007/978-3-540-89394-3"
}

@article{Atiyah:1982fa,
    author = "Atiyah, Michael F. and Bott, Raoul",
    title = "{The Yang-Mills equations over Riemann surfaces}",
    journal = "Phil. Trans. Roy. Soc. Lond. A",
    volume = "308",
    number = "1505",
    pages = "523--615",
    year = "1983",
    doi = "10.1098/rsta.1983.0017"
}

@article{Goldman:1988,
    author = "Goldman, William M.",
    title = "{Topological components of spaces of representations}",
    journal = "Invent. Math.",
    volume = "93",
    number = "3",
    pages = "557--607",
    year = "1988",
    doi = "10.1007/BF01410200"
}

@article{Witten:1997bs,
    author = "Witten, Edward",
    title = "{Toroidal compactification without vector structure}",
    eprint = "hep-th/9712028",
    archivePrefix = "arXiv",
    reportNumber = "IASSNS-HEP-97-124",
    doi = "10.1088/1126-6708/1998/02/006",
    journal = "JHEP",
    volume = "02",
    pages = "006",
    year = "1998"
}

@article{Borel:1999bx,
    author = "Borel, Armand and Friedman, Robert and Morgan, John W.",
    title = "{Almost commuting elements in compact Lie groups}",
    eprint = "math/9907007",
    archivePrefix = "arXiv",
    doi = "10.4310/MRL.2002.v9.n2.a6",
    journal = "Math. Res. Lett.",
    volume = "9",
    pages = "153--161",
    year = "2002"
}

@article{Vassilevich:2003xt,
    author = "Vassilevich, D. V.",
    title = "{Heat kernel expansion: user's manual}",
    eprint = "hep-th/0306138",
    archivePrefix = "arXiv",
    reportNumber = "ITP-UH-11-03",
    doi = "10.1016/j.physrep.2003.09.002",
    journal = "Phys. Rept.",
    volume = "388",
    number = "5-6",
    pages = "279--360",
    year = "2003"
}

@article{DrydenEtAl:2008,
    author = "Dryden, Emily B. and Gordon, Carolyn S. and Greenwald, Sarah J. and Webb, David L.",
    title = "{Asymptotic expansion of the heat kernel for orbifolds}",
    eprint = "0805.3148",
    archivePrefix = "arXiv",
    primaryClass = "math.DG",
    doi = "10.1307/mmj/1213972406",
    journal = "Mich. Math. J.",
    volume = "56",
    number = "1",
    pages = "205--238",
    year = "2008"
}

@article{Georgi:2000ks,
    author = "Georgi, Howard and Grant, Aaron K. and Hailu, Girma",
    title = "{Brane couplings from bulk loops}",
    eprint = "hep-ph/0012379",
    archivePrefix = "arXiv",
    reportNumber = "HUTP-00-A030",
    doi = "10.1016/S0370-2693(01)00408-7",
    journal = "Phys. Lett. B",
    volume = "506",
    pages = "207--214",
    year = "2001"
}

@article{vonGersdorff:2002as,
    author = "von Gersdorff, Gero and Irges, Nikolas and Quir{\'o}s, Mariano",
    title = "{Bulk and brane radiative effects in gauge theories on orbifolds}",
    eprint = "hep-ph/0204223",
    archivePrefix = "arXiv",
    reportNumber = "CERN-TH-2002-090, IEM-FT-224-02",
    doi = "10.1016/S0550-3213(02)00392-1",
    journal = "Nucl. Phys. B",
    volume = "635",
    number = "1-2",
    pages = "127--157",
    year = "2002"
}

@article{Antoniadis:1993jp,
    author = "Antoniadis, Ignatios and Benakli, K.",
    title = "{Limits on extra dimensions in orbifold compactifications of superstrings}",
    eprint = "hep-th/9310151",
    archivePrefix = "arXiv",
    reportNumber = "CPTH-A257-0793",
    doi = "10.1016/0370-2693(94)91194-0",
    journal = "Phys. Lett. B",
    volume = "326",
    pages = "69--78",
    year = "1994"
}

@article{Antoniadis:1998sd,
    author = "Antoniadis, Ignatios and Dimopoulos, S. and Pomarol, A. and Quiros, M.",
    title = "{Soft masses in theories with supersymmetry breaking by TeV compactification}",
    eprint = "hep-ph/9810410",
    archivePrefix = "arXiv",
    reportNumber = "CERN-TH-98-332, CPTH-S686-1098, SU-ITP-98-58, IEM-FT-184-98",
    doi = "10.1016/S0550-3213(99)00037-1",
    journal = "Nucl. Phys. B",
    volume = "544",
    pages = "503--519",
    year = "1999"
}

@article{Barbieri:2000vh,
    author = "Barbieri, Riccardo and Hall, Lawrence J. and Nomura, Yasunori",
    title = "{A Constrained standard model from a compact extra dimension}",
    eprint = "hep-ph/0011311",
    archivePrefix = "arXiv",
    reportNumber = "SNS-PH-00-16, UCB-PTH-00-38, LBNL-47128",
    doi = "10.1103/PhysRevD.63.105007",
    journal = "Phys. Rev. D",
    volume = "63",
    pages = "105007",
    year = "2001"
}

@article{Antoniadis:2015chx,
    author = "Antoniadis, Ignatios and Benakli, Karim and Quiros, Mariano",
    title = "{Sequestered gravity in gauge mediation}",
    eprint = "1512.00029",
    archivePrefix = "arXiv",
    primaryClass = "hep-ph",
    doi = "10.1140/epjc/s10052-016-4192-8",
    journal = "Eur. Phys. J. C",
    volume = "76",
    number = "7",
    pages = "363",
    year = "2016"
}

@article{Atiyah:1975jf,
    author = "Atiyah, M. F. and Patodi, V. K. and Singer, I. M.",
    title = "{Spectral asymmetry and Riemannian Geometry 1}",
    doi = "10.1017/S0305004100049410",
    journal = "Math. Proc. Cambridge Phil. Soc.",
    volume = "77",
    pages = "43",
    year = "1975"
}

@article{Dai:1994kq,
    author = "Dai, Xian-zhe and Freed, Daniel S.",
    title = "{eta invariants and determinant lines}",
    eprint = "hep-th/9405012",
    archivePrefix = "arXiv",
    doi = "10.1063/1.530747",
    journal = "J. Math. Phys.",
    volume = "35",
    pages = "5155--5194",
    year = "1994",
    note = "[Erratum: J.Math.Phys. 42, 2343--2344 (2001)]"
}

@article{Freed:2014iua,
    author = "Freed, Daniel S.",
    editor = "Donagi, Ron and Douglas, Michael R. and Kamenova, Ljudmila and Rocek, Martin",
    title = "{Anomalies and Invertible Field Theories}",
    eprint = "1404.7224",
    archivePrefix = "arXiv",
    primaryClass = "hep-th",
    doi = "10.1090/pspum/088/01462",
    journal = "Proc. Symp. Pure Math.",
    volume = "88",
    pages = "25--46",
    year = "2014"
}

@article{Cheng:2026abk,
    author = "Cheng, Peng and De Freitas, Hector Parra",
    title = "{Dai-Freed anomalies and level matching in heterotic asymmetric orbifolds}",
    eprint = "2604.19634",
    archivePrefix = "arXiv",
    primaryClass = "hep-th",
    month = "4",
    year = "2026"
}

@article{Garcia-Etxebarria:2018ajm,
    author = "Garc{\'\i}a-Etxebarria, I{\~n}aki and Montero, Miguel",
    title = "{Dai-Freed anomalies in particle physics}",
    eprint = "1808.00009",
    archivePrefix = "arXiv",
    primaryClass = "hep-th",
    reportNumber = "MPP-2018-188",
    doi = "10.1007/JHEP08(2019)003",
    journal = "JHEP",
    volume = "08",
    pages = "003",
    year = "2019"
}

@article{Witten:1982fp,
    author = "Witten, Edward",
    title = "{An SU(2) Anomaly}",
    journal = "Phys. Lett. B",
    volume = "117",
    pages = "324--328",
    year = "1982",
    doi = "10.1016/0370-2693(82)90728-6"
}

@article{Vafa:1986wx,
    author = "Vafa, Cumrun",
    title = "{Modular Invariance and Discrete Torsion on Orbifolds}",
    reportNumber = "HUTP-86/A011",
    doi = "10.1016/0550-3213(86)90379-2",
    journal = "Nucl. Phys. B",
    volume = "273",
    pages = "592--606",
    year = "1986"
}

@article{Narain:1985jj,
    author = "Narain, K. S.",
    title = "{New Heterotic String Theories in Uncompactified Dimensions {\ensuremath{<}} 10}",
    reportNumber = "RAL-85-097",
    doi = "10.1016/0370-2693(86)90682-9",
    journal = "Phys. Lett. B",
    volume = "169",
    pages = "41--46",
    year = "1986"
}

@article{Narain:1986am,
    author = "Narain, K. S. and Sarmadi, M. H. and Witten, Edward",
    title = "{A Note on Toroidal Compactification of Heterotic String Theory}",
    reportNumber = "Print-86-0870 (RUTHERFORD), RAL-86-022",
    doi = "10.1016/0550-3213(87)90001-0",
    journal = "Nucl. Phys. B",
    volume = "279",
    pages = "369--379",
    year = "1987"
}

@article{Benakli:1995ut,
    author = "Benakli, Karim",
    title = "{Perturbative supersymmetry breaking in orbifolds with Wilson line backgrounds}",
    eprint = "hep-th/9509115",
    archivePrefix = "arXiv",
    reportNumber = "IC-95-306",
    doi = "10.1016/0370-2693(96)00902-1",
    journal = "Phys. Lett. B",
    volume = "386",
    pages = "106--114",
    year = "1996"
}

@article{Abel:2017vos,
    author = "Abel, Steven and Dienes, Keith R. and Mavroudi, Eirini",
    title = "{GUT precursors and entwined SUSY: The phenomenology of stable nonsupersymmetric strings}",
    eprint = "1712.06894",
    archivePrefix = "arXiv",
    primaryClass = "hep-th",
    doi = "10.1103/PhysRevD.97.126017",
    journal = "Phys. Rev. D",
    volume = "97",
    number = "12",
    pages = "126017",
    year = "2018"
}

@article{Antoniadis:1990ew,
    author = "Antoniadis, Ignatios",
    title = "{A Possible new dimension at a few TeV}",
    reportNumber = "EP-CPTH-A978-0690",
    doi = "10.1016/0370-2693(90)90617-F",
    journal = "Phys. Lett. B",
    volume = "246",
    pages = "377--384",
    year = "1990"
}

@article{Antoniadis:1998ki,
    author = "Antoniadis, Ignatios and Dudas, E. and Sagnotti, A.",
    title = "{Supersymmetry breaking, open strings and M theory}",
    eprint = "hep-th/9807011",
    archivePrefix = "arXiv",
    reportNumber = "CERN-TH-98-212, CPTH-S616-0698, LPTHE-ORSAY-98-42, ROM2F-98-21",
    doi = "10.1016/S0550-3213(98)00806-2",
    journal = "Nucl. Phys. B",
    volume = "544",
    pages = "469--502",
    year = "1999"
}

@article{Antoniadis:1998ep,
    author = "Antoniadis, Ignatios and D'Appollonio, G. and Dudas, E. and Sagnotti, A.",
    title = "{Partial breaking of supersymmetry, open strings and M theory}",
    eprint = "hep-th/9812118",
    archivePrefix = "arXiv",
    reportNumber = "CERN-TH-98-382, CPTH-S691-1198, LPTHE-ORSAY-98-70, ROM2F-98-41",
    doi = "10.1016/S0550-3213(99)00232-1",
    journal = "Nucl. Phys. B",
    volume = "553",
    pages = "133--154",
    year = "1999"
}

@article{Antoniadis:1999xk,
    author = "Antoniadis, Ignatios and Dudas, E. and Sagnotti, A.",
    title = "{Brane supersymmetry breaking}",
    eprint = "hep-th/9908023",
    archivePrefix = "arXiv",
    reportNumber = "CPHT-S727-0799, LPT-ORSAY-99-60, ROM2F-99-23",
    doi = "10.1016/S0370-2693(99)01023-0",
    journal = "Phys. Lett. B",
    volume = "464",
    pages = "38--45",
    year = "1999"
}

@article{Gaiotto:2014kfa,
    author = "Gaiotto, Davide and Kapustin, Anton and Seiberg, Nathan and Willett, Brian",
    title = "{Generalized Global Symmetries}",
    eprint = "1412.5148",
    archivePrefix = "arXiv",
    primaryClass = "hep-th",
    doi = "10.1007/JHEP02(2015)172",
    journal = "JHEP",
    volume = "02",
    pages = "172",
    year = "2015"
}

@article{Murray1996,
    author = "Murray, Michael K.",
    title = "{Bundle Gerbes}",
    doi = "10.1112/jlms/54.2.403",
    journal = "J. Lond. Math. Soc.",
    volume = "54",
    number = "2",
    pages = "403--416",
    year = "1996"
}

@incollection{CheegerSimons1985,
    author = "Cheeger, Jeff and Simons, James",
    title = "{Differential Characters and Geometric Invariants}",
    booktitle = "{Geometry and Topology}",
    editor = "Alexander, J. and Harer, J.",
    series = "Lecture Notes in Mathematics",
    volume = "1167",
    pages = "50--80",
    publisher = "Springer",
    address = "Berlin",
    year = "1985",
    doi = "10.1007/BFb0075216"
}

@article{HopkinsSinger2005,
    author = "Hopkins, Michael J. and Singer, Isadore M.",
    title = "{Quadratic Functions in Geometry, Topology, and M-Theory}",
    eprint = "math/0211216",
    archivePrefix = "arXiv",
    doi = "10.4310/jdg/1143642908",
    journal = "J. Differential Geom.",
    volume = "70",
    number = "3",
    pages = "329--452",
    year = "2005"
}

@article{Dabholkar:2002sy,
    author = "Dabholkar, Atish and Hull, Christopher",
    title = "{Duality twists, orbifolds, and fluxes}",
    eprint = "hep-th/0210209",
    archivePrefix = "arXiv",
    reportNumber = "TIFR-TH-02-38, QMUL-PH-02-18",
    doi = "10.1088/1126-6708/2003/09/054",
    journal = "JHEP",
    volume = "09",
    pages = "054",
    year = "2003"
}

@article{DAuria:2005npq,
    author = "D'Auria, Riccardo and Ferrara, Sergio and Trigiante, Mario",
    title = "{Curvatures and potential of M-theory in $D=4$ with fluxes and twist}",
    eprint = "hep-th/0507225",
    archivePrefix = "arXiv",
    reportNumber = "CERN-PH-TH-2005-131",
    doi = "10.1088/1126-6708/2005/09/035",
    journal = "JHEP",
    volume = "09",
    pages = "035",
    year = "2005"
}

@article{DallAgata:2005zff,
    author = "Dall'Agata, Gianguido and Prezas, Nikolaos",
    title = "{Scherk-Schwarz reduction of M-theory on $G_2$-manifolds with fluxes}",
    eprint = "hep-th/0509052",
    archivePrefix = "arXiv",
    reportNumber = "CERN-PH-TH-2005-149",
    doi = "10.1088/1126-6708/2005/10/103",
    journal = "JHEP",
    volume = "10",
    pages = "103",
    year = "2005"
}

@article{Fabinger:2000jd,
    author = "Fabinger, Michal and Horava, Petr",
    title = "{Casimir effect between world-branes in heterotic M-theory}",
    eprint = "hep-th/0002073",
    archivePrefix = "arXiv",
    reportNumber = "CALT-68-2255, CITUSC-00-004, PRA-HEP-00-02",
    doi = "10.1016/S0550-3213(00)00421-5",
    journal = "Nucl. Phys. B",
    volume = "580",
    pages = "243--263",
    year = "2000"
}

@article{Bossard:2024mls,
    author = "Bossard, Guillaume and Casagrande, Gabriele and Dudas, Emilian",
    title = "{Twisted orientifold planes and S-duality without supersymmetry}",
    eprint = "2411.00955",
    archivePrefix = "arXiv",
    primaryClass = "hep-th",
    reportNumber = "CPHT-RR012.072024",
    doi = "10.1007/JHEP02(2025)062",
    journal = "JHEP",
    volume = "02",
    pages = "062",
    year = "2025"
}

@article{Kiritsis:1997ca,
    author = "Kiritsis, Elias and Kounnas, Costas",
    title = "{Perturbative and nonperturbative partial supersymmetry breaking: {$N=4 \to N=2 \to N=1$}}",
    eprint = "hep-th/9703059",
    archivePrefix = "arXiv",
    reportNumber = "CERN-TH-97-039, CERN-TH-97-39, LPTENS-97-10",
    doi = "10.1016/S0550-3213(97)00430-6",
    journal = "Nucl. Phys. B",
    volume = "503",
    pages = "117--156",
    year = "1997"
}

@article{Antoniadis:1999ux,
    author = "Antoniadis, Ignatios and D'Appollonio, G. and Dudas, E. and Sagnotti, A.",
    title = "{Open descendants of Z(2) x Z(2) freely acting orbifolds}",
    eprint = "hep-th/9907184",
    archivePrefix = "arXiv",
    reportNumber = "CPTH-S725-0799, DFF-340-07-99, LPT-ORSAY-99-58, ROM2F-99-22",
    doi = "10.1016/S0550-3213(99)00616-1",
    journal = "Nucl. Phys. B",
    volume = "565",
    pages = "123--156",
    year = "2000"
}

@article{Cotrone:1999xs,
    author = "Cotrone, A. L.",
    title = "{A Z(2) x Z(2) orientifold with spontaneously broken supersymmetry}",
    eprint = "hep-th/9909116",
    archivePrefix = "arXiv",
    reportNumber = "IFUM-647-FT",
    doi = "10.1142/S0217732399002595",
    journal = "Mod. Phys. Lett. A",
    volume = "14",
    pages = "2487--2497",
    year = "1999"
}

@article{Angelantonj:2003hr,
    author = "Angelantonj, Carlo and Antoniadis, Ignatios",
    title = "{Suppressing the cosmological constant in nonsupersymmetric type I strings}",
    eprint = "hep-th/0307254",
    archivePrefix = "arXiv",
    reportNumber = "CERN-TH-2003-165",
    doi = "10.1016/j.nuclphysb.2003.09.047",
    journal = "Nucl. Phys. B",
    volume = "676",
    pages = "129--148",
    year = "2004"
}

@article{Angelantonj:2006ut,
    author = "Angelantonj, Carlo and Cardella, Matteo and Irges, Nikos",
    title = "{An Alternative for Moduli Stabilisation}",
    eprint = "hep-th/0608022",
    archivePrefix = "arXiv",
    reportNumber = "DFTT-16-2006, IFUM-869-FT",
    doi = "10.1016/j.physletb.2006.08.072",
    journal = "Phys. Lett. B",
    volume = "641",
    pages = "474--480",
    year = "2006"
}

@article{Abel:2018zyt,
    author = "Abel, Steven and Dudas, Emilian and Lewis, Daniel and Partouche, Herv{\'e}",
    title = "{Stability and vacuum energy in open string models with broken supersymmetry}",
    eprint = "1812.09714",
    archivePrefix = "arXiv",
    primaryClass = "hep-th",
    reportNumber = "CPHT-RR117.122018, DCPT-18/37, IPPP/18/112",
    doi = "10.1007/JHEP10(2019)226",
    journal = "JHEP",
    volume = "10",
    pages = "226",
    year = "2019"
}

@article{Kounnas:2016gmz,
    author = "Kounnas, Costas and Partouche, Herve",
    title = "{Super no-scale models in string theory}",
    eprint = "1607.01767",
    archivePrefix = "arXiv",
    primaryClass = "hep-th",
    reportNumber = "LPTENS-16-04, CPHT-RR033.062016",
    doi = "10.1016/j.nuclphysb.2016.10.001",
    journal = "Nucl. Phys. B",
    volume = "913",
    pages = "593--626",
    year = "2016"
}

@article{Benakli:2026operator,
    author = "Benakli, Karim",
    title = "{Anti-periodic Gravitino and Dirac Effective Operators}",
    eprint = "2611.16047",
    archivePrefix = "arXiv",
    primaryClass = "hep-th",
    year = "2026",
    month = "11"
}

@article{Linch:2002wg,
    author = "Linch III, William D. and Luty, Markus A. and Phillips, J.",
    title = "{Five dimensional supergravity in N=1 superspace}",
    journal = "Phys. Rev. D",
    volume = "68",
    pages = "025008",
    year = "2003",
    doi = "10.1103/PhysRevD.68.025008",
    eprint = "hep-th/0209060",
    archivePrefix = "arXiv"
}

@article{Becker:2019ywd,
    author = "Becker, Katrin and Becker, Melanie and Butter, Daniel and Linch III, William D. and Randall, Stephen",
    title = "{Five-dimensional supergravity in N=1/2 superspace}",
    journal = "JHEP",
    volume = "03",
    pages = "098",
    year = "2020",
    doi = "10.1007/JHEP03(2020)098",
    eprint = "1909.09208",
    archivePrefix = "arXiv",
    primaryClass = "hep-th"
}
\end{document}